\documentclass[11pt,oneside,english]{book}
\usepackage{graphics}
\usepackage{psfrag}
\usepackage{makeidx}
\usepackage{nomencl}

\input epsf.tex

\usepackage[T1]{fontenc}
\usepackage[latin1]{inputenc}
\usepackage{geometry}
\usepackage{amssymb,amsmath}
\usepackage{babel}

\def\beq{\begin{equation}}
\def\eeq{\end{equation}}
\def\ba{\begin{eqnarray}}
\def\ea{\end{eqnarray}}

\def\bda#1{{${\cal{D}}_{#1}$}}

\def\a{\alpha}
\def\b{\beta}

\def\te{\theta}

\def\La{\Lambda}

\def\t{\tau}

\def\beq{\begin{equation}}
\def\eeq{\end{equation}}
\def\ba{\begin{eqnarray}}
\def\ea{\end{eqnarray}}

\def\beq{\begin{equation}}
\def\eeq{\end{equation}}
\def\beqn{\begin{eqnarray}}
\def\eeqn{\end{eqnarray}}

\def\matrix{\begin{matrix}}

\makeindex
\makeglossary
\begin{document}

\baselineskip = 17pt

\title{{\bf{Semi-realistic heterotic ${\mathbb Z}_2 \times {\mathbb Z}_2$ orbifold models}}}
\author{Thesis submitted in accordance with the requirements of \\
the University of Liverpool for the degree of Doctor in Philosophy \\
by \\
Elisa Manno}
\date{June 2009}

\maketitle 
\frontmatter

\chapter{Abstract}

Superstring phenomenology explores classes of vacua which can reproduce
 the low energy data provided by the Standard Model. 
We consider the 
heterotic $E_8\times E_8$ string theory, which gives  
rise to four-dimensional Standard-like
 Models and 
allows for their $SO(10)$ embedding.
The exploration of realistic vacua 
consists of finding compactifications of the 
heterotic string from ten to four dimensions.
We investigate two different schemes of compactification:
the free fermionic
formulation and the orbifold construction. 
The relation of free fermion models to ${\mathbb Z}_2 \times {\mathbb Z}_2$ 
orbifold compactifications implies that they produce
three pairs of untwisted Higgs multiplets.  
In the examples presented in this dissertation we
explore the removal of the extra Higgs representations by using the 
free fermion boundary conditions directly at the string level, rather
than in the effective low energy field theory.
Moreover,
by employing the standard analysis of
flat directions 
we present a quasi--realistic three generation string model
in which stringent $F$-- and $D$-- flat solutions do not appear to exist
to all orders in the superpotential. 
We speculate that this result is indicative of the non--existence 
of supersymmetric $F$-- and $D$-- flat solutions in this model and discuss
its potential implications. 
By continuing our search of semi-realistic models in different string
compactifications we present a simple, yet rich, set up: the orbifold.
The simplest examples of orbifold compactifications generally
produce a large number of families, which are clearly
unappealing for experimental reasons. We show that, by choosing
a non-factorisable compactification lattice, defined by skewing
its standard simple roots, we decrease the total number of generations.
Although we do not provide a semi-realistic model 
 in this framework,
the method represents an intermediate
step to the final realisation of 
phenomenologically viable three generation models.
Moreover, we mention other possible tools which may be applied
in the search of Standard Model-like solutions. 
Finally, the construction
of modular invariant 
partition functions for $E_8\times E_8$ orbifold
compactifications is presented. 
Several interesting examples
are derived with this formalism, such as the case of
a ${\mathbb Z}_2 \times {\mathbb Z}_2$ shift orbifold model, in order to provide
a more technical approach in the construction of consistent string models.

\tableofcontents


\addcontentsline{toc}{chapter}{Contents}

\listoffigures


\addcontentsline{toc}{chapter}{List of Figures}

\chapter{Acknowledgement}
\begin{flushright}
{\textit{Non basta guardare, occorre guardare 
con gli occhi che vogliono vedere,}}\\
{\textit{che credono in quello che vedono.}}\quad {Galileo Galilei.}
\end{flushright}
I would like to express my gratitude to my supervisor,
Prof. Alon Faraggi, for his valuable suggestions and constructive
advises through this research work. His guidance into the complicated and
at the same time fascinating world 
of String Phenomenology enabled me to complete my work 
successfully.
I am immensely grateful to Dr. Cristina Timirgaziu for her precious
instructions in the study of string models and her willingness
to answer all my questions without hesitation.
Her warm encouragement and thoughtful guidance have been
crucial in the completion of this project. 
I wish to acknowledge the String Phenomenology Group which
provided numerous ideas and useful discussions during the weekly
meetings, as well
as the useful seminars which presented new aspects
and outlooks on the 
progress of string theory in the rest of the scientific world.
In particular,
I would like to thank
Dr. Thomas Mohaupt for helpful suggestions in several occasions
and Dr. Mirian Tsulaia for the interesting collaboration
 of the last few months, whose help and knowledge
 has been significant in the solutions of 
numerous questions.
My great appreciation goes to Prof. Carlo Angelantonj,
whose kindness and availability allowed me to answer
critical doubts in the analysis of the partition function 
constructions.
I wish to thank Prof. Gerald Cleaver for the collaboration
which produced the second paper presented in this thesis.
I am thankful to Prof. Ian Jack for his 
concrete, and moral, support during these years of study at
the University of Liverpool.
I wish to thank Prof. Claudio Coriano who persuaded
 me to apply for
this PhD and first got me interested in the research of
Theoretical Physics.
My PhD would not have been so exciting and enjoyable without
my great friends Ben, Cathy and Chris, who shared with me
an office and the difficulties of the scientific research.
Among my many friends in Liverpool, I want to thank Paola,
Linda and Laura for
the pleasant evenings spent together. Their precious friendship and advises 
during these years have been fundamental to me.
A special thanks to Adriano and Andrea for their delicious italian dinners, 
that were a nice break from the writing up.
In the last year of my PhD the constant and loving 
support of my boyfriend
Gary has been decisive 
to overcome moments of frustration and tiredness. 
The great time we have spent together gave me the enthusiasm and
drive to accomplish my work. I will cherish all the memories
of the last days for the rest of my life.
I am forever grateful to my family, my mother
Filomena, my father Luciano and my sister Paola, since they taught me
the way to reach my goals in life.


\mainmatter

\chapter{Introduction}
\section{Motivation}

The Standard Model (SM) of Particle Physics 
describes correctly the physics of the elementary particles
and their interactions, as confirmed by
the experiments up to the electroweak scale $M_W=246$ GeV.
It combines three of the four fundamental
forces in nature, the weak, 
the strong and the electromagnetic interaction, 
into a unique theoretical framework, which is 
a Yang-Mills gauge theory based on the symmetry group 
$SU(3)_C\times SU(2)_L\times U(1)_Y$ ($C$, $L$ and $Y$ denote
the colour, the weak isospin and the hypercharge quantum number respectively).
In particular, the weak and the electromagnetic interactions are
described by the $SU(2)_L\times U(1)_Y$ gauge symmetry, which is spontaneously
broken to a $U(1)_{em}$ by the Higgs mechanism \cite{Higgs:1964ia}.
The resulting massive gauge bosons, $W^{\pm}$ and $Z^0$, mediate the 
weak interactions, while the massless boson $\gamma$, the photon, is the 
carrier of the electromagnetic force.
The Quantum Chromodynamics is described by the $SU(3)_C$ sector,
which remains unbroken, where the messengers of the strong
interaction are eight massless gluons.
The Standard Model content consists
of three generations of leptons and
three generations of quarks, 
in agreement with the observed experiments.
The predictability of the Standard Model  
is a consequence of its renormalizability, which assures
a consistent perturbative analysis of quantities related
to the particle physics (infinities that may appear in the 
calculations are consistently absorbed into a finite number
of physical parameters).
Despite the achievements accomplished in this set up,
 several issues
have not been resolved yet.
We list below some among the most important shortcomings 
of the Standard Model \cite{Kounnas:1985cj}.

$\bullet$ Absence of gravity: the Standard Model 
does not include in its description 
the Newtonian force, which is $42$ orders
of magnitude smaller than the nuclear forces. 
Although General Relativity describes its
infrared properties consistently, gravity is  
characterised by non-renormalizable operators which 
produce ultraviolet divergences. 

$\bullet$ The hierarchy problem: the Higgs boson, 
responsible for the
electroweak symmetry breaking and
 for the generations of the masses
for the elementary particles, 
has a mass of the order of $100$ GeV 
(if correctly predicted by the Standard 
Model). This mass receives
radiative corrections which can make the Higgs 
very heavy ($\sim 10^{19}$ GeV), while its
vacuum expectation value is of the order 
of the electroweak scale.
The hierarchy between the two energy 
scales in the 
physics of the Higgs boson appears very unnatural, 
and certainly unappealing for a fundamental theory.
The 
introduction of supersymmetry (a symmetry
between fermionic and bosonic degrees of freedom in the theory) 
solves this problem by preventing the scalar particle to acquire
the dangerous contributions 
from the perturbation theory, thus 
 stabilising its mass.

$\bullet$ The grand unification: the coupling 
constants for the 
electromagnetic and nuclear forces
 are parameters which depend on the
energy scale. If their behaviour is 
extrapolated at high energy,
 roughly $10^{16}$ GeV, these values 
approach to one point 
but do not coincide. If supersymmetry is included,
 the final theory provides 
a unified description of the forces
of the Standard Model at high energy.

$\bullet$ The arbitrariness: more than 
twenty free parameters
describe the physics of the Standard 
Model and their values are
completely arbitrary. For instance, 
the fermion masses, 
the gauge and Yukawa couplings, 
the Kobayashi-Maskawa parameters
and many others have to be fixed by the 
experiments and put by hand
into the theory.

There are many other open questions related to the
physics of the Standard Model, such as the problem of 
the cosmological constant, whose small value cannot be
explained in this set up. Also, the number of families 
does not find a reasonable explanation. Moreover, we  
mention the non-zero
neutrino masses, due to their oscillations, which does not
fit into the description of the leptonic physics of the Standard Model. 
The attempts of surmounting all these inconsistencies 
lead to several different theoretical solutions in
the physics beyond the 
Standard Model, for instance the introduction of 
grand unification theories (GUTs) and supersymmetry.
The main target of GUTs theories \cite{Kounnas:1985cj,Langacker:1980js} 
is solving the unification problem
 previously mentioned,
by extending the gauge symmetry group 
of the SM to a $G_{GUT}$ characterised by
 only one gauge coupling. In principle,
the strong, the weak and the electromagnetic
 interaction merge
together at some higher energy 
scale $M_{GUT}$ where the theory has 
the larger gauge symmetry $G_{GUT}$. 
When the energy decreases
below $M_{GUT}$ then the GUT symmetry breaks 
to the SM gauge group $SU(3)\times SU(2)\times U(1)$
 and the
couplings associated with different 
factors evolve at different rate.
The smallest simple group which
accommodates the SM is the 
 $SU(5)$ with $M_{GUT}\simeq 10^{15}$ GeV \cite{Georgi:1974sy}.
A typical feature of grand unified theories
is the mixing of 
quarks and leptons into the same group representation.
Thus, in the case
of $SU(5)$ gauge group, a matter generation 
is contained into the two irreducible
representations $\{{\bf{10}}\,,\,{\bf{\bar 5\}}}\in SU(5)$.
By considering a larger $G_{GUT}$, for example
an $SO(10)$ symmetry \cite{Georgi:1974my}, 
it is possible to combine 
one generation into only one irreducible 
representation, precisely the ${\bf{16}}$ of $SO(10)$.
In the last case, the presence of a singlet state, 
the right-handed neutrino, 
and the absence of exotic particles makes 
the model very predictive.
Unfortunately, there are several unsolved questions
appearing in grand unified theories, most of which originated
from the quark-lepton mixing.
A first example 
is given by the existence of new interactions that violate 
lepton and baryon number, which are responsible for 
the instability of the proton.
Another typical problem is the presence of colour-triplet 
Higgs states which we do not expect to see in the low energy spectrum
(the so-called doublet-triplet splitting problem). 
Additionally,
the hierarchy problem, 
which affects already the physics of the SM,
does not find a solution in GUTs theories.
Finally, they still suffer from the lack 
of gravity.

Several answers to the previous problems 
are presented by supersymmetric theories.
In particular the hierarchy problem is solved with 
the introduction of supersymmetry (SUSY), as anticipated earlier,
which associates to each 
boson of the theory a fermionic
superpartner with the same quantum numbers 
(since any internal symmetry
commutes with SUSY). 
This symmetry is an extension of 
the Poincar\'e algebra which includes the 
fermionic generators $Q^i$, $i=1,..N$,
 satisfying anticommutation relations.
The way supersymmetry overcomes 
the hierarchy problem is by
"doubling" the spectrum,
where each scalar coexists with
its fermionic partner. 
Basically,
the radiative corrections of the scalar Higgs
at one-loop include a divergent scalar self-energy term.
In supersymmetric theories a quadratically divergent
term from the bosonic superpartner arises, 
giving exactly an opposite contribution.
 Hence, we assist to 
a cancellation of terms which stabilises 
the scalar masses of the theory.
At low energies 
there is no experimental evidence
of supersymmetric particles, implying 
that SUSY has to be broken at a 
relatively low scale, while
 being an exact symmetry 
at high energies.

\section{String theory as a theory of unification}

As mentioned before, the non-renormalizability of General Relativity
makes a consistent description of quantum gravity problematic.
Therefore, the formulation of a quantum theory that includes
gravity and the other forces is very important.
String theory seems to be the most successful candidate for 
a unified theory of all forces in nature, as we explain in the following. 
The regularization of 
the gravitational interactions is realised thanks to 
the introduction of an extended object, the string.
The known particles are identified with massless excitations
of the string. Beside these particles there is an infinite tower
of fields with increasing masses and spins 
\cite{Veneziano:1968yb,Virasoro:1969pd}
with typical mass of the 
order of the Plank scale $M_P\sim 10^{19}$ GeV.
Among all excitation modes the graviton,
the quantum of the gravitational field, 
arises in the spectrum, and suggests the interpretation
of string theory
as a quantum theory of gravity.
Moreover, the presence of only one parameter 
(the string coupling $g_s$) used in the description of
all phenomena, is considered a key feature in the prospective 
of an unifying picture.
From a more technical point of view, 
string theory contains gauge symmetries which may 
incorporate the SM symmetry. 
Finally, supersymmetry arises in a natural way in this set up, 
despite the existence of consistent modular invariant string theories
which are not supersymmetric.  
In the quantization procedure, 
the consistency of the 
string theory requires
spacetime to have the critical dimension, which corresponds to $D=10$ 
for supersymmetric
strings.
In the table below we present the five 10-dimensional perturbative
superstring theories and some of their most important properties.
\vspace{0.3cm}
\begin{center}
\begin{tabular}{cccc}
Type & $N_{susy}$ & String & Massless bosonic content \\
\hline
\hline
\vspace{0.3cm}
$H_{E_8\times E_8}$ & 1 & closed and oriented &  
$g_{\mu\nu},\varphi, B_{\mu\nu}, A_{\mu}$ of $E_8\times E_8$\\
\hline
\vspace{0.3cm}
$H_{SO(32)}$ & 1& closed and oriented & 
$g_{\mu\nu},\varphi, B_{\mu\nu}, A_{\mu}$ of $SO(32)$ \\
\hline
\vspace{0.3cm}
$I-SO(32)$ & 1 &open+ closed unoriented &  
$g_{\mu\nu},\varphi, A_{\mu\nu}, A_{\mu}$ of $SO(32)$\\
\hline
\vspace{0.3cm}
$IIA$ & 2 & closed and oriented &  
$g_{\mu\nu},\varphi, B_{\mu\nu}, C_{\mu\nu\rho}, A_{\mu}$ of $U(1)$ \\
\hline
\vspace{0.3cm}
$IIB$ & 2 & closed and oriented & 
$g_{\mu\nu},\varphi, B_{\mu\nu}^N, \varphi',B_{\mu\nu}^R, D_{\mu\nu\rho\sigma}^{\dagger}$\\
\hline
\end{tabular}
\end{center}
\vspace{0.3cm}
In the table above, 
$g_{\mu\nu},\varphi, B_{\mu\nu}, A_{\mu}$ represent the graviton, the dilaton,
the antisymmetric tensor and the gauge bosons respectively.
The bosons $A_{\mu}$ belong to the adjoint
representation of $E_8\times E_8$ or $SO(32)$ for the first three
cases, while they are bosons of  $U(1)$ symmetries for the type IIA case.  
$C_{\mu\nu\rho}, \varphi', B_{\mu\nu}^R$ and $D_{\mu\nu\rho\sigma}^{\dagger} $
 are respectively a three-index tensor potential,
a zero-form, a two-form and a four-form potential, the
latter with self-dual field strength.  
The five superstring models are considered
as different manifestations (in different regimes),
 of an unique theory, known as {\bf{$M$-theory}}, and
they are connected by some kind
 of equivalences, the so-called string dualities \cite{Sen:1996yy}.
The underlying fundamental 
theory, whose low energy limit 
is 11 dimensional SUGRA 
\cite{Hull:1995mz}, is unfortunately still poorly understood. 
 
\vspace{0.3cm}
\begin{figure}[htp]
\centering
\scalebox{0.4}{\includegraphics{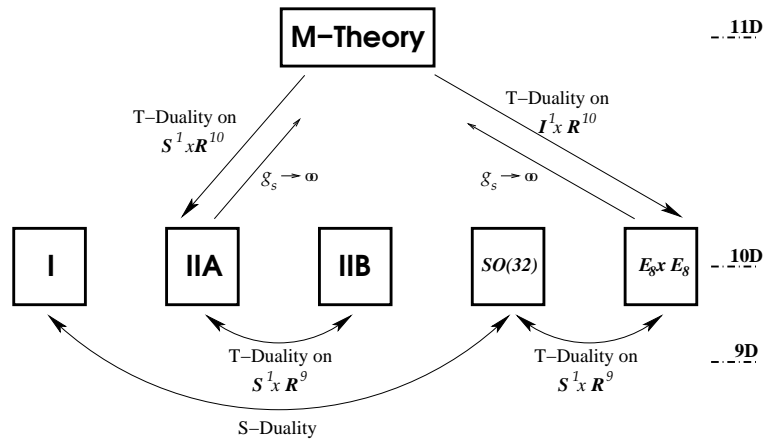}}
\caption{Supersymmetric perturbative consistent string theories in 10 dimensions.}
\label{fig:strinthy}
\end{figure}

As we can see from fig.\ref{fig:strinthy},
the duality transformations relate the superstring theories
in nine and ten dimensions. $T$ duality inverts the radius $R$
of the circle $S^1$, along which a space
direction is compactified, $R\rightarrow \frac{1}{R}$.
In particular, this duality relates
the weak-coupling limit of a theory compactified on a space with
large volume to the correspondent weak-coupling limit of 
another theory compactified on a small volume. 
$S$ duality instead provides the quantum equivalence
of two theories which are perturbatively distinct.
In fact, it inverts the string coupling 
$g_s\rightarrow \frac{1}{g_s}$. The perturbative excitations
of a theory are mapped to non-perturbative excitations of
the dual theory and viceversa.  
Fig.{\ref{fig:strinthy}} summarises the relevant information
of the perturbative string theories and their web of dualities.

In order to make contact with the real world, the
compactification of the six extra dimensions is needed.
This procedure follows the Kaluza-Klein dimensional 
reduction of quantum field theory
and is generalised to the case where a certain number
of spacetime dimensions give rise to a compact manifold, 
invisible at low-energy \cite{Scherk:1978ta,Cremmer:1973mg}.
Demanding four-dimensional $N=1$ supersymmetric 
models leads us to a special choice of internal manifolds, the so-called
Calabi-Yau manifolds \cite{Yau:1977ms}.
Compactifications of this kind are characterised 
by some free parameters, the moduli,
 generally related to the size and shape of the
extra dimensions. The low energy parameters often depend 
on these free values which spoil
 the predictivity of the theory.
The moduli describe possible deformations of the theory
and their continuous changes allow to go from one vacuum to
another. So far, the problem of fixing the moduli has not
been solved yet, since no fundamental principle is able to
single out a unique physical vacuum.
The study of Calabi-Yau manifolds is, 
unfortunately, fairly complicated since
the computation of properties which are not of topological
nature is very difficult.
A simpler class of compact manifolds is given by the
toroidal compactification, although the resulting theory 
is not chiral. 
Hence, combining
the desirable pictures of Calabi-Yau manifolds and 
toroidal compactifications, we finally arrive to
the orbifold construction. The orbifold seems to provide
a simple framework for the realisation of $N=1$ supersymmetric
models in four dimensions, with chiral particles in the spectrum.

In this thesis we discuss two main compactification schemes
which offer complementary advantages in the understanding 
of semi-realistic heterotic string models. 
The first approach is the free fermionic 
construction, which is based on an algebraic method to build 
consistent string vacua directly in four dimensions.
In the fermionic formalism  
all the worldsheet degrees of freedom, required to 
cancel the conformal anomaly, are given by free fermions on the
string worldsheet. This set up offers
a very convenient setting for experimentation of models,
 allowing a systematic classification
of free fermion vacua and their phenomenological properties. 
Moreover, this set up provided the most semi-realistic models
to date.
On the other hand, the orbifold compactification, previously
mentioned, leads to the analysis of other interesting features of heterotic
models. For instance, the geometric 
picture provided by the orbifold construction
may be instrumental for examining other
questions of interest, such as the 
dynamical stabilisation of the moduli
fields and the moduli dependence of the Yukawa couplings.
The correspondence of free fermionic models 
\cite{Faraggi:1993pr,Faraggi:1995yd,Donagi:2004ht} to 
${\mathbb Z}_2 \times {\mathbb Z}_2$ orbifold compactification is a key point 
of this thesis. In fact,
the phenomenologically appealing properties of the
free fermionic models and their relation to ${\mathbb Z}_2\times
{\mathbb Z}_2$ orbifolds 
provide the clue that we might gain further insight into
the properties of this class of quasi--realistic string compactifications 
by constructing
${\mathbb Z}_2\times {\mathbb Z}_2$ orbifolds on enhanced
non-factorisable lattices 
(the point at which the internal dimensions are realised as
free fermions on the worldsheet is a maximally symmetric point with 
an enhanced $SO(12)$ lattice, which is in principle non-factorisable). 

In this thesis we produced the following results.
We presented two semi-realistic models in the free fermionic 
formulation with a reduced Higgs spectrum. The truncation
of the Higgs content 
is realised for the first time in this set up
 at the level of the string scale,
by the assignment of asymmetric boundary conditions
to the internal right- and left-moving fermions of the theory. 
Moreover, the analysis of flat directions, performed with the standard
methods, leads to an
unexpected result. The Fayet-Iliopoulos D-term which breaks 
supersymmetry perturbatively in our models is not compensate
by the existence of 
D- and F- flat solutions, which would restore supersymmetry.
The Bose-Fermi degeneracy of the spectrum implies that the models 
are supersymmetric at tree level. Thus, the models presented 
may provide a new interpretation of 
the supersymmetry breaking in string theory.
In the framework of the orbifold construction, we 
built a ${\mathbb Z}_2 \times {\mathbb Z}_2$ orbifold with a skewed $SO(4)^3$
compactification lattice and analysed its spectrum and
symmetry group. Our main goal initially was reproducing
a three generation free fermionic model \cite{Faraggi:2004rq}
with gauge symmetry $E_6\times U(1)^2\times SO(8)^2_H$.
Unfortunately we could not obtain the wished features, 
not even after the introduction of Wilson lines. Nevertheless,
several interesting properties are discussed concerning 
 the compactification lattice and its possible tools to realise
semi-realistic four dimensional models in the
construction of orbifold models.
Finally, we concluded this thesis with
the construction of modular invariant partition functions
for heterotic shift orbifolds. In this context we presented  
different examples of consistent vacua with the
derivation of the full perturbative spectrum. In particular,
we discussed the details of a ${\mathbb Z}_2 \times {\mathbb Z}_2$
 shift orbifold
model which contains some technical subtleties due to the
elements of the orbifold group, and presented in detail
its massless spectrum.

\newpage

\section{Organisation of the chapters}

The topics of this thesis are organised as follows.

$\bullet$ \underline{\bf{Chapter 2}}
 
A general introduction on the bosonic and fermionic string
is presented in order to provide perturbative superstring
constructions. A brief overview on the partition function
which encodes the modular invariant properties of the 
theory is discussed. We
explain the bosonization procedure necessary for
 the correspondence between fermionic
 and bosonic conformal field theories. 
We close the chapter with some generalities on the
heterotic string, which will be analysed in great detail
in the next chapters.

$\bullet$ \underline{\bf{Chapter 3}}

We present the main features of four-dimensional semi-realistic 
models in the free fermionic construction and 
show the advantages of using this compactification scheme.
We fix the formalism to provide the consistency constraints
and the model building rules for this framework and explain
the general derivation of the spectrum.
In the second part of the chapter we present two very peculiar
examples of semi-realistic free fermionic models, 
where the reduction of the Higgs content is, for the first time, realised
at the string scale. Moreover, the standard
 analysis of flat directions is in both cases 
unable to restore supersymmetry perturbatively, although the models are 
supersymmetric at the classical level. This point opens new interpretations
for the supersymmetry breaking mechanism in string theory.  

$\bullet$ \underline{\bf{Chapter 4}}

We start by introducing the heterotic string in its
bosonic formulation, followed by the description
of the toroidal compactification. We proceed by providing
 the generalities of orbifold constructions.
The discussion of the spectrum is initially performed 
at an abstract level to find in
 the last part of the chapter 
a concrete application, in the case of a ${\mathbb Z}_2 \times {\mathbb Z}_2$
orbifold with $SO(4)^3$ compactification lattice. 
In our example
we seize the opportunity to 
present the explicit derivation
of the fixed tori for a non-factorisable lattice and
investigate possible ways to control the number of 
families, for example by considering Wilson lines.

$\bullet$ \underline{\bf{Chapter 5}}

Some interesting examples of heterotic strings compactified on
shift orbifolds are presented, providing the technical 
details on the derivation of ${\mathbb Z}_2$
 and ${\mathbb Z}_2 \times {\mathbb Z}_2$ 
orbifold partition functions. As an example is obtained,
a consistent modular invariant string vacuum
with no graviton. This model is in a way 
reminiscent of string vacua without gravity - "little string" models.

$\bullet$ \underline{\bf{Chapter 6}} 

We conclude this thesis underlining the main aim of our research, 
the semi-realistic heterotic string
constructions in different compactification schemes. 
We present the main results obtained and finally provide 
possible interesting outlooks.

\chapter{Background notions on consistent perturbative superstring theories}

In this chapter we briefly present aspects of the perturbative 
formulation of string theory and introduce the necessary
 tools for the construction of 
semi-realistic four dimensional superstring models.
The sources of the introductory part are given by
 \cite{Green:1987sp,Becker:2007zj,Polchinski:1998rq,Lust:1989tj,Szabo:2002ca,Mohaupt:2002py,Kiritsis:2007zz,Kiritsis:1997hj}.

We start by presenting the bosonic string, which is the simplest
instance of a string theory. This two-dimensional conformal 
theory at the classical level is consistent only
 at the critical dimension D=26.
 In its low energy spectrum, 
provided by the massless excitation modes, 
the presence of a symmetric metric tensor $g_{\mu\nu}$,
the candidate of the graviton field,
gives the
 main motivation for interpreting string theory 
as a quantum theory of gravity.
Two main reasons make the bosonic string inadequate
for a complete description of the fundamental interactions, such as
the existence of tachyonic states,   
a sign of instability for the theory, and the
 absence of fermionic excitations in the perturbative spectrum. 
The solution to these problems leads to 
the introduction of the superstring,
a superconformal theory with critical dimension D=10.
After presenting the classical action for the 
bosonic and fermionic string, we will discuss the quantization procedure
of the theory. 
The concepts of conformal invariance and modular invariance are 
explained in detail.
We concisely mention how to calculate string interactions 
whilst giving a detailed 
overview on the partition function for the closed
bosonic string, the torus amplitude,
 since this quantity  represents 
one of the main topics treated in the following chapters
 \cite{Angelantonj:2002ct,Bianchi:1988ux}.
In the last section
we introduce the concept of toroidal compactification 
which will be considered extensively in chapter 4, before 
the orbifold constructions of semirealistic models.
In most cases we restrict our discussion to the closed strings
since our target is the construction of the heterotic string.

\section{Bosonic strings}

\begin{figure}[htp]
\epsfxsize=2 in
\centering
\centerline{\epsffile{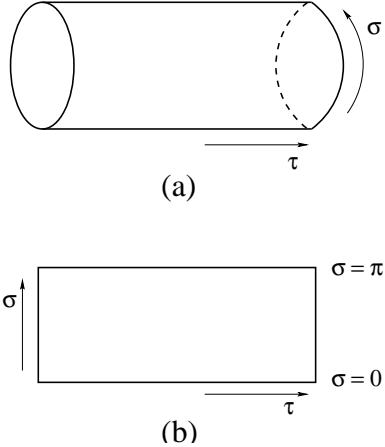}}
\caption{a) closed string worldsheet. b)open string worldsheet.}
\label{fig:ws}
\end{figure}

Strings are one dimensional finite objects whose
 propagation in a D dimensional spacetime gives rise to
 a two dimensional worldsheet $X^{\mu}(\sigma,\tau)$, 
$\mu=0,..D-1$.
In fig. \ref{fig:ws}
 this surface is shown
in both cases of free closed and free open strings.
The worldsheet is parametrized by the two real 
independent coordinates, $\tau$ and $\sigma$, where the first variable 
is a time-like parameter while the second is space-like and belongs to the interval  
$[0,\pi]$.

The physics of the 
string\footnote{To be more precise, the simplest action which describes
the motion of the string is the Nambu-Goto action, $S_{NG}=
- T \int d^2\sigma\sqrt{-\gamma}$, where $\gamma$ is the determinant of the
induced metric on the worldsheet, $\gamma_{\alpha\beta}= \partial_{\alpha}
X^{\mu}\partial_{\beta}X^{\nu}g_{\mu\nu}$. This action is proportional 
to the area swept from the worldsheet, thus it provides
 a more geometric and intuitive
meaning of the string action. The Polyakov action,
which supplies in a simpler way the equations of motion, is equivalent 
to the Nambu-Goto action and can be obtained by introducing
the independent metric on the worldsheet $h^{\alpha\beta}$.}
is described by the Polyakov action
that, in a flat Minkowski D dimensional spacetime, assumes the form \cite{Polyakov:1981rd,Polyakov:1981re}
\beq
S=-\frac{T}{2}\int d^2\sigma\sqrt{-h}h^{\alpha\beta}
\eta_{\mu\nu}\partial_\alpha X^{\mu}
\partial_\beta X^{\nu} ,
\label{Paction}
\eeq
where $T$ is the string tension,
$h^{\alpha\beta}$ is the worldsheet metric and $h=det(h^{\alpha\beta})$, 
while $d^2{\sigma}$ implies the equivalent notation
 $\sigma=(\sigma^0,\sigma^1)=(\tau, \sigma)$.

For a general background we can simply replace the flat
metric $\eta_{\mu\nu}$ by $g_{\mu\nu}(X)$ and 
eq.(\ref{Paction}) becomes the worldsheet action of D dimensional 
scalar fields $X^{\mu}$ coupled to the dynamical two-dimensional metric 
(theory of quantum gravity coupled to matter).

The Polyakov action has three symmetries:
\begin{itemize}
\item[1)] Poincar\'e invariance in the target space $X^{\mu}$.
\item[2)]Local reparametrization invariance.
\item[3)]Conformal (Weyl) invariance.
\end{itemize} 

The last two properties are local symmetries
 which can be used to fix the worldsheet metric in the conformal 
gauge, $h_{\alpha\beta}=e^{\phi(\tau,\sigma)}\eta_{\alpha\beta}$, obtaining
 a flat metric 
up to a scaling function.
The equations of motion (e.o.m.) for the bosonic fields $X^{\mu}$ and
for the metric $h^{\alpha\beta}$ are obtained in the usual procedure, 
as the variation of the action with respect to each of these fields
respectively. At this point it is convenient to introduce 
 the two-dimensional  
stress tensor $T_{\alpha\beta}$ 
which provides the constraints for the string theory.
We define $T_{\alpha\beta}$ (also known as
 energy-momentum tensor) 
as the variation of the Polyakov action with respect to the 
world-sheet metric 
\beq
T_{\alpha\beta}=-\frac{2}{T\sqrt{-h}}\frac{\delta S}{\delta h^{\alpha\beta}}
=\partial_{\alpha}X^{\mu}\partial_{\beta}X_{\mu}-\frac{1}{2}h_{\alpha\beta}
h^{\rho\gamma}
\partial_{\rho}X^{\mu}\partial_{\gamma}X_{\mu} ,
\label{stress}
\eeq
then the request that
the energy-momentum tensor 
vanishes,
\beq
T_{\alpha\beta}=0 , 
\label{vir}
\eeq
corresponds exactly  to the e.o.m. for
$h^{\alpha\beta}$. This condition is called the 
Virasoro constraint and
represents a very important ingredient  
when considering the physical states of the model under consideration.
The stress tensor is symmetric, traceless ($T_{\alpha \alpha}=0$),
 as consequence of the Weyl
invariance and conserved.

It is very convenient to 
rewrite the Virasoro conditions
in the light-cone coordinates 
$\sigma^+=\tau+\sigma$, $\sigma^-=\tau-\sigma$ , 
where $\partial_{\pm}=\frac{1}{2}(\partial_{\tau}\pm \partial_{\sigma})$. 
Then eq.(\ref{vir}) would simply become 
\beq
T_{--}=\frac{1}{2}(\partial_-X)^2 = 0 \,\,;
\quad T_{++}=\frac{1}{2}(\partial_+X)^2 = 0\,\,; \quad T_{\pm\mp}=0.
\eeq
The equations of motion for the fields  $X^{\mu}$
take the form $\partial_+ \partial_- X^{\mu}=0$, whose 
general solution can be written as 
the sum of a ``right-moving'' solution 
plus a ``left-moving'' solution,
\beq
X^{\mu}(\tau,\sigma)=X^{\mu}_R(\tau-\sigma)+X^{\mu}_L(\tau+\sigma).
\label{llrr}
\eeq
Together with the periodicity constraint 
$X^{\mu}(\sigma,\tau)
=X^{\mu}(\sigma+2\pi,\tau)$, eq.(\ref{llrr}) leads to the
 mode expansion
\ba
X^{\mu}_R(\tau-\sigma)&=&\frac{1}{2}x^{\mu}+\alpha'p^{\mu}(\tau-\sigma)+
i\sqrt{\frac{\alpha'}{2}}\sum_{n\neq0}\frac{1}{n}
\alpha^{\mu}_n e ^{-2in(\tau-\sigma)},\nonumber\\
X^{\mu}_L(\tau+\sigma)&=&\frac{1}{2}x^{\mu}+\alpha'p^{\mu}(\tau+\sigma)+
i\sqrt{\frac{\alpha'}{2}}\sum_{n\neq0}\frac{1}{n}
\tilde\alpha^{\mu}_n e ^{-2in(\tau+\sigma)} ,
\label{e}
\ea
where the Regge slope parameter $\alpha'$ 
is defined in terms of the string
tension as  $\alpha'=1/2 \pi T$.
From (\ref{e}) we see that
 the classical motion of the string is described by 
the centre of mass position $x^{\mu}$, the momentum $p^{\mu}$ and
the oscillator modes.

For later convenience we define the 
Virasoro operators as Fourier modes of the stress tensor, that in 
the right-moving sector become
\[L_m=\frac{T}{2}\int_0^{\pi}d\sigma e^{2im(\tau-\sigma)}\,\,T_{--}=\frac{1}{2}
\sum_{n=-\infty}^{\infty}\alpha^{\mu}_{m-n}\cdot\alpha_{\mu n}\,\, (m\neq 0).
\]
The Virasoro operators satisfy the constraints $L_m=0$, $\forall n\in Z$ 
and for the case $n=0$ we obtain the mass equation for the right 
oscillation modes, discussed 
in the following section. Moreover 
$\alpha^{\mu}_0=\sqrt{\frac{\alpha'}{2}}p^{\mu}_0$.
The correspondent left-moving expression 
$\tilde L_m$ is given by 
 the substitutions 
$T_{--}\rightarrow T_{++}$, $\sigma^- \rightarrow \sigma^+$ 
and the complex conjugate oscillators and similar conditions
to the right sector hold in the left sector as well.

{\bf{Quantization of the bosonic string}}

The oscillators, the centre of mass position and the momentum presented 
in eq.(\ref{e}) satisfy the standard commutation relations, while
the Virasoro operators form the so-called Virasoro algebra.
In the covariant canonical 
quantization procedure the previous 
conditions are translated into the following commutators
\ba
[x^{\mu},p^{\mu}]&=& i\eta^{\mu\nu}, \nonumber\\
\left[ \alpha^{\mu}_{m} ,\alpha^{\nu }_{n} \right] &=& m \delta_{m+n}
 \eta^{\mu\nu} ,\nonumber\\
\left[ \tilde\alpha^{\mu}_m,\tilde\alpha^{\nu }_n \right]
&=&m\delta_{m+n}\eta^{\mu\nu},\nonumber\\
\left[ L_m,L_n \right]&=& (m-n)L_{m+n}+\frac{D}{12}m(m^2-1)\delta_{m+n}.
\ea
The other commutators between
different combination of operators are zero. 
The Hermiticity of $X^{\mu}$ gives 
$(\alpha^{\mu}_n)^{\dagger}=\alpha^{\mu }_{-n}\,;\,
(\tilde\alpha^{\mu}_n)^{\dagger}=\tilde\alpha^{\mu }_{-n}$.
$D$ represents the central charge and for the bosonic string 
$D=\eta^{\mu\nu}\eta_{\mu\nu}$. 
The same algebra holds for the left operator $\tilde L_m$.
 From now on, when defining properties of operators in the right sector,
we will assume implicitly that analogous relations hold in the left sector.
In the quantization of a classical system 
an ambiguity is introduced in the definition of the operators.
This can be solved if we consider the corresponding 
normal-ordered expressions. In the case of the Virasoro 
operators the correct definition is given by $L_m=
\sum_{n=-\infty}^{\infty}:\alpha^{\mu}_{m-n}\alpha_{\mu n}:$. 
The only term sensitive to normal ordering is $L_0$ where
a normal ordering constant $a$ is introduced.

In the covariant quantization we obtain states with negative norm
which destroy the unitarity of the theory, but we can 
discharge those by imposing the following 
constraints
\beq
L_{m>0}|phys\rangle = 0\,\,\,,\,\,\,(L_0 - a)|phys\rangle = 0.
\label{physical}
\eeq
It has been shown that the subset of positive norm states 
exists only for $D\le 26$ and $a \le 1$ \cite{Green:1987mn}.

It is easier to 
solve the Virasoro constraints in the light-cone quantization 
(we have already defined the operators in terms of light-cone
 coordinates) where the states, obtained by solving the mass-shell equation,
 are always positive.
But if unitarity is guaranteed in this procedure, we will need 
to verify the Lorentz invariance, which is not manifest. 
We have already mentioned that
for $D=26$ and $a=1$ Lorentz invariance
is preserved. $D=26$ is  thus a very special choice of
spacetime dimensions, called the critical 
dimension of the bosonic string.
   
We use now a residual invariance, leftover after imposing
the conformal gauge, which is 
a reparametrization invariance up to scaling, generally defined as
\[
\sigma'_{+}\rightarrow f(\sigma_+)\,\,\,,\,\,\,\sigma'_{-}\rightarrow f(\sigma_-).
\]
This invariance allows to fix the value of $X^+$ as follows, leading to
the light cone gauge,
\[ X^+ = x^+ +2 \alpha' p^+ \tau.
\]
 The light-cone coordinates are 
given by $X^{\pm}=(X^0\pm X^{D-1})/\sqrt{2}$ and by 
using the Virasoro constraints we can express $X^-$ in terms of the transverse 
coordinates $X^i$, where $i$ takes values in the transverse directions.
 This means that we are left only with the 
transverse oscillators, while the 
light-cone ones are given by
\ba
\alpha^-_n &=& \frac{1}{\sqrt{2\alpha'}p^+}\{ \sum_{m\in Z}:\alpha^i_{n-m}
\alpha^i_{m}: -2a\delta_{n0}\} ,\nonumber\\
\alpha^+_n &=&  \sqrt{\frac{\alpha'}{2}p^+ 
\delta_{n0}} ,
\ea
and analogous expressions hold for $\tilde \alpha^{\pm}_n$.
The Virasoro constraints in the light-cone gauge
define the mass-shell condition for the physical states
\beq
2p^+p^-=\frac{2}{\alpha'}(L_0+\tilde{L}_0-\frac{D-12}{12})
\quad ;\quad L_0= \tilde{L}_0.
\label{elisa}
\eeq
In the first equation of (\ref{elisa}) 
the Riemann $\zeta$ function\footnote{The infinite 
sum due to the zero-point energy is calculated
by a regularisation procedure introducing the Riemann $\zeta$ function:
$\zeta(s)=\sum_{k=1}^{\infty}k^{-s}$. It provides the value of $a$ in terms of
the space-time dimension $D$, which is exactly 
$a=\frac{D-2}{24}$, as shown in formula
 (\ref{mass}) for $\zeta(-1)=-1/12$ \cite{Brink:1986ja}.}
$\zeta(-1)=-1/12$ has been used,
as a result of the divergent sums of zero-points energies due to 
the normal ordering $a$ of $L_0$ and $\tilde{L}_0$ \cite{Brink:1986ja}.
The second equation in (\ref{elisa}) is the level matching condition,
a relation which connects the left with the right excitation modes
of the closed string. This constraint has to be imposed for 
the consistency of every closed string model and contains
an important information concerning the physical states of the model,
the right and the left modes provide the same contribution to the mass 
of the physical states.
The masses of the string excitations are obtained
 by the contributions of the 
transverse momenta, which for the right sector are provided
by the formula 
$L_0=\frac{\alpha'}{4}p^ip^i+N$. 
The mass operator is
\beq
M^2=\frac{2}{\alpha'}(N+\tilde{N}-\frac{D-2}{12})
\label{mass}
\eeq
and $N=\sum_{m>0}\alpha_{-m}\cdot\alpha_{m}$. In the case at hand $D=26$, thus
the first state obtained from eq.(\ref{mass}) is 
the ground state $|p^{\mu}\rangle$, with $N=\tilde N =0$. 
Its mass is given by $M^2=-4a/\alpha'$, 
where $a$ takes the value $1$ for consistency, as we said before.
This state is the tachyon.
The first excited state is the tensor 
$\alpha^i_{-1}\tilde\alpha^j_{-1}|p^{\mu}\rangle$.
If we decompose it into irreducible representations 
of the group $SO(24)$ we obtain a symmetric tensor $g_{\mu\nu}$ 
(a spin-2 particle, the graviton), the antisymmetric tensor
$B_{\mu \nu}$ and a scalar $\varphi$, the dilaton.

At the next level we obtain states which are organised in representations 
of $SO(25)$ and which are massive.

\section{Vertex operators and string interactions}
\label{vvsba}
A local unitary quantum field theory
has an operator-state correspondence which 
associates to each field a quantum state created 
from the vacuum. In string theory the same correspondence is 
realised by mapping the worldsheet cylinder 
to the complex plane. 

\begin{figure}[htp]
\centering
\scalebox{0.5}{\centerline{\epsffile{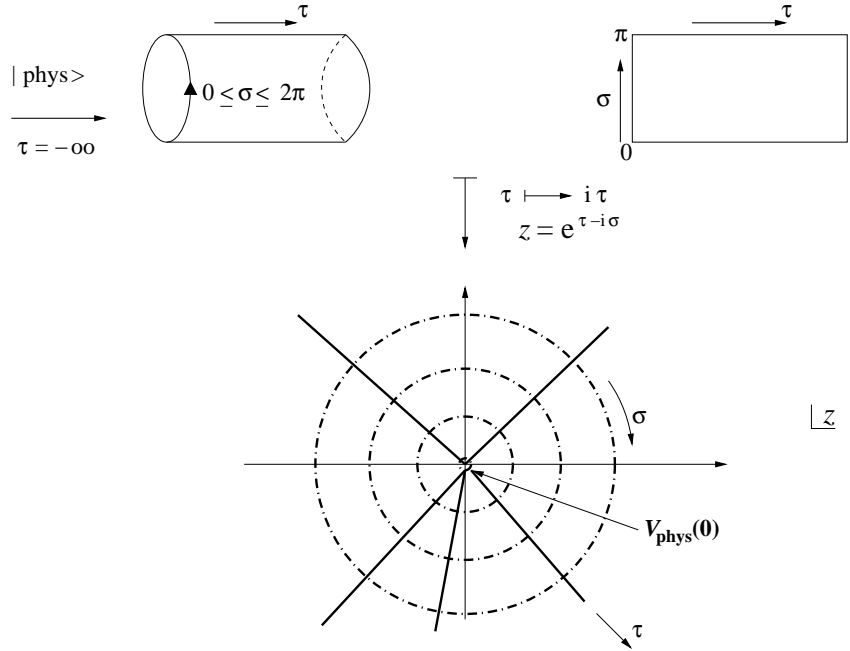}}}
\caption{Mapping of the worldsheet cylinder
into the complex plane. The dotted lines of constant $\tau$ are 
concentric circles while the lines of constant $\sigma$ follow radial directions from the origin.}
\label{fig:cylinder}
\end{figure}

In this context it is possible to 
build the so-called vertex operators 
which give rise to a spectrum generating algebra.
By using this formalism, for instance, 
an incoming physical state $|phys\rangle$ in the 
infinite worldsheet past $(\tau =-\infty)$
is given by the insertion of a vertex operator $V(z)$
at the origin $z=0$, see fig.(\ref{fig:cylinder}).

In this thesis we will not go into further
details concerning the vertex operators,
but it is important to stress their role
in the construction of string amplitudes 
and in the description
of strings interactions.

In quantum field theory the perturbative 
expansion of Feynman diagrams describes the interacting particles at 
well defined points. The worldline of particles in spacetime 
is described by propagators that meet in a vertex, singular point
which is responsible
for ultraviolet divergences in loop amplitudes. 
The string Polyakov perturbation theory is
 given by the sum of two-dimensional surfaces
which correspond to the worldsheets.
When considering all contributions of the infinite tower of massive 
particles of the string spectrum, the ultraviolet divergences of
quantum gravity loop amplitudes cancel out.
The reason why 
the non-renormalizability of quantum field theory 
is solved in string theory is because its interactions 
are described by
smooth surfaces  with no singular points.
The main consequence of 
this property is that string 
interactions are completely determined by the worldsheet topology.
In oriented closed strings the perturbative expansion
 is given by only one contribution 
at each order of perturbation theory. 
This contribution corresponds to closed 
orientable Riemann surfaces with increasing number of handles $h$
and the perturbative series is hence weighted by $g_s^{-\chi}$,
where $\chi$ is 
 the Euler character, defined as $\chi=2-2h$, while
the string coupling $g_s$ is dynamically determined by the vacuum expectation
value of the dilaton field $\varphi$, $g_s=e^{<\varphi>}$. 

A generic string scattering amplitude is given by a path integral
of the form
\beq
A=\int\mathcal{ D}h_{\alpha\beta}
\mathcal{D}X^{\mu}e^{-S_P}\prod_{i=1} ^{n}\int_{M}d^2\sigma_i
V_{\alpha_i},
\label{amplitude}
\eeq
where $h_{\alpha\beta}$ is the metric on the worldsheet $M$,
$S_P$ is the Polyakov action and $V_{\alpha}$ is the vertex operator 
that describes the emission or absorption of a closed string state 
of type $\alpha_i $ from the worldsheet.
The conformal invariance reduces these expressions to 
integrals on non-equivalent worldsheets which are described 
by some complex parameters, the moduli. The amplitudes 
in eq.(\ref{amplitude}) are then finite dimensional integrals over 
the moduli space of $M$.

\section{The superstring}

As we have mentioned at the beginning,
the bosonic string suffers of two main problems: the absence of 
spacetime fermions (necessary for a realistic description
of nature) and the presence of tachyons (sign of 
an incorrect identification of the vacuum).
The solution to these problems
 leads us to the construction of the superstring.
The new theory is constructed by the introduction of 
worldsheet supersymmetry, realised by including $D$
two-dimensional Majorana fermions
 $\Psi^{\mu}=(\psi_-^{\mu},\psi_+^{\mu})$, 
$\mu=0,..D-1,$ on the worldsheet. These fields are
vectors from the spacetime point of view but when
combined with appropriate boundary conditions will provide
spacetime fermions. 
In the following we will work in 
the RNS (Ramond-Nevew-Schwarz) 
formalism \cite{Neveu:1971rx,Ramond:1971gb},
 where the GSO (Gliozzi-Scherck-Olive)
 projections are introduced in order to 
obtain supersymmetry \cite{Gliozzi:1976jf}. 
The generalised action $S_T$ 
in the conformal gauge 
\beq
S_T=-\frac{T}{2}\int d^2\sigma(\partial_{\alpha}X^{\mu}
\partial^{\alpha}X_{\mu}-i\overline{\psi}^{\mu}\rho^{\alpha}
\partial_{\alpha}\psi_{\mu})
\label{azione}
\eeq
is invariant under 
worldsheet global supersymmetric transformations
\[ \delta_{\epsilon}X^{\mu}=\bar{\epsilon}\psi^{\mu} \quad, \quad
\delta_{\epsilon}\psi^{\mu}=-i\rho^{\alpha}\partial_{\alpha}X^{\mu}\epsilon ,
\]
with $\epsilon$ constant spinor and $\rho^{\alpha}$, $\alpha=0,1$,
Dirac matrices which can be chosen as follows
\[\rho^0 = \left( \begin{array}{cc}
0 & -i  \\
i & 0  \end{array} \right) 
 ,\quad
\rho^1 = \left( \begin{array}{cc}
0 & i  \\
i & 0  \end{array} \right) .  \]
In the light-cone coordinates the fermionic contribution 
of eq.(\ref{azione}) is simply
\beq
\psi_{-} {\partial}_{+} \psi_{-} +\psi_{+} {\partial}_{-} \psi_{+} ,
\eeq
where the space-time index $\mu$ has been suppressed.

The equations of motion are simply the Dirac equations 
$\partial_{\pm}\psi_{\mp}=0$. Their solutions are of the form 
$\psi_{-}=\psi_{-}(\sigma_+)$ and $\psi_{+}=\psi_{+}(\sigma_-)$, hence we 
can say that 
$\psi_-$ represents the right-moving field while  $\psi_+$ is the 
left-moving one.
The boundary conditions arise by requiring that 
\beq(\psi_+\delta\psi_++\psi_-\delta\psi_-)|_{\sigma=0}^{\sigma=\pi} = 0.
\label{vani}
\eeq 
Equation (\ref{vani}) is satisfied if $\psi_{+}$ and $\psi_-$ are periodic or anti-periodic
\ba
\psi^{\mu}_{+}(\sigma+\pi,\tau)&=&\pm\psi^{\mu}_{+}(\sigma,\tau),\nonumber\\
\psi^{\mu}_{-}(\sigma+\pi,\tau)&=&\pm\psi^{\mu}_{-}(\sigma,\tau).
\ea
The periodic case is called Ramond (R) boundary condition while the 
anti-periodic is known as Neveu Schwarz (NS).
The general solution in terms of mode expansion is given by
\beq
\psi_-^{\mu} =\sum_r b^{\mu}_r e^{-2i\pi(\sigma_-)} ,
\eeq
for the right-moving states and an 
analogous expression applies for the left-movers 
$\psi_+^{\mu}$ (by replacing $\sigma_-$ by 
 $\sigma_+$ and $b^{\mu}_r$ by $\tilde b^{\mu}_r$ ).
As a result of the boundary conditions, 
the frequency $r$ is integer 
for R boundary conditions and half-integer 
for the NS case.

The Ramond boundary conditions and the integer modes 
will describe string states that are spacetime fermions.
In fact, if we consider the fundamental state $b_0^{i}|0;p^{\mu}>$,
we see that it is massless and degenerate, as $b_0$ satisfies
the Clifford algebra $\{ b_0^i,b_0^j\}=\delta^{ij}$. This
means that the Ramond vacuum is a spinor of $SO(8)$ and
all the states obtained from the vacuum with the 
creation operators are fermionic as well.
Instead the NS boundary conditions with the half-integer 
excitations give
bosons. The fundamental state
$|0;p^{\mu}>$ has negative mass (tachyon)
 and is a scalar. The first excited massless state 
$b^i_{-\frac{1}{2}}|0;p^{\mu}>$ is a vector of $SO(8)$
and all the states in this sector, created by half-integer
modes, provide bosons.

Since the superstring is an extension of the bosonic case,
it is necessary to enlarge the algebra which describes the theory. 
Thus, the classical Virasoro constraints are now generalised to
\beq
J_{\pm}=0\quad ,\quad T_{\pm\pm}=0 ,
\label{vv}
\eeq
where the supercurrents and the energy-momentum tensors
are given in their light-cone gauge coordinates
\[J_+ = \psi^{\mu}_+\partial_+X_{\mu} ,\quad T_{++}=\partial_+X^{\mu}\partial_+X_{\mu}+
\frac{i}{2}\psi_+^{\mu}\partial_+\psi_{+\mu} ,
\]
\[J_- = \psi^{\mu}_-\partial_- X_{\mu} ,
\quad T_{--}=\partial_-X^{\mu}\partial_-X_{\mu}+
\frac{i}{2}\psi_-^{\mu}\partial_-\psi_{-\mu}.
\]

{\bf{Quantization of the superstring}}

The quantization of the fermionic fields is obtained by
imposing the anticommutation relations 
\[\{ b_r^{\mu},b_s^{\nu} \}=
\eta^{\mu\nu}
\delta_{r+s}\quad ,\quad
\{\tilde b_r^{\mu},\tilde b_s^{\nu} \}=
\eta^{\mu\nu}
\delta_{r+s}. \]
The anticommutator of left and right oscillators vanishes.
For $r<0$ ($r>0$) $b_r$ denotes creation (annihilation) operators. 
The complete spectrum is provided by the action of the creation operators
on the vacuum. 

The mass-shell condition in eq.(\ref{mass}) is now generalised
by redefining 
$N$ as the number of right bosonic plus 
right fermionic oscillators acting on the vacuum. Same
redefinition applies to $\tilde{N}$.
We have to take into account that fermions can assume 
 R or NS 
boundary conditions and this will change the 
contribution to the zero point energy $a$. 
Each fermionic coordinate contributes with a $-1/48$
in the NS sector and $1/24$ in the R sector, while each boson
gives a contribution of $-1/24$. In $D$ dimensions, if we are 
in the light-cone gauge, we have $D-2$ transverse bosons and $D-2$
transverse fermions which give $a=0$ in the Ramond sector while
$a=-1/16(D-2)$ in the Neveu-Schwarz.

After quantizing the supersymmetric
 theory, the Virasoro constraints become
\ba
\left[ L_m,L_n \right]&=& (m-n)L_{m+n}+\frac{D}{8}m(m^2-1)\delta_{m+n},\nonumber\\
\left[ L_m,G_r \right]&=&(\frac{m}{2}-r)G_{m+r},\nonumber\\
\{G_r, G_s\}&=& 2 L_{r+s}+\frac{D}{2}(r^2-\frac{a}{2})\delta_{r+s},
\ea
where the operators are defined by their normal ordered expressions
\ba L_m&=&L_m^{a'}+L_m^{b'},\nonumber\\
L_m^{a'}&=&\frac{1}{2}\sum_{n\in Z}:\alpha_{-n}\cdot\alpha_{m+n}: ,\nonumber\\
L_m^{b'}&=&\frac{1}{2}\sum_{n\in Z+a}:(r-\frac{m}{2})b_{m-r}\cdot b_{r}: ,
\nonumber\\
\ G_r &=&\sum_{n\in Z}: b_{r-n} \cdot\alpha_{n} : .
\ea
For completeness with respect to 
the bosonic case, we shall provide the 
light-cone quantization for the superstring case.
The theory is ghost-free but not explicitly covariant, but we can assure
Lorentz invariance if $D=10$ and $a=1/2$ \cite{Green:1987sp}.

The gauge is fixed with the relation 
$\psi^+=0$ and $X^+ =\alpha' p^+ \tau$ and since we 
are fixing the longitudinal oscillator modes, the only 
independent degrees of freedom are the transverse ones.

A supersymmetric non-tachyonic theory is obtained 
when the spectrum is truncated by some GSO (Gliozzi, 
Scherk and Olive) projections \cite{Gliozzi:1976qd}. 
We will explain this truncation separately
in the NS and in the R sector.
In the Neveu-Schwarz sector the GSO projections $P_{GSO}$ 
is defined by keeping  
states with an odd number of $b^{i}_{-r}$ oscillator excitations 
and removing those with even number. We define below
the projection operator in the NS sector and the fermion number,
$$
P^{NS}_{GSO}=\frac{1}{2}(1-(-1)^F)\,\,,\,\,
F=\sum_{r=1/2}^{\infty}b^i_{-r}\cdot b^i_r .
$$
Thus, the bosonic ground state is now massless and the
spectrum no longer contains a tachyon (which has fermion number $F=0$).
In the Ramond sector, the fundamental state (a Majorana spinor)
lives in the spinorial representation of $SO(8)$, as mentioned before.
If we introduce the projector operator
$$P_{GSO}^R=\frac{1}{2}(1+(-1)^F\Gamma_9),$$
where $\Gamma_9=b_0^1\cdot\cdot b_0^8$
 is the chiral operator in the transverse dimensions,
then the fundamental state becomes a Majorana-Weyl spinor
of definite chirality.
$P_{GSO}^R$, while projecting onto spinors of opposite chirality, 
guarantees spacetime 
supersymmetry of the physical superstring spectrum (we 
note that the choice of sign of $(-1)^F\Gamma_9=\pm 1$, corresponding to 
different chirality projections on the spinors, is a matter of convention).

The general procedure to obtain the massless spectrum 
is to solve 
the massless equations for 
right and left sector, apply level matching condition and the
particular GSO 
projections depending on the perturbative
superstring model considered, finally tensor the left with the right states.
If we want to proceed with the explicit calculation 
of the spectrum we need to specify the string theory 
we want to analyse.
Supersymmetric theories with only closed 
strings are type IIA, type IIB and heterotic models.
For type IIA and type IIB (where supersymmetry is realised in 
the left and right sector), by taking the tensor products 
of right and left movers we get four distinct sectors:
NS-NS, R-R, NS-R, RN-R, where the first two sets give bosons 
and the last two sectors provide fermion fields in the target space.  
The features and differences among these two models
have been given in the introduction. 
In this thesis we are interested 
in the
heterotic string hence we will focus on the technicalities 
concerning the heterotic case starting from section \ref{het}. 

\section{One loop amplitude and modular invariance}

The one loop vacuum amplitude, also known as genus-one partition
function, represents a fundamental quantity of the theory since
it encodes the full perturbative spectrum.
Differently from quantum field theory, in the string theory
this is a finite quantity that makes the theory modular invariant.
The modular invariant constraints 
are in fact derived
from the calculation of the 
one-loop vacuum amplitude. 
The Feynman diagram, which describes a closed string 
propagating in time and returning 
to its initial state, is a donut-shaped surface, 
equivalent to a two-dimensional torus.
\begin{figure}[htp]
\centering
\scalebox{0.5}{\centerline{\epsffile{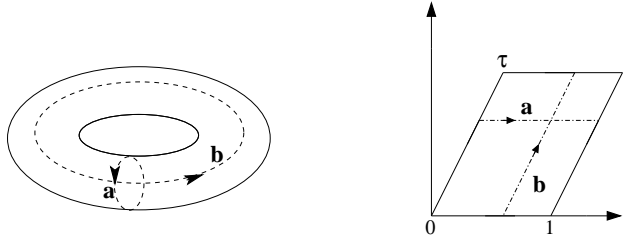}}}
\caption{1)Torus diagram. 2)The flat torus as 
a two dimensional lattice. $a$ and $b$ represent 
the two non-contractible cycles of the Riemann surface.}
\label{fig:torus2}
\end{figure}
We can parametrize the torus 
 by a complex parameter $\tau=\tau_1+i \tau_2$ ,
$\tau_2>0$.
If we define in the complex plane 
a lattice by identifying $z=z+1$ , 
$z=z+\tau$, then
the torus is obtained by identifying the opposite sides of this 
parallelogram (see fig.\ref{fig:torus2}).

\begin{figure}[htp]
\centering
\scalebox{0.3}{\centerline{\epsffile{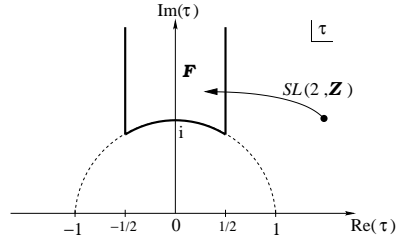}}}
\caption{Fundamental domain.}
\label{fig:F}
\end{figure}

The full family of equivalent tori is obtained by the
transformations 
\beq
S : \tau \rightarrow -\frac{1}{\tau}\quad, \quad 
T : \tau \rightarrow \tau + 1 \quad ,
\label{modular}
\eeq
that are the generators of the modular invariant group, 
whose most general transformation is given by
\beq
\tau \rightarrow \frac{a \tau + b}{c \tau + d} \quad  ad-bc=1
\quad a,b,c,d
\in {\mathbb Z}.
\label{invm}
\eeq
The formula (\ref{invm}) generates the modular group $PSL(2,{\mathbb Z})$.
The non-equivalent tori are contained in the so-called
 fundamental region 
$$\mathcal{F}={\mathbb C}^1/PSL(2,{\mathbb Z}) =\{ |\tau|\ge 1 \, 
,\,-\frac{1}{2}\le\tau_1 < \frac{1}{2} \, , \,\tau_2 >0\}\,$$
 (see fig.\ref{fig:F}).
Any point outside the modular domain can be
 mapped by a modular transformation
inside $\mathcal{F}$.
We calculate now the vacuum amplitude for the bosonic string
 in analogy
with the quantum field theory approach.
In the case of a single scalar particle 
the vacuum energy $\Gamma$ is 
defined by the path integral 
\beq
e^{-\Gamma}=\int\mathcal{D} \,\phi e^{-S} ,
\label{ee}
\eeq
where $S$ is the action of the boson in $D$ dimensions.
If we want to make explicit the dependence 
of the integral on the particle mass $M$
we can rewrite it in terms of the Schwinger parameter $t$ and
eq.(\ref{ee}) assumes the form
\beq
\Gamma=-\frac{V}{2}\int_{\epsilon}^{\infty}\frac{dt}{t}e^{-t M^2}\int
\frac{d^D p}{(2\pi)^{D}}e^{-tp^2} ,
\label{pf}
\eeq
where $V$ is the volume of the spacetime
 and $p$ the momentum of the particle.
 The parameter $\epsilon$ is an ultraviolet cutoff 
that will disappear when we restrict the integration region to 
the fundamental region of the torus. 
If we calculate the Gaussian momentum integral and generalise 
formula (\ref{pf}) for bosonic and/or fermionic fields
then we obtain 
\beq
\Gamma=-\frac{V}{2(4\pi)^{D/2}}\int_{\epsilon}^{\infty}\frac{dt}
{t^{D/2+1}}Str (e^{-t M^2}),
\label{pf2}
\eeq
where the Supertrace $Str$ takes into account the Bose-Fermi statistics.

Let us now consider the case of the bosonic string for which
we want to derive the one-loop amplitude.
For the bosonic theory we have $D=26$
 and $M^2=\frac{2}{\alpha'}(L_o+\tilde L_o-2)$.
At this point we need to take into account the level matching condition that
can be implemented by a constraint given in terms of a 
real variable $s$. 
Subsequently, we rearrange
the $t$ and $s$ parameters in the new complex "Schwinger" parameter 
$\tau=\tau_1+i\tau_2=s+i\frac{t}{\alpha' \pi}$. Since the closed string 
sweeps a torus at one loop then we identify $\tau$ as 
the Teichmuller parameter parametrizing the torus 
(see for example \cite{Angelantonj:2002ct}).

Defining $q = e^{2i\pi\tau}$ and $\overline q = e^{-2i\pi\overline\tau}$ and 
calculating the integral in the fundamental domain gives
the partition function of the torus amplitude
\beq
\mathcal{T}=\int_{\mathcal{F}}\frac{d^2\tau}{\tau_2^2}\frac{1}{\tau_2^{12}}
tr q^{L_0-1} \overline q^{\tilde L_0-1}.
\label{T}
\eeq
The same expression can be obtained by some geometric considerations. 
A point on the string propagates in the time direction 
as $2\pi \tau_2$ and in space as $2\pi \tau_1$. The time translation is 
given by the Hamiltonian $H= L_0+\tilde L_0-2$ and the shift along the
string is given by the momentum operator $P= L_0-\tilde L_0$. 
The path integral is then
\[\mathcal{T} \propto tr(e^{-2\pi\tau_2H}e^{2i\pi\tau_1P})\sim
tr(q^{L_0-1}\overline q^{\tilde{L}_0-1}).
\]

The expansion of the operator $L_0$ and the calculation of the trace
will transform equation (\ref{T}) into
\beq
\mathcal{T}=\int_{\mathcal{F}}\frac{d^2\tau}{\tau_2^2}\frac{1}{\tau_2^{12}}
\frac{1}{|\eta(\tau)|^{48}} ,
\label{T2}
\eeq
where the Dedekind $\eta$ function is defined in Appendix A,
as well as its properties under modular transformations. 
Each bosonic mode then gives a contribution to the partition function
equal to $\frac{1}{|\eta|^2}$.
The integrand of eq.(\ref{T2}) is modular invariant, as we can prove
by using the formulae in Appendix A. 
\newpage

\section{Spin structures}

When we consider the parallel transport properties of
spinors on a two dimensional surface, for example on the torus,
we need to introduce the so-called spin structures.
They provide the fermionic contributions to 
the partition function and have to be defined 
in both Ramond and Neveu-Schwarz sectors.
Some kind of GSO
projections enter in the game to ensure the 
consistency of the theory. 

A fermion
moving around the two non-contractible loops
of the torus gives rise to 
four possible spin structures, indicated 
as following: $A^{(++)}(\tau)$, $A^{(+-)}(\tau)$, $A^{(-+)}(\tau)$
 and $A^{(--)}(\tau)$. The first entry in the exponent 
represents the boundary condition in the $\sigma^1$ direction while
the second gives the boundary condition in time direction  $\sigma^0$.
The $"+"$ and $"-"$ signs label the Ramond 
and Nevew-Schwarz boundary conditions respectively.
 For brevity we 
focus our discussion on the spin structures 
of the right sector of the string.

The NS sector provides
states with anti-periodic boundary conditions in the $\sigma$ direction and
if we implement the periodicity in the time direction we
will need to introduce the Klein operator $(-1)^F$ in the trace.
The fermionic contributions to the path integral are given, in the 
R and NS sector Hilbert space, by the following expressions
\ba
A^{(+-)}&\sim& Tr_R (e^{-2\pi\tau_2H})\quad\,\,\,,\,\, 
A^{(++)}\sim Tr_R ((-1)^F e^{-2\pi\tau_2H}),\nonumber\\
A^{(--)}&\sim& Tr_{NS} (e^{-2\pi\tau_2H})\quad,\,
A^{(-+)}\sim Tr_{NS} ((-1)^F e^{-2\pi\tau_2H}).
\label{A}
\ea
The modular transformations change the boundary conditions, 
thus it is possible to obtain a spin structure from another by
applying T and S transformations.
We note that $A^{(++)}$ is modular invariant while for the 
other expressions the following relations hold 
$
A^{(+-)}\,\,\,\,\underrightarrow{S}\,\,\,\,
A^{(-+)}\,\,\,\,\underrightarrow{T}\,\,\,\, A^{(--)}.
$
Each of these contributions is multiplied by a phase which 
can be derived by imposing modular invariance of the total 
partition function of the model under consideration.
 A detailed explanation on the derivation
of the phases can be found in \cite{Lust:1989tj}.

The one-loop modular invariant partition function for the 
right-moving sector is given by 
\beq
Z=\frac{1}{2}Tr_{NS}[(1-(-1)^F)q^{L_0-\frac{1}{2}}]
+\frac{1}{2}Tr_R[(1+(-1)^F)q^{L_0}] .
\label{a}
\eeq
The total superstring amplitude is obtained by combining 
eq.(\ref{a}) with the left-moving fermionic contribution
 and multiply the whole expression by the bosonic part.

If we calculate the traces in eqs.(\ref{A}) we 
can rewrite the spin structures in terms of
 the Jacobi $\theta$-functions
\ba
A^{(++)}&=&16q^{1/3}\prod_{n=1}^{\infty}(1-q^{n})^8=
\frac{\theta_1^4(0|\tau)}
{\eta^4(\tau)} ,\nonumber\\
A^{(+-)}&=&16q^{1/3}\prod_{n=1}^{\infty}(1+q^{n})^8=
\frac{\theta_2^4(0|\tau)}
{\eta^4(\tau)} ,\nonumber\\
A^{(--)}&=&q^{-1/3}\prod_{n=1}^{\infty}(1+q^{n+1/2})^8=
\frac{\theta_3^4(0|\tau)}
{\eta^4(\tau)} ,\nonumber\\
A^{(-+)}&=&q^{-1/3}\prod_{n=1}^{\infty}(1-q^{n-1/2})^8=
\frac{\theta_4^4(0|\tau)}
{\eta^4(\tau)} .
\ea
Eq.(\ref{a}) corresponds to the famous Jacobi identity
\[ \theta_4^4-\theta_3^4-\theta_2^4=0 , 
\]
which tells us that the superstring amplitude vanishes.
 The meaning of the previous result is that the contribution
of NS spacetime bosons and R fermions is the same 
(but the two contributions have
 opposite statistics). This is considered
an indication of supersymmetry.
The general definition of $\theta$-functions as Gaussian sums and 
in product representations are given in Appendix A, along with
 their modular transformation properties. 

\section{Partition functions of 10D superstrings}


In this section we present the partition function
for the five perturbative superstring theories and the case of
the heterotic $E_8 \times E_8$ string with orbifold actions
 will be discussed widely
in the chapter 5.
A very convenient and compact way of writing the fermionic
contributions (in the previous section they were given in terms
of $\theta$-functions) is by defining the characters 
$O_{2n}, V_{2n}, S_{2n} $ and $C_{2n}$, representations
 of the $SO(2n)$ group.
Their general definitions and modular transformations 
are presented in Appendix A.
Here we give as an example the characters of the little group $SO(8)$
\ba
O_8= \frac{\theta^4_3+\theta^4_4}{2\eta^4}&,\quad& 
V_8= \frac{\theta^4_3-\theta^4_4}{2\eta^4},\nonumber\\
S_8= \frac{\theta^4_2+\theta^4_1}{2\eta^4}&,\quad& 
C_8= \frac{\theta^4_2-\theta^4_1}{2\eta^4}.
\label{esau}
\ea 
Each definition in eqs.(\ref{esau})
represents a conjugacy class of the $SO(8)$ group, 
in particular, $O_8$ is the scalar representation,
$V_8$ the vectorial, $S_8$ and $C_8$ are spinors with opposite chirality.
The characters
 $V_8$ and $O_8$ provide a decomposition of the NS sector, while 
the $C_8$ and $S_8$ give the R spectrum.
We are finally ready to present the partition functions
for the $10 D$ spectra of type II and type 0
\ba
\mathcal{T}_{IIA}&=&(\overline V_8-\overline C_8)|(V_8- S_8)|,
\quad\mathcal{T}_{0A}=|O_8|^2+|V_8|^2+
\overline C_8| S_8|+ \overline S_8|C_8|,\nonumber\\
\mathcal{T}_{IIB}&=&|V_8-S_8|^2 ,
\quad \quad\quad\,\,\,\,\,\,\,\,\,\,\,\quad\mathcal{T}_{0B}= |O_8|^2+|V_8|^2+|S_8|^2+|C_8|^2.
\ea
The spectrum can be read by expanding the characters 
in powers of $q$ and $\bar q$, as indicated in Appendix B.

For the heterotic case we need to introduce
$SO(16)$ and $SO(32)$ characters in the partition function,
in order to include the gauge degrees of freedom
of the theory.
The only two supersymmetric modular invariant heterotic models in
10 dimensions are those
where the  $E_8\times E_8$ and the $Spin(32) $ symmetries
are realised and their torus amplitude is respectively
\ba
\mathcal{T}_{E_8\times E_8}&=&(\overline V_8- \overline S_8)
( O_{16}+ S_{16})( O_{16}+ S_{16})\nonumber\\
\mathcal{T}_{S_{32}}&=&( \overline V_8-\overline S_8)(O_{32}+ S_{32}).
\ea

\section{Bosonization}\label{bosonization}

In this section we present the equivalence 
between fermionic and bosonic
conformal field theories in two dimensions,
a correspondence which allows the consistent
construction of free fermionic models.

Before entering into the details we will
 give the definition of
 operator product expansions (OPEs) 
in conformal theories in two dimensions.

\subsection{Product expansion operator}

In quantum field theory, the infinitesimal conformal
transformations 
\[z\rightarrow z + \epsilon(z)\quad ,\quad 
\bar z\rightarrow \bar z + \bar\epsilon(\bar z)\]
produce a variation of a field $\Phi(z,\bar z)$ 
given by the equal time commutator
with the conserved charge $Q=\frac{1}{2\pi i}
\oint (dz T(z)\epsilon(z)+d\bar z
\bar T(\bar z)\bar \epsilon(\bar z))$, where $T$ and $\bar T$ are the 
stress-energy tensors in complex coordinates. 
 The products of the operators is well defined only if time-ordered.  
The radial quantization introduced in section \ref{vvsba} is
an example of the construction of a quantum theory of conformal fields
on the complex plane. In this set up the time-ordered product is
replaced by the so called radial-ordering\footnote{The radial ordering
operator R for two fields A and B is given by \begin{equation*}
R(A(z)B(w))=\left\{
\begin{array}{rl}
A(z)B(w) & |z|>|w|\\
B(w)A(z) & |z|<|w| ,
\end{array} \right.
\end{equation*}
 where a minus sign appears if we interchange two fermions.},
 realised by the operator $R$. 
A complete treatment of the complex tensor analysis 
can be found in \cite{Kiritsis:2007zz,Kiritsis:1997hj}.
Here we only mention the main results which will be useful
for our purpose.  

The commutator of an operator A with a spacial integral of an
operator B corresponds to
\beq
 \Big[\int d\sigma B, A\Big]=\oint dz R(B(z)A(z)).
\label{integ}
\eeq 
This result leads \cite{Kiritsis:1997hj} to the operator
product expansions (OPEs) of the stress energy tensors $T(z)$ and $\bar T(\bar z)$ 
with the field $\Phi(w,\bar w)$ 
\ba
R(T(z)\Phi(w,\bar w))&=&\frac{h}{(z-w)^2}\Phi+\frac{1}{z-w}\partial_w\Phi+... ,\nonumber\\
R(\bar T(\bar z)\Phi(w,\bar w))&=&\frac{\bar h}
{(\bar z-\bar w)^2}\Phi+\frac{1}{\bar z-\bar w}\partial_{\bar w}\Phi+...\,.\nonumber\\
\label{pro}
\ea
Eqs.(\ref{pro}) contain the conformal transformation properties
of the field $\Phi$, hence they can be used as a definition of primary 
field\footnote{Its definition is given in $^{5}$.}
for $\Phi$ with conformal weight $(h, \bar h)$.
We observe that the above products are given by the
expansion of poles (singularities that contribute to
integrals of the type (\ref{integ})) plus regular terms, which we can omit.
From now on we assume that the operator product expansion is always 
radially ordered. 

\subsection{Free bosons and free fermions}

We start by considering a massless free boson $X(z,\bar z)$,
where we can split the holomorphic and anti-holomorphic components into  
$X_L(z)$ and $X_R(\bar z)$. 
For our purpose it is sufficient to consider the holomorphic
part only.
The propagator of the left component corresponds to 
$<X_L(z) X_L(w)>=-log(z-w)$, which says that it is not a
conformal field, but its derivative
 $\partial X_L(z)$ is a (1,0) conformal field.
This is showed by taking the OPE with 
the stress tensor, that is defined as $T=-\frac{1}{2}:\partial X_L^2:$,
and comparing with eq.(\ref{pro}) one obtains
\beq
T(z)\partial X_L(w)\sim\frac{1}{(z-w)^2}\partial X_L(w)+\frac{1}{z-w}\partial^2 X_L(w)+...\,.
\eeq 

We now 
consider two Majorana-Weyl
fermions $\psi^i(z)$, $i=1,2$, where a change of
basis rearranges the fermions into the complex form
\[\psi=\frac{1}{\sqrt{2}}(\psi^1 + i \psi^2)\quad ,\quad 
\bar\psi=\frac{1}{\sqrt{2}}(\psi^1 - i \psi^2).
\] 
The theory contains a $U(1)$ current algebra (see following section) generated by the (1,0)
current $J(z)=:\psi\bar\psi:$. The OPE for $\psi \bar \psi$ and the holomorphic
energy tensor are defined as
\beq
\psi(z)\bar\psi(w)=-\frac{1}{z-w}\quad , 
\quad T(z)=\frac{1}{2}:\psi(z)\partial \psi(z):. 
\eeq
If we calculate the product expansion $T(z)\psi(w)$ with
the above definitions, we see that $\psi$ is an affine primary 
field\footnote{The formal definition of primary field is the 
following: $\Phi$ is primary of conformal weight ($h$,$\bar h$) 
if it satisfies
 the transformation law $\Phi(z,\bar z)\rightarrow \left(
\frac{\partial f}{\partial z}\right)^h
\left(\frac{\partial \bar f}{\partial \bar z}\right)^{\bar h}
\Phi(f(z),\bar f(\bar z))$, where $h$ and $\bar h$ are real values.\label{che}}
of conformal weight $(1/2,0)$.

We present the boson-fermion correspondence
by showing that the same operator algebra is produced
by two Majorana-Weyl fermions on one hand and a chiral boson
 on the other hand.
In fact, in the fermionic case \[T(z)=\frac{1}{2}:J^2: ,
\]
formula that says that the stress tensor has central charge $c=1$.
We can produce the same operator algebra by using a single chiral boson
$X(z)$, whose current is provided by 
\[J(z)=i\partial X(z) ,
\]
where is the stress-energy tensor
 $T=-\frac{1}{2}:\partial X^2:$, as presented at 
the beginning of the section. 
The definitions below thus contain explicitly the 
boson-fermion equivalence 
\beq
\psi=:e^{iX(z)}:\quad , \quad \bar\psi=:e^{-iX(z)}:.
\label{coco}
\eeq
Further details can be found in \cite{Polchinski:1998rq,
Kiritsis:1997hj,Ginsparg:1988ui}.

\section{The heterotic string}\label{het}

The heterotic string \cite{Gross:1984dd} 
was constructed
after the famous work of Green and Schwarz \cite{Green:1984sg} had 
shown that the consistency of 
an $N=1$ supersymmetric string theory
requires the
presence of an $E_8\times E_8$ or $Spin(32)$ gauge symmetry. 
10 dimensions supergravity with these gauge groups
is free of gravitational and gauge anomalies.
This observation fuelled an increased activity 
in heterotic models. 
Before this discovery, the standard procedure
to introduce gauge groups in string theory
 consisted of attaching the Chan-Paton charges at the endpoints
of open strings \cite{Paton:1969je}. 
This prescription does not produce 
the exceptional $E_8\times E_8$ \cite{Schwarz:1982md,Marcus:1982fr},
a non-abelian GUT gauge group which allows a more
natural embedding of the Standard Model spectrum at low energy.

In this section we describe the basics
of the heterotic superstring, an orientable closed-string 
theory in ten dimensions with $N=1$ supersymmetry and
 with gauge group 
$E_8 \times E_8$ or $Spin(32)/Z_2$
 \cite{Green:1987sp}. Its low-energy
limit is supergravity coupled with Yang-Mills theory.
This theory is an hybrid of the $D=10$ fermionic string and
the $D=26$ bosonic string and the resulting
 spectrum is supersymmetric,
 tachyon free, Lorentz invariant and unitary.
The absence of gauge and gravitational anomalies is obtained
by the compactification of the extra sixteen bosonic coordinates
on a maximal torus of determined radius. All these properties make the 
heterotic string one of the most appealing candidates for 
an unified field theory.

{\bf{Current algebra on the string worldsheet}}

In heterotic models the gauge symmetries 
are introduced 
 by distributing
symmetry charges on the closed strings.
These charges are not localised, so we obtain a continuous
charge distribution throughout the string.
A way to describe their currents is to introduce,
on the worldsheet,  fermions with internal
quantum number, which are singlets under the Lorentz group.
If we take $n$ real Majorana fermions $\lambda^a$, $a=1,..n$, and
we split them into right- and left-moving modes ($\lambda^a_{\pm}$),
then we can write the bosonic action on the worldsheet,
including the new internal symmetries, as
\beq
S=-\frac{T}{2}\int d^2 \sigma(\partial_{\alpha}
X_{\mu}\partial^{\alpha}X^{\mu}-\lambda_-^a\partial_+\lambda^a_-
-\lambda_+^a\partial_-\lambda^a_+).
\label{h}
\eeq
The equivalence of bosons and fermions 
in two dimensions (see eq.(\ref{coco}))
allows us to convert two 
Majorana fermions on the worldsheet into a real boson. We can then 
obtain  
$\frac{n}{2}$ bosons $\phi^i$ in the place of $n$ fermions $\lambda^a$.
With this substitution 
the theory contains $D+n/2$ free bosons and
 has a 
$SO(D-1,1)$ Lorentz symmetry plus an internal 
$SO(n)\times SO(n)$ symmetry.
Its consistency
requires $D+n/2=26$, and in the case of a supersymmetric 
theory ($D=10$) it means that $n=32$.
Let us go back to eq.(\ref{h}) and
consider for our purposes 
only a $SO(n)_R$ symmetry.
The right-fermion currents are given by
\beq
J^{\alpha}_+(\sigma)=\frac{1}{2\pi}T^{\alpha}_{ab}\lambda_+^{a}(\sigma)
\lambda_+^{b}(\sigma).
\eeq
The $T^{\alpha}$ generators satisfy the algebra $[T^{\alpha},
T^{\beta}]=if^{\alpha \beta \gamma}T^{\gamma}$ and this relation fixes 
the commutation relation for the currents
\beq
[J^{\alpha}_+(\sigma),J^{\beta}_+(\sigma')]= i f^{\alpha \beta \gamma}
J^{\gamma}_+(\sigma)\delta(\sigma-\sigma')+
\frac{i k }{4 \pi}\delta^{\alpha \beta}\delta'(\sigma-\sigma').
\eeq
The previous formula describes
 the affine Lie algebra $\hat{SO}(n)$ with 
central extension represented by the
 second term (anomaly contribution).
If this algebra is built up from $n$ fermions in the 
fundamental representation of $SO(n)$ then $k=1$.
If the fermions are not in the fundamental representation 
we would obtain a different (quantized) value of $k$.
We are interested in obtaining the extended
 algebra for the exceptional group $E_8$ 
but it turns out that the task is unrealisable in terms of
 free fermions with a minimal 
value of $k$.
It has been shown \cite{Green:1987sp} that 
this realisation is possible by using 
eight free bosons.

We are now ready to describe the heterotic string as 
it was first formulated by Gross, Harvey, Martinec and Rohm.
As we said already, the left moving modes are described 
in a bosonic string theory (D=26) while the right movers
are supersymmetric (D=10). Specific GSO projections
ensure supersymmetry for our model.
 The gauge degrees of freedom are included in 
the left sector with an appropriate current algebra.

The general action of this theory is
\beq
S=-\frac{T}{2}\int d^2 \sigma\left(\sum_{\mu}^9(\partial_{\alpha}
X_{\mu}\partial^{\alpha}X^{\mu}-2\psi^{\mu}_+\partial_-\psi_{+\mu }
)-2\sum_{a=1}^n\lambda_-^a\partial_+\lambda^a_-\right).
\eeq

We observe here that the spacetime fermions $\psi^{\mu}$ have
only right-moving components, superpartners of $X_R^{\mu}$.
The content therefore differs from the
type IIB, where supersymmetry is realised in both left and right sectors.
The left-moving sector contains the space-time fields
$X^{\mu}_L$ and the internal
Majorana fermions $\lambda^a_-$.  

If the boundary conditions for $\lambda^a_-$ are all the same,
we obtain the $Spin(32)$ heterotic theory; choosing different
boundary conditions for the internal fermions will provide the 
$E_8 \times E_8$ heterotic string. In this thesis we want to analyse
the second possibility. 
It can be shown that the two theories
are continuously related \cite{Ginsparg:1986bx}. 
In fact an equal number of states at every mass level appear
in the two heterotic string theories.

The $E_8\times E_8$ heterotic theory
 is obtained when we split the internal fermions 
into two groups and assign different boundary conditions to each set.
In this case the gauge group would be
$SO(n)\times SO(32-n)$. The interesting case for us is 
when n is a multiple of 8 and in particular $n=16$. The massless
left-moving states are of the form
\[\lambda^i_{1/2}\lambda^j_{1/2}|\Omega> \quad i,j=1,..32.
\]
These combinations give rise
 to the vector and the adjoint representations for each $SO(16)$ 
present in the current algebra of the theory.
 We also obtain the spinorial 
representation of $SO(16)$. The introduction of 
appropriate GSO projections produces the final
content given by 
the adjoint and the spinorial
representations of $SO(16)$. This sum enhances the Lie algebra 
of $SO(16)$ to the 
exceptional group $E_8$. Since we started with 
an $SO(16)\times SO(16)$ symmetry we conclude that
the enlarged current algebra obtained is $E_8 \times E_8$.
In the next section we consider the 
toroidal compactification, fundamental in the
description of the bosonic formalism. 

\section{Toroidal compactifications}

The current algebra can be realised in the bosonic formulation
by introducing a toroidal compactification. We can start with
a bosonic theory in 26 dimensions and compactify one
dimension on a circle. In this simple case we only get one
toroidal boson while if the compactification includes $d$ of
these bosons the space-time is reduced from 26 to $26-d$ dimensions.

In this section we 
describes the simple compactification on a circle,
leaving the explanation on how gauge groups are created in this setup
for the case of higher dimensional compactifications in chapter 4.

The coordinate compactified on the circle satisfies
the condition $x\equiv x + 2\pi R n$. $R$ is the radius 
of the circle and $n$ an integer which defines the winding number,
 a quantity 
that gives the number of times the string wraps around the circle.
The winding represents a stringy new feature which arises in 
the compactification procedure.

The general expansion for the compact boson becomes
\beq
X=x + 2\alpha'\frac{m}{R}\tau
+2nR\sigma + (oscillators).
\label{com}
\eeq 
The expression (\ref{com}) can be rewritten in terms of
the chiral components $p_L$ and $p_R$ of the compact coordinate
as
\beq
X_{L,R}=\frac{1}{2}x+\alpha'p_{L,R}(\tau\mp\sigma)+(oscillators)_{L,R} ,
\eeq
where the chiral momenta are defined as 
\beq
p_{L,R}=\frac{m}{R}\pm\frac{nR}{\alpha'}.
\label{plr}
\eeq
The invariance under $x\rightarrow x+2\pi R$ requires 
$m$ to be integer. The presence of a $n\neq0$ describes a soliton state
that does not exist in the uncompactified theory, since its energy
would diverge for $R\rightarrow\infty$. This means that the spectrum 
of a compactified theory can in general be larger than
the non compact corresponding case. 
When a non-compact boson is compactified,
its contribution to the partition function becomes
a discrete sum, given below
\beq
\frac{1}{\sqrt{\tau_2}\eta\overline\eta}\rightarrow
\sum_{m,n}\frac{q^{\alpha' p_L^2/4}\overline q^{\alpha' p_R^2/4}}
{\eta\overline\eta}.
\eeq

We can underline here the presence of a symmetry which relates
$m$ and $n$ quantum numbers, the so-called T-duality, one of the
symmetries relating the five perturbative string models \cite{Hull:1994ys,
Witten:1995ex}.
\beq
n\leftrightarrow m \quad \quad R\leftrightarrow \alpha'/R.
\eeq
The previous formula tells us that the closed bosonic 
string compactified on a radius $R$ is equivalent
to the theory with radius $\alpha'/R$.
$T$ duality is an exact symmetry of the perturbative theory
for the closed bosonic string and it relates type 0A with type 0B,
type IIA with type IIB, as mentioned in the introduction.  
As we announced before,
the generalisation to higher dimensional tori will be 
considered in chapter 4. 
We will introduce the compactification
on a 16 dimensional tori that leads to the $E_8\times E_8$
symmetry, as expected.

\chapter{Free Fermionic Models}

In this chapter we describe the free fermionic formulation of 
the heterotic superstring and mainly focus on a subset of these models
 which are called 
semi-realistic free fermionic models.
 Moreover, we provide some indicative examples among this class of 
string compactifications, whose results are published in
\cite{Faraggi:2006qa,Cleaver:2008mu}.

In the first part of our discussion we will describe the 
consistency rules necessary for the construction of the theory.
The interested reader can find further details in the original papers
\cite{Antoniadis:1987wp,Antoniadis:1985az,Antoniadis:1986rn,Kawai:1986ah,Kawai:1987ew}.

In the second part of this chapter we present
some examples of semi-realistic models in the free fermionic formulation
produced in the past, in which the only Standard Model
charged states are the MSSM states 
\cite{Cleaver:1999cj,Cleaver:1999mw}.
Therefore we revisit some of their properties.
The presence of three Higgs doublets in the untwisted spectrum
is another feature of semi-realistic free fermionic models
and the general procedure to reduce them to one pair is 
given by the analysis of
 the supersymmetric flat directions. This method consists
 in
giving heavy masses to some 
of the Higgs doublets
 in the low energy field theory \cite{Cleaver:1998saa, Cleaver:2001ab}.
The two models largely discussed in this chapter introduce
 instead a new mechanism that achieves the same reduction by
an appropriate choice of boundary conditions, in particular, 
asymmetric boundary conditions among left and right internal fermions. 
An additional effect related to this choice is
the reduction of the supersymmetric moduli space.
The procedure, explained in detail later on,
 represents a selection 
mechanism useful to pick phenomenologically
 interesting string vacua. 
We will present some generalities
on the analysis of flat directions and introduce the concept of
stringent flat directions, since this will allow the
investigation of the
low energy properties
of free fermionic models.
The flat direction analysis is needed because of 
an anomalous $U(1)$ which 
generally appears in this set up.
Its presence gives rise to a Fayet-Illiopulos 
D-term which breaks supersymmetry but,
by looking at supersymmetric flat directions 
 and imposing F and D flatness on the vacuum, 
supersymmetry can be restored.
In the last example presented in this chapter
an extensive search could not provide any flat solution, raising 
the question on the perturbatively broken supersymmetry.
At the tree level the Bose-Fermi degeneracy of the spectrum implies that
 the theory is instead supersymmetric,  
yielding a vanishing cosmological
constant. Therefore, this unconventional result may lead to
an interesting new interpretation
of the supersymmetry breaking mechanism in string theory.

\section{The free fermionic formulation}

In contrast with the ten dimensional superstrings,
 where the compactification
of the "extra-dimensions" is needed to reduce
 the spacetime to four dimensions, 
the free fermionic formulation provides
 directly a four-dimensional theory with 
a certain number of internal degrees of freedom.
In fact, an internal sector of two-dimensional conformal field
theories is required in order to fulfil
\begin{itemize}
\item conformal invariance, 
\item worldsheet supersymmetry,
\item modular invariance.
\end{itemize}
In this approach all internal degrees of freedom are fermionised,
thus producing worldsheet fermions. Requiring anomaly cancellation fixes
the number of fields in the left and right sector, obtaining
18 left-moving Majorana fermions $\chi^a$, $(a=1,..18)$, and
44 right-moving Majorana fermions $\bar{\Phi}^I$, $(I=1,..44)$.
The spacetime is described by the 
left-moving coordinates $(X^{\mu},\psi^{\mu})$ and
the right-moving bosons $\overline{X}^{\mu}$. 
Since the heterotic string is $N=1$ spacetime supersymmetric (we choose 
here a different convention w.r.t. the bosonic approach by fixing
the supersymmetry in the left sector), then we require
left-moving local supersymmetry. This is realised non-linearly \cite{Antoniadis:1985az}
 among all
the fields in the left sector, spacetime and internal ones, by the 
supercurrent
\beq
T_{F}=\psi^{\mu}\partial X^{\mu}+ f_{abc}\chi^{a}\chi^b\chi^c,
\eeq
where $f_{abc}$ are the structure constants of a semi-simple
Lie group $G$ of dimension $18$. The $\chi^a$ transform in the 
adjoint representation of $G$. 
In \cite{Dreiner:1988zf} it is shown that $N=1$ spacetime 
supersymmetry can be obtained in four dimensions when the Lie algebra 
$G=SU(2)^6$. In this case it is convenient to
group the $\chi^a$ into six triplets $(\chi^i,y^i,w^i)$, 
$(i=1,..6)$. Each of them transforms as the adjoint representation
of $SU(2)$.
So far we have ensured superconformal invariance of the theory.
We still need to verify its modular invariance to get a consistent theory.
The target is achieved by investigating the properties of
the partition function.
In this prescription, a modular invariant partition
function must be the sum over all different boundary conditions
for the worldsheet fermions, with appropriate weights.
For a genus-$g$ worldsheet $\Sigma_g$, fermions moving 
around a non trivial loop $\alpha\in \pi_1(\Sigma_g)$
transform as
\ba
&&\overline\Phi^I\rightarrow R_g(\alpha)^I_J\overline\Phi^J,\nonumber\\
&&\psi^{\mu}\rightarrow -\delta_{\alpha}\psi^{\mu},\nonumber\\
&&\chi^a\rightarrow L_g(\alpha)^a_b\chi^b,
\label{spin}
\ea
where the first transformation refers to the
right-moving fields,
$L^a_{g a'}L^b_{g b'}L^c_{g c'} f_{abc}= - \delta_{\alpha}
f_{a'b'c'}$ and $\delta_{\alpha}=\pm 1$.
 The spin structure of each fermion is a 
representation of the first homotopy group $\pi_1(\Sigma_g)$
\cite{AlvarezGaume:1986es}. 
The transformations (\ref{spin}) ensure the 
invariance of the supercurrent. We need to require
the orthogonality of $ R_g(\alpha)$ to leave the energy-tensor
invariant in the right sector.
In order to keep the theory tractable, commutativity of the boundary conditions
has been assumed \cite{Antoniadis:1987wp}, implying the
following restrictions on $L_g(\alpha)$ and $R_g(\alpha)$:
they have to be abelian matrix representations of $\pi_1(\Sigma_g)$; 
it is assumed commutativity between the boundary conditions
on surfaces of different genus. The previous constraints
allow the diagonalization
of the matrices $R(\alpha)$ and $L(\alpha)$, simplifying
the equations (\ref{spin}) into 
\beq
f\rightarrow -e^{i\pi \alpha(f)}\,f,
\label{ffff}
\eeq
where $f$ is any fermion $(\psi^{\mu},\chi^a,\overline\Phi^I)$ and
$\alpha(f)$ is the phase acquired by $f$ when moving 
around the non contractible loop $\alpha$.

Thus, the spin structure for a non contractible loop can
be expressed as a vector
\beq
\alpha=\{\alpha(f_1^r),..\alpha(f_k^r);
\hat{\alpha}(f_1^c),..\hat{\alpha}(f_{k'}^c) \},
\label{bv}
\eeq
where $\alpha(f^r)$ is the phase for a real fermion while
$\hat{\alpha}(f^c)$ corresponds to a complex one. By convention, 
$\alpha(f)\in(-1,1]$. Obviously for the complex conjugate 
fermion $\alpha(f^*)\in[-1,1)$. 
We set the notation 
\[ \delta_{\a} = \left\{ \begin{array}
{r@{\quad \text{if} \quad}l}
1 & \a(\psi^{\mu}) =0\\
-1 & \a(\psi^{\mu})=1
\end{array} \right.\]
where, according to eq.(\ref{ffff}), the entry $1$ represents a periodic 
boundary condition and $0$ is the anti-periodic boundary condition.
Since there are $2g$ non-contractible loops for a genus $g$
Riemann surface, we have to specify two sets of phases
$\alpha_1,..\alpha_g$, $\beta_1,..\beta_g$ to obtain
the full partition function.
 In its general form it 
can be written as
\beq
Z=\sum_{genus}\sum_{i,j=1}^g c \left(\begin{array}{c}  
\a_i\\
\b_j
\end{array}\right) z\left(\begin{array}{c}  
\a_i \\
\b_j
\end{array}\right) ,
\eeq
where $z\left(\begin{array}{c}  
\a_i \\
\b_j
\end{array}\right)$ can be expressed in terms of 
$\theta$-functions. The modular
invariance imposes constraints onto the coefficients 
$ c \left(\begin{array}{c}  
\a_i \\
\b_j
\end{array}\right)$.
It was shown \cite{Seiberg:1986by}
that modular invariance and unitarity imply that these coefficients
for higher genus surfaces factorise into the form
$$ c \left(\begin{array}{c}  
\a_1,..\a_g \\
\b_1,..\b_g
\end{array}\right)= c \left(\begin{array}{c}  
\a_1\\
\b_1
\end{array}\right) c \left(\begin{array}{c}  
\a_2\\
\b_2
\end{array}\right)... c \left(\begin{array}{c}  
\a_g\\
\b_g
\end{array}\right).$$
For this reason it is sufficient to consider
only the
one-loop coefficients.

\subsection{Model building rules and physical spectrum}

In the free fermionic framework,
the construction of consistent string 
vacua in four dimensions
 is achieved by 
applying two sets of rules, namely, the constraints
for the boundary condition vectors (we restrict to the case
of rational spin structure \cite{Antoniadis:1987wp}) and the rules for
the one-loop phases.

A set of consistent boundary condition vectors
 form an additive group \[ \Xi \sim Z_{N_1}\otimes...\otimes Z_{N_k},\]
generated by the basis 
$B=\{b_1,..b_k\}$, where each $b_i$ is in the form of eq.(\ref{bv}).

This basis has to satisfy the following conditions
\begin{itemize}
\item$ \sum m_i b_i = 0 \Longleftrightarrow m_i = 0$({mod $N_i$}), 
\,$\forall i ,$
\item$ N_{ij} b_i\cdot  b_j=0 \,\text{mod 4},$
\item $N_i b_i \cdot b_i = 0 \,\text{mod 8}, $
\item $  b_1 =  1, $
\item the number of periodic real fermions must be even in each $b_i$,
\end{itemize}
where $N_i$ is the smallest integer for which $N_ib_i=0$(mod2) and
 $N_{ij}$ is the least common multiplier between $N_i$ and $N_j$.
The inner Lorentz product is defined by 
\[ b_i\cdot  b
_j = \left\{\frac{1}{2}\sum_{real\,\, left} 
+ \sum_{complex\,\, left} -\,\, \frac{1}{2}\sum_{real\,\, right}-
\sum_{complex\,\, right}  \right\}b_i(f)b_j(f)\,\,.\]
For a consistent basis $B$ there are several different modular invariant
choices of phases, each one leading to a consistent string theory.
The phases under consideration have to satisfy the requirements,
which provide the 
second group of constraints below 
\begin{itemize}
\item $c \left(\begin{array}{c}  
 b_i \\
 b_j
\end{array}\right)=\delta_{ b_i} e^{\frac{2\pi i n_i}{N_j}}=
\delta_{ b_j}e^{\frac{2\pi i m_i}{N_i}}e^{\frac{i \pi  b_i\cdot  b_j}{2}},$ 
\item $c \left(\begin{array}{c}  
 b_i \\
 b_i
\end{array}\right)=-\,e^{\frac{i\pi  b_i\cdot  b_i}{4}}c \left(\begin{array}{c}  
 b_i \\
 1
\end{array}\right) , $
\item $c \left(\begin{array}{c}  
 b_i \\
 b_j
\end{array}\right)=e^{\frac{i\pi  b_i\cdot  b_j }{2}} c ^* \left(\begin{array}{c}  
 b_j \\
 b_i
\end{array}\right) , $
\item $c \left(\begin{array}{c}  
 b_i \\
 b_j+ b_k
\end{array}\right)=\delta_{ b_i}c \left(\begin{array}{c}  
 b_i \\
 b_j
\end{array}\right) c\left(\begin{array}{c}  
 b_i \\
 b_k
\end{array}\right), $
\end{itemize}
where $1 < n_i < N_j$ and $1 < m_i < N_i$.
Moreover, there is some freedom for the phase $c \left(\begin{array}{c}  
 b_1 \\
 b_1
\end{array}\right)=\pm e^{\frac{i\pi  b_1\cdot  b_1}{4}}$, 
while by convention $c \left(\begin{array}{c}  
0 \\
0
\end{array}\right)=1$ and  $c \left(\begin{array}{c}  
\a \\
0
\end{array}\right)=\delta_{\a}$, condition which assures the presence
of the graviton in the spectrum.

If we indicate by\footnote{The notation can seem confusing
since we indicate by $\a$ a generic boundary condition
vector and at the same time the generic sector
in the Hilbert space. We assure that from the context it is always
clear to understand which quantity we are referring to.}$\a$ a generic sector in $\Xi$,
the corresponding Hilbert space $H_{\a}$ contributes to the
partition function of the model.
We adopt the notation $\a=\{\a_L|\a_R\}$ to separate
the left and the right phases. 
The states
in $H_{\a}$ have to satisfy the Virasoro conditions and the
level matching condition, that, in our
formulation, appear as
\beq
M_L^2=-\frac{1}{2}+\frac{\alpha_L\cdot \alpha_L}{8}+N_L=
-1+\frac{\alpha_R\cdot \alpha_R}{8}+N_R=M^2_R\,\,,
\label{si}
\eeq
where $N_L$ and $N_R$ are respectively
 the total left and the total right oscillator number
acting on the vacuum $|0\!>_{\a}$.
The frequencies are given respectively for
a fermion $f$ and its conjugate $f^*$ by
\[\nu_{f}=\frac{1+\alpha(f)}{2},\,\,\,\,\,\nu_{f^*}=\frac{1-\alpha(f)}{2}
\,\,.\]
The physical states contributing to the partition
function are those satisfying the GSO conditions
\beq
e^{i\pi b_i \cdot F_{\a}} |s>_{\a}=\delta_{\a}\,\, c
\left(\begin{array}{c}  
\a \\
b_i
\end{array}\right)^*
|s>_{\a} ,
\eeq
where $|s>_{\a} $ is a generic state in the sector $\alpha$,
given by bosonic and fermionic oscillators acting on the vacuum.
The operator $(b_i\cdot F_{\a})$ is given by
\beq
b_i\cdot F_{\a}=\Big\{\sum_{left} - \sum_{right} \Big\}b_i(f)F_{\a}(f) ,
\eeq
where $F$ is the fermion number 
operator. 
$F$ gets the following values
\[ F(f) = \left\{ \begin{array}
{r@{\quad \rightarrow \quad}l}
1 & \text{for $f$} \\
-1 & \text{for $f*$} .
\end{array} \right.\]
If the sector $\a$ contains periodic fermions, 
then the vacuum is degenerate
and transforms in the representation of a $SO(2n)$ Clifford algebra.
Hence, if $f$ is such a periodic fermion, it will
be indicated as $|\pm>$ and $F$ assumes the value below 
\[ F(f) = \left\{ \begin{array}
{r@{\quad \rightarrow \quad}l}
0 & \text{for}\,\, |+> \\
-1 & \text{for}\,\, |-> .
\end{array} \right.\]
The $U(1)$ charges for the physical states correspond 
to the currents $f^*f$ and are calculated by the following expression
\[ Q(f)= \frac{1}{2}\alpha(f) + F(f) .\]

\subsection{Construction of semi-realistic models}
\label{semisemi}

The construction of semi-realistic free fermionic models is related
to a particular choice of boundary condition basis vectors and
the general procedure of the construction is based 
on two principal steps.
The first stage is considering the NAHE 
(Nanopoulos-Antoniadis-Hagelin-Ellis) 
set \cite{Faraggi:1990ac,Faraggi:1992fa,Ferrara:1987jr} of boundary 
condition basis vectors $B=\{1,S,b_1,b_2,b_3 \}$, which 
corresponds to ${\mathbb Z}_2 \times {\mathbb Z}_2$ compactification
with the standard embedding of the 
gauge connection \cite{Faraggi:1993pr,Ellis:1997ec}.
The basis $B$ is explicitly given below
\ba
1 &=& \{\psi^{1,2}, \chi^{1,..6}y^{1,..6},w^{1,..6}
|\bar{y}^{3,..6},\bar w^{1,..6},\bar\psi^{1,..5},
\bar\eta^{1,2,3},\bar\phi^{1,..8} \},\nonumber\\
S &=& \{\psi^{1,2}, \chi^{1,..6} \},\nonumber\\
b_1 &=& \{\psi^{1,2}, \chi^{1,2},y^{3,..6}|\bar{y}^{3,..6},
\bar\psi^{1,..5},\bar\eta^1 \},\nonumber\\
b_2 &=& \{\psi^{1,2}, \chi^{3,4},y^{1,2},\omega^{5,6}|\bar{y}^{1,2},
\bar \omega^{5,6},
\bar\psi^{1,..5},\bar\eta^2 \},\nonumber\\
b_3 &=& \{\psi^{1,2}, \chi^{5,6},\omega^{1,..4}|\bar{\omega}^{1,..4},
\bar\psi^{1,..5},\bar\eta^3 \},
\ea
where the notation means that only periodic fermions 
are listed in the vectors. The left-moving internal coordinates
are fermionised by the relation $e^{iX^i}=1/\sqrt{2}
(y^i+iw^i)$, as explained in section \ref{bosonization} and a similar prescription holds for the right-moving internal
coordinates. 
The superpartners of the left-moving bosons
are indicated by $\chi^i$.
The extra 16 degrees of freedom $\bar\psi^{1,..5},
\bar\eta^{1,2,3},\bar\phi^{1,..8}$ are complex fermions.
The GSO one-loop phases for the NAHE set are given below 
\[c \left(\begin{array}{c}  
 b_i \\
 b_j
\end{array}\right)=-1 ,\quad c \left(\begin{array}{c}  
 1 \\
 S
\end{array}\right)= 1 , \quad c \left(\begin{array}{c}  
 b_i \\
 1,S
\end{array}\right)=-1 .
\]
The gauge group induced by the NAHE set is 
$SO(10)\times SO(6)^3\times E_8$ and $N=1$
supersymmetry. The spacetime vector bosons generating
the symmetry group arise in the Neveu-Schwarz sector and in 
the sector $\xi_2=1+b_1+b_2+b_3$. In particular,
the $\bar\psi^{1,..5}$ are responsible for the $SO(10)$
symmetry, the $\bar\phi^{1,..8}$ generate the hidden $E_8$
and the internal fermions  $\{{\bar y}^{3,\cdots,6},{\bar\eta}^1\}$,
$\{{\bar y}^1,{\bar
y}^2,{\bar\omega}^5,{\bar\omega}^6,{\bar\eta}^2\}$,
$\{{\bar\omega}^{1,\cdots,4},{\bar\eta}^3\}$ generate the three
horizontal ${\rm SO}(6)$ symmetries.
In the untwisted sector we note the presence
of states in the ${\bf{10}}$ vectorial representation
of $SO(10)$, that represent the best
 candidates for the Higgs doublets.
The three twisted sectors 
$b_1, b_2$ and $b_3$ produce 48 multiplets in the ${\bf{16}}$
representation of $SO(10)$, which carry $SO(6)^3$ charges
but are singlets under the hidden gauge group. 

In the second stage of the construction we
consider additional basis vectors (generally indicated
by $\a,\b,\gamma$) which reduce the number
of generations to three and simultaneously break 
the four dimensional
gauge group. This breaking is implemented by the assignment
of boundary conditions, in the new vectors,
 corresponding to the generators of
the subgroup considered. For instance, the breaking of 
$SO(10)$ is due to the 
boundary conditions of $\bar\psi^{1,..5}$ in $\a,\b,\gamma$,
which can provide $SU(5)\times U(1)$ \cite{Antoniadis:1989zy},
 $SO(6)\times SO(4)$ \cite{Antoniadis:1990hb}, $SU(3)\times SU(2)\times U(1)^2$ 
gauge groups 
\cite{Faraggi:1989ka,Faraggi:1991jr,Faraggi:1992fa,Cleaver:1998saa}. 
Further attempts in the construction of
realistic models can be found in \cite{Cleaver:2000ds,Cleaver:2001fd}.
The $SO(6)^3$ symmetries are also broken to
flavour $U(1)$ symmetries.
The worldsheet currents $\eta^i\bar\eta^i$, $i=1,2,3$, 
produce $U(1)$ charges in the visible sector and further 
$U(1)^n$ symmetries arise by the pairing of real fermions among the 
right internal sector. If a left moving real fermion
is paired with a right real fermion then 
the right gauge group has rank reduced by one.
The pairing of the left and right movers is a key point
in the phenomenology of free fermionic models, for example 
it is strictly related to the reduction of the untwisted
Higgs states, as we will discuss widely in the following.

The correspondence of the free fermionic models
with the orbifold construction is illustrated
by extending the NAHE set, $\{ 1,S,b_1,b_2,b_3\}$, by at least 
one additional boundary condition basis vector \cite{Faraggi:1993pr,Faraggi:1995yd,Donagi:2004ht}
\beq
\xi_1 = \{\bar\psi^{1,\cdots,5},   
{\bar\eta^{1,2,3}}\}.
\label{vectorx}
\eeq
With a suitable choice of the GSO projection coefficients the   
model possesses an ${\rm SO}(4)^3\times {\rm E}_6\times {\rm U}(1)^2
\times {\rm E}_8$ gauge group
and ${ N}=1$ space-time supersymmetry. The matter fields
include 24 generations in the 27 representation of
${\rm E}_6$, eight from each of the sectors $b_1\oplus b_1+\xi_1$,
$b_2\oplus b_2+\xi_1$ and $b_3\oplus b_3+\xi_1$.
Three additional ${\bf{27}}$ and ${\bf{\overline{27}}}$ pairs are obtained
from the Neveu-Schwarz $\oplus~\xi_1$ sector.

To construct the model in the orbifold formulation one starts
with the compactification on a torus with nontrivial background
fields \cite{Narain:1985jj,Narain:1986am}.
The subset of basis vectors
\beq
\{ 1,S,\xi_1,\xi_2\},
\label{neq4set}
\eeq
where $\xi_2 = \{\bar\phi^{1,\cdots,8}\}$,
generates a toroidally-compactified model with ${ N}=4$ spacetime
supersymmetry and ${\rm SO}(12)\times {\rm E}_8\times {\rm E}_8$ gauge 
group.
The same model is obtained in the geometric (bosonic) language
by tuning the background fields to the values corresponding to
the SO(12) lattice. The
metric of the six-dimensional compactified
manifold is then the Cartan matrix of SO(12),
while the antisymmetric tensor is given by
\begin{equation}
b_{ij}=\begin{cases}
g_{ij}&;\ i>j,\cr
0&;\ i=j,\cr
-g_{ij}&;\ i<j.\cr\end{cases}
\label{bso12}
\end{equation}
When all the radii of the six-dimensional compactified
manifold are fixed at $R_I=\sqrt2$, it is seen that the
left- and right-moving momenta  
\beq
P^I_{R,L}=[m_i-{\frac{1}{2}}(B_{ij}{\pm}G_{ij})n_j]{e_i^I}^*
\label{lrmomenta}
\eeq
reproduce the massless root vectors in the lattice of
SO(12). Here $e^i=\{e_i^I\}$ are six linearly-independent
vielbeins normalised so that $(e_i)^2=2$.
The ${e_i^I}^*$ are dual to the $e_i$, with
$e_i^*\cdot e_j=\delta_{ij}$.

Adding the two basis vectors $b_1$ and $b_2$ to the set
(\ref{neq4set}) corresponds to the ${\mathbb Z}_2\times {\mathbb Z}_2$
orbifold model with standard embedding.
Starting from the $N=4$ model with ${\rm SO}(12)\times
{\rm E}_8\times {\rm E}_8$
symmetry, and applying the ${\mathbb Z}_2\times {\mathbb Z}_2$
twist on the
internal coordinates, reproduces
the spectrum of the free-fermion model
with the six-dimensional basis set
$\{ 1,S,\xi_1,\xi_2,b_1,b_2\}$ 
\cite{Faraggi:1993pr,Faraggi:1995yd,Donagi:2004ht}.
The Euler characteristic of this model is 48 with $h_{11}=27$ and
$h_{21}=3$.

It is noted that the effect of the additional basis vector $\xi_1$ of eq.\
(\ref{vectorx}) is to separate the gauge degrees of freedom, spanned by
the world-sheet fermions $\{{\bar\psi}^{1,\cdots,5},
{\bar\eta}^{1,2,3},{\bar\phi}^{1,\cdots,8}\}$,
from the internal compactified degrees of freedom $\{y,\omega\vert
{\bar y},{\bar\omega}\}^{1,\cdots,6}$.
In the realistic free fermionic
models this is achieved by the vector $2\gamma$
 \cite{Faraggi:1993pr,Faraggi:1995yd,Donagi:2004ht}, with
\beq
2\gamma=\{\bar\psi^{1,\cdots,5},
{\bar\eta}^{1,2,3}, {\bar\phi}^{1,\cdots,4}\},
\label{vector2gamma}
\eeq
which breaks the ${\rm E}_8\times {\rm E}_8$ symmetry to ${\rm 
SO}(16)\times
{\rm SO}(16)$.
The ${\mathbb Z}_2\times {\mathbb Z}_2$ twist induced by $b_1$ and $b_2$
breaks the gauge symmetry to
${\rm SO}(4)^3\times {\rm SO}(10)\times {\rm U}(1)^3\times {\rm SO}(16)$.
The orbifold still yields a model with 24 generations,
eight from each twisted sector,
but now the generations are in the chiral 16 representation
of SO(10), rather than in the ${\bf{27}}$ of ${\rm E}_6$. The same model can
be realised \cite{Faraggi:2002qh} with the set
$\{ 1,S,\xi_1,\xi_2,b_1,b_2\}$,
by projecting out the ${\bf{16}}\oplus{\bf{{\overline{16}}}}$
from the $\xi_1$-sector taking
\beq\label{changec}
  c \left(\begin{array}{c}  
\xi_1\\
\xi_2
\end{array}\right) \rightarrow -  c \left(\begin{array}{c}  
\xi_1\\
\xi_2
\end{array}\right).
\eeq
This choice also projects out the massless vector bosons in the
128 of SO(16) in the hidden-sector ${\rm E}_8$ gauge group, thereby
breaking the ${\rm E}_6\times {\rm E}_8$ symmetry to
${\rm SO}(10)\times {\rm U}(1)\times {\rm SO}(16)$.
We can define two ${ N}=4$ models generated by the set
(\ref{neq4set}), ${ Z}_+$ and ${ Z}_-$, depending on the sign
in eq.\ (\ref{changec}). The first, say ${ Z}_+$,
produces the ${\rm E}_8\times {\rm E}_8$ model, whereas the second, say
${ Z}_-$, produces the ${\rm SO}(16)\times {\rm SO}(16)$ model.
However, the ${\mathbb Z}_2\times
{\mathbb Z}_2$
twist acts identically in the two models, and their physical 
characteristics
differ only due to the discrete torsion eq.\ (\ref{changec}).

The free fermionic formalism provides useful means to classify and
analyse ${\mathbb Z}_2\times {\mathbb Z}_2$ heterotic orbifolds at
special points in the  
moduli space. The drawbacks of this approach is that the geometric
view of the underlying compactifications is lost. On the other 
hand, the geometric picture may be instrumental for examining other
questions of interest, such as the dynamical stabilisation of the moduli
fields and the moduli dependence of the Yukawa couplings.
In chapter 4 we will analyse ${\mathbb Z}_2\times {\mathbb
Z}_2$ orbifolds  
on non-factorisable toroidal manifolds.

Once we extract the massless spectrum of a particular free fermion model,
the next step is the analysis of its superpotential. We postpone
the explanation of this topic since it will be treated in the next sections.
Further details concerning the construction of free fermionic models
carried on step by step can be found in \cite{Nooij:2004cz}.

\newpage

\section{Minimal Standard Heterotic String Models}



After providing the main tools on the construction of the theory,
we would like to revisit some of the properties
of semi--realistic Standard Model--like free fermionic models.
One of their remarkable successes 
has been the fact that they can accommodate
the right top quark mass
\cite{Faraggi:1995bv,Faraggi:1996pa,Faraggi:1991mu,Faraggi:1991be}.
The models offered
an explanation why only the top quark mass is characterised by the
electroweak scale, whereas the masses of the lighter quarks and
leptons are suppressed 
\cite{Faraggi:1991jr,Faraggi:1992rd,Faraggi:1992yz,Faraggi:1993su}. The reason is that only the top quark
Yukawa coupling is obtained at the cubic level of the superpotential,
whereas the Yukawa couplings of the
lighter quarks and leptons are obtained from nonrenormalizable terms
which are suppressed relative to the leading order term. 
As we explained before, the three generations
arise from the three twisted sectors,
 whereas the Higgs doublets, to which
they couple in leading order, arise from the untwisted sector.
 At leading
order each twisted generation couples to a separate pair of untwisted
Higgs doublets. Analysis of supersymmetric flat directions implied
that at low energies only one pair of Higgs doublets remains light and
other Higgs doublets obtain heavy mass from VEVs of Standard Model singlet
fields. Hence, in the low energy effective field theory, only the
coupling of the twisted generation that couples to the light Higgs
remains at leading order. The consequence is that only the top quark
mass is obtained at leading order, whereas the masses of the remaining
quarks and leptons are obtained at subleading orders. Evolution of the
calculated Yukawa couplings from the string to electroweak scale
 then yields a prediction for the top quark mass.
The analysis of the top quark mass therefore relies on the analysis
of supersymmetric flat directions and the decoupling of the additional
untwisted electroweak Higgs doublets, that couple to the twisted generations
at leading order. In the examples presented 
in the following an alternative
construction is given, where
only one pair of untwisted Higgs doublets remains in the massless
spectrum after the application of the Generalised GSO (GGSO) projections.
Therefore, the massless string spectrum contains a single
electroweak Higgs doublet pair, without relying on analysis of supersymmetric
flat directions in the effective low energy field theory.
Although the Higgs reduction is obtained by applying the new procedure,
 the flat direction analysis is still necessary to investigate
the supersymmetric properties of the model.
The existence of an ``anomalous'' $U(1)$
symmetry is a common feature
of free fermionic models \cite{Cleaver:1997rk}.
 The anomalous $U(1)_A$ is broken by the
Green--Schwarz--Dine--Seiberg--Witten mechanism \cite{Dine:1987xk}
in which a potentially large Fayet--Iliopoulos $D$--term
$\xi$ is generated by the VEV of the dilaton field.
Such a $D$--term would, in general, break supersymmetry, unless
there is a direction $\hat\phi=\sum\alpha_i\phi_i$ in the scalar
potential for which $\sum Q_A^i\vert\alpha_i\vert^2$ is of opposite sign to
$\xi$ and that
is $D$--flat with respect to all the non--anomalous gauge symmetries,
as well as $F$--flat. 
If such a direction
exists, it will acquire a VEV, cancelling the Fayet--Iliopoulos
$\xi$--term, restoring supersymmetry and stabilising the vacuum.
The set of $D$- and $F$-flat constraints is given by
\beqn
&& \langle D_A\rangle=\langle D_\alpha\rangle= 0~;\quad
\langle F_i\equiv
{{\partial W}\over{\partial\eta_i}}\rangle=0~~;\label{dterms}\\
\nonumber\\
&& D_A=\left[K_A+
\sum Q_A^k\vert\chi_k\vert^2+\xi\right]~~;\label{da}\\
&& D_\alpha=\left[K_\alpha+
\sum Q_\alpha^k\vert\chi_k\vert^2\right]~,~\alpha\ne A~~;\label{dalpha}\\
&& \xi={{g^2({\rm Tr} Q_A)}\over{192\pi^2}}M_{\rm Pl}^2~~;
\label{dxi}
\eeqn
where $\chi_k$ are the fields which acquire VEVs of order
$\sqrt\xi$, while the $K$--terms contain fields
like squarks, sleptons and Higgs bosons whose
VEVs vanish at this scale. $Q_A^k$ and $Q_\alpha^k$ denote the anomalous
and non--anomalous charges,
and $M_{\rm Pl}\approx2\times 10^{18}$ GeV denotes the
reduced Planck mass. The solution ({\it i.e.}\  the choice of fields
with non--vanishing VEVs) to the set of
eqs.(\ref{dterms})--(\ref{dalpha}),
though nontrivial, is not unique. Therefore in a typical model there exist
a moduli space of solutions to the $F$ and $D$ flatness constraints,
which are supersymmetric and degenerate in energy \cite{Font:1988tp}. Much of
the study of the superstring models phenomenology 
(as well as non--string supersymmetric models)
involves the analysis and classification of these flat directions.
The methods for this analysis in string models have been systematised
in \cite{Buccella:1982nx,Cleaver:1997jb,Cleaver:2001ab,Cleaver:1997cr,Cleaver:1997rk}.

In general it has been assumed in the past that in a given string model
there should exist a supersymmetric solution to the $F$ and $D$
flatness constraints. The simpler type of solutions utilise only
fields that are singlets of all the non--Abelian groups in a given
model (type I solutions). More involved solutions (type II solutions),
that utilise also non--abelian fields, have also been considered
\cite{Cleaver:1997rk}, as well as inclusion of non--abelian fields
in systematic methods of analysis \cite{Cleaver:1997rk}.
The general expectation that a given model admits a supersymmetric 
solution arises from analysis of supersymmetric point quantum field theories.
In these cases it is known that if supersymmetry is preserved at the 
classical level, 
then there exist index theorems that forbid supersymmetry breaking 
at the perturbative quantum level \cite{Witten:1982df}.
Therefore in point quantum field theories
supersymmetry breaking
may only be induced by non--perturbative effects \cite{Intriligator:2007cp}.

In the model of table \ref{secondexample}
the reduction of the Higgs states is obtained 
by imposing asymmetric boundary conditions in a boundary condition
basis vector that does not break the $SO(10)$ symmetry. 
Another consequence of the Higgs reduction mechanism 
is the simultaneous projection of untwisted $SO(10)$ singlet fields,
provoking a vast reduction of the moduli space of supersymmetric flat
solutions. 
The model under investigation does not 
contain supersymmetric flat directions that do not break some
of the Standard Model symmetries.
Thus, by continuing the 
search of semirealistic models 
with reduced Higgs spectrum
we are lead to the second model proposed in table \ref{stringmodel}, where 
the Higgs reduction mechanism
 utilises boundary 
conditions that are both symmetric and asymmetric
in the basis vectors that break $SO(10)$ to $SO(6)\times SO(4)$,
with respect to two 
of the twisted sectors of the ${\mathbb Z}_2\times {\mathbb Z}_2$ orbifold.
The consequence is that two of the untwisted Higgs multiplets,
associated with two of the twisted sectors, are projected entirely
from the massless spectrum. As a result, the string model
contains a single pair of untwisted electroweak Higgs doublets.

In the process of seeking supersymmetric flat direction, 
we arrive to the unexpected
conclusion that the model may not contain any 
supersymmetric flat directions at all.
In the least, this model 
appears to have no $D$-flat directions that can
be proved
to be $F$-flat to all order, other than through order-by-order analysis. 
That is, there does not appear to be any $D$-flat directions with {\it 
stringent} $F$-flatness (as defined in \cite{Cleaver:2000ey,Cleaver:2007ek}).
In the analysis of the flat directions we include all the fields in the
string model, {\it i.e.} Standard Model singlet states as well as
Standard Model charged states. The model therefore does not contain a
$D$--flat direction that is also stringently $F$--flat to all order
of non--renormalizable terms. 
The model may of course still admit non-stringent flat directions that rely 
on cancellations between superpotential terms. However, past experience
suggests that non--stringent flat directions
can only hold order by order, and are not maintained to all orders
\cite{Cleaver:2000ds}.
We therefore speculate that in this case supersymmetry is not exact, 
but is in general broken at some order.
If this finding remains true after the entire parameter 
space of possible all-order non--stringent flat
directions has been examined,
we must ask what are the implications. 
If a model without all-order
$F$-flatness were to be found, then supersymmetry would remain broken by the
Fayet--Iliopoulos term at a finite order, which is generated at the one--loop 
level in string perturbation theory, rather than be cancelled by a $D$-flat 
direction with anomalous charge. If so, then this would imply, although 
supersymmetry is unbroken at the classical level and the string spectrum is 
Bose--Fermi degenerate, that supersymmetry may be broken at the perturbative
quantum level. Nevertheless, since the spectrum is Bose--Fermi degenerate,
the one--loop cosmological constant still vanishes. 
The details of this model are given in section \ref{stringmodelsection}.

Below we provide the details of the Yukawa mechanism and
the Higgs doublet-triplet splitting which are realised in the examples
proposed in the next sections.


\subsection{Yukawa Selection Mechanism}\label{ysm}

At the cubic level of the superpotential
the boundary condition
basis vectors fix the Yukawa couplings for the quarks
and leptons \cite{Faraggi:1991be}.
These Yukawa couplings are fixed by the vector $\gamma$
which breaks the $SO(10)$ symmetry to $SU(5)\times U(1)$.
Each sector
$b_i$ gives rise to an up--like or down--like cubic level Yukawa coupling.
We can define three quantities $\Delta_i$, $i=1,2,3$, in the vector $\gamma$,
which measures the difference
of the left-- and right--moving
boundary conditions assigned to the internal fermions from the set
$\{y,w\vert{\bar y},{\bar\omega}\}$ and which are periodic in the vector
$b_i$,
\begin{equation}
\Delta_i=\vert\gamma_L({\rm internal})-
\gamma_R({\rm internal})\vert=0,1~~(i=1,2,3)
\label{udyc}.
\end{equation}
If $\Delta_i=0$ then the sector $b_i$ gives rise to a
down--like Yukawa coupling while the
up--type Yukawa coupling vanishes. The opposite occurs if $\Delta_i=1$.
In models that produce $\Delta_i=1$ for $i=1,2,3$
the down--quark type cubic--level Yukawa couplings vanish
and the models produce only up--quark type Yukawa couplings
at the cubic level of the superpotential. Models with these characteristics
were presented in refs. \cite{Faraggi:1992fa,Faraggi:1991be}.

\subsection{Higgs Doublet--Triplet Splitting}\label{hdts}

The Higgs doublet--triplet splitting operates as follows \cite{Faraggi:2001ry,Faraggi:1994cv}.
The Neveu--Schwarz sector gives rise to three fields in the
10 representation of $SO(10)$.  These contain the  Higgs electroweak
doublets and colour triplets when breaking
the gauge group to the SM symmetry. Each of those is charged with respect to one
of the horizontal $U(1)$ symmetries $U(1)_{1,2,3}$ generated by 
$\bar\eta^{1}$, $\bar\eta^{2}$ and $\bar\eta^{3}$.  Each one of these
multiplets is associated, by the horizontal symmetries, with one of the
twisted sectors, $b_1$, $b_2$ and $b_3$. The doublet--triplet
splitting results from the boundary condition basis vectors which break
the $SO(10)$ symmetry to $SO(6)\times SO(4)$. We can define a quantity
$\Delta_i$ in these basis vectors which measures the difference between the
boundary conditions assigned to the internal fermions from the set
$\{y,\omega\vert{\bar y},{\bar\omega}\}$ and which are periodic in the vector
$b_i$,
\begin{equation}
\Delta_i=\vert\alpha_L({\rm internal})-
\alpha_R({\rm internal})\vert=0,1~~(i=1,2,3)
\label{dts}.
\end{equation}
If $\Delta_i=0$ then the Higgs triplets, $D_i$ and ${\bar D}_i$,
remain in the massless spectrum while the Higgs doublets, $h_i$ and ${\bar
h}_i$ are projected out
and the opposite occurs for $\Delta_i=1$.
The rule in eq.(\ref{dts}) is a generic rule that operates in NAHE--based
free fermionic models.

Another relevant question with regard to the Higgs doublet--triplet
splitting mechanism is whether it is possible to construct models in which
both the Higgs colour triplets and electroweak doublets associated
to a given twisted sector $b_j$ from the
Neveu--Schwarz sector are projected out by the GSO projections.
This is a viable possibility as we can choose for example
$$\Delta_j^{(\alpha)}=1 ~{\rm and}~ \Delta_j^{(\beta)}=0,$$
where $\Delta^{(\alpha,\beta)}$ are the projections due
to the basis vectors $\alpha$ and $\beta$ respectively.
This is a relevant question as the number of Higgs representations,
which generically appear in the massless spectrum,
is larger than what is allowed by the low energy phenomenology.
Attempts to construct such models were discussed in ref. \cite{Faraggi:1997dc}.
In section \ref{higgsreduced} we present 
three generation models with reduced untwisted Higgs spectrum,
without resorting to analysis of supersymmetric flat directions.

\section{Models with reduced untwisted Higgs spectrum}\label{higgsreduced}

As an illustration of the Higgs reduction mechanism we consider the
model in table \ref{firstexample}.
\beqn
 &\begin{tabular}{c|c|ccc|c|ccc|c}
 ~ & $\psi^\mu$ & $\chi^{12}$ & $\chi^{34}$ & $\chi^{56}$ &
        $\bar{\psi}^{1,...,5} $ &
        $\bar{\eta}^1 $&
        $\bar{\eta}^2 $&
        $\bar{\eta}^3 $&
        $\bar{\phi}^{1,...,8} $ \\
\hline
\hline
  ${\alpha}$  &  1 & 1&0&0 & 1~1~1~0~0 & 0 & 1 & 0 & 0~1~1~0~0~0~0~0 \\
  ${\beta}$   &  1 & 0&1&0 & 1~1~1~0~0 & 1 & 1 & 1 & 0~1~1~0~0~0~0~0 \\
  ${\gamma}$  &  1 & 0&0&1 &
		${1\over2}$~${1\over2}$~${1\over2}$~${1\over2}$~${1\over2}$
	      & ${1\over2}$ & ${1\over2}$ & ${1\over2}$ &
                ${1\over2}$~0~0~0~${1\over2}$~${1\over2}$~${1\over2}$~0 \\
\end{tabular}
   \nonumber\\
   ~  &  ~ \nonumber\\
   ~  &  ~ \nonumber\\
     &\begin{tabular}{c|c|c|c}
 ~&   $y^3{y}^6$
      $y^4{\bar y}^4$
      $y^5{\bar y}^5$
      ${\bar y}^3{\bar y}^6$
  &   $y^1{\omega}^6$
      $y^2{\bar y}^2$
      $\omega^5{\bar\omega}^5$
      ${\bar y}^1{\bar\omega}^6$
  &   $\omega^1{\omega}^3$
      $\omega^2{\bar\omega}^2$
      $\omega^4{\bar\omega}^4$
      ${\bar\omega}^1{\bar\omega}^3$ \\
\hline
\hline
$\alpha$ & 1 ~~~ 0 ~~~ 0 ~~~ 0  & 0 ~~~ 0 ~~~ 1 ~~~ 1  & 0 ~~~ 0 ~~~ 1 ~~~ 0 \\
$\beta$  & 0 ~~~ 0 ~~~ 1 ~~~ 1  & 1 ~~~ 0 ~~~ 0 ~~~ 1  & 0 ~~~ 1 ~~~ 0 ~~~ 1 \\
$\gamma$ & 0 ~~~ 1 ~~~ 0 ~~~ 1  & 0 ~~~ 1 ~~~ 0 ~~~ 0  & 1 ~~~ 0 ~~~ 0 ~~~ 1 \\
\end{tabular}
\label{firstexample}
\eeqn
with the choice of generalised GSO coefficients:
\[c \left(\begin{array}{c}  
\a,\b \\
\a
\end{array}\right)= c \left(\begin{array}{c}  
\b,\gamma \\
\b
\end{array}\right)= -c \left(\begin{array}{c}  
\gamma \\
1,\a
\end{array}\right)= 
c \left(\begin{array}{c}  
\a \\
b_3
\end{array}\right)=\]\[ c \left(\begin{array}{c}  
\gamma \\
b_1
\end{array}\right)= - c \left(\begin{array}{c}  
\b \\
b_j
\end{array}\right)=-
 c \left(\begin{array}{c}  
\a \\
b_1,b_2
\end{array}\right)= - c \left(\begin{array}{c}  
\gamma \\
b_2, b_3
\end{array}\right)=1
\]
(j=1,2,3), with the others specified by modular invariance and spacetime
supersymmetry. As noted from the table, in this model the boundary conditions
with respect to $b_2$ and $b_3$ in the basis vector $\alpha$
are asymmetric and symmetric, respectively, while the opposite occurs for the basis vector $\beta$.
At the same time, the boundary conditions with respect to the sector
$b_1$ are asymmetric in both $\alpha$ and $\beta$. Therefore,
in this model
$\Delta_1^{(\alpha)}=\Delta_1^{(\beta)}=1$;
$~\Delta_2^{(\alpha)}=1,~\Delta_2^{(\beta)}=0$ and
$\Delta_3^{(\alpha)}=0,~\Delta_3^{(\beta)}=1$.
Consequently, irrespective of the choice of the generalised GSO
projection coefficients, both the Higgs colour triplets and electroweak
doublets associated with $b_2$ and $b_3$ are projected out by the
GSO projections, whereas the electroweak Higgs doublets that are associated
with the sector $b_1$ remain in the spectrum.
However, the sector $\alpha$ produces chiral fractionally
charged
exotics, and is therefore not viable. We also note that in this model the
non--vanishing cubic level Yukawa couplings produce a down--quark type mass
term, and not a potential top--quark mass term.

An alternative model is presented in table \ref{secondexample}.
\beqn
 &\begin{tabular}{c|c|ccc|c|ccc|c}
 ~ & $\psi^\mu$ & $\chi^{12}$ & $\chi^{34}$ & $\chi^{56}$ &
        $\bar{\psi}^{1,...,5} $ &
        $\bar{\eta}^1 $&
        $\bar{\eta}^2 $&
        $\bar{\eta}^3 $&
        $\bar{\phi}^{1,...,8} $ \\
\hline
\hline
  $b_4$  &  1 & 1&0&0 & 1~1~1~1~1 & 0 & 1 & 0 & 1~1~1~1~0~0~0~0 \\
  ${\beta}$   &  1 & 0&1&0 & 1~1~1~0~0 & 1 & 1 & 1 & 0~0~0~0~1~1~0~0 \\
  ${\gamma}$  &  1 & 0&0&1 &
		${1\over2}$~${1\over2}$~${1\over2}$~${1\over2}$~${1\over2}$
	      & ${1\over2}$ & ${1\over2}$ & ${1\over2}$ &
                0~0~${1\over2}$~${1\over2}$~0~0~${1\over2}$~${1\over2}$ \\
\end{tabular}
   \nonumber\\
   ~  &  ~ \nonumber\\
   ~  &  ~ \nonumber\\
     &\begin{tabular}{c|c|c|c}
 ~&   $y^3{y}^6$
      $y^4{\bar y}^4$
      $y^5{\bar y}^5$
      ${\bar y}^3{\bar y}^6$
  &   $y^1{\omega}^6$
      $y^2{\bar y}^2$
      $\omega^5{\bar\omega}^5$
      ${\bar y}^1{\bar\omega}^6$
  &   $\omega^1{\omega}^3$
      $\omega^2{\bar\omega}^2$
      $\omega^4{\bar\omega}^4$
      ${\bar\omega}^1{\bar\omega}^3$ \\
\hline
\hline
$b_4$ & 1 ~~~ 0 ~~~ 0 ~~~ 0  & 0 ~~~ 0 ~~~ 1 ~~~ 1  & 0 ~~~ 0 ~~~ 1 ~~~ 0 \\
$\beta$  & 0 ~~~ 0 ~~~ 1 ~~~ 1  & 1 ~~~ 0 ~~~ 0 ~~~ 1  & 0 ~~~ 1 ~~~ 0 ~~~ 1 \\
$\gamma$ & 0 ~~~ 1 ~~~ 0 ~~~ 1  & 0 ~~~ 0 ~~~ 0 ~~~ 1  & 1 ~~~ 1 ~~~ 0 ~~~ 0 \\
\end{tabular}
\label{secondexample}
\eeqn
with the choice of generalised GSO coefficients:
\[c \left(\begin{array}{c}  
b_4 \\
b_4,\b,\gamma
\end{array}\right)= c \left(\begin{array}{c}  
\b \\
\b, \gamma
\end{array}\right)= c \left(\begin{array}{c}  
b_4,\gamma \\
b_j
\end{array}\right)= 
-c \left(\begin{array}{c}  
\gamma \\
1
\end{array}\right)=- c \left(\begin{array}{c}  
\b \\
b_j
\end{array}\right)=1 ,
\]
(j=1,2,3), with the others specified by modular invariance and spacetime
supersymmetry. In this model the basis vector\footnote{We use a different notation here for the boundary condition vector $\a$, which is now called $b_4$,
since in the literature $\a$ breaks the $SO(10)$ gauge group while in this case 
the boundary conditions w.r.t. $\bar\psi^{1,..5}$ leave intact the $SO(10)$ symmetry.} 
$b_4$ preserves the $SO(10)$
symmetry, which is broken by the basis vectors $\beta$ and $\gamma$ to
$SU(3)\times SU(2)\times U(1)^2$. The $b_4$ projection is asymmetric
with respect to the internal fermions that are periodic in
the sectors $b_1$ and $b_2$ and, therefore, projects out the entire
untwisted vectorial representations of $SO(10)$,
that couple to the sectors $b_1$ and $b_2$, irrespective of the
$\beta$ projection. On the other hand, it is symmetric with respect to $b_3$,
while the basis
vector $\beta$, that breaks $SO(10)\rightarrow SO(6)\times SO(4)$,
is asymmetric with respect to $b_3$. Therefore, the Higgs doublets that
couple to $b_3$ remain in the massless spectrum.
We note also that the boundary conditions
in the vector $\gamma$, that breaks $SO(10)\rightarrow SU(5)\times U(1)$, are
asymmetric with respect to the internal fermions that are periodic in the
sector $b_3$. Therefore, this model will select an up--quark type Yukawa
couplings at the cubic level of the superpotential.
The gauge group of this model is generated entirely from the untwisted vector
bosons and there is no gauge symmetry enhancement from additional sectors.
The four dimensional gauge group is
$SU(3)_C\times SU(2)_L\times U(1)_{B-L}\times U(1)_{T_{3_R}}\times
U(1)_{1,\cdots,6}\times SU(2)^H_{1,\cdots,6}\times U(1)^H_{7,8}$.

The spectrum of the model is detailed in the Table $3.a$ in
Appendix B.
The cubic level superpotential, including states from the observable and hidden
sectors, is straightforwardly calculated following the rules given in
\cite{Kalara:1990sq} and reads:
\ba
W&=&N^c_{L_3} L_3\bar{h}+u^c_{L_3} Q_3\bar{h} +
C_+^{-+}D_-\bar{h}+C_-^{+-}D_+h+\nonumber\\
 &+&(\phi_1{\phi_3}'+\phi_1'\phi_3)\phi_2+
(C_+^{-+}C_-^{+-}+C_-^{-+}C_+^{+-})\phi_{3}'\nonumber\\
&+& (D_+D_-+C_+C_-+T_+T_-+D_{+-}^{(6)}D_{-+}^{(6)}+
D_{--}^{(6)}D_{++}^{(6)}){\phi_{3}}\nonumber\\
&+&(D^{(3,4)}_{+-}D^{(3,4)}_{-+}+D_+^{(5)}D_-^{(5)}+
D_{++}^{(3)}D_{--}^{(3)}+D_{+-}^{(3)}D_{-+}^{(3)}){\phi_{1}}
\nonumber\\
&+&A_+A_-\phi_{1}'.\nonumber
\ea

As expected, we obtain a Yukawa coupling for the top quark,
but also couplings of the Higgs with exotic states.
 One can also see that not all the fractionally charged 
\footnote{The hypercharge is defined as $Q_Y=1/3~Q_C+1/2~Q_L$ and
the electric charge is given by $Q_e=T_{3L}+Q_Y$, with $T_{3L}$ the
electroweak isospin.}states in the spectrum appear in the cubic level
superpotential, which means that they remain massless at the trilinear level.
However,
this does not exclude the possibility of giving them masses at higher orders.

\subsection{Flat directions}

In this section we investigate the flat directions of the model 
of table (\ref{secondexample}). The model contains 6 anomalous $U(1)$'s with
\ba &&{\rm Tr} \ Q_1={\rm Tr}\ Q_2=-{\rm Tr}\ Q_3={\rm Tr}\ Q_5=-24,\nonumber\\
&&{\rm Tr}\ Q_4=-{\rm Tr}\ Q_6=12.\ea
The total anomaly can be rotated into a single $U(1)_A$ and the new basis reads
\ba Q_1'&=&Q_1-Q_2,\nonumber\\
Q_2'&=&Q_3+Q_5,\nonumber\\
Q_3'&=&Q_4+Q_6,\nonumber\\
Q_4'&=&Q_1+Q_2+Q_3-Q_5,\nonumber\\
Q_5'&=&Q_1+Q_2-Q_3+Q_5+4(Q_4-Q_6),\nonumber\\
Q_A&=&2(Q_1+Q_2-Q_3+Q_5)-Q_4+Q_6.
\ea
In the following we will call $Q_i'$, i=1,...,5,  simply $Q_i$.

To search for flat directions we use the methodology developed in
\cite{Cleaver:1998gc}.
We start  by constructing a basis of D-flat directions
under $Q_{1...5}$ and then we investigate the existence of D-flat directions
in the anomalous $U(1)_A$. Subsequently we will have to impose D-flatness
under the remaining gauge groups and F-flatness. To generate the basis of
flat directions under $Q_{1...5}$ we start by forming a basis of gauge
invariant monomials under $U(1)_1$, then we use these invariants to
construct a basis of invariant monomials under $U(1)_2$ and so forth.

We include in the analysis only the fields with vanishing
hypercharge and which are singlets under the
Standard Model gauge group.  The $Q_{1...5,A}$ charges of these fields
are detailed in table \ref{flat},
where, following the notation of \cite{Cleaver:1998gc}, we signal by
$^{(')}$($^{('')}$) the presence in the spectrum of a second (third)
field with the same $U(1)_{1...5,A}$ charges and by $\surd$ the presence
of a field with opposite $U(1)_{1...5,A}$ charges.
For instance, the field $\phi$ stands for $\phi_1$, while $\phi'$ stands for
$\phi_3$ and the two fields with opposite charges are $\phi_1'$ and
$\phi_3'$. The fields with opposite charges to $A_+$ and
$A_-$ are $D^{(5)}_-$ and $D^{(5)}_+$, respectively, while the field with
opposite charges to $D_2$ is $D^{(3,4)}_{+-}$  and $\tilde{D}_2''$ stands
for $D^{(3,4)}_{-+}$, in the notation of the Table 3.a in Appendix B.
 We did not include in table \ref{flat} the fields
$\tilde{\phi}_1$, $\phi_2$ and $\tilde{\phi}_3$, which have vanishing
charges. These fields are trivially flat directions in the
$U(1)_{1...5}$, but they are not flat under the anomalous $U(1)$.

For simplicity we rescaled the charges $Q_1$, $Q_3$ and $Q_A$ by a
factor 2 and the charges  $Q_2$, $Q_4$ and $Q_5$ by a factor 4.
The seventh column is given by
\beq
\hat{Q}=\frac{1}{18}(Q_A-Q_5+9\ Q_3)
\eeq
and, as explained in \cite{Cleaver:1998gc}, it will be useful for the
search of flat directions in the anomalous $U(1)$.
\beqn
 &\begin{tabular}{l|rrrrrrr}
 ~ & $\quad Q_1$ & $\quad Q_2$ & $\quad Q_3$ & $\quad Q_4$ &
$\quad Q_5$ & $\quad Q_A$ & $\quad \hat{Q}$\\
\hline
\hline
  $\phi^{(')}~ \surd^{(')}$  & 0 & 4& 0 & -4 & 4 & 4 & 0 \\
 $S_1^{(')},D_1$ & 1 & 2 & -1 & 0 & -12 & -3 & 0\\
$\tilde{S}_1^{(')}, \tilde{D}_1^{(')}$ & 1 & 2 & 1 & 0 & 4 & -5 & 0 \\
$S_2^{(')}, D_2 ~\surd$ & -1 & 4 & 0 & -2 & -2 & -2 & 0\\
$\tilde{S}_2^{(')}, \tilde{D}_2^{(')('')}$ & -1 & 0 & 0 & 2 & -6 & -6 & 0 \\
$S_3^{(')},D_3$ & 0 & 0 & 1 & -4 & -12 &-3 & 1\\
$\tilde{S}_3^{(')}, \tilde{D}_3^{(')}$ & 0 & 0 & -1 & -4 & 4 & -5 & -1\\
$N_1$ & -1 & 0 & -1 & -2 & -10 & -1 & 0\\
$N_2$ & 1 & -2 & 0 & 0 & -4 & -4 & 0\\
$N_3$ & 0 & 2 & -1 & 2 & 6 & -3 & -1\\
$A_+ ~\surd$ & -1 & 0 & 0 & -6 & 2 & 2 & 0\\
$A_- ~\surd$ &1 & 4 & 0 & 2 & 2 &2 & 0\\
$F^{(')}$ &1 & -1 & 2 & -1 &1 &1 & 1\\
$\tilde{F}^{(')}$ & -1 &1 & 0 &1 &15 & -3 & -1\\
$F_1$ & 0 & 3 &1 &1 & 11 & 2 & 0\\
$F_2$ & 0 & -1 & 1 & 5 & 7 & -2 & 0\\
$F_3$ & 0 & 1 &1 & -5 & 9 & 0 & 0\\
$F_4$ & 0 &  -3 & 1 &  -1 & 5 & -4 & 0\\
\end{tabular}
\label{flat}
\eeqn

As a first step we investigate the existence of flat directions involving
vacuum expectation values only for the fields which are singlets under
both the visible and the hidden gauge groups. These fields are
$\phi^{(')}\surd^{(')}$, $S_1^{(')}$,  $\tilde{S}_1^{(')}$, $S_2^{(')}$,
$\tilde{S}_2^{(')}$, $S_3^{(')}$, $\tilde{S}_3^{(')}$, $N_1$, $N_2$ and
$N_3$. Bearing in mind the equivalence in the charges for some fields,
these count as 11 fields and so, given the fact that we have to impose 5
constraints, the basis of flat directions should contain 6 elements. But
a simple Mathematica program can show that it is impossible to incorporate
the fields  $S_1^{(')}$, $S_3^{(')}$, $N_1$, $N_2$ and $N_3$ into the flat
directions. This leave us with 6 fields, so we expect a basis with just one
element. It turns out that, in respect with the charges of the remaining
fields, $Q_4$ and $Q_5$ are a linear combination of the previous $U(1)$'s,
so there are actually only 3 independent constraints and, hence, we
obtain three basis elements
\beq \phi\bar{\phi}, ~~\bar{\phi}\tilde{S}_1^2\tilde{S}_2^2\tilde{S}_3^2,
{}~~\bar{\phi}^3\tilde{S}_1^2S_2^2\tilde{S}_3^2,\eeq
where we expressed the flat directions as gauge invariant monomials.
For example, the monomial
$\bar{\phi}\tilde{S}_1^2\tilde{S}_2^2\tilde{S}_3^2$
corresponds to the following choice of VEVs

\beq |\bar{\phi}|^2=|\psi|^2, ~ |\tilde{S}_1|^2=2|\psi|^2,
{}~ |\tilde{S}_2|^2=2|\psi|^2, ~ |\tilde{S}_3|^2=2|\psi|^2, \eeq
for an arbitrary $|\psi|$.

Note that in the precedent basis any field A can be replaced with its
copy A$^{'}$. Any flat direction, $P$,  can be obtained from the elements of
the basis as

\beq P^n = \prod_\alpha M_\alpha^{n_\alpha}, \eeq
where $M_\alpha$ stand for the elements of the basis, $n$ is a positive
integer and $n_\alpha$ are integers \cite{Cleaver:1998gc}.

In order to obtain D-flat directions in the anomalous $U(1)$ we need to
construct invariant monomials containing the field $S_3^{(')}$, since this
is the only field with a positive $\hat{Q}$ charge\footnote{The $\hat{Q}$
charge of an invariant monomial is equal, up to positive factors, with his
$Q_A$ charge, since the difference between the two is a linear combination
of $Q_{1...5}$, under which the invariant monomials have zero charge by
construction.}, necessary to cancel the negative Fayet-Iliopoulos term
generated by the anomalous $U(1)$ \footnote{In our model
${\rm Tr}~Q_A<0$.}. And, since none of the elements of the basis contains
this field, we conclude that there are no flat directions involving only
VEVs of the singlets.

Therefore, we proceed with the analysis including also non-abelian fields
under the hidden gauge group.  This amounts to including all the fields in
table (\ref{flat}), which contains 22 fields with non-equivalent charges.
Again, we look for a basis of gauge invariant monomials under $Q_{1...5}$.
Such a basis is given by
\ba &&\phi\bar{\phi}, ~~ D_2\bar{D_2}, ~~A_+\bar{A}_+, ~~A_-\bar{A}_-,
{}~~\bar{\phi}\tilde{S}_1^2\tilde{S}_2^2\tilde{S}_3^2,
{}~~\bar{\phi}^3\tilde{S}_1^2S_2^2\tilde{S}_3^2,~~\bar{\phi}A_+A_-,\nonumber\\
\nonumber\\
&&\bar{\phi}S_1^4N_1^2F^2\tilde{F}^4F_4^2,
{}~~\bar{\phi}S_1^2S_3^2N_3^2\tilde{F}^2F_4^2,
{}~~\bar{\phi}S_3^2N_1^2N_2^2N_3^4F^2\tilde{F}^2,\nonumber\\
\nonumber\\
&&\bar{\phi}S_3^2N_1^2N_2^2N_3^4F_1^2F_4^2,
{}~~\bar{\phi}S_3^2N_1^2N_2^2N_3^4F_2^2F_3^2,
{}~~S_1^2\tilde{S}_2S_3\tilde{S_3}\bar{A}_+\tilde{F}^2F_4^2,\nonumber\\
\nonumber\\
&&S_1^3\tilde{S}_2^3S_3\tilde{S_3}^2N_1N_2\bar{A}_+^3\tilde{F}^3F_3^2F_4^3,
{}~~\tilde{S}_2^5S_3\tilde{S_3}^5\bar{A}_+^5F_3^3F_4,
{}~~\bar{\phi}S_1^{10}\tilde{S}_2^2S_3^2\tilde{F}^8F_4^8,\nonumber\\
\nonumber\\
&&S_1^9\tilde{S}_2^2S_3^2N_1N_2\bar{A}_+\tilde{F}^8F_4^8 ,
\label{nabasis}
\ea
where, again, any field can be replaced with one of its copies with equal
$Q_{1...5}$ charges. All the elements of the basis have negative or vanishing
$\hat{Q}$ charges, but, since some of the elements contain the fields $S_3$ and
$F$, which have positive
$\hat{Q}$ charge, and, since flat directions can be obtained as a combination
of the basis elements with negative powers, we cannot conclude immediately that
there are no D-flat directions under the anomalous $U(1)$.  Nevertheless, a
simple Mathematica program shows that it is impossible to obtain viable
invariant monomials with positive $\hat{Q}$ charge, by viable meaning that
the fields that do not have a partner field with opposite charges should appear
with positive powers in the monomials. We conclude that there are no flat
directions involving only singlets of the visible gauge group.

Therefore, the only possibility
to obtain flat directions which do not break
electric charge
is to consider the option of giving a VEV also to
the neutral component of the Higgs field, in which case the flat directions
would break the electroweak symmetry.
The Higgs doublets in our model have the following charges:
\beqn
 &\begin{tabular}{l|rrrrrrr}
 ~ & $\quad Q_1$ & $\quad Q_2$ & $\quad Q_3$ & $\quad Q_4$ & $\quad Q_5$ &
$\quad Q_A$ & $\quad \hat{Q}$\\
\hline
\hline
  $h~ \surd$  & 0 & 4& 0 & 4 & -4 & -4 & 0
\end{tabular}
\eeqn
and including them into our analysis amounts to adding the invariant 
$\bar{\phi}hS_1^2S_3\tilde{S}_3\tilde{F}^2F_4^2$ to the basis (\ref{nabasis}).
The new basis element also has a negative $\hat{Q}$ charge and, again, it turns
out to be impossible to construct flat directions with positive  $\hat{Q}$
charge. This means that the only stable vacuum solutions of our model are the
ones that break the Standard Model gauge group.

Interested in the analysis of flat directions
in free fermionic models with reduced Higgs spectrum
we performed an extensive search in a similar case, where
we could not find any solutions.
Before providing the details of this model, the definition of stringent
 flat directions is introduced.

\section{Stringent flat directions}\label{stringent}

In general, systematic analysis of simultaneously $D$- and $F$-flat directions 
in anomalous models is a complicated, non-linear 
process 
\cite{Buchmuller:2006ik,Wess:1992cp}.
In weakly coupled heterotic string (WCHS) model-building, $F$-flatness of a
specific VEV direction in the low energy effective field theory may be proved
to a given order by cancellation of $F$-term components, only to be lost a
mere one order higher at which cancellation is not found. An exception is
directions with stringent $F$-flatness \cite{Cleaver:1999mw,Cleaver:2007ek}.
Rather than allowing cancellation between two or more components in an
$F$-term, stringent $F$-flatness requires that each possible component in an
$F$-term have zero vacuum expectation value. 

When only non-Abelian singlet fields acquire VEVs, stringent flatness
implies that two or more singlet fields in a given $F$-term cannot take
on VEVs. For example, in section \ref{abcabc}, which presents the third 
and forth
order superpotential for the model under consideration, the components of
the $F$-term for $\Phi_{45}$ are (through third order):
\beqn
F_{\Phi_{45}} &=& \bar{\Phi}_{46}\bar{\Phi}'_{56}+
\bar{\Phi}'_{46}\bar{\Phi}_{56}.
\label{fflat1}
\eeqn
For stringent $F$-flatness we require not just that $<F_{\Phi_{45}}> = 0$, but 
that each component within is zero, i.e., 
\beqn
<\bar{\Phi}_{46}\bar{\Phi}'_{56}> = 0,\, <\bar{\Phi}'_{46}\bar{\Phi}_{56}> = 0.
\label{fflat2}
\eeqn
Thus, by not allowing cancellation between components in a given $F$-term, 
stringent $F$-flatness imposes stronger constraints than generic $F$-flatness, 
but requires significantly less fine-tuning between the VEVs of fields.

The net effect of all stringent $F$-constraints on a given superpotential term 
is that at least two fields in the term must not take on VEVs. This condition 
can be relaxed when non-abelian fields acquire VEVs. Self-cancellation of a 
single component in a given $F$-term is possible between various VEVs within a 
given non-abelian representation. Self-cancellation was discussed
in \cite{Cleaver:2000aa} for $SU(2)$ and $SO(2n)$ states.

A given set of stringent $F$ flatness constraints are not independent and 
solutions to a set can be expressed in the language of Boolean algebra (logic) 
and applied as constraints to linear combinations of $D$-flat basis
directions.The Boolean algebra language makes clear that the effect
of stringent $F$-flat constraints is strongest for low order superpotential
terms and lessens with increasing order. In particular, for the model
presented in the following,
 stringent flatness is extremely constraining on VEVs of
the reduced number of (untwisted) singlet fields appearing in the third
through fifth order superpotential, in comparison to its constraints on the
larger number of singlets in the model of table \ref{secondexample}
\cite{Faraggi:2006qa}.  

One might imagine that 
stringent $F$-flatness constraints requires order-by-order testing of 
superpotential terms.
This is, in fact, not necessary. All-order stringent $F$-flatness can actually
be proved or disproved by examining only a small finite set of possible dangerous
(i.e., $F$-flatness breaking) superpotential terms. 
Through a process such as matrix 
singular value decomposition (SVD)\footnote{A SVD FORTRAN subroutine is provided in
\cite{Press_numericalrecipes:}.},
a finite set of superpotential terms can be constructed that generates all possible 
dangerous superpotential terms for a specific $D$-flat direction. This basis of
gauge-invariants can always be formed with particular attributes: 
(1) each basis element term contains at most one unVEVed field (since to threaten
$F$-flatness, a gauge-invariant term, necessarily without anomalous charge, 
can contain no more than one unVEVed field);
(2) there is at most one basis term for each unVEVed field in the model; and
(3) when an unVEVed field appears in a basis term, it appears only to the first power.
The SVD process generated a possibly threading basis of superpotential terms for several
models (see for example \cite{Cleaver:1999mw,Cleaver:2000ds,Perkins:2005zh}).

To appear in a string-based superpotential, a gauge invariant term must also  
follow Ramond-Neveu-Schwarz worldsheet charge conservation rules.
For free fermionic models these rules were generalised from finite order in \cite{Kalara:1990sq,Rizos:1991bm}
to all-order in \cite{Cleaver:2001ab}. The generic all order rules can be applied to systematically
determine if any product of SVD-generated $F$-flatness threatening superpotential basis
elements survive in the corresponding string-generated superpotential. If none survive,
then $F$-flatness is proved to all finite order. This technique has been used to prove
$F$-flatness to all finite order for various directions in several models 
\cite{Cleaver:1999mw,Cleaver:2000ds,Perkins:2005zh}.
Alternately, if any terms do survive, the lowest order is determined at which stringent
$F$-flatness is broken.

How should stringent (especially all-order) flat directions be interpreted in 
comparison to general (perhaps finite order) flat directions? All-order 
stringent flat directions contain a minimum
number of VEVs and appear in models 
as the roots of more fine-tuned (generally finite-order) flat directions that
require specific cancellations between $F$-term components. The latter may 
involve cancellations between sets of components of different orders in the 
superpotential. 

All-order stringent flat directions have indeed been
discovered to be such roots 
in all prior free fermionic heterotic models for which we have performed 
systematic flat direction classifications. However, the model presented 
in the next section  
appears to lack any stringent flat directions, at least within the expected 
range of VEV parameter space. We have reached this conclusion after employing 
our standard systematic methodology for $D$- and $F$-flat direction analysis.

\section{The string model with no stringent flat-directions}\label{stringmodelsection}

The string model that we present here contains three chiral generations,
charged under the Standard Model gauge group and with the canonical
$SO(10)$ embedding of the weak--hypercharge; one pair of untwisted
electroweak Higgs doublets; a cubic level top--quark Yukawa coupling.
The string model therefore shares some of the phenomenological
characteristics of the quasi--realistic free fermionic string models.
The boundary condition
basis vectors beyond the NAHE--set and the one--loop GSO projection 
coefficients are shown in table \ref{stringmodel} and in table 
\ref{phasesmodel1}, respectively. 
\beqn
 &\begin{tabular}{c|c|ccc|c|ccc|c}
 ~ & $\psi^\mu$ & $\chi^{12}$ & $\chi^{34}$ & $\chi^{56}$ &
        $\bar{\psi}^{1,...,5} $ &
        $\bar{\eta}^1 $&
        $\bar{\eta}^2 $&
        $\bar{\eta}^3 $&
        $\bar{\phi}^{1,...,8} $ \\
\hline
\hline
  ${\alpha}$  &  0 & 0&0&0 & 1~1~1~0~0 & 1 & 0 & 0 & 1~1~0~0~0~0~0~0 \\
  ${\beta}$   &  0 & 0&0&0 & 1~1~1~0~0 & 0 & 1 & 0 & 0~0~1~1~0~0~0~0 \\
  ${\gamma}$  &  0 & 0&0&0 &
		${1\over2}$~${1\over2}$~${1\over2}$~${1\over2}$~${1\over2}$
	      & ${1\over2}$ & ${1\over2}$ & ${1\over2}$ &
                0~0~0~0~$1\over2$~$1\over2$~${1\over2}$~${1\over2}$ \\
\end{tabular}
   \nonumber\\
   ~  &  ~ \nonumber\\
   ~  &  ~ \nonumber\\
     &\begin{tabular}{c|c|c|c}
 ~&   $y^3{y}^6$
      $y^4{\bar y}^4$
      $y^5{\bar y}^5$
      ${\bar y}^3{\bar y}^6$
  &   $y^1{\omega}^5$
      $y^2{\bar y}^2$
      $\omega^6{\bar\omega}^6$
      ${\bar y}^1{\bar\omega}^5$
  &   $\omega^2{\omega}^4$
      $\omega^1{\bar\omega}^1$
      $\omega^3{\bar\omega}^3$
      ${\bar\omega}^2{\bar\omega}^4$ \\
\hline
\hline
$\alpha$ & 1 ~~~ 0 ~~~ 0 ~~~ 1  & 0 ~~~ 0 ~~~ 1 ~~~ 1  & 0 ~~~ 0 ~~~ 1 ~~~ 1 \\
$\beta$  & 0 ~~~ 0 ~~~ 1 ~~~ 1  & 1 ~~~ 0 ~~~ 0 ~~~ 1  & 0 ~~~ 1 ~~~ 0 ~~~ 1 \\
$\gamma$ & 0 ~~~ 1 ~~~ 0 ~~~ 0  & 0 ~~~ 1 ~~~ 0 ~~~ 0  & 1 ~~~ 0 ~~~ 0 ~~~ 0 \\
\end{tabular}
\label{stringmodel}
\eeqn
with the choice of generalised GSO coefficients:
\begin{equation}
{\bordermatrix{
        &{\bf 1}&  S & &{b_1}&{b_2}&{b_3}& &{\alpha}&{\beta}&{\gamma}\cr
 {\bf 1}&   ~~1 &~~1 & & -1  &  -1 & -1  & &  -1    &  -1   & ~~i   \cr
       S&   ~~1 &~~1 & &~~1  & ~~1 &~~1  & &  -1    &  -1   &  -1   \cr
        &       &    & &     &     &     & &        &       &       \cr
   {b_1}&    -1 & -1 & & -1  &  -1 & -1  & &  -1    &  -1   & ~~i   \cr
   {b_2}&    -1 & -1 & & -1  &  -1 & -1  & &  -1    & ~~1   & ~~i   \cr
   {b_3}&    -1 & -1 & & -1  &  -1 & -1  & & ~~1    &  -1   & ~~1   \cr
	&       &    & &     &     &     & &        &       &       \cr
{\alpha}&    -1 & -1 & & -1  &  -1 &~~1  & & ~~1    & ~~1   & ~~1   \cr
 {\beta}&    -1 & -1 & & -1  & ~~1 & -1  & &  -1    & ~~1   & ~~1   \cr
{\gamma}&    -1 & -1 & &~~1  & ~~1 & -1  & &  -1    &  -1   &  -i   \cr}}
\label{phasesmodel1}
\end{equation}
Both the basis vectors $\alpha$ and $\beta$ break the 
$SO(10)$ symmetry to $SO(6)\times SO(4)$ and the basis vector
$\gamma$ breaks it further to $SU(3)\times U(1)_C\times SU(2)\times U(1)_L$.  
The basis vector $\alpha$ is symmetric with respect to the sector
$b_1$ and asymmetric with respect to the sectors $b_2$ and $b_3$, 
whereas the basis vector $\beta$ is symmetric with respect to $b_2$ 
and asymmetric with respect to $b_1$ and $b_3$. As a consequence of these
assignments and of the
string doublet--triplet splitting mechanism \cite{Faraggi:1994cv}, both the untwisted
Higgs colour triplets and electroweak doublets, with leading coupling
to the matter states from the sectors $b_1$ and $b_2$, are projected
out by the generalised GSO projections. 
At the same time the untwisted colour Higgs triplets 
that couple at leading order to the states from the sector $b_3$ are projected out, 
whereas the untwisted electroweak Higgs doublets
remain in the massless spectrum. Due to the asymmetric boundary 
conditions in the sector $\gamma$ with respect to the sector
$b_3$, the leading Yukawa coupling is that of the up--type quark from the
sector $b_3$ to the untwisted electroweak Higgs doublet \cite{Faraggi:1991be}.
Hence, the leading Yukawa term is that of the top quark and only its
mass is characterised by the electroweak VEV. The lighter
quarks and leptons couple to the light Higgs doublet through
higher order nonrenormalizable operators that become effective 
renormalizable operators by the VEVs that are used to cancel the
anomalous $U(1)_A$ $D$--term equation \cite{Faraggi:1991be}. 
We remind once again that the novelty
in the construction of the model in \cite{Faraggi:2006qa}, and in the model of table \ref{stringmodel}, is that the reduction of the untwisted
Higgs spectrum is obtained by the choice of the boundary
condition basis vectors in table \ref{stringmodel}, whereas
in previous models it was obtained by the choice of
flat directions and analysis of the superpotential \cite{Cleaver:1997rk}.

The final gauge group of the string model arises
as follows: in the observable sector the NS boundary conditions 
produce gauge group generators for 
\beq
SU(3)_C\times SU(2)_L\times U(1)_C\times U(1)_L\times U(1)_{1,2,3}\times
U(1)_{4,5,6}~~~ .
\label{observablegg}
\eeq
Thus, the $SO(10)$ symmetry is broken to
$SU(3)\times SU(2)_L\times U(1)_C\times U(1)_L$,
where, 
\begin{eqnarray}
U(1)_C
&\Rightarrow&~Q_C=
			 \sum_{i=1}^3Q({\bar\psi}^i)~,\label{u1c}\\
U(1)_L 
&\Rightarrow&~Q_L=
			 \sum_{i=4}^5Q({\bar\psi}^i)~.\label{u1l}
\end{eqnarray}
The flavour $SO(6)^3$ symmetries are broken to $U(1)^{3+n}$ with
$(n=0,\cdots,6)$. The first three, denoted by $U(1)_{j}$ $(j=1,2,3)$, arise 
{}from the worldsheet currents ${\bar\eta}^j{\bar\eta}^{j^*}$, 
as mentioned previously.
The additional horizontal $U(1)$ symmetries, denoted by $U(1)_{j}$ 
$(j=4,5,...)$, arise by pairing two real fermions from the sets
$\{{\bar y}^{3,\cdots,6}\}$, 
$\{{\bar y}^{1,2},{\bar\omega}^{5,6}\}$ and
$\{{\bar\omega}^{1,\cdots,4}\}$. 
The final observable gauge group depends on
the number of such pairings. In this model there are the 
pairings ${\bar y}^3{\bar y}^6$, ${\bar y}^1{\bar\omega}^5$
and ${\bar\omega}^2{\bar\omega}^4$, which generate three additional 
$U(1)$ symmetries, denoted by $U(1)_{4,5,6}$. 

It is important to note that the existence of these three additional 
$U(1)$ currents is correlated with the assignment of asymmetric
boundary conditions with respect to the set of internal
worldsheet fermions $\{y,\omega|{\bar y},{\bar\omega}\}^{1,\cdots,6}$,
in the basis vectors that extend the 
NAHE--set, $\{\alpha, \beta,\gamma\}$.
This assignment of asymmetric boundary conditions in the basis
vector that breaks the $SO(10)$ symmetry to $SO(6)\times SO(4)$
results in the projection of the untwisted Higgs colour--triplet fields
and preservation of the  corresponding electroweak--doublet Higgs
representations \cite{Faraggi:1994cv}.

In the hidden sector, 
which arises from the complex
worldsheet fermions ${\bar\phi}^{1\cdots8}$,
the NS boundary conditions produce the generators of
\beq
SU(2)_{1,2,3,4}\times SU(4)_{H_1}\times U(1)_{H_1}\, .
\label{hiddengg}
\eeq
$U(1)_{H_1}$ 
corresponds to the combinations of the worldsheet charges
\begin{equation}
Q_{H_1}=\sum_{i=5}^8Q({\bar\phi}^i)~.\label{qh1}
\end{equation}

The model contains several additional sectors that may a priori 
produce spacetime vector bosons and enhance the gauge symmetry, which include
the sectors
${\bf 1}+b_1+b_2+b_3$ and ${\bf 1}+S+\alpha+\beta+\gamma$.
Additional spacetime vector bosons from these sectors would enhance
the gauge symmetry that arise from the spacetime vector bosons produced
in the Neveu--Schwarz sector.
However, with the choice of generalised GSO projection coefficients 
given in table \ref{phasesmodel1} all of the extra gauge bosons from these
sectors are projected out and the four dimensional gauge group is 
given by eqs. (\ref{observablegg}) and (\ref{hiddengg}).

In addition to the graviton, dilaton,
antisymmetric sector and spin--1 gauge bosons, 
the Neveu--Schwarz  sector gives one pair of electroweak Higgs 
doublets $h_3$ and $\bar h_3$; six pairs of $SO(10)$ singlets, which 
are charged with respect to $U(1)_{4,5,6}$;
three singlets of the entire four dimensional gauge group.
A notable difference as compared to models with unreduced
untwisted Higgs spectrum, like the model of ref.\  \cite{Faraggi:1991jr},
is that the $SO(10)$ singlet fields, which are charged under
$U(1)_{1,2,3}$, are projected out from the massless spectrum.
The three generations are obtained from the sectors $b_1$, $b_2$ and 
$b_3$, as usual. The model contains states that are vector--like with respect to the
Standard Model and all non--abelian group factors, but may be chiral with
respect to the $U(1)$ symmetries that are orthogonal to the $SO(10)$ group. 
The full massless spectrum of the model is detailed in Table 3.b
in Appendix B.

As a final note we remark that the boundary conditions with respect to the 
internal worldsheet fermions of the set
$\{y,\omega|{\bar y},{\bar\omega}\}^{1,\cdots,6}$
in the basis vectors $\alpha$, $\beta$ and $\gamma$,
that extend the NAHE--set, are similar to those
in the basis vectors that generate the string 
model of ref.\ \  \cite{Faraggi:1991jr}, with the replacements
\beqn
\alpha({\bar y}^3{\bar y}^6)& \longleftrightarrow & 
\gamma({\bar y}^3{\bar y}^6) \nonumber\\
\beta({\bar y}^1{\bar\omega}^5)& \longleftrightarrow & 
\gamma({\bar y}^1{\bar\omega}^5) . \label{278substitutions}
\eeqn
The worldsheet 
fermions $\{y,\omega|{\bar y},{\bar\omega}\}^{1,\cdots,6}$
correspond to the compactified dimensions in a corresponding
bosonic formulation.
The substitutions in eqs.(\ref{278substitutions})
are augmented with suitable modifications
of the boundary conditions of the worldsheet fermions
$\{{\bar\psi}^{1,\cdots,5},{\bar\eta}^{1,\cdots,3},{\bar\phi}^{1,\cdots,8}\}$,
which correspond to the gauge degrees of freedom.  
The effect of these additional 
modifications is to alter the hidden sector gauge group.
While the substitutions in eqs.(\ref{278substitutions})
look innocuous enough, they in fact produce substantial 
changes in the massless spectrum and, as a consequence, in the
physical characteristics of the models. With regard to the
flat directions of the superpotential, the effect of these
changes on the untwisted states will be particularly noted.

\subsection{Third and Fourth Order Superpotential}
\label{abcabc}
The three singlets of the entire four dimensional gauge group are obtained from: 
\beqn
\xi_{1}&=&\chi^{12*}\bar{\omega}^3\bar{\omega}^6|0>~~,\nonumber\\
\xi_{2}&=&\chi^{34*}\bar{\omega}^1\bar{y}^5|0>~~,\nonumber\\
\xi_{3}&=&\chi^{56*}\bar{y}^2\bar{y}^4|0>~~.\nonumber
\eeqn

We show below the cubic and fourth order superpotential terms. 

\noindent Trilinear superpotential:
\beqn
W_3&=&N^c_3 L_3\bar{h}+u^c_3 Q_3\bar{h} + 
		H_4\bar{H}_7 h+\bar{H}_4H_7\bar{h}+\nonumber\\
&+&{\xi_1}(H_1\bar{H}_1+H_8\bar{H}_8+H_9\bar{H}_9)\nonumber\\
&+&{\xi_2}(H_2\bar{H}_2+H_{10}\bar{H}_{10}+H_{11}\bar{H}_{11})\nonumber\\
&+&{\xi_3}(H_3\bar{H}_3+H_4\bar{H}_4+H_5\bar{H}_5+H_6\bar{H}_6+
		H_7\bar{H}_7)\nonumber\\
&+&{\xi_3}(\Phi_1^{\alpha\beta}\bar{\Phi}_1^{\alpha\beta}+
		\Phi_2^{\alpha\beta}\bar{\Phi}_2^{\alpha\beta})\nonumber\\
&+&\Phi_{45}(\bar{\Phi}_{46}\bar{\Phi}'_{56}+\bar{\Phi}'_{46}\bar{\Phi}_{56})+
	\bar{\Phi}_{45}(\Phi_{46}\Phi'_{56}+\Phi'_{46}\Phi_{56})\nonumber\\
&+&\Phi'_{45}(\bar{\Phi}_{46}\Phi_{56}+
	\bar{\Phi}'_{46}\Phi'_{56})+\bar{\Phi}'_{45}(\Phi_{46}\bar{\Phi}_{56}+
	\Phi'_{46}\bar{\Phi}'_{56})\nonumber\\
&+&\Phi'_{45}\left((\Phi_1^{\alpha\beta})^2+(\Phi_2^{\alpha\beta})^2\right)+
	\bar{\Phi}'_{45}\left((\bar{\Phi}_1^{\alpha\beta})^2+
	(\bar{\Phi}_2^{\alpha\beta})^2\right)\nonumber\\
&+&\bar{\Phi}'_{45}H_{12}H_{13}+\Phi_{46}H_{14}H_{15}+
	\bar{\Phi}'_{56}H_{16}H_{17}\nonumber\\
&+&\Phi'_{56}(H_1)^2+\bar{\Phi}'_{56}(\bar{H}_1)^2+
	\bar{\Phi}'_{46}(H_2)^2+ \Phi'_{46}(\bar{H}_2)^2\nonumber\\
&+&\Phi_1^{\alpha\beta}H_9H_{11}+
	\bar{\Phi}_2^{\alpha\beta}(\bar{H}_1\bar{H}_2+\bar{H}_8\bar{H}_{10})+
	 \bar H_1 \bar H_4 H_{10} + H_2 \bar H_4 \bar H_8~~.
\label{w3all}
\eeqn
\noindent Quartic superpotential:
\beqn
W_4&=&  Q_1 u_1 H_4 \bar{H}_5
      + Q_2 u_2 H_4 \bar{H}_6
      + L_1 N^c_1 H_4 \bar{H}_5
      + L_2 N^c_2 H_4 \bar{H}_6~~.
\label{w4all}
\eeqn		
We provide the expression of the quintic order superpotential in (\ref{w5all}) 
in Appendix B.

\subsection{Flat directions}\label{flatdir}

The model in table \ref{stringmodel} possesses nine local $U(1)$ symmetries, 
eight in the observable part
and one in the hidden part. Six of these are anomalous:
\beq
{\rm Tr}{U_1}= {\rm Tr}{U_2}= - {\rm Tr}{U_3}=  
2 {\rm Tr}{U_4}= -2 {\rm Tr}{U_5}= 2 {\rm Tr}{U_6}= -24.
\label{tru1s}
\eeq
$U(1)_L$ and $U(1)_C$ of the $SO(10)$ subgroup are anomaly free.  
Consequently, the weak hypercharge and the orthogonal combination,
$U(1)_{Z^{\prime}}$, are  anomaly free.
The hidden sector $U(1)_{H_1}$ is also anomaly free.\

Of the six anomalous $U(1)$s, five can be rotated by
an orthogonal transformation to become anomaly free. 
The unique combination that remains anomalous is: 
${U_A}=k\sum_j [{{\rm Tr} {U(1)_j}}]U(1)_j$, 
where $j$ runs over all the anomalous $U(1)$s and $k$ is
a normalisation constant. 
For convenience, we take $k={1\over{12}}$ and therefore 
the anomalous combination is given by:
\beq
U_A=-2U_1-2U_2+2U_3-U_4+U_5-U_6,{\hskip .5cm}{\rm Tr}Q_A=180.\label{u1a}
\eeq

The five rotated non-anomalous orthogonal combinations are not unique,
with different
choices related by orthogonal transformations. One choice is given by:
\beqn
 {U^\prime}_1 & = & U_1-U_2{\hskip .5cm},{\hskip .5cm}
 {U^\prime}_2  ~=~  U_1+U_2+2U_3,\label{u1pu2p}\\
 {U^\prime}_3 & = & U_4+U_5{\hskip .5cm},{\hskip .5cm}
 {U^\prime}_4  ~=~  U_4-U_5-2U_6,\label{u3pu4p}\\
 {U^\prime}_5 & = & U_1+U_2-U_3-2U_4+2U_5-2U_6.\label{u5p}
\eeqn
Thus, after this rotation there are a total of eight $U(1)$s
free from gauge and gravitational anomalies. 
In the following we use a different method to
calculate D- and F- flatness, which is suitable 
for the implementation of a FORTRAN program. 
A basis set of (norm-squares of) VEVs of scalar fields
satisfying the non-anomalous 
$D$-flatness constraints (\ref{dxi}) can be created en masse 
\cite{ Cleaver:1997cr, Cleaver:2001ab}. The basis directions can have
positive, negative, or zero anomalous charge. 
In the maximally orthogonal basis used in the singular value decomposition 
approach of \cite{Cleaver:1997cr, Cleaver:2001ab}, 
each basis direction is uniquely identified with a particular VEV. That is, 
although each basis direction generally contains many VEVs, each basis direction 
contains at least one particular VEV that only appears in it.

A physical $D$-flat direction $D_{\rm phys}$, with anomalous charge of sign 
opposite that of the FI term  $\xi$, is formed from linear combinations of the 
basis directions, 
\beqn
D_{\rm phys} = \sum_{\rm i = 1}^{\#~{\rm basis~dirs.}} a_i D_i,
\label{phydir}
\eeqn
where the integer coefficients $a_i$ are normalised to have no non-trivial 
common factor.

In our notation, a physical flat direction (\ref{phydir}) may have a negative 
norm-square for a vector-like field. This denotes that it is the oppositely 
charged vector-partner field that acquires the VEV, rather than the field. 
Basis directions themselves may have vector-like partner directions if all 
associated fields are vector-like. On the other hand, if in particular, the 
field generating the VEV uniquely associated with a basis direction does not 
have a vector-like partner, that basis direction cannot have a vector-like 
partner direction. 

In pursuit of physical all-order flat directions for this model, we first 
examined directions formed solely from the VEVs of non-abelian singlet fields. An 
associated maximally orthogonal basis set, denoted by
$\{{\cal{D}}^{'}_{i=1\ {\rm to}\ 13}\}$, containing only non-abelian singlet 
VEVs is shown in Table 3.c in Appendix B. 
The respective unique VEV fields of these basis 
directions are identified in Table 3.d in the same Appendix.
 Examination of Tables 3.c and 3.d 
reveals that no physical $D$-flat directions can be formed solely from VEVs
of non-abelian singlet fields.
Since the FI term $\xi$ in eq.(\ref{dxi}) is positive for this model, with ${\rm 
Tr}Q_A=180$, a physical flat direction must carry a negative anomalous charge. 
However, of the 13 singlet $D$-flat basis directions, three carry anomalous 
charge of $+15$, $+30$, $+30$ while the remaining ten do not carry anomalous charge. Further,
the unique VEVed fields for the 3 basis directions with positive anomalous charge 
do not have corresponding vector-like partner fields. Hence, there are no 
vector-like paired basis directions 
with negative anomalous charge. Thus, Tables 
3.c and 3.d imply that one or more fields carrying non-abelian charges must also 
acquire VEVs in physical $D$-flat directions. This result is, in itself, not 
necessarily unexpected, as non-abelian VEVs have been required for physical 
(all-order) flat directions in other quasi-realistic free fermionic heterotic 
models in the past, for example \cite{Cleaver:2000ds}. 

Thus, we expanded our flat direction search to include VEVs of both non-Abelian
singlet fields and non-abelian charged fields. Our chosen set of 50 maximally
orthogonal $D$-flat basis directions for both non-abelian singlet VEVs and
non-abelian charged VEVs, denoted by $\{{\cal{D}}_{i=1\ {\rm to}\ 50}\}$, is
presented in Table 3.e. The respective unique field VEVs identified with these
basis directions are given in Table 3.f. In this enlarged basis the anomalous 
charges are given in units of ($\frac{Q^{(A)}}{15}$) and the directions 
containing only singlet VEVs are rotations of those in Table 3.c.

Nine of the 50 directions, denoted $D_{i=1,...,9}$, carry one or two units of
negative anomalous charge. Twenty basis directions, denoted $D_{10}$ through
$D_{29}$, carry no anomalous charge. Twenty-one basis directions, denoted
$D_{30}$ through $D_{50}$, carry one or two units of positive anomalous charge.  
All basis directions possessing negative anomalous charge contain $SU(3)_C\otimes
SU(2)_L$ charges or hidden sector $SU(4)\otimes \prod^{4}_{j=1} SU(2)_j$ charges.
(Thus, this basis set also reveals that anomaly cancellation will necessarily
break one or more non-abelian local symmetries.) All of the $\Phi$ fields, the
$H_{1~{\rm to}~11}$ fields and $h$ have vector-like pairs. Thus, physical flat
directions can have negative components for any of these. A subset of these
fields, specifically $\Phi_{46}$, $\Phi^{'}_{45}$, $\bar{\Phi}^{'}_{56}$,
and $H_{4,5,6,7}$,
has VEVs appearing in multiple basis directions. The only non-vector-like field
with a VEV that appears in multiple directions is $e^c_{3}$.
 
$D_{10}$ through $D_{17}$ and $D_{22}$ are composed solely of varying
combinations of the vector-like fields. Hence, all of these basis directions have
corresponding vector-like partner basis directions, $\bar{D}_{i} \equiv -D_{i}$, 
for which the VEV of each field is replaced by the VEV of the vector-like partner
field. Thus, in a physical flat direction in eq.(\ref{phydir}), each of the respective
integer coefficients $a_{10}$ through $a_{17}$ and $a_{22}$, may be
negative, positive, or zero.

Note that $D_{7}$, $D_{8}$, $D_{9}$ and $D_{20}$ are vector-like except for their 
$e^c_{3}$ components. Thus, each of $a_{7}$, $a_{8}$, $a_9$ and $a_{20}$ may be 
negative, positive, or zero in a physical $D$-flat direction, so long as the net
norm-square VEV of $e^c_{3}$ is non-negative.\footnote{Note that non-vector-like
fields, such as $e^c_3$, that appear in multiple directions with some basis directions
having positive and some having negative norm-square components,
are common in this process.
Further, some models explored in the past have had (at least) one basis direction with two
(or more) field VEVs unique to it and with norm-square VEVs with differing signs.
This latter type of basis direction can never appear in a physical direction and, hence,
implies that the fields unique to it can never appear in a $D$-flat
direction. (If all of the
norm-squares of the fields unique to a basis direction were initially negative, then these
signs, along with those of the norm-squares of any vector-like field VEVs in that basis
direction, could all be changed together to allow the basis direction
to appear in a physical
direction.)} The remaining basis directions
contain at least one unique non-vector-like field VEV. Thus, in a physical flat
direction, the coefficients of the remaining basis directions must be 
non-negative.

What does this mean for a physical $D$-flat direction formed as a linear
combination of the basis directions? For a physical flat direction there are,
thus, two specific constraints on the $a_i$ coefficients and one general set of
non-negative norm-square constraints on a subset of the $a_i$. First, negative
anomalous charge for a flat direction requires
\beqn
-2 \sum_{i=1}^2 a_i     -   \sum_{i=3}^{9} a_i 
+  \sum_{i=30}^{44} a_i + 2 \sum_{i=45}^{50} a_i < 0.
\label{physdc1} 
\eeqn 
Second, a non-negative norm-square VEV for $e^c_3$ requires 
\beqn
&&-6 \sum_{i=1}^{2} a_i -3 a_3 - 6\sum_{i=4}^{6} a_i -2 a_7 -6 \sum_{i=8}^{9} a_i 
-2 \sum_{i=18}^{19} a_i - a_{20} + a_{21}\nonumber\\ 
&&-2 \sum_{i=23}^{24} a_i - a_{25} -2 a_{26} +2 a_{27} - 2 a_{28} +2 a_{29}
+6 a_{30} +6 a_{32}+ a_{38} \nonumber\\
&&+ 3\sum_{i=39}^{40} a_i
+6 a_{42} + 6 \sum_{i=45}^{47} a_i
+2\sum_{i=48}^{49} a_i + 6 a_{50}\geq 0.
\label{physdc2} 
\eeqn 
Last, for the set of 
non-vector-like fields that are each identified with a respective unique
$D$-flat direction, the general set of non-negative norm-square VEV constraints 
is \beqn
a_i~\geq 0~{\rm for}~i=1~{\rm to}~6,~18,~19,~21,~23~{\rm to}~50.
\label{physdc3}
\eeqn

At low orders, each individual superpotential term also induces several
stringent $F$-term  constraints on the $a_i$ coefficients of physical flat
directions. As stated prior, 
the set of constraints from superpotential terms with only singlet
fields translate into the requirement that two or more singlet fields in a given
superpotential term cannot take on VEVs. For the model under investigation,
constraints from third order superpotential terms are especially severe. For 
this model, all six $\Phi$ singlet fields and their vector-like partners appear in
third order superpotential terms (specifically, the sixth and seventh lines) of 
eq.(\ref{w3all}). Stringent $F$-flatness from these terms forbids at least 8 of the
12 singlet fields from acquiring VEVs.

For example, when solely third order stringent $F$-flatness constraints are
applied to the six pairs of $\Phi$ vector-like singlets (and no $F$-flatness
constraints are applied to the non-abelian states), there are just nine solution
classes that allow the
maximum of 4 singlet VEVs. (Flat directions in any of these nine classes are
defined by their respective non-abelian VEVs.)

For three of these nine singlet third order flatness classes, the VEVs are of two fields
and their respective vector-like partners: either,
\beqn
<{\Phi}_{45}>,\, <{\Phi}'_{45}>,\, <\bar{\Phi}_{45}>,\,
<\bar{\Phi}'_{45}> &\ne 0,&\, {\rm or}
\label{fflat3sa}\\
<\bar{\Phi}_{46}>,\, <\bar{\Phi}'_{46}>,\, 
<{\Phi}_{46}>,\, <{\Phi}'_{46}> &\ne 0,&\, 
{\rm or}
\label{fflat3sb}\\
<\bar{\Phi}'_{56}>,\, <\bar{\Phi}_{56}>,\, 
<{\Phi}'_{56}>,\, <{\Phi}_{56}> &\ne 0.&
\label{fflat3sc}
\eeqn
Higher order stringent flatness constraints can further reduce the allowed number
of singlet VEVs of each of these solutions. Further, a component of a $D$-flat
basis direction in Table 3.a in Appendix B only specifies the difference between the 
norm-squares of the VEV of a given field and of the given vector-like partner
field (if it exists). Completely chargeless VEVs solely involving a field 
$\Phi_i$ and its vector-like partner $\bar{\Phi}_i$ such that $|<\!\!\!\Phi_i\!\!\!>|^2 = 
|<\!\!\!\bar{\Phi}_i\!\!\!>|^2$ can always be added to a physical $D$-flat direction. 
However, it is preferable for higher order $F$-flatness to impose that a field
and its vector-partner do not simultaneously acquire VEVs. Hence, these three
solutions effectively allow only two unique singlet fields to acquire VEVs.

The next three classes of singlet solutions do allow up to four distinct
singlet fields to
acquire VEVs: either,
\beqn
&&<{\Phi}_{45}>,\, <{\Phi}'_{45}>,\, <{\Phi}_{46}>,\, <{\Phi}'_{46}> \ne 0,
{\rm or},
\label{fflat3sd}\\
&&<{\Phi}_{45}>,\, <{\Phi}'_{56}>,\, 
<{\Phi}_{56}>\, <\bar{\Phi}'_{45}>   \ne 0,\, 
{\rm or},
\label{fflat3se}\\
&&<\bar{\Phi}_{46}>,\, <\bar{\Phi}_{56}>,\, <{\Phi}'_{56}>,\, <{\Phi}'_{46}>
\ne 0.
\label{fflat3sf}
\eeqn

For the three remaining solution classes, the fields in (\ref{fflat3sd}),
(\ref{fflat3se}) 
and (\ref{fflat3sf}), are respectively replaced with their vector-like partner
fields. For any of these nine stringent $F$-flat choices, no other $\Phi$ singlet 
fields can acquire VEVs.

Any of the constraints on allowed and disallowed VEVs, such as the above, can be
re-expressed in terms of constraints on the $a_i$ coefficients specifying the
basis directions contributions to a physical $D$-flat direction. For example,
setting $<\Phi_{46}>=0$ would require
\beqn
&&4 a_1 + a_2  +2 \sum_{i=3}^{4} a_i +8 a_5 +2 a_6 + 
a_7 - a_8 - a_9 + a_{10} + a_{16}  \nonumber\\ 
&&- a_{18} +2 a_{19} + a_{20} - a_{21} + a_{23}-2 a_{27} - a_{28} + 
a_{29} + 4 a_{30} \nonumber\\
&&+ a_{31}  - a_{32} +\sum_{i=33}^{35} a_i -2 \sum_{i=36}^{37} a_i -
2 a_{39} + a_{40} 
-2 \sum_{i=41}^{43} a_i + a_{44}\nonumber\\
&& -4 a_{45} + 2 a_{46} - a_{47} -3 a_{49} - a_{50} = 0 ~ ~ .
\label{examp1} 
\eeqn

To systematically investigate physical $D$-flat directions with non-abelian VEVs,
over a course of several months we generated and examined physical $D$-flat
directions composed of from 1 to 6 basis directions. Under the assumption that
all VEVs of physical flat directions are nearly of the same order of magnitude,
we allowed coefficients of 0 to 20 for the non-vector-like basis directions and
coefficients of -20 to 20 for the vector-like basis directions.

To be classified as a physical $D$-flat direction, a linear combinations of basis
directions needed to obey eqs.(\ref{physdc1}-\ref{physdc3}) and was, of course, 
also required to have non-abelian $D$-flatness. (The general process by
which we enforced non-abelian $D$-flatness followed that presented in
\cite{Cleaver:1997cr,Cleaver:2001ab}.) Each resulting physical $D$-flat direction was then
tested for stringent $F$-flatness from all third order through fifth order
superpotential terms and additionally for some key sixth order superpotential
terms.\footnote{While only the third through fifth order superpotential is given
in section (\ref{abcabc}), we have generated the complete superpotential to eighth order
and can generate it to any required order.}
 
Following the SVD method discussed earlier in section 3.4 and described in
\cite{Cleaver:1999cj,Cleaver:2007ek},
we had planned to then test for possible all-order stringent $F$-flatness, the 
subset of physical $D$-flat directions that had proved stringently $F$-flat to
at least fifth or sixth order. Based on all of the prior models we had
investigated, we had expected to find around four to six physical $D$-flat
directions that were, in fact, stringently $F$-flat to all finite order.
However, in contrast we discovered that no physical $D$-flat directions that we
had generated even kept stringent
$F$-flatness through sixth order. So there were no
physical $D$-flat directions to examine for all-order testing. For this model,
with its reduced 
set of singlet fields from the untwisted sector, not even self-cancellation of 
non-abelian terms could provide stringent $F$-flatness through sixth order for 
any of these physical $D$-flat directions.

We will continue a search for $F$-flatness past sixth order
for physical $D$-flat directions in this model that are comprised of seven or
more basis directions.
However, a continued null result is likely: since each of our
basis directions contains a unique field VEV, increasing the number of non-zero
$a_i$ coefficients linearly increases the minimum number of unique field VEVs.
With each increase in number of basis directions composing a physical $D$-flat
direction, the probability of obtaining stringent $F$-flatness much beyond sixth
order further decreases.

In this model no physical D-flat direction that we generated 
kept F-flatness through six order. We said that only
stringent flat directions can be flat to all orders of 
nonrenormalizable terms. This would indicate that this model
has no D-flat directions that can be proved to be F-flat
to all order.
If a non-vanishing F-term does exist, then supersymmetry remains unbroken
at finite order. The Fayet-Iliopoulos term that breaks supersymmetry
is generated at one-loop level
in the perturbative string expansion. On the other hand the string spectrum
is Bose-Fermi degenerate and possesses $N=1$ spacetime
supersymmetry at the classical level.
This would suggest that, contrary to the expectation from
supersymmetric quantum field theories, 
perturbative supersymmetry breaking may ensue in string theory.
Futhermore, the modular invariant one-loop partition function
vanishes, giving a vanishing one-loop cosmological constant. 
This model may therefore represent an example of a quasi-realistic
string vacuum with vanishing one-loop cosmological constant
and perturbatively broken supersymmetry.

\chapter{${\mathbb Z}_2\times {\mathbb Z}_2$ orbifold constructions}

In the previous chapter we have largely 
discussed the free fermionic models,
which correspond to ${\mathbb Z}_2\times {\mathbb Z}_2$ orbifolds
at special points of the moduli space (see section \ref{semisemi}).
In this chapter we want to present the orbifold construction as 
it provides complementary information on
heterotic models away from the special points.

We first consider the heterotic 
superstring compactified on a flat torus,
where the physical dimensions are reduced from ten to four.
In order to obtain models with appealing phenomenology,
for instance with $N=1$ supersymmetry, the initial
toroidal compactification is modified by modding-out 
a discrete symmetry described by a point group $P$
and giving rise to an orbifold 
\cite{Dixon:1985jw,Dixon:1986jc}, for review see \cite{Bailin:1999nk}. 
We briefly present the orbifold construction rules, 
the derivation of the massless spectrum and the 
projection conditions required for modular invariance.
 We mainly follow \cite{Ibanez:1986tp,Ibanez:1987pj,Nilles:2004ej,Forste:2004ie,wingerter,vaudregange,Ibanez:1987dw},
where an extensive treatment of the topic can be found.

 An indicative example of orbifold compactification 
is presented in the second part of this chapter.
Our model is a six dimensional torus
defined by the $SO(4)^3$ root lattice, with ${\mathbb Z}_2\times {\mathbb Z}_2$
discrete symmetry. The derivation of its fixed tori, 
their centralisers and the introduction
of the Wilson lines is explained in details.
The main motivation for considering the skewed model analysis
was the attempt of reproducing the three generation free fermionic model
\cite{Faraggi:2004rq}, with $E_6\times U(1)^2\times SO(8)^2_H$ gauge symmetry.
A model with these properties was not found in the classification
by Donagi and Wendland \cite{Donagi:2008xy}, which extended
the analysis of Donagi and Faraggi \cite{Donagi:2004ht}.
The aim of the skewed model analysis is to try
to build an orbifold model with similar characteristics to the free fermionic
model. While unsuccessful, the inclusion of this analysis in the thesis
aims to provide details of the complementary orbifold construction.
In particular, we explain the implications of using 
factorisable or non-factorisable lattices and how the 
presence of Wilson lines may change the phenomenology
of the model. A detailed study concerning ${\mathbb Z}_2\times {\mathbb Z}_2$
orbifolds with different compactification
 lattices is given in \cite{Faraggi:2006bs}.
Several semi-realistic orbifold models have been presented 
in the literature with different discrete symmetry 
\cite{Lebedev:2006kn,Lebedev:2008un},
although we are mainly interested in the ${\mathbb Z}_2\times {\mathbb Z}_2$ case.
This is mainly because we believe that 
the correspondence with free fermionic models can provide some intuition
in the selection of phenomenological interesting vacua,
since the number of modular invariant orbifold models is huge and 
a complete classification under general physical properties
represents an incredible feat.
This chapter describes the general procedure
of the orbifold construction and suggests some technical tricks
in choosing the most favourable lattices to construct
possible semi-realistic orbifold vacua.

\section{Heterotic string and toroidal compactification}
\label{4.1}

The ten dimensional heterotic superstring can provide 
a realistic four-dimensional theory if six of the nine spatial 
dimensions are compactified to a ``sufficiently small'' scale,
unobservable in nowadays experiments.
The simplest compactification scheme is on a
torus that, being a flat surface,
assures no modifications in the equations of motion.
We start this section by revisiting the content of the heterotic string
in ten dimensions in the bosonic construction, since in chapter 2
we have presented the correspondent fermionic description,
where the compact bosons are substituted by internal 
degrees of freedom (32 real left-moving fermions
 with a precise choice of boundary conditions).
In the bosonic formalism, the heterotic string is a right-moving
superstring combined with a bosonic left-moving string. In the light-cone
gauge the eight fermionic and eight bosonic right coordinates 
are given respectively by $\Psi^i_R$ and $X^i_R$, $i=1,..8$.
The indices $i=1,2$ denote the two transverse spacetime dimensions,
 while the other six refer to the compact spatial dimensions. 
The left movers are given by the bosonic 
$X^i_L$ and sixteen further bosons $X^I_L$, $I=1,..16$, compactified
on a 16-torus. The anomaly cancellation requirement imposes
that the 16-torus is either the root lattice of $E_8\times E_8$ or the one
of $Spin(32)/{\mathbb Z}_2$ \cite{Green:1984sg}.
In this thesis we are interested in the $E_8\times E_8$ symmetry,
then the equations given below 
will refer to the first case.
The compactification procedure does not affect
the mode expansion of the fields, whose expressions 
have been provided in chapter 2.
We specify here the expansion of the gauge degrees of freedom
\beq
X^I_L(\tau+\sigma)= x^I_L+p_L^I(\tau+\sigma)+\frac{i}{2}
\sum_{n\neq 0}\frac{\tilde \a_n^I}{n}e^{-2in(\tau+\sigma)} ,
\eeq 
where we fixed $\a'=1/2$ and the momenta 
$p^I_L$ lay on the $E_8\times E_8'$
 lattice.
In the canonical basis, any element of the $E_8$ lattice
can be written as eight-dimensional vectors
\[(n_1\,,...,\, n_8)\quad,\quad (n_1+1/2,...,n_8+1/2)\quad,
\]
where $\sum n_i =$0(mod2). 
The first notation labels the adjoint representation
of $SO(16)$, while the second vector represents
 the spinorial of the same 
symmetry group.

The compactification of the internal coordinates
on the 6-torus, namely $X^i=X^i_L+X^i_R$ with $i=3,..8$, identifies
 the centre of mass coordinates $x^i$ with 
points that are separated by lattice vectors of the torus
\[x^i=x^i+2\pi L^i ,
\]
where $\vec L=(L^3,...,L^8)$ belongs to a six-dimensional lattice 
$\Lambda=\{\sum_{t=3}^{8} r_t\vec e_t\,\,|\,\, r_t\in{\mathbb Z} \}$ 
and $\vec e_t$ are
the basis vectors of the lattice. 
This implies that the boundary
conditions for the compact spatial bosons
are also satisfied if 
$X^i(\tau,\pi)=X^i(\tau,0)+2\pi L^i$, which
correspond to winding states around the torus.
The compactification also requires the quantization of the momenta
$p^i$ and this result is achieved
by imposing the condition $\sum_{i=3}^{8}p^i L^i \in {\mathbb Z}$.
Thus, the momenta are quantised on the dual lattice $\Lambda^*$, defined
as
\[\Lambda^*=\{\sum_{t=3}^{8} m_t\vec {e_t}^*\,\,|\,\, m_t\in {\mathbb Z} \} ,
\]
where the basis vectors $\vec {e_{t'}}^*$ satisfy the relation
 $\vec {e_{t'}}^*\cdot \vec e_t=\delta_{t't}$.

After the compactification to four dimensions,
the mass formula for the right movers takes the form 
\beq
\frac{1}{4}m_R^2= N_R+\frac{1}{2}p_R^ip_R^i - a_R,
\label{mr}
\eeq 
where $N_R$ is the number operator which counts the bosonic 
and fermionic (both R and NS) oscillators.
The constants $a_{R,L}$ are the normal ordering
for the Virasoro operators $\tilde L_0$ and $L_0$, introduced
in chapter 2. There we have showed that 
they get different values when considering
the Ramond or the Neveu-Schwarz sector
(we notice that these values were determined for the non-compactified
theory, while different values will be
calculated in the next section
for twisted states arising in orbifold constructions).

For the left movers in four dimensions 
the mass formula is given by
\beq
\frac{1}{4}m_L^2= \tilde N_L+\frac{1}{2}p_L^Ip_L^I - a_L ,
\label{ml}
\eeq
where the left number operator $\tilde N_L$
includes the spatial oscillators 
$\tilde\a_{-n}^i\tilde\a_n^i$ and the left gauge contributions
$\tilde\a_{-n}^I\tilde\a_n^I$.

In eq.(\ref{mr}) and (\ref{ml}) the contribution 
from the momenta
$p_{L,R}^i$ can give rise to massless states for particular 
values of the parameters of the lattice $\Lambda$, 
such as the length of the basis vectors, the angles between them,
a scale factor. Apart from these isolated values,
massless states arise when momenta and winding numbers
are zero 
\[p_R^i=p_L^i=0 ,
\] as we can see from their definition in eq.(\ref{plr}).
The toroidal compactification described so far provides a 
$N=4$ supersymmetric theory in four dimensions.

In fact, let us show explicitly how
four gravitinos are generated in this set up.
In the massless spectrum, we notice the presence 
of the states
\beq
b^i_{-1/2}|0>_R\otimes\,\, \tilde\a^j_{-1}|0>_L \,\,,\,\,
\quad b^i_{0}|0>_R\otimes \,\,\tilde\a^j_{-1}|0>_L \,\,,
\label{asz}
\eeq
where $i=1,2$ and $j$ takes values in the compact space.
The first combination provides spacetime vectors, the
second state is in the Ramond groundstate and 
transforms as an $SO(8)$ chiral spinor, the opposite chirality
spinor being deleted by GSO projections 
used in the superstring construction.
The weight vector notation for such a spinor
is given by
$q=(\pm\frac{1}{2}\pm\frac{1}{2}\pm\frac{1}{2}\pm\frac{1}{2})$,
with an even number of ``$+$''.
The $SO(8)$ chiral spinor can be decomposed
into representations of $SO(2)\times SO(6)\in SO(8)$,
with the $SO(2)$ corresponding to the two transverse spacetime
coordinates and the $SO(6)$ referring to the six compactified coordinates.
Hence, there are four spacetime spinors of each chirality, providing
four gravitinos. 
The analogous notation is used for the NS
right moving state, corresponding to the first
entry in eq.(\ref{asz}) and indicated by $q=(\underline{1,0,0,0})$
(the underscore denotes that all permutations are included).

For completeness we provide 
the massless physical states of the heterotic string in $D=10$.

\underline{{\bf{Spectrum of the heterotic string}}}

\ba
&&|q>_R \times \tilde\a^i_{-1}|0>_L\quad :\quad i=1,..8\quad {\text{supergravity multiplet}} ,
\nonumber\\
&&|q>_R \times \tilde\a^I_{-1}|0>_L\quad : \quad I=1,..16\quad {\text{uncharged gauge bosons of $E_8\times E_8$}} ,\nonumber\\
&&|q>_R \times |p>^I_L\quad :\quad 240 + 240\quad {\text{charged gauge bosons of $E_8\times E_8$}} ,
\ea
where $|q>_R$ indicates both R and NS solutions, meaning that
the bosonic and its correspondent
fermionic state are
 present in the spectrum at the same time (susy superpartners).

\section{Orbifold construction}

So far we have shown
that the toroidal compactification reduces our ten dimensional
heterotic string to four dimensions, but the theory is not chiral.
In order to obtain a phenomenological interesting $N=1$ supersymmetric
theory, we consider the orbifold construction by starting with
the toroidal case. 
A torus is created by the identification of points $\vec x$
of the underline space that differs by a lattice vector
$\vec l\in\Gamma=2\pi\Lambda$
\beq
\vec x\sim \vec x+\vec l .
\label{q}
\eeq
In the toroidal compactification
six spatial internal dimensions are compactified
on the torus $T^6$ and the sixteen left-moving coordinates,
corresponding to the gauge degrees of freedom,
are compactified on the self-dual lattice $T_{E_8\times E_8'}$.
$T^6$ is generated by the lattice $\Lambda$
defined in the previous section, while
$T_{E_8\times E_8'}$ is given by the root lattice of 
the group $E_8\times E_8'$.
An orbifold is obtained when we identify points on the torus
which are related by the action of an isometry $\theta$,
more precisely, an automorphism of the lattice 
($\theta \vec l \in 2\pi\Lambda
$) that preserves the scalar products among 
the basis vectors $\vec e_a \in \Lambda$, $a=1,..6$,
where the vector $\vec l=\vec e_a n_a$.
In the following we indicate the lattice roots simply as
$e_a$, specifying the entries of the vector with
a new label $i$ when necessary. 
The orbifold is defined as
\beq
\Omega=T^6/P \times T_{E_8\times E_8}/G  ,
\eeq
where $P$ is the point (isometry) group, $G$ its embedding
in the gauge degrees of freedom. 
The construction of an orbifold depends 
on the choice of the point group P, its embedding G
and the lattice $T^6$. In particular,
the requirement of $N=1$
spacetime supersymmetry is achieved by imposing $P\subset SU(3)$.
We restrict our discussion to an abelian $P$.
In this case the point group is discrete and there
are two possible choices :
\ba
&\bullet& P \equiv {\mathbb Z}_N = \{ \theta^k\,\, |\,\, k = 0,..N-1\}\nonumber\\
&\bullet& P \equiv {\mathbb Z}_N\times {\mathbb Z}_M = \{ \theta^k_1\circ \theta_2^l\,\,
 |\,\, k = 0,..N-1\,\, {\text{and}}\,\, l=0,..M-1\}
\ea
where $\theta$ can be seen as a rotation of $2\pi /N$, with $N$
being the order of the twist.
The gauge
twisting group $G$ is an automorphism
of the $E_8\times E_8$ Lie algebra and its action is required 
in order to satisfy 
modular invariance.
The six-dimensional torus
can be written in the equivalent notation
 $T^6={\mathbb R}^6/\Gamma$ when considering the
identification\[
\vec x\sim \theta \vec x+\vec l .\]
The previous expression is 
useful when we define the space group, given by
the set of elements
\[S=\{(\theta,\vec l)|\quad\theta\in P,\,\,\, \vec l\in 2\pi \Lambda \} .
\]
By using the previous definition the following equivalence holds: 
$T^6/P \equiv {\mathbb R}^6/S$.

The inner automorphism of the $E_8\times E_8$ algebra
 can be realised \cite{hel}
by a shift $V^I$ in the root lattice and 
the embedding of a generic element of $S$ is implemented by
\[(\theta,n_ae_a) \rightarrow (\sigma_{V^I}, n_a\sigma_{A_a^I}),
\]
where $\sigma_{A_a^I}$ corresponds to the action of the shifts
$A_a$ in the gauge lattice. These shifts 
are the gauge transformations
associated with the non-contractible loops
given by $e_{a}$ and they are called Wilson lines.

We can finally present the orbifold action on 
the spatial compact coordinates and on the 
gauge degrees of freedom 
\beq
X^i\rightarrow (\theta X)^i + n_a e^i_a \,\,, \quad X^I_L\rightarrow
X^I_L+ V^I + n_aA_a^I.
\eeq
In particular, if we use the complex notation $Z^a$, $a=1,2,3$,
for the compact dimensions $X^{3,..8}$, the action of $\theta$
is simply
\beq
\theta^k : Z^a \rightarrow e^{2\pi i k v^a} Z^a
\eeq
where the vector $\vec v=(v_1,v_2, v_3)$ corresponds to the
twist action 
\[\theta^k \rightarrow k v .
\]
Since the number of independent cycles on a six-torus is six,
we could initially think that there are six independent Wilson lines.
However, the lattice vectors defining the torus are generally
related by the point group symmetry, thus some of the 
Wilson lines are identified.
We will clarify this statement 
when we consider
our example in section (\ref{so43}).
Differently from the toroidal compactification,
in orbifold backgrounds there are singular points,
the so-called fixed points, where
the metric is not isomorphic to $R^6$.
This is a crucial feature of orbifold models, 
related to the presence of twisted sectors
in the spectrum. 
A fixed point is defined by $X_f^i=(\theta^kX_f)^i+n_a e_a^i$,
 for $i=3,..8$.
We will show the explicit derivation of the fixed points
for the ${\mathbb Z}_2\times {\mathbb Z}_2$ orbifold with a given compactification lattice
in the second part of this chapter.

At this point we show how
the twist vector $v^a$ has to be fixed to achieve $N=1$ supersymmetry.
Since P is abelian, it 
must belong to the Cartan subalgebra of $SO(6)$
associated with the coordinates $X^{3,..8}$. 
If the generators of this subalgebra are indicated as
$M^{34}$, $M^{56}$ and $M^{78}$, then
the action of the point group element acts
on the complex basis $Z^a$ as
\beq
\theta = exp[2\pi i (v_1M^{34}+v_2M^{56}+v_3M^{78})],
\eeq
where $|v_a|<1$, $a=1,2,3$. The condition $P\subset SU(3)$
thus requires
\beq \pm v_1 \pm v_2 \pm v_3 = 0 .\label{vvv}
\eeq
The condition (\ref{vvv}) and the fact
 that the twist is a symmetry of the torus
restrict the choices of $P$ to the following possibilities:
 it has to be a ${\mathbb Z}_N$ symmetry with $N=3,4,6,7,8,12$ 
or a ${\mathbb Z}_N\times {\mathbb Z}_M$ symmetry, with $N$ multiple of $M$ and
$N=2,3,4,6$ \cite{Dixon:1985jw,Dixon:1986jc}.
In general there can be several lattices for a given $P$.
The massless spectrum and the gauge symmetries are determined
by the point group and not by the choice of the lattice.
We point out that when the space group is taken into account, then
the embedding into the gauge lattice $E_8\times E_8$ provides properties
depending  on the lattice.
A complete list of point group generators for 
${\mathbb Z}_N$ and ${\mathbb Z}_N\times {\mathbb Z}_M$  $\subset SU(3)$ 
orbifolds can be found in \cite{Bailin:1999nk}.

\subsection{Consistency conditions}

The embedding of the point group $P$ into 
the twist gauge group $G$ is an homomorphism of the lattice, thus
for a $N$ order twist $\theta$ the action of $NV^I$ corresponds to
the identity on the root lattice. The same principle holds for
the Wilson lines and these conditions
are translated into the equations below
\beq
NV\in T_{E_8\times E_8}\quad , \quad NA_a\in T_{E_8\times E_8}.
\eeq
Modular invariance has to be required in order 
to guarantee anomaly freedom and in the orbifold 
construction 
this requirement is implemented by the following conditions
\ba
N(V^2-v^2)=0{\text{(mod2)}} ,\nonumber\\
NV\cdot A_a=0{\text{(mod1)}} ,\nonumber\\
N A_a\cdot A_b=0{\text{(mod1)}} ,\nonumber\\
N A_a^2=0 {\text{(mod2)}} .
\label{modcon}
\ea 
For ${\mathbb Z}_N\times {\mathbb Z}_M$ orbifolds the previous relations
can be generalised. For instance, in the second part
of this chapter, when the
${\mathbb Z}_2\times {\mathbb Z}_2$ orbifold is introduced,
it will be defined by two independent
twist vectors $\vec v_1$ and $\vec v_2$,
while the standard embedding
is realised by the shifts 
$V^I_1$ and $V^I_2$.
The first two formulae in eqs.(\ref{modcon}) must
hold for both of these vectors.
Moreover, the Wilson lines conditions in eqs.(\ref{modcon})
must be fulfilled by both these vectors as well.

\subsection{Generalities on the spectrum}

There are different ways in which
the closed boundary conditions can be satisfied on an orbifold.
This leads to the conclusion that there are two types
of strings, the untwisted string closed on the torus 
before the identification of points by the twist,
and the twisted string which is closed on the torus after
imposing the point group symmetry.
This is simply resumed in the following expression
\beq
X^{3,..8}(\tau,\sigma)=\theta^k X^{3,..8}(\tau, 0)+ n_ae_a ,
\label{bbcc}
\eeq
where the untwisted sector ($k=0$) corresponds to the 
toroidal compactification, while the additional twisted sectors
generate all new string states, localised at the points left
fixed under the action of the
 elements $(\theta^k, n_a e_a)$ of the space group S.
A generic element $h\in S\otimes G$ has a correspondent operator $\hat h$
which implements the action of $h$ on the Hilbert space.
We call $\hat h$ a constructing element and denote
the states localised at the corresponding fixed point by $H_h$.
Hence, since the orbifold is defined by modding out 
the action of $S\otimes G$, then physical states must be invariant
 under the projection $S \otimes G$. 
We will explain this concept on an explicit example in section 
\ref{so43}.

\underline{{\bf{UNTWISTED SECTOR}}}

The untwisted states are those obtained by the heterotic string
compactified on a torus which survive the $S \otimes G$
projections.
Below we rewrite eqs.(\ref{mr}--\ref{ml}) demanding the level
matching condition and 
using the 
weight vector notation for the right movers, as introduced previously,
\beq
\frac{1}{2}q^2-\frac{1}{2}=\frac{1}{4}m_R^2=
\frac{1}{4}m_L^2=\frac{1}{2}p^2+N_L-1=0.\label{qq}
\eeq
Under the action of $S \otimes G$ the left and the right states
transform respectively as
\[|p>= e^{(2\pi i p\cdot V)}|p>\quad ;\quad 
|q>= e^{(2\pi i q\cdot v)}|q>. 
\]
Invariant states are created when the product
of these eigenvalues is 1.
We obtain two kinds of states, the gauge bosons providing the
unbroken gauge group, and the charged matter states.
The first set of solutions satisfies the conditions
\beq
p\cdot V=0{\text{(mod 1)}}\,\,, \quad p\cdot A_a={\text{(0 mod 1)}}\,\, ,
\eeq 
combined with right movers which are invariant under $S$.
When considering right movers transforming non trivially,
we get the second set of solutions.
In this case, in fact,
the only possible surviving states are those tensored with
left states transforming as
\beq
p\cdot V=k/N {\text{(mod 1)}}\,\,,\,\,
 k=1,..N-1 , \quad p\cdot A_a=0{\text{(mod 1)}} .
\eeq
The most important result
in the untwisted spectrum
is that three of the four gravitinos present in the
toroidal compactification are projected out,
giving a four-dimensional $N=1$ supergravity theory.


\underline{{\bf{TWISTED SECTORS}}}

The boundary conditions in eq.(\ref{bbcc}) for $k\neq 0$ 
provide the massless states of the 
twisted sectors. Each twisted sector corresponds to 
a constructing element, previously called $h$.
Obviously, the new boundary conditions change
the mode expansions of the bosonic and fermionic oscillators,
while the weight lattice has been shifted. In particular, we obtain
$|q>_R^{tw}=|q + k v >_R$ and shifted 
momenta $|p^I>^{tw}= |p^I + kV^I +n_a A_a>$.
The mass formula in each twisted sector reads as
\beq
\frac{1}{2}(q+v_i)^2-\frac{1}{2}+\delta_c=\frac{1}{4}m_R^2=
\frac{1}{4}m_L^2=\frac{1}{2}(p^I+V^I+n_aA_a)^2+N_L-1+\delta_c=0.\label{rr}
\eeq
In the formula above the quantity $\delta_c$ is the
zero point energy due to the  moded oscillators. 
It can be calculated by $\delta_c=\frac{1}{2}
\sum_{a=0}^3\eta^a (1-\eta^a)$, where $\eta^a = kv^a{\text{(mod 1)}}$
and $0 \le \eta^a < 1$. We anticipate here that for
the case of the ${\mathbb Z}_2\times {\mathbb Z}_2$ orbifold 
$$\delta_c=1/4 .$$
As mentioned already, in the twisted sector 
the oscillators are moded if they correspond
to a complex dimension $a$ where the twist acts non trivially,
giving for example $\tilde \a^a_{m-\eta^a}$ for a bosonic oscillator.
In this case the number operator $\tilde N$ can be fractional.
The physical spectrum is obtained after the projections
under each element of $S \otimes G$. If we indicate
with $h=(\theta,n_a e_a; V,n_a A_a)$ a constructing element 
of this group, the invariant states under $h$ define
the Hilbert space $H_h$, as
we stated at the start of this section.
 Now we consider a different element 
of the group, that we call 
$g=(\overline\theta,\overline n_a e_a; V,\overline n_a A_a)$.
If $g$ commutes with $h$, then, by using the definition
of twisted boundary conditions, we can see that the states
invariant under $g$ belong to $H_h$. 
Moreover, all states in $H_h$ which transform non 
trivially under $g$ have to be projected out.
This reasoning has to be applied for all commuting elements
of $S \otimes G$ and the whole set that contains these elements
is called centraliser
\beq
Z_h = \{ g\in S \otimes G \,\,{\text{such that}}\,\, [h,g]=0 \}.
\eeq 
Requiring that non invariant states are projected
out means that all the elements in the centraliser
act as the identity on $H_h$.
For each non commuting element $\tilde g$, $[\tilde g, h]\neq 0$,
the procedure to apply consists in building linear combinations
of states of Hilbert spaces $H_h$, $H_{\tilde gh\tilde g^{-1}}$, ...
 $H_{\tilde g^nh\tilde g^{-n}}$, with $\tilde g^n=1$. 
In the ${\mathbb Z}_2\times {\mathbb Z}_2$ case it
is always possible to restrict the previous procedure
to a reasonable finite number of elements of $S \otimes G$.

\section{${\mathbb Z}_2\times {\mathbb Z}_2$ orbifold with $SO(4)^3$ compactification lattice}
\label{so43}

There are several lattices with ${\mathbb Z}_2\times {\mathbb Z}_2$ symmetry
that can be  considered to describe the $T^6$ torus.
One of the simplest instances \cite{wingerter,vaudregange}
is the factorisable $T^6=T^2\times T^2\times T^2$, with
orthogonal roots $e_i=(0,0,..,i,..,0)$, $i=1,..,6$.
The action of the point group in the ${\mathbb Z}_2\times {\mathbb Z}_2$ orbifold
is given by $P_{{\mathbb Z}_2\times {\mathbb Z}_2}=\{ 1,\theta_1, \theta_2, \theta_3\}$,
where the trivial element $1$ generates the untwisted spectrum 
and $\theta_k$, $k=1,2,3$, generate the twisted sectors.
$\theta_3$ is the combination of the two independent twists $\theta_1$
and $\theta_2$.
 We present explicitly
the twist vectors associated to each twisted sector
\[1\rightarrow (0,0,0,0) \,\,,\quad \theta_1 \rightarrow
v_1=(0,1/2,-1/2,0) ,\]\[\theta_2\rightarrow v_2=(0,0,1/2,-1/2) \,\,,
\quad \theta_3 \rightarrow v_3=(0,1/2,0,-1/2), \]
where the four entries in the vectors $v_k$ refer to the spatial
dimensions in complex coordinates.
The space group is defined by $S=\{(kv_1+lv_2,\,\, 
n_ae_a \,\,|\,\,k,l=0,1,\,\, n_a\in {\mathbb Z}) \}$ and the twisted sectors are
obtained by the combinations of $k,l=0,1$.
It has been shown \cite{wingerter,vaudregange} that
 the factorisable $T^6=T^2\times T^2\times T^2$ lattice
needs proper Wilson lines to provide 
three standard model generations, although it still does 
not realise the standard embedding of the hypercharge. 
For this reason we can conclude that 
the factorisable lattice can be considered a toy model
in the class of orbifold constructions.
We need to introduce more challenging cases 
to implement some interesting
phenomenological properties. 
The next step
is to consider different compactification
 lattices for the six-dimensional torus
 that for instance generate an inferior number of generations before
even adding Wilson lines.
Hence, we rely on two mechanisms for the generation reduction,
such as the introduction of Wilson lines and the choice of the lattice. 
Before entering into the details, we specify the fact that
for the ${\mathbb Z}_2\times {\mathbb Z}_2$ orbifold the 
fixed points are actually two dimensional
objects, thus providing fixed tori. They give rise to 
generations or anti-generations (representations of
the symmetry group of the model under consideration in terms of 
multiplets which provide the Standard Model families, eventually after
the breaking of the gauge group).
The number of fixed tori depends on the compactification lattice and
for this reason and appropriate choice of the lattice
provides the 
options to decrease the 
net number of generations, often too many in orbifold compactifications.
In the standard embedding the net number of generations
is actually given by the Euler number, hence we compare
the result obtained by the explicit calculation of the fixed tori
for our model with its the Euler number.

As we have mentioned in the introduction,
 the $SO(4)^3$ orbifold example is far from being a 
semi-realistic model. Our point is showing how  
the presence of Wilson lines, that in general
 change drastically the outcome of 
a model, in this particular case do not modify
the number of generations, for any choice
 of Wilson lines. The proof of this 
statement is shown at the end of the chapter.   

We introduce now our example, where $T^6$ is obtained by
compactifying ${\mathbb R}^6$ on an $SO(4)^3$ root lattice, whose basis
vectors are given by the 
simple roots 
\begin{eqnarray}
e_1 & = & \left( 1 , 0 , 0,-1,0,0\right) , \nonumber \\
e_2 & = & \left( 1, 0, 0, 1,0,0\right) , \nonumber \\
e_3 & = & \left( 0 ,1 , 0, 0,-1,0\right) , \nonumber \\
e_4 & = & \left( 0, 1, 0, 0,1,0\right) , \nonumber \\
e_5 & = & \left( 0,0,1,0,0,-1\right) , \nonumber \\
e_6 & = & \left( 0,0,1,0,0,1\right) .\label{roots}
\end{eqnarray}
We remark here that the action of the orbifold on the 
$SO(4)^3$ compactification lattice is non-factorisable, as
it is obvious from the displacement of the entries
in the roots (\ref{roots}).
This choice produces 
interesting consequences for 
the spectrum of the model. In fact, 
the number of fixed tori 
is reduced from 48 in the standard $SO(4)^3$
to 12  for the case with skewed action on the compactification lattice, 
resulting into
a drastic reduction of the number of generations.
The derivation of the massless spectrum follows
the rules given in the previous sections. We find
convenient to obtain at this point some relevant information
which will be used in the calculation of the twisted 
states. In fact, from the analysis of the $SO(4)^3$ skew lattice,
we obtain the fixed tori and the 
centralisers which are necessary for the discussion 
of the massless twisted states.

\subsection{Analysis of the lattice}\label{sec1}

The study of a lattice consists of the following steps:

\begin{itemize}
\item find the generators of the lattice,
\item look at the symmetries of the roots under the orbifold action,
\item calculate the fixed tori and the centraliser,
\item analyse the consistency conditions for the Wilson lines.
\end{itemize}
We remind that we removed the vector symbol
on the roots and any general vector lattice to simplify the notation. 

\subsubsection{Generators and symmetries}\label{11}

The generators of the lattice are defined as
 the minimal shifts that, added to a fixed torus,
provide exactly the equivalent torus. 
In order to make this concept more understandable, we will illustrate
 the procedure to obtain the generators in the case of the trivial 
torus\footnote{The fixed tori for the $\theta_2$ sector are calculated in the next section.} of the $\theta_2$ twisted sector
\begin{equation}
\left\{ \left( x_1 , x_2 , 0,0,0,0\right) \left|\, x_1, x_2 \in {\mathbb R}^2/
\Lambda^2 \right. \right\} .
\end{equation}
The compactification lattice $\Lambda^2$ is generated by the vectors
$\left( 2 ,0 \right)$ and  $\left( 0 , 2 \right)$; in fact we need to 
satisfy the condition
\begin{equation}\label{generators}
\left( x_1,x_2,0,0,0,0\right) = 
\left( x_1+a,x_2+b,0,0,0,0\right) + \sum a_i e_i , 
\end{equation}
where the $e_i$ are the $SO(4)^3$ roots and $a$ and $b$ are the 
minimal shifts on the ($x_1$, $x_2$) coordinates of the
 2-torus. The constants $a_i$ have to be integer 
since we are looking for equivalent tori, meaning that they can 
differ only by  $SO(4)^3$ lattice shifts.
Eq. (\ref{generators}) can be written as
$$\frac{x_1}{2}(e_1+e_2)+\frac{x_2}{2}(e_3+e_4)=\frac{x_1+a}{2}(e_1+e_2)+
\frac{x_2+b}{2}(e_3+e_4)+ \sum a_i e_i  . $$
Requiring $a_i\in Z$ implies $(a,b)= (0 $(mod2)$ ,0 $(mod2)$) $.
Hence, we are lead to the conclusion that
the minimal shift is determined by the points 
$(0,0),(0,2),(2,0),(2,2)$ and we can 
choose the two independent generators to be $a=(2,0)$, $b=(0,2)$.

The symmetry of the roots (\ref{roots}) 
determines analogous results for the
fixed tori in the $\theta_1$ and the $\theta_3$
twisted sectors, although this is not a general 
property\footnote{The $SO(6)^2$ 
non factorisable lattice is an example where the fixed tori in the
three twisted sectors have different generating elements \cite{Faraggi:2006bs}.}. 
The symmetries of the lattice are derived 
by looking at the 
transformation properties 
of the roots under the elements $\theta_i$.
\begin{center}
\beq\label{theta123}
\begin{tabular}{|c|c|c|}\hline
$\theta_1$ & $\theta_2$ & $\theta_3$ \\
\hline
$e_1\rightarrow$ - $e_1$ & $e_1 \rightarrow$ $e_2$ & $e_1 \rightarrow$ - $e_2$\\
\hline
$e_2\rightarrow$ - $e_2$ & $e_2 \rightarrow$ $e_1$ & $e_2 \rightarrow $ - $e_1$\\
\hline
$e_3\rightarrow $ - $ e_4 $ & $e_3 \rightarrow $ $e_4$ & $e_3 \rightarrow $ - $ e_3$\\
\hline
$e_4\rightarrow $ - $ e_3 $& $e_4 \rightarrow $ $e_3 $ & $e_4 \rightarrow $ - $ e_4$\\
\hline
$e_5\rightarrow $ - $ e_6 $ & $e_5 \rightarrow $ - $ e_5$ & $e_5 \rightarrow $ $e_6$\\
\hline
$e_6\rightarrow $ - $ e_5$ & $e_6 \rightarrow $ - $ e_6$ & $e_6 \rightarrow $ $ e_5$\\
\hline
\end{tabular}
\eeq
\end{center}
We observe that there are three sets
of roots $\{e_1,e_2\}$, $\{e_3,e_4\}$ 
and $\{e_5,e_6\}$, which behave analogously under the twists. 
This means that the Wilson lines
associated to each group must be equal, in particular 
$A_1 $=$ A_2$ , $A_3 $=$ A_4$ , $A_5 $=$ A_6$\footnote{This result
gives rise to the consistency conditions for
the Wilson lines that we can possibly introduce
in the case of the $SO(4)^3$ skew lattice.}. 

\subsubsection{Fixed tori and centraliser}\label{tori}

In this section we present
the fixed tori for each twisted sector of the model.
The element in parenthesis
on the right-hand-side of a fixed torus 
represents the constructing 
element for its correspondent centraliser. 
We have also specified if the torus provides 
a generation or an anti-generation to the twisted spectrum,
a concept that will be explained later on.
 
\underline{The fixed tori for the sector $\theta_1$ } :
\begin{eqnarray}
\left\{ \left( 0 , 0 , 0,0,x_5,x_6\right) 
  \left|\, x_5,x_6 \in {\mathbb R}^2/ 
\    {\Lambda}^2 \right. \right\} , & & \mbox{1 generation},\label{1t1}\\ 
\left\{ \left( 
 1,0,0,0,x_5,x_6\right) 
  \left|\, x_5,x_6 \in {\mathbb R}^2/ 
\    {\Lambda}^2 \right. \right\} , & & \mbox{($e_1+e_2$) 1 generation},\label{2t1} \\
\left\{ \left( 1/2,0,0,1/2, x_5, x_6\right)
  \left|\, x_5,x_6 \in {\mathbb R}^2/ 
\    {\Lambda}^2 \right. \right\} , & & \mbox{($e_2$) 1 generation}, \label{3t1}\\ 
\left\{ \left( 1/2,0,0,-1/2, x_5, x_6\right)
  \left|\, x_5,x_6 \in {\mathbb R}^2/ 
\    {\Lambda}^2 \right. \right\} , & & \mbox{($e_1$) 1 anti-generation}.\label{4t1}
\end{eqnarray}

\underline{The fixed tori for sector $\theta_2$} :
\begin{eqnarray}\label{2}
\left\{ \left( x_1 , x_2 ,
  0,0,0,0\right) 
  \left|\, x_1,x_2 \in {\mathbb R}^2/ 
\Lambda^2 \right. \right\} , & & \mbox{ 1 generation},\label{1t2}\\
\left\{ \left( x_1 , x_2 ,
  1,0,0,0\right) 
  \left|\, x_1,x_2 \in {\mathbb R}^2/ 
\Lambda^2 \right. \right\} , & & \mbox{($e_5+e_6$) 1 generation},\label{2t2} \\ 
\left\{ \left( x_1 , x_2 ,
  1/2,0,0,1/2 \right)
  \left|\, x_1,x_2 \in {\mathbb R}^2/ 
\Lambda^2 \right. \right\} , & & \mbox{($e_6$) 1 generation}, \label{3t2}\\ 
\left\{ \left( x_1 , x_2 ,
  1/2,0,0,-1/2 \right) \right.
  \left|\, x,y \in {\mathbb R}^2/ 
\Lambda^2  \right\} , & & \mbox{($e_5$) 1 anti-generation}.\label{4t2}
\end{eqnarray}

\underline{The fixed tori for sector $\theta_3$ }:
\begin{eqnarray}
\left\{ \left( 0 , 0 , x_3,x_4,0,0\right) 
  \left|\, x,y \in {\mathbb R}^2/ 
\    {\Lambda}^2 \right. \right\} , & & \mbox{ 1 generation},\label{1t3}\\ 
\left\{ \left( 
  0,1,x_3,x_4,0,0\right) 
  \left|\, x_3,x_4 \in {\mathbb R}^2/ 
\    {\Lambda}^2 \right. \right\} , & & \mbox{($e_3+e_4$) 1 generation},\label{2t3} \\
\left\{ \left( 0,1/2,x_3,x_4, 1/2, 0\right)
  \left|\, x,y \in {\mathbb R}^2/ 
\    {\Lambda}^2 \right. \right\} , & & \mbox{($e_4$) 1 generation}, \label{3t3}\\ 
\left\{ \left(0,1/2,x_3,x_4,-1/2,0\right)
  \left|\, x,y \in {\mathbb R}^2/ 
\    {\Lambda}^2 \right. \right\} , & & \mbox{($e_3$) 1 anti-generation}.\label{4t3}
\end{eqnarray}

\subsubsection{Derivation of the fixed tori (\ref{1t2})-(\ref{4t2})}
Only the calculation of the fixed tori in $\theta_2$ twisted sector
is presented in detail, since the treatment
for the other twisted sectors is similar.
The mathematical condition which provides
 fixed tori in the $\theta_2$ sector is the following
\beq\label{torus}
\theta_2 T = T + \sum a_ie_i ,
\eeq
where we indicate the generic torus as $T = (x_1,x_2,x_3,x_4,x_5,x_6) .$
Equation (\ref{torus}) gives
\ba
(x_1,x_2,-x_3,-x_4,-x_5,-x_6)&=&(x_1,x_2,x_3,x_4,x_5,x_6)
+(a_1,0,0,-a_1,0,0)\nonumber\\
&&+(a_2,0,0,a_2,0,0)
+(0,a_3,0,0,-a_3,0)\nonumber\\
&&+(0,a_4,0,0,a_4,0)+(0,0,a_5,0,0,-a_5)\nonumber\\
&&+(0,0,a_6,0,0,a_6) ,
\ea
or equivalently
\begin{equation}
\left\{
\begin{array}{rl}
x_1 &= x_1+a_1+a_2\\
x_2 &= x_2+a_3+a_4\\
-x_3 &= x_3+a_5+a_6\\
-x_4 &= x_4-a_1+a_2\\
-x_5 &= x_5-a_3+a_4\\
-x_6 &= x_6-a_5+a_6\,\,.\\
\end{array}\right.
\end{equation}
The first two equations restrict some of the
coefficients by requiring the equivalence of fixed points
(see eq.(\ref{generators})) 
 $a_1+a_2 = a_3+a_4=0$(mod2).
We can distinguish several cases 
which give different solutions for the $x_i$ coordinates
\ba
\bullet&& a_1+a_2=0 ; a_3+a_4=0\label{a1}\,\,,\\
\bullet&& a_1+a_2=2 ; a_3+a_4=0\label{a2}\,\,,\\
\bullet&& a_1+a_2=0 ; a_3+a_4=2\label{a3}\,\,,\\
\bullet&& a_1+a_2=2 ; a_3+a_4=2\label{a4}\,\,.
\ea
If we take the case (\ref{a1}), 
for example, we would get 
\begin{equation*}
\left\{
\begin{array}{rl}
x_1 &= x_1\\
x_2 & = x_2\\
-2x_3 &= a_5+a_6\\
-2x_4 &= 2a_2\\
-2x_5 &= 2a_4\\
-2x_6 &= -a_5+a_6\,\,.\\
\end{array}\right.
\end{equation*}
Let us consider initially the case $a_1$=$a_2$=$a_3$=$a_4$=0, which implies
 $(x_4,x_5)=(0,0)$.
We are left with the equations
\begin{equation}\label{eq.}
\left\{
\begin{array}{rl}
-2x_3 &= a_5+a_6\\
-2x_6 &= -a_5+a_6\\
\end{array}\right.
\end{equation}
which means looking for all 
$(x_3,x_6)\in[0,2]$ such that $a_5,a_6$ are integers.
The possible options are  $(x_3,x_6)\in \{0,1/2,1,3/2\}$. 
We note here that $3/2 \sim -1/2 $(mod2).
It is easy to verify that the complete 
set of  solutions is given by
\[(x_3,x_6)=(0,0),(\underline{0,1}),(1/2,1/2),(1,1),
(\underline{1/2,3/2}),(3/2,3/2) ,\] 
where the underline script indicates any solution
 obtained by swapping the entries. 
We can finally collect the results corresponding to the first case analysed
 and write down the fixed tori
\begin{equation}
\begin{array}{rl}
T_1 = (x_1,x_2, 0,0,0,0);& T_5 = ( x_1,x_2, 1,0,0,1);\\
T_2 = (x_1,x_2, 1,0,0,0);& T_6 = ( x_1,x_2, 0,0,0,1);\\
T_3 = (x_1,x_2, 1/2,0,0,1/2);& T_7 = ( x_1,x_2, 3/2,0,0,3/2);\\
T_4 = (x_1,x_2, 1/2,0,0,3/2);& T_8 = ( x_1,x_2, 3/2,0,0,1/2).\\
\label{corre}
\end{array}
\end{equation}
Among these solutions we have to select only the independent 
ones, since there are identifications up to shift lattices:
\beq
\begin{array}{rl}
 T_1 = T_5 + e_5 \,\,&,\quad T_2 = T_6 + e_5\,\,\,\,, \\
 T_3 = T_7 + e_6\,\,&,\quad T_4 = T_8 + e_6\,\,\,\,. \\
\end{array}
\eeq
The total independent fixed tori are then $T_1,T_2,T_3,T_4$ as shown in eqs.(\ref{1t2}-\ref{4t2}).

If we apply the same procedure for
the other cases
in eqs.(\ref{a2}--\ref{a4})
we notice that the solutions are redundant, reproducing
equivalent fixed tori.
For instance, it is straightforward to check that eq.(\ref{a4}) may
fix the constants $a_1$=$a_2$=$a_3$=$a_4$=1 which provides the solutions
in eqs.(\ref{corre}).
Furthermore, we notice that for each case 
in (\ref{a1}-\ref{a4}) there are several choices to
fix the constants $a_i$. For example,
eq.(\ref{a1}) can fix $a_1$=1, $a_2$=$-1$, $a_3$=1, $a_4$=$-1$.
This choice provides $(x_4, x_5)$=$(1,1)$=$(0,0)+e_2+e_4$,
 yielding exactly the same solutions obtained for the choice
$a_1$=$a_2$=$a_3$=$a_4$=0.
An analogous calculation has been performed
 for the twisted sectors $\theta_1$ and $\theta_3$,
whose results are shown in eqs.(\ref{1t1}--\ref{4t1}, \ref{1t3}--\ref{4t3}).

In the analysis of the twisted sectors, the string states 
arising at the fixed points in general do provide a generation 
(or anti-generation) of fermions of the Standard Model,
after the breaking of the gauge symmetry group into the
Standard Model gauge group. For instance,
if the gauge bosons of the model provide an $E_6$ symmetry,
a generation is identified by the supermultiplet
which falls into the ${\bf{27}}$ representation of $E_6$,
while a ${\bf{\overline{27}}}$ would indicate the anti-generation (the choice
of generation/anti-generation w.r.t. the 
representation is a matter of convention).
We are interested in the net number of generations for our model,
thus we need to know what each fixed torus gives rise to.
Let us consider a fixed torus under $\theta_i$.
 If the fixed points of this torus under the action 
of $\theta_j$, where $i\ne j$, are mapped onto points
of the same torus, then that torus provides a generation.
In the case where this torus is mapped onto a different 
fixed torus in the same twisted sector, then these two tori 
give a generation and an anti-generation.
By applying this reasoning to each fixed torus, we get a total number of
nine generations and three anti-generations (hence a net number
of six generations) in the twisted sector.  

\subsubsection{Calculation for the centraliser}

The analysis of the compactification lattice 
proceeds with the calculation
of the centraliser. This information will provide 
 the projections 
under which the twisted states have to be invariant.
For brevity we give the details only for sector $\theta_2$, since
the calculation for the other ${\mathbb Z}_2\times {\mathbb Z}_2$ 
non-trivial elements $\theta_1$ and $\theta_3$
 is a straightforward modification of the following derivation.

The first step is to find the constructing element for each torus, which
 we call now $g=(\theta_2,\overline{e}_{inv})$. As mentioned
in the introductory part, the centraliser is the set of
all elements $h=(\theta_i,\sum a_i e_i)$ of
the orbifold group that commute with $g$.
This condition in the $\theta_2$ sector is translated by the formula 
\beq\label{cen}
 \sum a_i e_i - \theta_2 (\sum a_i e_i)= 
\overline{e}_{inv}- \theta_j ( \overline{e}_{inv})\,\,,\,\, j = 1,2,3.
\eeq
The invariant vector $ \overline{e}_{inv}$ is determined for each torus by the transformation
\ba
&&T_1=(x_1,x_2,0,0,0,0)\,\,\quad\underrightarrow{\theta_2}\,\,
(x_1,x_2,0,0,0,0)+ \overline{e}_{inv.1}\quad\,\, \,(\overline{e}_{inv.1}= 0) ,
\nonumber\\
&&T_2=(x_1,x_2,1,0,0,0)\,\,\quad\underrightarrow{\theta_2}\,\,
(x_1,x_2,1,0,0,0)+ \overline{e}_{inv.2}\quad \,\,\,(\overline{e}_{inv.2}=
 e_5+e_6) ,\nonumber\\
&&T_3=(x_1,x_2,\frac{1}{2},0,0,\frac{1}{2})\quad\underrightarrow{\theta_2}\,\,
(x_1,x_2,\frac{1}{2},0,0,\frac{1}{2})+ \overline{e}_{inv.3}\quad\, (\overline{e}_{inv.3}=
 e_6) ,\nonumber\\
&&T_4=(x_1,x_2,\frac{1}{2},0,0,-\frac{1}{2})\,\underrightarrow{\theta_2}\,\,
(x_1,x_2,\frac{1}{2},0,0,-\frac{1}{2}) +\overline{e}_{inv.4}\,\,\,
(\overline{e}_{inv.4}=e_5) ,\nonumber
\ea
obtaining for each of the fixed four tori above the respective
constructing elements $g_1 = (\theta_2,0),\,\, g_2 = (\theta_2, e_5+e_6)
,\,\, g_3 = (\theta_2, e_6) ,\,\, g_4 = (\theta_2, e_5)$.
Let us see explicitly how we get the centraliser
for the torus $T_2$, for instance, by applying eq.(\ref{cen}). 
\beq\sum a_i e_i - \theta_2 (\sum a_i e_i)= 
e_5 + e_6 - \theta_j ( e_5 + e_6)\,\,,\,\,j=1,2,3\,\,,
\label{cece}\eeq will give the solutions for $\theta_1$
and $\theta_2$ :
$a_5=a_6=1$; $a_1=a_2$; $a_3=a_4$ and for the $\theta_3$
the set of solutions : $a_1=a_2$; $a_3=a_4$.
The centraliser is then determined by all possible
linear combinations of the previous constants w.r.t.
the correspondent twisted sector. The final result is shown below
\begin{eqnarray}
Z_{ \small{g_2 = (\theta_2, e_5+e_6)}}&=&\big\{h_1=(\theta_1,e_5+e_6),
h_2=(\theta_1,e_5+e_6+e_1+e_2),\nonumber\\
&&h_3=(\theta_1,e_5+e_6+e_3+e_4),
h_4=(\theta_1,e_5+e_6+e_1+e_2+e_3+e_4),\nonumber\\
&&h_5=(\theta_2,e_5+e_6) ,
h_6=(\theta_2,e_5+e_6+e_1+e_2),\nonumber\\
&&h_7=(\theta_2,e_5+e_6+e_3+e_4),
h_8=(\theta_2,e_5+e_6+e_1+e_2+e_3+e_4),\nonumber\\
&&h_9=(\theta_3,e_1+e_2),
h_{10}=(\theta_3,e_3+e_4),\nonumber\\
&&h_{11}=(\theta_3,e_1+e_2+,e_3+e_4)
\big\} .
\end{eqnarray}
None of these elements induce a 
projection on the states from the $T_2$ torus because of the consistency
conditions in section 4.3.1.
For the trivial torus $T_1$ there are obviously no
projections at all induced by the Wilson lines and this condition implies
that the transformation laws of the massless states of $T_1$
(and $T_2$ for the same reason) are determined under $\theta_1$
only. By analysing the centralisers of $T_3$ and $T_4$ we
see that the transformations under the $\theta_1$ sector 
are not defined, meaning that it is impossible to 
create invariant states by tensoring with the twisted 
right movers in $\theta_2$ sector 
(in fact we will show later on that these transform 
as $e^{\pm \frac{i\pi}{2}}$ under $\theta_1$).
This particular result depends completely on the 
choice of the compactification lattice. 

\subsubsection{Fixed points}

A different way to calculate the net number of generations
for a given model is to find the Euler
 number $\chi$ of the orbifold under investigation.
In case of the standard embedding, $\chi$ gives 
the number of generations multiplied by 2.
We are now interested to check the validity of
 our previous result
by determining $\chi$.
The Euler number is provided by  the formula 
\begin{equation}
\chi = \frac{1}{\left| G \right|} \sum_{[\theta_i,\theta_j] 
= 0} \chi_{\theta_i,\theta_j}\,\,,\,\,i,j=1,2,3 \,\,, 
\end{equation}
where $\left| G \right|$ is the order of the orbifold group (in this case 2)
 with
elements $\theta_i$, $\theta_j$ and $\chi_{\theta_i,\theta_j}$ is the number of points which are
simultaneously fixed under the action of $\theta_i$ and $\theta_j$.
Again we decide to consider only the $\theta_2$
twisted sector where each fixed torus will 
provide certain fixed points under 
the action of $\theta_1$ and $\theta_3$.
The condition
\[(T_1 - \theta_1 T_1) =(2x_1,2x_2,0,0,0,0)
\]
is satisfied by the four points
\beq\label{so1}(0,0,0,0,0,0),(1,0,0,0,0,0),(0,1,0,0,0,0),(1,1,0,0,0,0).
\eeq
For the fixed torus $T_2$ 
\[(T_2 - \theta_1 T_2) =(2x_1,2x_2,2,0,0,0) ,
\]
which is satisfied by the four points
\beq\label{so2}(0,0,1,0,0,0),(1,0,1,0,0,0),(0,1,1,0,0,0),(1,1,1,0,0,0).
\eeq
Finally,
\[(T_3 - \theta_1 T_3) =(2x_1,2x_2,1,0,0,0)
\]
has no solutions, such as the torus $T_4$.
The solutions (\ref{so1}--\ref{so2}) 
are invariant under $\theta_1$, 
obviously invariant under $\theta_2$ (since we are investigating
the fixed tori under $\theta_2$ sector).
Therefore, invariance under $\theta_3$ is guaranteed.
We have identified eight fixed points 
of $\theta_2$ sector under all the three 
twisted sectors so $\chi_{\theta_2,\theta_1}=8$.
In the same way we find  the other 
contributions  $\chi_{\theta_1,\theta_3}$ and
 $\chi_{\theta_2,\theta_3}$ which will totally give 
\beq
\chi = \frac{\chi_{\theta_2 ,\theta_1}+
\chi_{\theta_1 ,\theta_3}+\chi_{\theta_2 ,\theta_3}}{2} = \frac{3 \cdot 8}{2} . 
\eeq
The number of generations is then $ N = \chi/2 = 6$. 
This confirms our previous result on the net number of generations.

\subsection{Introduction of Wilson lines}\label{WLWL}

A Wilson line is a vacuum expectation value for an internal
gauge field component $A_i$, where the index labels the direction along
the lattice vectors (\ref{roots}).
As we have mentioned already, the maximum number of
independent Wilson lines depends on the compactification lattice
and for the $SO(4)^3$ skew case we get only three possible independent
Wilson lines that can be added.
The orbifold action on the vectors generating the
$SO(4)^3$ lattice (see table \ref{theta123})
provides the consistency condition for these Wilson lines
\begin{equation}
2 A_i , \,\,\, A_1 + A_2,\,\,\, A_3 + A_4,\,\,\, A_5 + A_6  \in
\Lambda_{\mbox{\tiny E$_8\times$E$_8$}} ,\,\,\, i = 1, \ldots , 6 .
\end{equation}
We note that the first condition
holds for any Wilson lines, for any lattice.

The effects of Wilson lines in the orbifold construction
are threefold.
First, the modular invariant conditions are more
restricted for the choice of the embedding $V^I$ and new constraints
are introduced.
Secondly, in the untwisted sector they introduce new projections,
breaking the gauge group. 
Finally, in the twisted sectors, the massless equations
change with respect to each fixed point and this provides
different left states from the case with no Wilson lines.
Moreover, the transformation laws
of these states change, accordingly to the formula
\beq
|p+kV+n_a A_a>_L\rightarrow e^{2\pi i(p+kV+n_a A_a)\cdot (lV+m_a A_a)}
|p+kV+n_a A_a>_L ,
\eeq
where the fixed point considered here
 is given by the constructing element 
$(\theta^k, n_a e_a)$ and the projection is performed under
the elements of the centraliser $h=(\theta^l, m_a e_a)$.
The last step in the derivation of the spectrum is
tensoring left-moving and right-moving states to obtain
invariant objects under the full space group. The modification
 introduced by the Wilson lines is that now the 
states have to be invariant under the centraliser, which is
a subset of $S$.
The particular choice of our compactification lattice
does not allow us to reduce the total number of generations
with the introduction of Wilson lines, as we explain in detail
at the end of the chapter. The other interesting implication
due to the presence of Wilson lines 
is the breaking of the symmetry group and
we will show how this is realised in a particular case. 
In \cite{vaudregange,Faraggi:2006bs}
the visible gauge group has  been broken
into $SO(10)$ or $SU(5)$ or into the Standard Model gauge group
$SU(3)\times SU(2)\times U(1)$ plus additional $U(1)$s. 
The nice breaking pattern is not enough to get semi-realistic 
orbifolds, since in fact in the previous examples many phenomenological
requirements could not be implemented.

In this section we show how the breaking of the hidden 
$E'_8\rightarrow SO(8)'\times SO(8)'$ 
is realised, in order to explain some technical details
regarding this sort of calculation. Few remarks on the choice of
Wilson lines are listed below.

$\bullet$ We note that Wilson lines with entries $\in\{0, \pm 1/2, \pm 1 \}$
break the initial gauge symmetry to $SO(2n)$ subgroups, 
while entries $\sim \pm 1/4$ produce $SU(n)$ algebras.

$\bullet$ If we want to break only the hidden (observable) sector,
the Wilson lines have to have only non-zero entries in the second (first)
8 dimensional vector.

$\bullet$ Wilson lines containing a single entry equal to $1$
 project the spinorial
roots (\ref{xx}) or (\ref{zz}) in the untwisted sector
by the projection condition $p^I\cdot A^I=0$(mod 1).

$\bullet$ The modular invariant conditions in eqs.(\ref{modcon}) have to
hold for any choice of Wilson lines. 

Keeping in mind the previous observations,
we proceed by introducing 
the following Wilson lines 
\begin{eqnarray}
A_1  = A_2 & = & \left( 0^8 \right) \left(\left( \frac{1}{2}\right)^{4}, 1,0,0,0\right) ,\nonumber \\
A_3 = A_4 & = & \left( 0^8\right) \left( 1,1,0^6 \right)  ,\nonumber
\\ 
A_5  =A_6 & = & \left( 0^8\right)\left( 1,0,0,0,-\frac{1}{2} ,\left(\frac{1}{2}\right)^3\right) .\label{wi} 
\end{eqnarray}
It is easy to verify their modular invariance:
$V_{1,2}\cdot A_{\alpha}= 0$(mod 1);
 $A_{\alpha}\cdot A_{\beta}= 0$(mod 1), $\alpha \ne \beta$ ;
$A_{\alpha}^2 = 0$(mod 2), $\a,\b=1,..6$.

Each $SO(8)$ factor has rank four, thus the total initial rank
is not reduced. We know how many roots to expect for the
algebra of each $SO(8)$ by using the relation
\[D_{SO(8)}- R_{SO(8)} =  T.R._{SO(8)}\rightarrow 28 - 4 = 24 ,\]
where $D$ is the dimension of the group, $R$ its rank
and $T.R.$ the number of total root weights.
By applying the projections induced by the Wilson lines on
the initial roots of $E_8'$, in eqs.(\ref{z}) and (\ref{zz}),
only the following roots survive
\beq
p^I=(0^8)(\underline{\pm 1, \pm 1 , 0 , 0} , 0 , 0 , 0 , 0)\,\,,\,\,
p^I=(0^8)(0 , 0 , 0 , 0 ,\underline{\pm 1 ,\pm 1 , 0 , 0} )\,\,,
\eeq
providing in fact the algebra of a $SO(8)\times SO(8)$ group.

\subsection{Massless Spectrum}

The massless spectrum of the model 
is produced by the solutions of the eqs.(\ref{qq}) and (\ref{rr})
 in the 
untwisted and in the twisted sectors respectively.
An invariant solution is obtained by tensoring 
right- and left-moving solutions which
survive the orbifold projections.

\subsubsection{Untwisted spectrum}

The untwisted massless spectrum is derived by 
solving equation (\ref{qq}). Subsequently,
we have to look at the invariant states 
under the action of the orbifold group 
of ${\mathbb Z}_2\times {\mathbb Z}_2$, where a generic element
 is indicated by $G=(\theta_i, n_a e_a ;V_i^I,n_aA_a^I)$.
We write explicitly the definitions of the oscillator 
number operators $N$ and $\tilde N$ in the Neveu Schwarz
 (NS) and in the Ramond (R) sector.
 We remind that the right sector is 
supersymmetric while the left one
only contains bosonic oscillators.
\begin{equation*}
\tilde N_{NS}=\sum_{n=1}^{\infty}\tilde\alpha^i_{-n}\tilde\alpha^i_{n}
+\sum_{n=1/2}^{\infty} n \tilde b^i_{-n}\tilde b^i_{n}-\frac{1}{2}\,\,\,\,\,,\,\,\,\,
\tilde N_{R}=\sum_{n=1}^{\infty}\tilde\alpha^i_{-n}\tilde\alpha^i_{n}
+\sum_{n=0}^{\infty} n \tilde d^i_{-n}\tilde d^i_{n} \,\,,
\end{equation*}
\begin{equation*}
 N=\sum_{n=1}^{\infty}\alpha^i_{-n}\alpha^i_{n}+\alpha^I_{-n}\alpha^I_{n}-1 \,\,,
\end{equation*}
where we have called $\tilde d^i$ the fermionic oscillators
in the Ramond sector and included
the values of $a_{L,R}$.

The total set of bosonic and fermionic 
oscillators which can transform under the 
orbifold action is given below in the light-cone gauge.
For brevity we drop the tilde on the right 
oscillators and differentiate the bosonic
left and right oscillators with the label $L,R$ when needed. 
Moreover, the complex conjugate oscillators are indicated 
by a bar.
\[
\alpha^{\mu}_n \,\,\,,\,\,\,\alpha_n^i\,\,\,,\,\,\, \overline\alpha_n^i\,\,\,
,\,\,\,b_{-n}^{\mu}\,\,,\,\,\,b_{-n}^i\,\,\,,\,\,\,\overline b_{-n}^i\,\,\,,\,\,\,d_n^{\mu}\,\,\,
,\,\,\,d_{-n}^i\,\,\,,\,\,\,\overline d_{-n}^i.
\]
It is convenient to use here 
the complex notation $Z^i$ for the bosonic and $\Psi^i$ for the fermionic 
coordinates in the compact dimensions, $i=1,2,3$. As anticipated before, 
the transformation properties
 for these oscillators in the compact dimensions are
\[Z^i\rightarrow e^{2 \pi i v_i}Z^i\,\,\,\,,\,\,\,\, \Psi^i \rightarrow e^{2\pi i v_i}\Psi^i .\]
If we consider only the 
massless contributions, we obtain the terms
\[b_{-1/2}^{\mu}\,\,,\,\,\,b_{-1/2}^i\,\,\,,\,\,\,\overline b_{-1/2}^i\,\,\,,\,\,\,d_0^{\mu}\,\,\,
,\,\,\,d_{0}^i\,\,\,,\,\,\,\overline d_{0}^i\,\,\,,\,\,\,\alpha_{-1}^{\mu}\,\,\,,\,\,\,\alpha_{-1}^{i}\,\,\,,\,\,\,\overline\alpha_{-1}^{i}\,\,.
\]
The
right moving solutions are obtained from the massless
equation (\ref{qq}). The  
correspondence between $SO(8)$ weight roots and oscillators is given 
in the table below, where the transformation 
laws under $\theta_1$ and $\theta_2$ are also provided.
\begin{center}
\beq
\begin{tabular}{|c|c|c|c|}\hline
Right Oscillator & Weight & $\theta_1 $ &$\theta_2 $\\
\hline
$b_{-1/2}^{\mu=1,2}$ & $(\pm 1,0,0,0)$ & $1$ & $1$\\
\hline
$b_{-1/2}^{i=1}$ & $(0,1,0,0)$ & $e^{i\pi}$ & $1$\\
\hline
$b_{-1/2}^{i=2}$ & $(0,0,1,0)$ & $e^{-i\pi}$ & $e^{i\pi}$\\
\hline
$b_{-1/2}^{i+3}$ & $(0,0,0,1)$ & $1$ & $e^{-i\pi}$\\
\hline
\hline
$d_{0}^{\mu=1,2}$ & $\pm(1/2, 1/2, 1/2, 1/2)$ & $1$ & $1$\\
\hline
$d_{0}^{i=1}$ & $(1/2,-1/2,1/2,1/2)$ & $e^{-i\pi}$ & $1$\\
\hline
$d_{0}^{i=2}$ & $(1/2,1/2,-1/2,1/2)$ & $e^{i\pi}$ & $e^{-i\pi}$\\
\hline
$d_{0}^{i=3}$ & $(1/2,1/2,1/2,-1/2)$ & $1$ & $e^{i\pi}$\\
\hline
\end{tabular}
\label{tableoscillators}\eeq
\end{center}

The phases of $\alpha_{-1}^{\mu}\,\,\,,\,\,\,\alpha_{-1}^{i}\,\,\,,\,\,\,
\overline\alpha_{-1}^{i}\,\,$ in the right 
and in the left sector are analogous to 
$b_{-1/2}^{\mu}\,\,,\,\,\,b_{-1/2}^i\,\,\,,\,\,\,
\overline b_{-1/2}^i$.
The oscillators of the gauge 
degrees of freedom $\alpha_{-1}^{I=1}$ 
are invariant under the action of the twists.
The correspondent complex oscillators 
transform obviously with opposite phases. 
In the left sector 
the solutions of the massless equation can be oscillators
and momenta $p^I$,
 roots of $E_8\times E'_8$ lattice.
The orbifold projection for the $p^I$ is given by  
\beq
G(p)= e^{2i\pi p^I.V^I}=1 .
\label{proj}
\eeq
Its solutions give rise to the gauge bosons which
describe the symmetry of the theory.
Solutions $p^I$ which pick a phase under 
the previous projection can still survive the total projection
of the orbifold when they are tensored with non invariant right states,
transforming with opposite phase w.r.t. the left contribution.
These are the charged matter states.
We show now how the projections produce 
the bosons of the unbroken gauge group.
The roots of the $E_8\times E_8'$ lattice are of the form
\ba
&\bullet& p^I=\underline{(\pm 1,\pm 1,0,0,0,0,0,0)}(0)^8\label{x}\\
&\bullet& p^I=(\pm 1/2,\pm 1/2,\pm 1/2,\pm 1/2,
\pm 1/2,\pm 1/2,\pm 1/2,\pm 1/2)(0)^8\label{xx}\\
&\bullet& p^I=(0)^8\underline{(\pm 1,\pm 1,0,0,0,0,0,0)}\label{z}\\
&\bullet& p^I=(0)^8(\pm 1/2,\pm 1/2,\pm 1/2,\pm 1/2,\label{zz}
\pm 1/2,\pm 1/2,\pm 1/2,\pm 1/2)
\ea
The roots (\ref{x}--\ref{xx}) produce 
the observable sector while the vectors (\ref{z}--\ref{zz})
 give the hidden sector.
In the standard embedding the shift vectors $V_1$ and $V_2$ are 
\[V_1= (1/2,-1/2,(0)^6)(0)^8\,\,\,,\,\,\,V_2= (0, 1/2,-1/2,(0)^5)(0)^8 ,
\]
hence it is straightforward that no roots are projected 
out for the hidden sector, when applying condition (\ref{proj}).
The surviving roots from the observable sector are instead
\[p^I = (0,0,0,{\underline{\pm 1,\pm 1,0,0,0}})(0)^8\,\,,
\]\[
p^I=((1/2)^3(\pm 1/2)^5)(0)^8 ({{even}}-)
\,\,,\,\,p^I=((-1/2)^3(\pm 1/2)^5)(0)^8 ({{odd}}-) .
\]
Finally, we have obtained 
$240$ invariant roots for the hidden $E_8$ and $72$
invariant roots for the observable sector. The last ones represent 
the weight vectors of the exceptional Lie group $E_6$,
as it is derived by the analysis of the simple roots \cite{Slansky:1981yr}.
From the complete set of roots at hand, only fourteen are 
simple roots, corresponding to a rank $14$ algebra.
However, the rank $16$ of the gauge group is not reduced,
meaning that there are two additional $U(1)$ symmetries. 
The final gauge group is \[E_6\times U(1)^2\times E_8' .\]
In the table below we list all the invariant momenta 
and the matter states 
with their transformation laws.

\begin{center}
\begin{tabular}{|c|c|c|}
\hline
$P^I$ & $\theta_1 $& $\theta_2 $ \\
\hline
$(0,0,0,\underline{\pm 1,\pm 1,0,0,0})(0)^8$ & $1$ & $1$\\
\hline
 $((1/2)^3,(\pm 1/2)^5)(0)^8$,$({\text{even}}-)$ & $1$ & $1$\\
\hline
 $((-1/2)^3,(\pm 1/2)^5)(0)^8$,$({\text{odd}}-)$ & $1$ & $1$\\
\hline
$(a)(\pm 1,0,0, \underline{\pm 1,0,0,0,0})(0)^8$ & $e^{\pm i\pi}$ & $1$\\
\hline
$(b)(0,\pm 1,0, \underline{\pm 1,0,0,0,0})(0)^8$ & $e^{\pm i\pi}$ & $e^{\pm i\pi}$\\
\hline
$(c)(0,0,\pm 1, \underline{\pm 1,0,0,0,0})(0)^8$ & $1$ & $e^{\pm i\pi}$\\
\hline
$(a)(0,\pm 1,\pm 1, (0)^5)(0)^8$  & $e^{\pm i\pi}$& $1$\\
\hline
$(b)(\pm 1,0,\pm 1, (0)^5)(0)^8$  & $e^{\pm i\pi}$& $e^{\pm i\pi}$\\
\hline
$(c)(\pm 1,\pm 1, (0)^6)(0)^8$  & $1$& $e^{\pm i\pi}$\\
\hline
$(a)(\pm (1/2,1/2,-1/2),(\pm (1/2)^5)(0)^8$ *  & $1$& $e^{\pm i\pi}$\\
\hline
$(b)(\pm (1/2,-1/2,1/2),(\pm (1/2)^5)(0)^8$ *  & $e^{\pm i\pi}$& $e^{\pm i\pi}$\\
\hline
$(c)(\pm (-1/2,1/2,1/2),(\pm (1/2)^5)(0)^8$ *  & $e^{\pm i\pi}$& $1$\\
\hline
\end{tabular}
\end{center}
The $*$ indicates that we have to consider an odd number of ``$-$''
for the last five entries if the first three entries have
a ``$+$'' sign in front, in the other case we take an even number
of $-1/2$ entries.
The untwisted massless spectrum is summarised below. 
\vspace{0.5cm}

\begin{tabular}{cccc}
Right mover & &Left mover & Particle \\
\hline
\vspace{0.3cm}
$(\pm 1,0,0,0)$&$\otimes$ &$(\pm 1,0,0,0)$ & $G_{\mu \nu}, B_{\mu \nu}, \phi$ \\
\vspace{0.3cm}
$\pm( 1/2,1/2,1/2,1/2)$ &$\otimes$&$(\pm 1,0,0,0)$ &$\Psi^{\alpha}_{\mu}$ + h.c. \\
\vspace{0.3cm}
$(\pm 1,0,0,0)$&$\otimes$ &$(0,0,0,\underline{\pm 1,\pm 1,0,0,0})(0)^8$ & 
$ $ \\
\vspace{0.3cm}
&$\otimes$ &$((-1/2)^3,(\pm 1/2)^5)(0)^8$ & $A_{\mu}  $ \\
\vspace{0.3cm}
&$\otimes$ &$((+1/2)^3,(\pm 1/2)^5)(0)^8$ & $ $ \\
\vspace{0.3cm}
$\pm (1/2,1/2,1/2,1/2)$&$\otimes$ &$(0,0,0,\underline{\pm 1,\pm 1,0,0,0})(0)^8$ 
& $ $ \\
\vspace{0.3cm}
&$\otimes$ &$((-1/2)^3,(\pm 1/2)^5)(0)^8$ & $\lambda^{\alpha}  $ \\
\vspace{0.3cm}
&$\otimes$ &$((+1/2)^3,(\pm 1/2)^5)(0)^8$ & $  $ \\
\hline
\vspace{0.3cm}
$b^1_{-1/2}\,\,,\,\,d^1_0$&$\otimes $&$\tilde\a_{-1}^1$ & \\
\vspace{0.3cm}
$b^2_{-1/2}\,\,,\,\,d^2_0$&$\otimes $&$\tilde\a_{-1}^2$ & \\
\vspace{0.3cm}
$b^3_{-1/2}\,\,,\,\,d^3_0$&$\otimes $&$\tilde\a_{-1}^3$ & \\
\hline
\vspace{0.3cm}
$b^1_{-1/2}\,\,,\,\,d^1_0$&$\otimes $&$(a)$ & \\
\vspace{0.3cm}
$b^2_{-1/2}\,\,,\,\,d^2_0$&$\otimes $&$(b)$ & \\
\vspace{0.3cm}
$b^3_{-1/2}\,\,,\,\,d^3_0$&$\otimes $&$(c)$ & \\
\end{tabular}

In the table above we have used an analogous notation
for left compact oscillators, see table \ref{tableoscillators},
 where, as usual, the first
entry of the vector corresponds to the complexified transverse
spacetime dimension, while the last three are the complex 
compact dimensions.
The first set of states provides the supergravity multiplet and the super Yang Mills multiplet,
the second gives rise to the moduli and the last provides 
\footnote{The total number
of states provided by the set (a), for instance,
is 56. We note that 27 of them transform with a
certain phase, the other 27 pick exactly the opposite phase,
indicating two opposite helicities. There are two singlets which 
we are neglecting at the moment. 
Each chiral hypermultiplet with its CPT partner
is combined to give one chiral supermultiplet
of $N=1$ supersymmetry in four dimensions.} 3 $(27,1)$ $\in E_6\times E_8'$
as the solutions are given in three combinations $a$, $b$ and $c$.

\subsubsection{Twisted sectors}

We derive the spectrum for one twisted sector only since 
the analysis is analogous in the other cases.
For instance, we solve the massless 
equations for the fixed tori of $\theta_2$ sector
\begin{eqnarray}
&T_1&: \frac{m_L^2}{4}=\frac{1}{2}(p+V_2)^2 + N -\frac{3}{4} ,\label{t1}\\
&T_2&: \frac{m_L^2}{4}=\frac{1}{2}(p+V_2+A_5+A_6)^2 + N -\frac{3}{4} ,\label{t2}\\
&T_3&: \frac{m_L^2}{4}=\frac{1}{2}(p+V_2+A_6)^2 + N -\frac{3}{4} ,\label{t3}\\
&T_4&: \frac{m_L^2}{4}=\frac{1}{2}(p+V_2+A_5)^2 + N -\frac{3}{4} ,\label{t4}
\end{eqnarray} 
while the right massless equation does not change for the different tori
and it has been presented in eq.(\ref{rr}). As we said, for the ${\mathbb Z}_2\times {\mathbb Z}_2$
orbifold $\delta_c$= $1/4$, thus the twisted right movers
have to be solutions of $(q+v_2)^2=1/2$.  
These solutions are showed in the
table below with their transformation properties
\vspace{0.5cm}

\begin{center}
\begin{tabular}{ccc}
Right movers & $\theta_1$& $\theta_2$ \\
\hline
\vspace{0.3cm}
$q_{1,sh}=(0,0,-1/2,-1/2)$&$e^{\frac{i\pi}{2}}$ & $1$ \\
\hline
\vspace{0.3cm}
$q_{2,sh}=(0,0,1/2,1/2)$&$e^{-\frac{i\pi}{2}}$ & $1$ \\
\hline
\vspace{0.3cm}
$\bar q_{1,sh}=(-1/2,1/2,0,0)$&$e^{\frac{i\pi}{2}}$ & $1$ \\
\hline
\vspace{0.3cm}
$\bar q_{2,sh}=(1/2,-1/2,0,0)$&$e^{-\frac{i\pi}{2}}$ & $1$ \\
\end{tabular}
\end{center}
\vspace{0.3cm}

where $\bar q_{1,sh}$ and $\bar q_{2,sh}$ correspond to the Ramond
shifted oscillators. We only consider the $q_{1,sh}$ solution tensored
with the left twisted states since 
the $q_{2,sh}$ is exactly the right contribution
of the correspondent antiparticles. In fact, we will find
a certain number of left states, solutions of the left massless
equation, which transform with opposite phase of $q_{1,sh}$, providing
invariant states. At the same time the same number of left movers
is present in the massless spectrum with opposite transformation
phases, giving invariants if combined with the $q_{2,sh}$.
Only one set of these solutions has to be considered,
as anticipated before.   
Moreover, we note that the spectrum is supersymmetric
since any left solution tensored with $q_{1,sh}$, for instance,
and providing an invariant state, is also automatically invariant
when multiplied by the Ramond right $\bar q_{1,sh}$. 

For completeness, we provide 
the right oscillators with their transformations
in the twisted sector $\theta_2$
\beq
\alpha^i_n \rightarrow \alpha^1_n ;\,\,\overline\alpha^1_n ;\,\,
\alpha^2_{n-1/2} ;\,\,\overline\alpha^2_{n+1/2} ;\,\,
\alpha^3_{n-1/2} ;\,\,\overline\alpha^3_{n+1/2} ;
\label{1111}
\eeq
\beq
\Psi^1_{n+\rho}\rightarrow \Psi^1_{n+\rho} ;\,\,\overline\Psi^1_{n-\rho} ;\,\,
 \Psi^2_{n-1/2+\rho} ;\,\,\overline\Psi^2_{n+1/2-\rho} ;\,\, \Psi^3_{n-1/2+\rho};\overline\Psi^3_{n+1/2-\rho} ;
\eeq
where $\rho=0$ in the Ramond case ($d^i$ oscillators)
 and $\rho=1/2$ in the NS ($b^i$ oscillators) respectively.
For the left compact oscillators 
we have analogous expressions to eqs.(\ref{1111}), while 
the oscillators for the gauge degrees 
of freedom do not transform under the twists.

We solve eq.(\ref{t1}) to obtain 
the contribution to the
massless spectrum from the trivial torus of $\theta_2$ sector. 
We distinguish two cases, when $N=0$ and when $N=1/2$.
A remark is to be done at this point. When looking for
the $p^I_{shift}$, not only we consider the roots (\ref{x}), 
(\ref{xx}), (\ref{z}) and (\ref{zz}), but also all their
linear combinations, as long as they still satisfy the massless
equation. Keeping this in mind, we obtain the following results:
for $N=0$, 56 
$p_{shift}^I$ are found, half of which 
take a phase $e^{\frac{i\pi}{2}}$ while 
the others transform with opposite phases 
(as explained before,  
only one set of
 these solutions is considered);
if $N=1/2$, only one $p_{shift}^I$ satisfies the massless equation.
In total the trivial torus provides the states in the table below.

\begin{center}
\begin{tabular}{|c|c|c|c|}
\hline
 Oscillators & $P^I_{shift}$ & Right oscillator & number of solutions \\
\hline
$N_L=0$&$(\pm 1, -1/2, -1/2, 0^5)(0^8)$& $q_{1,sh}$&2\\
&$(0, 1/2, 1/2, \underline{\pm 1, 0^4})(0^8)$ &$q_{1,sh}$&10\\
&$(-1/2, 0, 0, (\pm 1/2)^5)(0)^8$ even&$q_{1,sh}$&16\\
\hline
{{$N_L=1/2$}} : $\alpha^{3}_L$&$(0, 1/2, -1/2, 0^5)(0^8)$&$q_{1,sh}$& 1\\
{{$N_L=1/2$}} : $\alpha^{2}_L$&$(0, -1/2, 1/2, 0^5)(0^8)$&$q_{1,sh}$& 1\\
\hline
\end{tabular}
\end{center}
\vspace{0.3cm}
To identify the representations of the twisted states
we rewrite these weights as Dynking labels, with respect to 
$E_6$ and $SO(8)'\times SO(8)'$. Each multiplet is identified by grouping
the states with same $U(1)$ charges. If we indicate with $\alpha_i$,
$i=1,..6$, the simple roots of $E_6$, given in (\ref{appc}) in Appendix C,
 and with $\alpha_j$,
$j=9,..12$ and  $\alpha_k$,
$k=13,..16$ , the simple roots of the two $SO(8)$ gauge groups 
(we are not interested here 
in classifying the states under the hidden gauge group, since the 
potential standard model particles are singlets under it),
then for every root we need to calculate
\ba
p^I_{DL_{E_6}}&=&(\a_1\cdot p^I,\a_2\cdot p^I,\a_3\cdot p^I,
\a_4\cdot p^I,\a_5\cdot p^I,\a_6\cdot p^I,)_{Q_1,Q_2} ,\nonumber\\
p^I_{DL_{SO(8)_1'}}&=&(\a_9\cdot p^I,...,\a_{13}\cdot p^I) ,\quad
p^I_{DL_{SO(8)_2'}}=(\a_{13}\cdot p^I,...,\a_{16}\cdot p^I) ,
\ea
where the $Q_1$ and $Q_2$ charges are obtained by $Q_1=H_1-H_2$ and
 $Q_2=H_1+H_2-2H_3$.
This procedure is shown in Table c.1 in Appendix C.

We provide below the final result
for the contribution of the massless states for the
trivial fixed torus $T_1$, where the notation indicates
the representation of the 
multiplets under the gauge group $E_6\times SO(8)'\times
SO(8)'$ and the apex gives the $Q_{1,2}$ charges:
\ba 
&&\tilde N=\frac{1}{2}\quad (1,1,1)^{-\frac{1}{2},\frac{1}{2}} ,
\quad (1,1,1)^{\frac{1}{2},-\frac{3}{2}} ,\nonumber\\
&&\tilde N=0\quad (1,1,1)^{\frac{3}{2},-\frac{3}{2}} ,
\quad ({\bf{27}},1,1)^{-\frac{1}{2},-\frac{3}{2}} .
\ea
 This torus provides a generation 
under the $E_6$ gauge group.
By performing the same calculation for the other fixed
tori of $\theta_2$ we find out that the
$T_2$ torus provides exactly the same content of $T_1$,
and this is due to the property
$A_5+A_6 \in \Lambda_{16\times 16}$. 
If no Wilson lines are introduced in our model, we expect
eqs.(\ref{t3}) and (\ref{t4}) to reduce to (\ref{t1}),
providing a generation and an anti-generation, plus
a certain number of singlet states. Then, the total
contribution from $T_3$ and $T_4$ to the net number of generations is zero.
When we switch on the Wilson line $A_5$ we automatically 
get a huge change in both eqs.(\ref{t3}) and (\ref{t4}),
giving obviously the same contribution. In this case the choice 
of Wilson lines (\ref{wi}) only produces hidden charged states,
projecting the generations under the observable gauge group. 

To conclude,
the particular choice of our compactification lattice 
reduces the number of fixed tori to four par each twisted 
sector, providing a total number of nine generations (from the fixed 
tori (\ref{1t1}),(\ref{2t1}), (\ref{3t1}), (\ref{1t2}), 
(\ref{2t2}), (\ref{3t2}), (\ref{1t3}), (\ref{2t3}), (\ref{3t3}))
and three anti-generations (from the fixed tori 
(\ref{4t1}), (\ref{4t2}), (\ref{4t3})). We showed that,
independently on the choice of Wilson lines, there is no way to 
project out any of these generations. This is in fact a limitation
of the $SO(4)^3$ lattice.

\chapter{Construction of partition functions 
in heterotic $E_8\times E_8$ models}

In this chapter we discuss some examples of heterotic 
superstring models compactified on shift orbifolds.
In particular the cases presented are four dimensional
 shift orbifolds on which a ${\mathbb Z}_2$ or a
${\mathbb Z}_2 \times {\mathbb Z}_2$ projection acts on the internal tori.

As we explained in chapter 4, in the standard orbifold 
compactification the string coordinates are identified
under internal inversion operations, for instance the ${\mathbb Z}_2$ 
generators correspond to $\pi$ rotations. 
The shift orbifolds are instead 
created by the action of discrete shifts
on the basis vectors of the compactification lattice. The result
of this operation can lead to the implementation of
 the Scherk-Schwarz mechanism 
for the spontaneous supersymmetry breaking.
In quantum field theory the same mechanism 
is obtained by shifts on the
internal Kaluza Klein momenta \cite{Scherk:1978ta,Cremmer:1979uq},
while in string theory a more general
procedure is given when introducing momentum or winding shifts
along the compact directions 
\cite{Ferrara:1987es,Kounnas:1988ye,Kounnas:1989dk},
 while preserving modular invariance. 
The different choice for the two types of
 the shift will produce the so-called
 Scherk-Schwarz breaking or
 the M-theory breaking \cite{Antoniadis:1998ki,Antoniadis:1998ep}.

In this thesis we will consider the simple case of a 
one-dimensional momentum shift-orbifold with ${\mathbb Z}_2$ 
or ${\mathbb Z}_2 \times {\mathbb Z}_2$ action.

\section{Shift orbifold}\label{Z_-}

In this section we are interested in
 looking at a simple example of shift orbifold
realised in heterotic models. Thus, we 
start from the partition function of the 
 $E_8\times E_8$ heterotic string in 10 dimensions.
\beq
\mathcal{Z^+}_{E_{8}\times E_{8}} =(\overline{V_8} 
-\overline{S_8})(O_{16}+S_{16})(O_{16}+S_{16}).
\label{Z+}
\eeq
The next step is to compactify on a factorisable six torus 
of the form $T^2\times T^2 \times T^2$ and 
introduce the shift in one compact dimension $x^9$
\beq
\delta:\quad x^9\rightarrow x^9 +\pi R, \quad \delta^2 = 1.
\eeq  
The shift orbifold is generated by the elements
\[(1, \epsilon_1(-1)^{F_{\xi_1}}\delta, \epsilon_2(-1)^{F_{\xi_2}}
\delta, \epsilon_1\epsilon_2(-1)^{F_{\xi_1}+F_{\xi_2}} )=
(1,\epsilon_1a,\epsilon_2b,\epsilon_1\epsilon_2ab),\]
where $F_{\xi_1}$ is an internal
 fermion number in the sector describing the
 first  $E_8$
gauge group and  $F_{\xi_2}$
  is an internal fermion number
 in the sector describing the second  $E_8$ gauge group.
 The parameters $\epsilon_{1,2}\in
\{\pm 1\}$ lead to different models.
In this section we consider the case
with the group elements $(1, a, b, ab)$, where $\epsilon_{1,2}=1$,
 and show in detail
the derivation of the resulting partition function.

An other interesting case is when 
the group elements are given by
$(1,-a,b, -ab)$, obtained when $\epsilon_{1}=-1$ and $\epsilon_{2}=1$,
 and show in detail
the derivation of the resulting partition function.
This result will be presented 
briefly in section \ref{ls}.

Let us note first that the action of the previously introduced operators
on the lattice and on the $SO(2n)$ characters is given by 
\ba
\delta&:&\La_{m,n}\rightarrow(-1)^m\La_{m,n}\nonumber\\
(-1)^{F_{\xi_i}}&:&( O_{16}/ V_{16})_i \rightarrow 
( O_{16}/ V_{16})_i\,\,\nonumber\\
& &( S_{16}/ C_{16})_i \rightarrow 
(- S_{16}/- C_{16})_i\,,\,{(i=1,2)}. 
\label{action}
\ea
Now let us introduce the projection operator
\beq
\frac{1 \mp (-1)^{F_{\xi_1}}\delta}{2}\times \frac{1+(-1)^{F_{\xi_2}}\delta}{2}=
\frac{1}{4}\{1 \mp (-1)^{F_{\xi_1}}\delta+(-1)^{F_{\xi_2}}\delta \mp
(-1)^{F_{\xi_1}+F_{\xi_2}} \} ,
\label{projections}
\eeq
where the sign $+$  refers to the first case , and the  $-$ refers
to the second case.

The partition function in eq.(\ref{Z+}) 
after the compactification on the six-torus becomes
\beq
\mathcal{Z^+}_{E_{8}\times E_{8}} 
=(\overline{V_8} 
-\overline{S_8})\La_1\La_2\La_{m',n'}\La_{m,n}( O_{16}+ S_{16})( O_{16}+ S_{16}).
\label{Z+Z+}
\eeq
$\La_1$ and $\La_2$ are the two lattices for two-dimensional tori, while
the third two-torus has been factorised into two circles to facilitate
the implementation of the shift.
The full partition function which is obtained from eq.(\ref{Z+Z+}) and is 
invariant under the orbifold group (\ref{projections}) 
is given by
\beq
\mathcal{Z}_{tot}= \mathcal{Z}_{00}+\sum_{i}\mathcal{Z}_{0i}
+\sum_{i}(\mathcal{Z}_{i0}+\mathcal{Z}_{ii})
+ c_0 \sum_{i\ne j} \mathcal{Z}_{ij},
\label{total}
\eeq
where $i,j\in \{a,b,ab\}$ and
the constant $c_0$, called the discrete torsion, multiplies
a modular invariant orbit. 
The first two terms in eq.(\ref{total}) correspond to 
the total contribution of the
 untwisted sector of the orbifold and are given by
\ba
\mathcal{Z}_0&=&
\mathcal{Z}_{0,0}+ \mathcal{ Z}_{0,ab} +  \mathcal{Z}_{0,a} 
+  \mathcal{Z}_{0,b}=\nonumber\\
&=&\frac{1}{4}(\overline{V_8} -\overline{S_8})\La_1\La_2\La_{m',n'}\La_{m,n}[(O_{16}+S_{16})(O_{16}+S_{16})+(O_{16}-S_{16})(O_{16}-S_{16})\nonumber\\
&&+ (-1)^{m}\{(O_{16}-S_{16})(O_{16}+S_{16})+(O_{16}+S_{16})(O_{16}-S_{16}) \}].
\label{untwisted}
\ea
The term $\mathcal{ Z}_{0,ab}$ is obtained by acting on the 
eq.(\ref{Z+Z+}) with the operator $ab$, and  the third and the forth contributions 
($ \mathcal{Z}_{0,a} $, $\mathcal{Z}_{0,b}$) are 
respectively given by acting with operators $a$ and $b$ on the eq.(\ref{Z+Z+}).
In a similar way the last 
two terms in (\ref{total}) correspond to the twisted sector
which contributions have to be calculated.
In our model we choose the value of $c_0$ to be $+1$.
We can rewrite the untwisted sector 
(\ref{untwisted}) as
\beq
\mathcal{Z}_{0}=
\frac{1}{2} (\overline{V_8} -\overline{S_8})\La_1\La_2\La_{m',n'}
\La_{m,n} [(O_{16}O_{16}+S_{16}S_{16})
+(-1)^{m}(O_{16}O_{16}-S_{16}S_{16})],
\label{non}
\eeq
that can be rearranged, by using the formula (\ref{lambda}) 
in Appendix A, into the form
\beq
\mathcal{Z}_{0}=
 (\overline{V_8} -\overline{S_8})\La_1\La_2\La_{m',n'}
[\La_{2m,n}(O_{16}O_{16})+\La_{2m+1,n}(S_{16}S_{16})].
\eeq
The derivation of the twisted sector, neglecting
for the moment the torsion contribution, is given by 
the action of T and S transformations of each term in (\ref{non}). 
We illustrate the procedure with a schematic picture below. 
These terms are given in (\ref{D1}) in Appendix D. 
\ba
&&\overbrace{\mathcal{Z}_{0,ab}}^{\mathit{T\,\, invariant}}\,\,\,\,\,\,
\underleftrightarrow{S}\,\,\,\,\,\, 
\mathcal{Z}_{ab,0}\,\,\,\,\,\,
\underleftrightarrow{T}\,\,\,\,\,\, \overbrace{\mathcal{Z}_{ab,ab}}^{\mathit{S\, invariant} }\nonumber\\
&&\overbrace{\mathcal{Z}_{0,a}}^{\mathit{T\,\, invariant}}\,\,\,\,\,\,
\underleftrightarrow{S}\,\,\,\,\,\, 
\mathcal{Z}_{a,0}\,\,\,\,\,\,
\underleftrightarrow{T}\,\,\,\,\,\, \overbrace{\mathcal{Z}_{a,a}}^{\mathit{S\, invariant} }\nonumber\\
&&\overbrace{\mathcal{Z}_{0,b}}^{\mathit{T\,\, invariant}}\,\,\,\,\,\,
\underleftrightarrow{S}\,\,\,\,\,\, 
\mathcal{Z}_{b,0}\,\,\,\,\,\,
\underleftrightarrow{T}\,\,\,\,\,\, \overbrace{\mathcal{Z}_{b,b}}^{\mathit{S\, invariant} }\nonumber\\
&&\nonumber\\
&&\,\,\,\,\,\,\mathcal{Z}_{a,b}\,\,\,\,\,\,
\,\,\,\,\,\,\,\underleftrightarrow{T}\,\,\,\, 
\mathcal{Z}_{a,ab}\,\,\,\,\,\,
\underleftrightarrow{S}\,\,\,\,\,\,\,\,\,\,\mathcal{Z}_{ab,a}\nonumber\\
&&\,\,\,\,\,\,\,\,\,\,\updownarrow S\,\,\,\,\,\,\,\,\,\,\,\,\,\,\,\,\,\,\,\,\,\,\,\,\,\,\,\,\,\,\,\,\,\,\,\,\,\,\,\,\,\,\,\,\,\,\,\,\,\,\,\,\,\,\,\,\,\,\,\,\,\,\ \updownarrow T \nonumber\\
&&\,\,\,\,\,\,{\mathcal{Z}_{b,a}}\,\,\,\,\,\,
\,\,\,\,\,\,\underleftrightarrow{T}\,\,\,\,\,\, 
\mathcal{Z}_{b,ba}\,\,\,\,\,\,
\underleftrightarrow{S}\,\,\,\,\,\,\,\,\,\, {\mathcal{Z}_{ba,b}}
\ea
We note that 
the calculation of the terms which 
contribute to the torsion is more subtle since
 we have to define the way the projections act in a twisted sector, 
while preserving modular invariance.
If, for instance, we take the element $\mathcal{Z}_{a,0}$, its $a$ projection
would provide a different result w.r.t. the element
 $\mathcal{Z}_{a,a}$, obtained by
a T transformation of $\mathcal{Z}_{a,0}$. 
This means that we have to reproduce the same pattern of action
when the projector $b$ acts onto $\mathcal{Z}_{a,0}$.
The $b$ operator contains the shift $\delta$ which, in the twisted 
sector $a$ produces a change on the lattice equal to
 $(-1)^m \La_{m,n+1/2}$. The first group of 
gauge characters, which transforms accordingly to a T transform for the 
$\mathcal{Z}_{a,0}$ element, in the $b$ projection is untouched, while
 $b$ acts on the second set of characters in the usual way, as if
we are considering an untwisted element. The formula below summarises this 
procedure
\ba
\mathcal{Z}_{a,b}&=&\big[ (-1)^{F_{\xi_2}}\delta \big]_{a}\{(\overline{V_8} 
-\overline{S_8})\La_1\La_2\La_{m',n'}
\La_{m,n+1/2}[(V_{16}+C_{16})(O_{16}+S_{16})]  \}\nonumber\\
&=& (\overline{V_8} -\overline{S_8})(-1)^{m}\La_1\La_2\La_{m',n'}
\La_{m,n+1/2}[(V_{16}+C_{16})(O_{16}-S_{16})].
\label{zab}
\ea
At this point the remaining contributions are simply derived
 by an S and T transformation chain
\beq
\mathcal{Z}_{a,b}\,\,\,\,\underrightarrow{T}\,\,\,\,\mathcal{Z}_{a,ab}\,\,\,\,
\underrightarrow{S}\,\,
\,\,\mathcal{Z}_{ab,a}\,\,\,\,\underrightarrow{T}\,\,
\,\,\mathcal{Z}_{ab,b}\,\,\,\,\underrightarrow{S}\,\,
\,\,\mathcal{Z}_{b,ab}\,\,\,\,\underrightarrow{T}\,\,\,\,\mathcal{Z}_{b,a}.
\eeq
These expressions complete the list of terms
to get the full twisted sectors.

\underline{ab twisted sector}
\ba
&&\mathcal{Z}_{ab,0}+\mathcal{Z}_{ab,ab}+\mathcal{Z}_{ab,a}+\mathcal{Z}_{ab,b}=
\frac{1}{4} (\overline{V_8} -\overline{S_8})\La_1\La_2\La_{m',n'}
\La_{m,n}\nonumber\\
&&[(V_{16}+C_{16})(V_{16}+C_{16})
+(V_{16}-C_{16})(V_{16}-C_{16})]+\nonumber\\
&&c_0 (-1)^{m}
 [(-V_{16}+C_{16})(V_{16}+C_{16})+(V_{16}+C_{16})(-V_{16}+C_{16})].\nonumber\\
\ea

\underline{a twisted sector}
\ba
&&\mathcal{Z}_{a,0}+\mathcal{Z}_{a,a}+\mathcal{Z}_{a,b}+\mathcal{Z}_{a,ab}=
\frac{1}{4} (\overline{V_8} -\overline{S_8})\La_1\La_2\La_{m',n'}
\La_{{{m,n+1/2}}}\nonumber\\
&&[(V_{16}+C_{16})(O_{16}+S_{16})
+(-1)^{m}(-V_{16}+C_{16})(O_{16}+S_{16})]+\nonumber\\
&&c_0
 [ (-1)^{m}(V_{16}+C_{16})
(O_{16}-S_{16})+(-V_{16}+C_{16})(O_{16}-S_{16})].\nonumber\\
\ea

\underline{b twisted sector}
\ba
&&\mathcal{Z}_{b,0}+\mathcal{Z}_{b,b}+\mathcal{Z}_{b,a}+\mathcal{Z}_{b,ab}=
\frac{1}{4} (\overline{V_8} -\overline{S_8})\La_1\La_2\La_{m',n'}
\La_{m,n+1/2}\nonumber\\
&&[(O_{16}+S_{16})(V_{16}+C_{16})
+(-1)^{m}(O_{16}+S_{16})(-V_{16}+C_{16})]+\nonumber\\
&&c_0
 [(-1)^{m}(O_{16}-S_{16})(V_{16}+C_{16})+(O_{16}-S_{16})(-V_{16}+C_{16})].\nonumber\\
\ea
The S and T transformations used to derive the previous terms are given in
Appendix (\ref{S,T}).
Putting these results into (\ref{total}) we finally obtain 
\ba
\mathcal{Z_-}=&& (\overline{V_8}-\overline{S_8})\La_1\La_2\La_{m',n'}
\big[\La_{2m,n}(O_{16}O_{16}+C_{16}C_{16})+\La_{2m+1,n}(S_{16}S_{16}+V_{16}V_{16})\nonumber\\
&&+\,\, \La_{2m,n+\frac{1}{2}}(O_{16}C_{16}+C_{16}O_{16})+
\La_{2m+1,n+\frac{1}{2}}(V_{16}S_{16}+S_{16}V_{16})\big].\nonumber\\
\label{Z-}
\ea
At this level, the model presents $N=4$ supersymmetry in four dimensions
 and a 
$SO(16)\times SO(16)$ gauge group.
This model 
contains 
gravity and Yang Mills fields as its massless excitations.
In the next section we examine the $Z_2$ 
orbifold of (\ref{Z-}) and discuss 
its massless spectrum.
\newpage

\section{Partition function of the heterotic $E_8\times E_8$ shift orbifold superstring with ${\mathbb Z}_2$ action}\label{Z}

In this section we consider the ${\mathbb Z}_2$ orbifold of the 
partition function (\ref{Z-}). The model obtained by 
this further action
has $N=2$ supersymmetry in four dimensions
 and $SO(4)\times SO(12)\times SO(16)$ 
gauge symmetry.

The ${\mathbb Z}_2$ is generated by the elements $(1,h)$ where $h$ 
acts on the (complex) coordinates of the internal factorised torus
 $T^6=T^2\times T^2\times T^2$ as
\[Z_1\rightarrow e^{i\pi}Z_1\quad,\quad 
Z_2\rightarrow e^{i\pi}Z_2\quad,\quad
Z_3\rightarrow Z_3.
\]
We consider the standard embedding, thus the element $h$ 
acts non-trivially on the gauge degrees of freedom of the 
heterotic string as well.
For this reason it is convenient to decompose the $SO(2n)$ characters
in such a way to keep $O_4$, $V_4$, $S_4$ and $C_4$ factors 
(on which the element $h$ acts non-trivially) explicit.
The new partition function reads like
\beq
\mathcal{Z}_{Tot}= \mathcal{Z}_{00} + 
\mathcal{Z}_{0h} + \mathcal{Z}_{h0} + \mathcal{Z}_{hh},
\label{q}
\eeq
where $\mathcal{Z}_{00} $ is the untwisted 
term with no projection that corresponds 
exactly to (\ref{Z-}). The following 
term $\mathcal{Z}_{0h}$ is obtained by acting with $h$
onto the previous, while an S transformation produces the third term which, 
after a T transformation, provides $\mathcal{Z}_{hh}$.
If we decompose the characters by applying formula (\ref{decomp}),
 then the first term in (\ref{q}) becomes 
\ba
\mathcal{Z}_{00}=&& \frac{1}{4}\big[
\overline V_4 \overline O_4+\overline O_4 \overline V_4-\overline S_4 \overline S_4-\overline C_4 
\overline C_4\big]
\,\La_1\,\La_2\,\La_{m',n'}\,\times\nonumber\\
&&\big[(\La_{2m,n}+\La_{2m,n+\frac{1}{2}})
(O_4O_{12}+V_4V_{12}+C_4S_{12}+S_4C_{12})(O_{16}
+C_{16})\nonumber\\
&&+(\La_{2m,n}-\La_{2m,n+\frac{1}{2}})
(O_4O_{12}+V_4V_{12}-C_4S_{12}-
S_4C_{12})(O_{16}-C_{16})\nonumber\\
&&+(\La_{2m+1,n}+\La_{2m+1,n+\frac{1}{2}})
(V_4O_{12}+O_4V_{12}+S_4S_{12}+C_4C_{12})
(V_{16}+S_{16})\nonumber\\
&&+ (\La_{2m+1,n}-
\La_{2m,n+\frac{1}{2}})
(V_4O_{12}+O_4V_{12}
-C_4C_{12}-S_4S_{12})(V_{16}-S_{16})
\big].\nonumber\\
\ea
The action of the twist, imposed by 
the ${\mathbb Z}_2$ action on the characters,
has to be 
consistent with worldsheet supersymmetry 
\cite{Antoniadis:1985az,Kawai:1986va,Dixon:1985jw,Dixon:1986jc} 
and can be shown explicitly by applying the properties
 of the $\te$-functions
 into the definitions of the characters in (\ref{Char}).
These properties hold for the spacetime degrees 
of freedom
\ba
&&O_4\rightarrow O_4 ,\hspace{1cm} V_4\rightarrow -V_4 ,\nonumber\\
&&S_4\rightarrow -S_4 ,\hspace{1cm} C_4\rightarrow C_4 ,
\ea
 and for the gauge degrees of freedom as well, 
\ba
O_{16}&=& O_4O_{12}+V_4V_{12} \rightarrow O_4O_{12}-V_4V_{12} , \nonumber\\
V_{16}&=& V_4O_{12}+O_4V_{12} \rightarrow -V_4O_{12}+O_4V_{12}, \nonumber\\
S_{16}&=& S_4S_{12}+C_4C_{12} \rightarrow -S_4S_{12}+C_4C_{12}, \nonumber\\
C_{16}&=& S_4C_{12}+C_4S_{12} \rightarrow -S_4C_{12}+C_4S_{12},
\ea
where we have used the $SO(4)\times SO(12)$ decomposition
of the $SO(16)$ characters.
We finally get
\ba
\mathcal{Z}_{0h}=&& \frac{1}{4}\big[
-\overline V_4 \overline O_4+\overline O_4 \overline V_4+\overline S_4 \overline S_4-\overline C_4 \overline C_4\big]
\La_{m',n'}|\frac{2 \eta}{\theta_2}|^4\times\nonumber\\
&&\big\{(\La_{2m,n}+\La_{2m,n+\frac{1}{2}})
(O_4O_{12}-V_4V_{12}+C_4S_{12}-S_4C_{12})(O_{16}+C_{16})\nonumber\\
&&+(\La_{2m,n}-\La_{2m,n+\frac{1}{2}})(O_4O_{12}-V_4V_{12}-C_4S_{12}+S_4C_{12})(O_{16}
-C_{16})\nonumber\\
&&+(\La_{2m+1,n}+\La_{2m+1,n+\frac{1}{2}})(-V_4O_{12}+O_4V_{12}-S_4S_{12}+C_4C_{12})
(V_{16}+S_{16})\nonumber\\
&&+ (\La_{2m+1,n}-\La_{2m,n+\frac{1}{2}})(-V_4O_{12}+O_4V_{12}
-C_4C_{12}+S_4S_{12})(V_{16}-S_{16})
\big\}.\nonumber\\
\label{11}
\ea
The twisted sector is obtained by performing S and T transformations on each term of the previous expression. 
In particular the S transformation of $\mathcal{Z}_{0h}$ gives $\mathcal{Z}_{h0}$ while the T transform of the
last one provides
the $\mathcal{Z}_{hh}$.
It is indicative at this point to show explicitly the procedure for at least the first contribution of (\ref{11}).

From section \ref{SSS} in Appendix A
 we get the following S transformation laws:
\ba
&&(O+V)_{4,12,16}\rightarrow (O+V)_{4,12,16}\quad,\quad(O-V)_{4,12,16}\rightarrow(S+C)_{4,12,16}
,\nonumber\\
&&(S-C)_{4,12}\rightarrow(-S+C)_{4,12}\quad\quad\,\,\,
,\quad(S-C)_{16}
\rightarrow(S-C)_{16},\nonumber\\
&&(O+C)_{16}\rightarrow(O+C)_{16},
\ea
where the indices refer to the characters of $SO(4)$,
$SO(12)$ and $SO(16)$ respectively, and provides
\ba
&&(-\overline V_4 \overline O_4+\overline O_4 \overline V_4+\overline S_4 \overline S_4-\overline C_4 \overline C_4)
=\frac{1}{2}[-(\overline O +\overline V)_{4,4}(\overline O-\overline V)_{4,4}
+(\overline O-\overline V)_{4,4}(\overline O+\overline V)_{4,4}\nonumber\\
&&+(\overline S+\overline C)_{4,4}(\overline S-\overline C)_{4,4}+(\overline S-\overline C)_{4,4}
(\overline S+\overline C)_{4,4}]
\underrightarrow{S} \frac{1}{2}
[-(\overline O+\overline V)(\overline S+\overline C)
+(\overline S+\overline C)(\overline O+ \overline V)\nonumber\\
&&-(\overline O-\overline V)(\overline S-\overline C)-
(\overline S-\overline C)(\overline O-\overline V)]_{4,4}= (-\overline O\overline S
-\overline V\overline C+\overline S\overline V +\overline C\overline O)_{4,4}.
\ea
In section (\ref{bos}) we will show the transformation laws of the bosonic
contributions and in Appendix A their modular transformations are presented.
By applying those expressions we can write 
\[\La_{m',n'}|\frac{2 \eta}{\theta_2}|^4\times
(\La_{2m,n}+\La_{2m,n+\frac{1}{2}})\underrightarrow{S}\La_{m',n'}
|\frac{2\eta}{\theta_4}|^4
\times \big( \La_{2m,n}+\La_{2m,n+\frac{1}{2}}\big).
\]
The gauge degrees of freedom contribution becomes 
\ba
&&(O_4O_{12}-V_4V_{12}+C_4S_{12}-S_4C_{12})=
\frac{1}{2}[(O+V)(O-V)+(O-V)(O+V)\nonumber\\
&&-(C+S)(C-S)+(C-S)(C+S)]_{4,12}
\underrightarrow{S} \frac{1}{2}[(O+V)(S+C)+(S+C)(O+V)\nonumber\\
&&+(O-V)(C-S)-(C-S)(O-V)]_{4,12}=(OC+VS+CV+SO)_{4,12},
\ea
where the notation has been explained before.

We now apply the same procedure to all the other terms contained
in $\mathcal{Z}_{0h}$, obtaining the expression
\ba
\mathcal{Z}_{h0}=&& \frac{1}{4}\big[
-\overline V_4 \overline C_4+\overline S_4 \overline 
V_4-\overline O_4 \overline S_4+\overline C_4 \overline O_4\big]
\La_{m',n'}\times 16|\frac{ \eta}{\theta_4}|^4\times\nonumber\\
&&\big\{(\La_{2m,n}+\La_{2m,n+\frac{1}{2}})
(V_4S_{12}+S_4O_{12}+O_4C_{12}+C_4V_{12})(O_{16}+C_{16})\nonumber\\
&&+(\La_{2m+1,n}+\La_{2m+1,n+\frac{1}{2}})
(O_4S_{12}+S_4V_{12}+V_4C_{12}+C_4O_{12})(V_{16}+S_{16})\nonumber\\
&&+(\La_{2m,n}-\La_{2m,n+\frac{1}{2}})
(-V_4S_{12}+S_4O_{12}-O_4C_{12}+C_4V_{12})(O_{16}-C_{16})\nonumber\\
&&+(\La_{2m+1,n}-\La_{2m+1,n+\frac{1}{2}})
(-O_4S_{12}+S_4V_{12}-V_4C_{12}+C_4O_{12})(V_{16}-S_{16})
\big\}.\nonumber\\
\ea
The calculation which provides $\mathcal{Z}_{hh}$ consists in 
applying the T transformation for each term in the above result.
As the procedure is analogous for every contribution, we 
show only the T action on the first term, which 
we reproduce here again 
\ba
&&(-\overline V_4 \overline C_4+\overline S_4 \overline V_4-\overline O_4 \overline S_4+\overline C_4 \overline O_4)
\La_{m',n'}\times 16|\frac{ \eta}{\theta_4}|^4\times(\La_{2m,n}+\La_{2m,n+\frac{1}{2}})\nonumber\\
&&(V_4S_{12}+S_4O_{12}+O_4C_{12}+C_4V_{12})(O_{16}+C_{16}).
\ea
In the formulae below we 
factorise a global sign obtained from the phase-prefactor of the 
$T$ transformation, given in (\ref{S,T})
\ba
&&T_{SO(4)}= e^{-i\pi/6}diag(1,-1,i,i),\nonumber\\
&&T_{SO(12)}= e^{-i\pi/2}diag(1,-1,-i,-i),\nonumber\\
&&T_{SO(16)}= e^{-2i\pi/3}diag(1,-1,1,1),\nonumber\\ 
&&X_8X_{16}X_{16}\quad \underrightarrow{T} 
\quad e^{i\pi/3} e^{-4\pi i/3} \quad X_8X_{16}X_{16}= - X_8X_{16}X_{16}.
\label{TTT}
\ea
The spacetime factors transform as
\[-\overline V_4 \overline C_4+\overline S_4 \overline V_4-\overline O_4 \overline S_4+
\overline C_4 \overline O_4\rightarrow i(-\overline V_4 \overline C_4+\overline S_4 
\overline V_4+\overline O_4 \overline S_4-\overline C_4 \overline O_4),\]
while the bosonic contribution is given by the expression below
 (here we are omitting the
transverse bosons) 
\[|\frac{2\eta}{\theta_4}|^4
\times(\La_{2m,n}+\La_{2m,n+\frac{1}{2}})\rightarrow|\frac{2\eta}{\theta_3}|^4
\times(\La_{2m,n}+\La_{2m,n+\frac{1}{2}}).\]
Finally, the contribution from the gauge degrees of freedom gives
\ba
&& V_4 S_{12}+S_4  O_{12}+O_4  C_{12}+ C_4 V_{12}\rightarrow -i(- V_4 S_{12}+O_4 C_{12}- S_4  O_{12}+ C_4 
V_{12}).\nonumber\\
\ea
We observe that the term $(O+C)_{16}\rightarrow (O+C)_{16}$
remains invariant also under T. The combination of these results
leads to the final expression
\ba
\mathcal{Z}_{hh}=&& \frac{1}{4}\big[
-\overline V_4 \overline C_4+\overline S_4 
\overline V_4+\overline O_4 \overline S_4-\overline C_4 \overline O_4\big]
\La_{m',n'}\times 16|\frac{ \eta}{\theta_3}|^4\times\nonumber\\
&&\big\{(\La_{2m,n}+\La_{2m,n+\frac{1}{2}})
(V_4S_{12}+S_4O_{12}-O_4C_{12}-C_4V_{12})(O_{16}+C_{16})\nonumber\\
&&+(\La_{2m+1,n}-\La_{2m+1,n+\frac{1}{2}})
(-O_4S_{12}-S_4V_{12}+V_4C_{12}+C_4O_{12})(-V_{16}+S_{16})\nonumber\\
&&+(\La_{2m,n}-\La_{2m,n+\frac{1}{2}})
(-V_4S_{12}+S_4O_{12}+O_4C_{12}-C_4V_{12})
(O_{16}-C_{16})\nonumber\\
&&+ (\La_{2m+1,n}+\La_{2m+1,n+\frac{1}{2}})
(O_4S_{12}-S_4V_{12}-V_4C_{12}+C_4O_{12})(-V_{16}-S_{16})
\big\}.\nonumber\\
\ea
\subsection{The bosonic contribution}
\label{bos}
In this section we present some useful details, used 
already in the construction of the untwisted and twisted
 heterotic ${\mathbb Z}_2$ partition function, concerning the
bosonic contribution to each of these sectors.

\underline{Untwisted sector}:
\[
\underbrace{\frac{1}{|\eta|^2}\times\frac{1}{|\eta|^2}}_{\mu=2,3\,\, \text{spacetime}}
 \underbrace{\times\frac{1}{|\eta|^2}\times\frac{1}{|\eta|^2}\times\frac{1}{|\eta|^2}
\times\frac{1}{|\eta|^2}\times\frac{1}{|\eta|^2}
\times\frac{1}{|\eta|^2}}_{6\,\,\,\,\text{compactified dimensions}},
\]
where 
\[
\frac{1}{|\eta|^2}=\frac{1}{\eta}\times\frac{1}{\overline{\eta}}.
\]
Now, since the $Z_2$ action gives
 $\frac{1}{\eta^2}\rightarrow \frac{2\eta}{\te_2}$ when a $e^{i\pi}$
 acts on each
complex dimension, we obtain
\[
\mathcal{Z}_{00} \rightarrow \frac{1}{\eta^8}\frac{1}{\overline{\eta}^8}\,\,\,\,\,\,\,;\,\,\,\,\,\,\,
\mathcal{Z}_{0h}\rightarrow
 \frac{4}{\eta^2 \te^2_2}\frac{4}{\overline{\eta}^2 \overline{\te}^2_2}.
\]

\underline{Twisted sector}:
It is sufficient to apply S and T transformations on the previous result
\[(\frac{4}{\eta^2 \te^2_2})\,\,\,\,\,\,\,
\underrightarrow {{S}}\,\,\,\,\,\,\, (\frac{4}{\eta^2
\te^2_4})\,\,\,\,\,\,\,\underrightarrow{{T}}\,\,\,\,\,\,\,
 (\frac{4}{\eta^2
\te^2_3}),
\]
for the right contribution, reminding that the analogous
result holds for the left bosonic part. Combining both sectors, the bosonic 
contribution resulting in the twisted contributions is given respectively by
\[
\mathcal{Z}_{h0} \rightarrow  \frac{4}{\eta^2 \te^2_4}
\frac{4}{\overline{\eta}^2 \overline{\te}^2_4}
\,\,\,\,\,\,\,;\,\,\,\,\,\,\,
\mathcal{Z}_{hh}\rightarrow
 \frac{4}{\eta^2 \te^2_3}\frac{4}{\overline{\eta}^2 \overline{\te}^2_3}\,\,.
\]
It is useful for later purposes
 to expand in powers of $q$ the right 
( and in $\bar q$ for the left) bosonic contribution. 
In first approximation
\ba
&&\mathcal{Z}_{00}\rightarrow\frac{1}{\eta^8}\sim
q^{-1/3}(1+8q+..)
\,,\,\,
\mathcal{Z}_{h0}\rightarrow\frac{1}{\eta^2\te_4^2}\sim
q^{-1/12}(1-4q^{1/2}+..),
\nonumber\\
&&\mathcal{Z}_{0h}\rightarrow \frac{4}{\eta^2 \te^2_2}\sim
q^{-1/3}(1+..)
\,\quad,\,\,\mathcal{Z}_{hh}\rightarrow\frac{1}{\eta^2\te_3^2}\sim
q^{-1/12}(1+4q^{1/2}+..).
\ea
\subsection{Spectrum}
We have now all the ingredients to provide the untwisted 
spectrum of our model, which is obtained by the following sum  
\ba
\mathcal{Z}_{00}+\mathcal{Z}_{0h}\sim \La_{m',n'}\times\Big[&\La_{2m,n}&\big\{\big(
\overline O_4 \overline V_4-\overline C_4 \overline C_4\big)\big[
(O_4O_{12}O_{16}+C_4S_{12}C_{16})\big]\nonumber\\
& &+(\overline V_4 \overline O_4-\overline S_4 \overline S_4\big)\big[
(V_4V_{12}O_{16}+S_4C_{12}C_{16})\big]\big\}\nonumber\\
&+\La_{2m,n+\frac{1}{2}}&\big\{\big(
\overline O_4 \overline V_4-\overline C_4 \overline C_4\big)\big[
(O_4O_{12}C_{16}+C_4S_{12}O_{16})\big]\nonumber\\
& &+(\overline V_4 \overline O_4-\overline S_4 \overline S_4\big)\big[
(V_4V_{12}C_{16}+S_4C_{12}O_{16})\big]\big\}\nonumber\\
&+\La_{2m+1,n}&\big\{\big(
\overline O_4 \overline V_4-\overline C_4 \overline C_4\big)\big[
(O_4V_{12}V_{16}+C_4C_{12}S_{16})\big]\nonumber\\
& &+(\overline V_4 \overline O_4-\overline S_4 \overline S_4\big)\big[
(V_4O_{12}V_{16}+S_4S_{12}S_{16})\big]\big\}\nonumber\\
&+\La_{2m+1,n+\frac{1}{2}}&\big\{\big(
\overline O_4 \overline V_4-\overline C_4 \overline C_4\big)\big[
(O_4V_{12}S_{16}+C_4C_{12}V_{16})\big]\nonumber\\
&&+(\overline V_4 \overline O_4-\overline S_4 \overline S_4\big)\big[
(V_4O_{12}S_{16}+S_4S_{12}V_{16})\big]\big\}\Big].\nonumber\\
\ea
As announced previously, we are interested 
in the massless states, whose expansion is provided below. The first 
two terms give the untwisted right contributions, the last four provide
the untwisted left massless terms.
\ba
\frac{\overline O_4 \overline V_4/\overline C_4 \overline C_4}{\overline \eta^8}&&\sim 
\overline  q^{1/3}\overline q^{1/12}(1+6 \overline q^{-1} +..)\overline q^{1/12}(4\overline q^{-1/2}+..)\sim4 \overline
q^0+..\nonumber\\
\frac{\overline V_4 \overline O_4/\overline S_4 \overline S_4}{\overline \eta^8}&&\sim 4\overline
q^0+..\nonumber\\
 \frac{O_4O_{12}O_{16}}{\eta^8}&&\sim q^{-1/3}(1+8q+..)q^{-1/12}(1+6q+..)q^{-1/4}(1+66q+..)
q^{-1/3}(1+120q+..)\nonumber\\
&& \sim 4q^{0}+6q^0+66q^0+120q^0+..\nonumber\\
\frac{V_4V_{12}O_{16}}{\eta^8}&&\sim q^{-1/3}q^{-1/12}(4q^{1/2}+..)q^{-1/4}(4q^{1/2}+..)
q^{-1/3}(1+120q+..)\nonumber\\
&& \sim 16 q^0 +..\nonumber\\
\ea
Summing up, in the untwisted sector
 one has $N=(1,0)$, $D=6$ SUGRA multiplet and the Yang--Mills 
multiplet, partially
projected by the ${\mathbb Z}_2$ action,
provided by the terms
\beq
 \frac{ O_4 O_{12} O_{16}}
{\eta^8} \times \frac{ \overline O_4 \overline V_4 - 
\overline C_4 \overline  C_4 }{\overline \eta^8}, \quad
 \frac{V_4 V_{12} O_{16}}{\eta^8}
 \times \frac{\overline V_4 \overline O_4 -
\overline S_4 \overline S_4 }{\overline \eta^8}.
\eeq
A similar calculation is performed for the twisted sector, where
we present the full spectrum 
\ba
\mathcal{Z}_{h0}+\mathcal{Z}_{hh}&\sim& \nonumber\\
&&\nonumber\\
\frac{16}{2}\La_{m',n'}&\times&
\Big[\La_{2m,n}\big\{\big[\big(
-\overline V_4 \overline C_4+\overline S_4 \overline V_4\big)
(V_4S_{12}C_{16}+S_4O_{12}O_{16})\nonumber\\
&&+(-\overline O_4 \overline S_4+\overline C_4 \overline O_4\big)
(O_4C_{12}C_{16}+C_4V_{12}O_{16})\big]\times
(\frac{1}{|\eta|^4|\te_4|^4}+\frac{1}{|\eta|^4|\te_3|^4})     \nonumber\\
&&+\big[\big(
-\overline V_4 \overline C_4+\overline S_4 \overline V_4\big)
(O_4C_{12}C_{16}+C_4V_{12}O_{16})\nonumber\\
&&+(-\overline O_4 \overline S_4+\overline C_4 \overline O_4\big)
(V_4S_{12}C_{16}+S_4O_{12}O_{16})\big]\times
(\frac{1}{|\eta|^4|\te_4|^4}-\frac{1}{|\eta|^4|\te_3|^4})\big\}\nonumber\\
&&\nonumber\\
&&+ \La_{2m,n+\frac{1}{2}}\big\{ \big[\big(
-\overline V_4 \overline C_4+\overline S_4 \overline V_4\big)
(V_4S_{12}O_{16}+S_4O_{12}C_{16})\nonumber\\
&&+(-\overline O_4 \overline S_4+\overline C_4 \overline O_4\big)
(O_4C_{12}O_{16}+C_4V_{12}C_{16})\big]\times
(\frac{1}{|\eta|^4|\te_4|^4}+\frac{1}{|\eta|^4|\te_3|^4})
\nonumber\\
&&+\big[\big(
-\overline V_4 \overline C_4+\overline S_4 \overline V_4\big)
(O_4C_{12}O_{16}+C_4V_{12}C_{16})\nonumber\\
&&+(-\overline O_4 \overline S_4+\overline C_4 \overline O_4\big)
(V_4S_{12}O_{16}+S_4O_{12}C_{16})\big]\times
(\frac{1}{|\eta|^4|\te_4|^4}-\frac{1}{|\eta|^4|\te_3|^4})\big\}
\nonumber\\
&&\nonumber\\
&&+ \La_{2m+1,n}\big\{\big[\big(
-\overline V_4 \overline C_4+\overline S_4 \overline V_4\big)
(S_4V_{12}V_{16}+V_4C_{12}S_{16})\nonumber\\
&&+(-\overline O_4 \overline S_4+\overline C_4 \overline O_4\big)
(O_4S_{12}S_{16}+C_4O_{12}V_{16})\big]\times
(\frac{1}{|\eta|^4|\te_4|^4}+\frac{1}{|\eta|^4|\te_3|^4})\nonumber\\
&&+\big[\big(
-\overline V_4 \overline C_4+\overline S_4 \overline V_4\big)
(O_4S_{12}S_{16}+C_4O_{12}V_{16})\nonumber\\
&&+(-\overline O_4 \overline S_4+\overline C_4 \overline O_4\big)
(S_4V_{12}V_{16}+V_4C_{12}S_{16})\big]\times
(\frac{1}{|\eta|^4|\te_4|^4}-\frac{1}{|\eta|^4|\te_3|^4})\big\}\nonumber\\
&&\nonumber\\
&&+ \La_{2m+1,n+\frac{1}{2}}\big\{\big[\big(
-\overline V_4 \overline C_4+\overline S_4 \overline V_4\big)
(S_4V_{12}S_{16}+V_4C_{12}V_{16})\nonumber\\
&&+(-\overline O_4 \overline S_4+\overline C_4 \overline O_4\big)
(O_4S_{12}V_{16}+C_4O_{12}S_{16})\big]\times
(\frac{1}{|\eta|^4|\te_4|^4}+\frac{1}{|\eta|^4|\te_3|^4})\nonumber\\
&&+\big[\big(
-\overline V_4 \overline C_4+\overline S_4 \overline V_4\big)
(O_4S_{12}V_{16}+C_4O_{12}S_{16})\nonumber\\
&&+(-\overline O_4 \overline S_4+\overline C_4 \overline O_4\big)
(S_4V_{12}S_{16}+V_4C_{12}V_{16})\big]\times
(\frac{1}{|\eta|^4|\te_4|^4}-\frac{1}{|\eta|^4|\te_3|^4})\big\}\Big].\nonumber\\
\ea
The twisted massless states form a $N=1$, $D=6$ half hypermultiplet
are provided by
\beq
 \frac{ C_4 V_{12} O_{16}}
{\eta^2\theta^2_4} \times \frac{\overline O_4
\overline  S_4 - \overline C_4 \overline O_4 }{\overline \eta^2\bar\theta_4^2}, \quad
 \frac{ S_4 O_{12} O_{16}}{\eta^2\theta^2_4}
 \times \frac{\overline O_4 \overline S_4 - \overline C_4\overline  O_4 }{\overline \eta^2\bar\theta^2_4} .
\label{aw}
\eeq
The relevant expansions for the massless contributions in eq.(\ref{aw})
 are presented below,
for the right and left contributions respectively
\ba
\frac{\overline O_4 \overline S_4/
\overline C_4 \overline O_4}{\overline \eta^2\te_4^2 }&&\sim 
\overline q^{1/12}\overline q^{1/12}(1+6 \overline q^{-1} +..)2\overline q^{-1/6}(1+
4\overline q^{-1}+..)\sim 2 \overline
q^0+..\nonumber\\
\frac{\overline V_4 \overline C_4/\overline S_4 \overline V_4}{\overline
 \eta^2\te_4^2}&&\rightarrow
{massive}\nonumber\\
\frac{C_4V_{12}O_{16}}{ \eta^2\te_4^2}&&\sim q^{-1/12}2q^{1/6}(1+4q+..)q^{-1/4}4q^{1/2}
q^{-1/3}(1+120q+..)
 \sim 8 q^{0} +..\nonumber\\
\frac{S_4O_{12}O_{16}}{ \eta^2\te_4^2}&&\sim q^{-1/12}4q^{1/2}2q^{1/6}(1+4q+..)q^{-1/4}(1+66q+..)
q^{-1/3}(1+120q+..)\nonumber\\
&& \sim 8 q^{0} +.. .
\ea

\section{A string model with no gravity}\label{ls}

It is interesting to consider a variation of 
the previous model, obtained by a the shift orbifold 
with group elements $(1,-a,b,-ab)$, with
the choice of the torsion constant $c_0=1$.
The modular invariant string theory derived 
from this orbifold action is characterised by the 
absence of the graviton in its full spectrum.
This result leads us 
to the possible interpretation 
of a little heterotic string, 
in connection with \cite{Losev:1997hx,Kutasov:2000jp,Giveon:1999pxGiveon:1999px,Kutasov:2001uf}.
Many interesting properties of this kind of model
can be investigated by the string thermodynamics
at nonzero temperature \cite{McClain:1986id,Atick:1988si,Deo:1988jj,Bowick:1992qu}.
The basic idea is to generalise the partition function
by adding the temperature dependence and 
obtaining (\ref{Z-}) and (\ref{Z_-}) as particular cases \cite{nostro}.

We quote the expression of the partition
function after the orbifold action, which is indicated
by $\mathcal{Z}'_-$ to distinguish 
from the partition (\ref{Z-}).
\ba
\mathcal{Z'}_-=&& (\overline{V_8}-\overline{S_8})\La_1\La_2\La_{m',n'}
\big[\La_{2m,n}(S_{16}O_{16}-C_{16}C_{16})+\La_{2m+1,n}(O_{16}S_{16}-
V_{16}V_{16})\nonumber\\
&&+\,\, \La_{2m,n+\frac{1}{2}}(S_{16}C_{16}-C_{16}O_{16})+
\La_{2m+1,n+\frac{1}{2}}(O_{16}V_{16}-V_{16}S_{16})\big].
\label{Z_-}
\ea
We observe that, although (\ref{Z_-}) is a modular invariant string vacuum,
it does not respect the spin-statistics, since the bosonic contributions
to the partition function should arise with positive terms 
while the fermionic contribute with negative terms. This 
principle obviously does not hold for the partition function (\ref{Z_-}). 
 
From the expression above one can see that the zero mass spectrum
of the model contains no gravity because of the absence of the term 
$O_{16}O_{16}$ in the left sector. Moreover the adjoint representation
is missing as well so the gauge group cannot be defined either.
We consider a ${\mathbb Z}_2$ action of the orbifold $\mathcal{Z'}_-$ and provide 
its spectrum.
The details of the techniques used for the ${\mathbb Z}_2$
projections have been widely explained in section \ref{Z}.

\underline{Untwisted sector}
\ba
\mathcal{Z'}_{00}+\mathcal{Z'}_{0h}\sim \La_{m,n}\times\Big[&\La_{2m,n}&\big\{\big(
\overline O_4 \overline V_4-\overline C_4 \overline C_4\big)\big[
(C_4C_{12}O_{16}-C_4S_{12}C_{16})\big]\nonumber\\
& &+(\overline V_4 \overline O_4-\overline S_4 \overline S_4\big)\big[
(S_4S_{12}O_{16}-S_4C_{12}C_{16})\big]\big\}\nonumber\\
&+\La_{2m,n+\frac{1}{2}}&\big\{\big(
\overline O_4 \overline V_4-\overline C_4 \overline C_4\big)\big[
(C_4C_{12}C_{16}-C_4S_{12}O_{16})\big]\nonumber\\
& &+(\overline V_4 \overline O_4-\overline S_4 \overline S_4\big)\big[
(S_4S_{12}C_{16}-S_4C_{12}O_{16})\big]\big\}\nonumber\\
&+\La_{2m+1,n}&\big\{\big(
\overline O_4 \overline V_4-\overline C_4 \overline C_4\big)\big[
(O_4O_{12}V_{16}-O_4V_{12}S_{16})\big]\nonumber\\
& &+(\overline V_4 \overline O_4-\overline S_4 \overline S_4\big)\big[
(V_4V_{12}V_{16}-V_4O_{12}S_{16})\big]\big\}\nonumber\\
&+\La_{2m+1,n+\frac{1}{2}}&\big\{\big(
\overline O_4 \overline V_4-\overline C_4 \overline C_4\big)\big[
(O_4O_{12}S_{16}-O_4V_{12}V_{16})\big]\nonumber\\
&&+(\overline V_4 \overline O_4-\overline S_4 \overline S_4\big)\big[
(V_4V_{12}S_{16}-V_4O_{12}V_{16})\big]\big\}\Big]. \nonumber\\
\ea 
The massless untwisted contributions are given by
\beq
 \frac{ C_4 C_{12} O_{16}}
{\eta^8} \times \frac{\overline O_4\overline  V_4 -
\overline C_4\overline  C_4 }{\overline \eta^8}, \quad
 \frac{ S_4 S_{12} O_{16}}{\eta^8}
 \times \frac{\overline V_4\overline  O_4 -
\overline S_4\overline  S_4 }{\overline \eta^8},
\eeq
since right and left contributions give
$$
 \frac{ C_4 C_{12} O_{16}}{\eta^8}
\sim 2^6 q^0  ,\quad  \frac{ S_4 S_{12} O_{16}}
{\eta^8} \sim 2^6  q^0 , 
$$
$$
\frac{\overline O_4\overline V_4  }{\overline \eta^8} \sim 4\overline q^0, \quad
\frac{\overline V_4 \overline O_4  }{\overline \eta^8} \sim 4\overline  q^0,$$
$$ 
\frac{\overline S_4\overline  S_4  }{\overline \eta^8} \sim 4\overline q^0, \quad
\frac{\overline C_4 \overline C_4  }{\overline \eta^8} \sim 4 \overline q^0. 
$$

From the expressions above one can read the content of the massless spectrum in terms of the 
six-dimensional $N=(1,1)$ (which gives in $D=4$ $N=4$ supersymmetry upon 
the dimensional reduction to four dimensions) SUSY multiplets. In particular one has $2^6$ massless $(1,1)$
multiplets, whose bosonic part contains one vector and four scalar fields.

We remind that this string solution is not physical, since the graviton
does not appear in the untwisted spectrum. However, it represents a 
consistent solution for its modular invariance. 
Thus, 
the question arises: what role this solution plays in the string theory,
if any?

For completeness, we proceed the calculation by presenting the twisted sector. 

\underline{Twisted sector}
\ba
\mathcal{Z'}_{h0}+\mathcal{Z'}_{hh}&\sim& \nonumber\\
&&\nonumber\\
\frac{16}{2}\La_{m',n'}&\times&
\Big[\La_{2m,n}\big\{\big[\big(
-\overline V_4 \overline C_4+\overline S_4 \overline V_4\big)
(V_4C_{12}O_{16}-V_4S_{12}C_{16})\nonumber\\
&&+(-\overline O_4 \overline S_4+\overline C_4 \overline O_4\big)
(O_4S_{12}O_{16}-O_4C_{12}C_{16})\big]\times
(\frac{1}{|\eta|^4|\te_4|^4}+\frac{1}{|\eta|^4|\te_3|^4})     \nonumber\\
&&+\big[\big(
-\overline V_4 \overline C_4+\overline S_4 \overline V_4\big)
(O_4S_{12}O_{16}-O_4C_{12}C_{16})\nonumber\\
&&+(-\overline O_4 \overline S_4+\overline C_4 \overline O_4\big)
(V_4C_{12}O_{16}-V_4S_{12}C_{16})\big]\times
(\frac{1}{|\eta|^4|\te_4|^4}-\frac{1}{|\eta|^4|\te_3|^4})\big\}\nonumber\\
&&\nonumber\\
&&+ \La_{2m,n+\frac{1}{2}}\big\{ \big[\big(
-\overline V_4 \overline C_4+\overline S_4 \overline V_4\big)
(V_4C_{12}C_{16}-V_4S_{12}O_{16})\nonumber\\
&&+(-\overline O_4 \overline S_4+\overline C_4 \overline O_4\big)
(O_4S_{12}C_{16}-O_4C_{12}O_{16})\big]\times
(\frac{1}{|\eta|^4|\te_4|^4}+\frac{1}{|\eta|^4|\te_3|^4})
\nonumber\\
&&+\big[\big(
-\overline V_4 \overline C_4+\overline S_4 \overline V_4\big)
(O_4S_{12}C_{16}-O_4C_{12}O_{16})\nonumber\\
&&+(-\overline O_4 \overline S_4+\overline C_4 \overline O_4\big)
(V_4C_{12}C_{16}-V_4S_{12}O_{16})\big]\times
(\frac{1}{|\eta|^4|\te_4|^4}-\frac{1}{|\eta|^4|\te_3|^4})\big\}
\nonumber\\
&&\nonumber\\
&&+ \La_{2m+1,n}\big\{\big[\big(
-\overline V_4 \overline C_4+\overline S_4 \overline V_4\big)
(C_4V_{12}V_{16}-C_4O_{12}S_{16})\nonumber\\
&&+(-\overline O_4 \overline S_4+\overline C_4 \overline O_4\big)
(S_4O_{12}V_{16}-S_4V_{12}S_{16})\big]\times
(\frac{1}{|\eta|^4|\te_4|^4}+\frac{1}{|\eta|^4|\te_3|^4})\nonumber\\
&&+\big[\big(
-\overline V_4 \overline C_4+\overline S_4 \overline V_4\big)
(S_4O_{12}V_{16}-S_4V_{12}S_{16})\nonumber\\
&&+(-\overline O_4 \overline S_4+\overline C_4 \overline O_4\big)
(C_4V_{12}V_{16}-C_4O_{12}S_{16})\big]\times
(\frac{1}{|\eta|^4|\te_4|^4}-\frac{1}{|\eta|^4|\te_3|^4})\big\}\nonumber\\
&&\nonumber\\
&&+ \La_{2m+1,n+\frac{1}{2}}\big\{\big[\big(
-\overline V_4 \overline C_4+\overline S_4 \overline V_4\big)
(C_4V_{12}S_{16}-C_4O_{12}V_{16})\nonumber\\
&&+(-\overline O_4 \overline S_4+\overline C_4 \overline O_4\big)
(S_4O_{12}S_{16}-S_4V_{12}V_{16})\big]\times
(\frac{1}{|\eta|^4|\te_4|^4}+\frac{1}{|\eta|^4|\te_3|^4})\nonumber\\
&&+\big[\big(
-\overline V_4 \overline C_4+\overline S_4 \overline V_4\big)
(S_4O_{12}S_{16}-S_4V_{12}V_{16})\nonumber\\
&&+(-\overline O_4 \overline S_4+\overline C_4 \overline O_4\big)
(C_4V_{12}S_{16}-C_4O_{12}V_{16})\big]\times
(\frac{1}{|\eta|^4|\te_4|^4}-\frac{1}{|\eta|^4|\te_3|^4})\big\}\Big] .\nonumber\\
\ea

The massless twisted contributions are given by 
\beq
 \frac{ O_4 S_{12} O_{16}}
{\eta^2 \theta_4^2} \times \frac{\overline O_4\overline  S_4 -\overline C_4\overline  O_4 }
{\overline\eta^2\bar\theta^2_4} ,
 \eeq
since right and left contributions give
$$
 \frac{C_4 C_{12} O_{16}}
{\eta^2\theta_4^2}\sim 2^5  q^0, \quad  
\frac{\overline O_4 \overline S_4  }{\overline\eta^2\bar\theta^2_4} \sim 2 \overline q^0, \quad
\frac{\overline C_4 \overline O_4  }{\overline \eta^2\bar\theta^2_4} \sim 2  \overline q^0. 
$$
As it was for the case of the untwisted sector, one can group the massless spectrum in terms of six dimensional supersymmetry
multiplets. In the twisted sector  one has $D=6$ $N=1$ supersymmetry (which gives in $D=4$ $N=2$ upon
the reduction to four dimensions)  and the massless spectrum  forms  $2^5$ half-hypermultiplets.


\section{Supersymmetric ${\mathbb Z}_2 \times {\mathbb Z}_2$ shift orbifold model}\label{ZZ}
In this section we present the 
partition function for 
the shift orbifold (\ref{Z-})
 with the action of ${\mathbb Z}_2 \times {\mathbb Z}_2$ orbifold. 
The main difference w.r.t.
the case treated in section \ref{Z} is that the spectrum is not 
anymore completely determined by the modular invariance of one-loop
torus amplitude and an ambiguity is present when 
projecting the twisted sectors. 

The fact that many 
different choices (consistent with modular invariance) can be made   
is described by a phase $\epsilon$,
 called the discrete torsion, which disconnects 
the modular orbits \cite{Vafa:1986wx,Vafa:1994rv}.
An analogous situation was presented in the derivation of eq.(\ref{zab}).
In the case of ${\mathbb Z}_2 \times {\mathbb Z}_2$ orbifold,
 the elements acting on the torus $T^6$
are given by
\[1=(+++)\quad ,\quad g=(+--)\quad ,\quad f=(-+-)\quad ,\quad h=(--+),\]
where the notation means that each "$+$" or "$-$" acts on the complex coordinates
of each two-torus.
The elements $g$, $f$ and $h$ generate three independent 
twisted sectors.
The action of the orbifold group elements on the
$SO(2)$ characters corresponding 
to the three two-tori is given by the table below.
\vspace{0.7cm}

\begin{tabular}{|c|c|c|c|}
\hline
$T_1\times T_2\times T_3$ &$O_2\,\,\,$ $V_2\,\,\,$ $S_2\,\,\,$ $C_2$& 
$O_2\,\,\,$ $V_2\,\,\,$ $S_2\,\,\,$ $C_2$&
 $O_2\,\,\,$ $V_2\,\,\,$ $S_2\,\,\,$ $C_2$\\
\hline
g: &\quad +\quad +\quad +\quad +\quad 
& \quad +\quad $-$\quad $i$\quad $-i$\quad&
\quad +\quad $-$\quad $-i$\quad $i$\quad\\
\hline
h: &\quad +\quad $-$\quad $i$\quad $-i$\quad 
& \quad +\quad $-$\quad $-i$\quad $i$\quad&
\quad +\quad +\quad +\quad +\quad\\
\hline
f: &\quad +\quad $-$\quad $i$\quad $-i$\quad 
& \quad +\quad +\quad +\quad +\quad&
\quad +\quad $-$\quad $-i$\quad $i$\quad\\
\hline
\end{tabular}

\vspace{0.8cm}

\begin{figure}[htp]
\epsfxsize=2 in
\centering
\scalebox{1.2}{{\epsffile{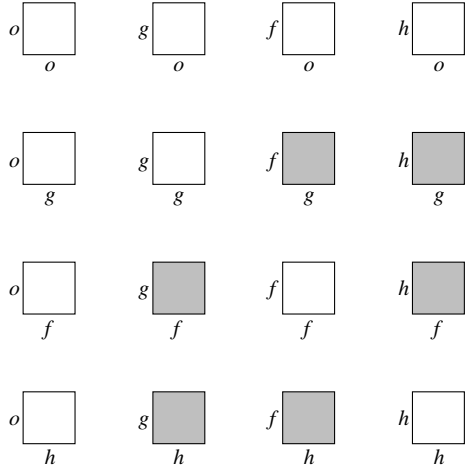}}}
\caption{Modular orbits in ${\mathbb Z}_2 \times {\mathbb Z}_2$ orbifolds.}
\label{fig:mod1}
\end{figure}

In chapter 2 we presented the spin structures. They 
represent the building blocks for 
the partition function in orbifolds models. 
Their modular transformation properties
can be presented with the schematic picture below 
\begin{figure}[htp]
\epsfxsize=2 in
\centering
\scalebox{0.7}{\centerline{\epsffile{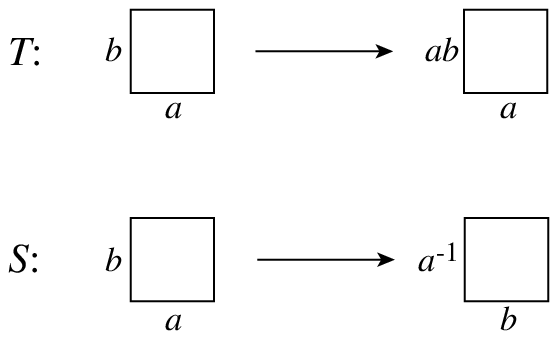}}}
\caption{Modular transformations for a generic amplitude in orbifold models,
$\newline$ where $a$, $b$ $\in\{{1,g,h,f}\}$ 
for a ${\mathbb Z}_2 \times {\mathbb Z}_2$ orbifold.}
\label{fig:mod}
\end{figure}

which shows that the ${\mathbb Z}_2 \times {\mathbb Z}_2$ orbifold needs at least
two independent modular orbits. For example, the element $(g,h)$ 
cannot be derived from any untwisted amplitude.   
Therefore, to obtain the full partition function, 
we have to calculate each of the contributions 
shown in fig.\ref{fig:mod1}, where the empty and the coloured
boxes are associated to two independent orbits. We remind that the full 
partition has to be modular invariant.
At the end of this section we
 provide the result of the partition function
in a compact form in
the case without discrete torsion ($\epsilon=1$).
In general, the value of the phase for the ${\mathbb Z}_2 \times {\mathbb Z}_2$ case 
can be $\epsilon=\pm 1$ since 
it has to be of the same order as the generators of the orbifold.
The explanations concerning the final form for the partition function
(\ref{z2xz2}) are presented 
in the next sections and the definitions concerning the 
terms $T_{ij}$ and $G_{ij}$ are given in Appendix D.

The generic expression for a ${\mathbb Z}_2 \times {\mathbb Z}_2$ 
orbifold partition function can be indicated as
\beq
\mathcal{Z}_{\text{Total}}= Tr_{(\text{un. + tw.})}
\frac{1+g+f+h}{4}\,\,Z_0 ,
\label{proev}
\eeq
where $Z_0$ in this case is given by eq.(\ref{Z-}).
The explicit calculation of (\ref{proev}) gives
\ba
\mathcal{Z}_{\text{Total}}\!\!\!&=&\!\!{1 \over 4} \biggl\{ 
T_{oo} \Lambda_1 \Lambda_2 \Lambda_{m',n'}
[\La_{2m,n}(O_{16}O_{16}+C_{16}C_{16})+\La_{2m+1,n}
(S_{16}S_{16}+V_{16}V_{16})\nonumber\\
&+&\,\, \La_{2m,n+\frac{1}{2}}(O_{16}C_{16}+C_{16}O_{16})+
\La_{2m+1,n+\frac{1}{2}}(V_{16}S_{16}+S_{16}V_{16})]\nonumber\\
&+&T_{og}  \Lambda_1 
|{2\eta \over \theta_2}|^4 G_{0g}
+ T_{of} \Lambda_2 
|{2\eta \over \theta_2}|^4 G_{of}
+T_{oh} \Lambda_{m',n'} 
|{2\eta \over \theta_2}|^4\big\{
\La_{2m,n} G_{oh} \nonumber\\
&+&\La_{2m+1,n} G'_{0h} +\La_{2m,n+1/2} G''_{0h} 
+\La_{2m+1,n+1/2} G'''_{0h}\big\} \nonumber\\ 
&+& T_{go}G_{g0} \Lambda_1
|{2\eta \over \theta_4}|^4+ T_{gg}G_{gg}  \Lambda_1 
|{2\eta \over \theta_3}|^4 
+ T_{fo} G_{fo}\Lambda_2
|{2\eta \over \theta_4}|^4 +T_{ff}G_{ff} 
\Lambda_2 |{2\eta \over \theta_3}|^4  \nonumber \\
&+& T_{ho}\Lambda_{mn} |{2 \eta \over \theta_4}|^4\big\{ 
\La_{2mn} G_{h0}+\La_{2m+1,n} G'_{h0} +\La_{2m,n+1/2} G''_{h0} 
+\La_{2m+1,n+1/2} G'''_{h0}\big\}\nonumber\\
&+&T_{hh} \Lambda_{mn} |{2 \eta \over \theta_3}|^4
\big\{ \La_{2mn} G_{hh}+\La_{2m+1,n} G'_{hh} +\La_{2m,n+1/2} G''_{hh} 
+\La_{2m+1,n+1/2} G'''_{hh}\big\}\nonumber\\
&+&  \left( T_{gh}G_{gh} + T_{gf}G_{gf} 
+ T_{fg}G_{fg} + T_{fh}G_{fh} + T_{hg}G_{hg}
+ T_{hf}G_{hf} \right) |{8\eta^3 \over {\theta_2 \theta_3 \theta_4}}|^2 
\biggr\}\quad , \nonumber\\ 
\label{z2xz2}
\ea
where $\Lambda_1$, $\Lambda_2$ and $\Lambda_{m'n'}\La_{mn}$ denote the three
lattice sums associated to the three internal tori, as usual.
The contributions of the transverse
bosons is implicit here.

The details for the derivation of eq.(\ref{z2xz2}) are 
presented in the following sections, where we provide the main steps.
In fact 
all the ingredients and the general methods have been
presented extensively
in the previous part of this chapter. 
The main difficulty for 
the ${\mathbb Z}_2 \times {\mathbb Z}_2$ orbifold consists in 
handling correctly
the numerous products of characters
 which have to be transformed under
$S$ transformation. In fact they generate huge sums of terms which need
some rearrangement to obtain a compact readable result. 
For this purpose a simple mathematica program has been used. 

\subsection{Untwisted spectrum}

As in the calculation of eq.(\ref{untwisted}), the untwisted contribution is
given by the sum of the projections w.r.t. the elements of the orbifold
group of the initial partition function. 
The first four rows of eq.(\ref{z2xz2}) indicate the total untwisted
sector.

The twisted sector is given by the sum of two pieces,
the first is the S and T transformation of the untwisted 
contribution, the second is the new independent modular
orbit with its S and T transforms.
We provide some further information concerning the 
derivation of the twisted sector in the next section.

\subsection{Twisted sector h}

We present in the following 
the details for one twisted sector only, in particular
the $h$ sector, where the element $G_{hg}$ fixes
the choice of the independent orbit for our model by a consistent
projection. 
In the other two twisted sectors there is no need for
such a choice since all elements are 
determined by modular transformations from the previous ones.
For the determination of a twisted sector $h$
 we proceed, as usual,
by taking the $S$ transform of the untwisted element projected
by $h$. An $S$ transformation of $G_{0h}$ gives 
\ba
G_{h0}&=&\big\{S_2C_2O_2O_{10}+ C_2S_2V_2V_{10}+
S_2S_2V_2O_{10}+C_2C_2O_2V_{10}\nonumber\\
&&+C_2S_2O_2O_{10}
+S_2C_2V_2V_{10}
+C_2C_2V_2O_{10}+S_2S_2O_2V_{10}\big\}O_{16}\nonumber\\
&&+\big\{V_2O_2S_2C_{10} + O_2V_2C_2S_{10}
+O_2O_2S_2S_{10} 
+ V_2V_2C_2C_{10}\nonumber\\
&&+V_2O_2C_2S_{10} + O_2V_2S_2C_{10}
+O_2O_2C_2C_{10} + V_2V_2S_2S_{10}\big\}C_{16}.
\label{Gh0}
\ea
The T transformation of (\ref{Gh0})
provides the contribution $G_{hh}$ and in its
expression we include the total phase
 arising from the following overall phases
\[G_{h0}\quad\underrightarrow{T}\quad iG_{hh}\quad,
\quad T_{h0}\quad\underrightarrow{T}\quad
 iT_{hh}\quad 
\Rightarrow  i\times i\times -1 = +1 ,
\]
where the last $-1$ in the formula
 is the global prefactor obtained 
in the T transformations (see eq.(\ref{TTT})).

\ba
G_{hh}&=&\big\{S_2C_2O_2O_{10}+ C_2S_2V_2V_{10}-
S_2S_2V_2O_{10}-C_2C_2O_2V_{10}\nonumber\\
&&+C_2S_2O_2O_{10}
+S_2C_2V_2V_{10}
-C_2C_2V_2O_{10}-S_2S_2O_2V_{10}\big\}O_{16}\nonumber\\
&&+\big\{V_2O_2S_2C_{10} + O_2V_2C_2S_{10}
-O_2O_2S_2S_{10} 
- V_2V_2C_2C_{10}\nonumber\\
&&+V_2O_2C_2S_{10} + O_2V_2S_2C_{10}
-O_2O_2C_2C_{10} - V_2V_2S_2S_{10}\big\}C_{16}.
\label{Ghh}
\ea
The independent orbit $G_{hg}$ is obtained by the action 
of the $g$ element onto $G_{h0}$. The overall
phase is included in the final expression
(\ref{Ghg}) and results from 
\[G_{h0}\quad\underrightarrow{g}\quad iG_{hg}\quad,
\quad T_{h0}\quad \underrightarrow{g}\quad iT_{hg}  \quad \Rightarrow i\times i = -1.
\]
\ba
G_{hg}&=&\big\{S_2C_2O_2O_{10}+ C_2S_2V_2V_{10}+
S_2S_2V_2O_{10}+C_2C_2O_2V_{10}\nonumber\\
&&-C_2S_2O_2O_{10}
-S_2C_2V_2V_{10}
-C_2C_2V_2O_{10}-S_2S_2O_2V_{10}\big\}O_{16}\nonumber\\
&&+(-1)\big\{-V_2O_2S_2C_{10} - O_2V_2C_2S_{10}
-O_2O_2S_2S_{10} 
- V_2V_2C_2C_{10}\nonumber\\
&&+V_2O_2C_2S_{10} + O_2V_2S_2C_{10}
+O_2O_2C_2C_{10} + V_2V_2S_2S_{10}\big\}C_{16}.
\label{Ghg}
\ea
We observe that the choice for our
projection is not the conventional one since after 
performing the $g$ action onto the gauge degrees of freedom we also 
added a minus sign in front of all the terms multiplying $C_{16}$.
This operation 
provides a natural result for $G_{hg}$, meaning that
the composition of the characters is analogous to $G_{h0}$ and
 $G_{hh}$ and
assures the modular invariance of the partition function.

The T transformation of eq.(\ref{Ghg}) provides
$G_{hf}$ which, as usual, includes the total phase from
\[G_{hg}\quad\underrightarrow{T} \quad iG_{hf}\quad,
\quad T_{hg}\quad\underrightarrow{T}\quad iT_{hf}\quad 
\Rightarrow  i\times i\times -1 = +1 .\]
\ba
G_{hf}&=&\big\{S_2C_2O_2O_{10}+ C_2S_2V_2V_{10}-
S_2S_2V_2O_{10}-C_2C_2O_2V_{10}\nonumber\\
&&-C_2S_2O_2O_{10}
-S_2C_2V_2V_{10}
+C_2C_2V_2O_{10}+S_2S_2O_2V_{10}\big\}O_{16}\nonumber\\
&&+\big\{V_2O_2S_2C_{10} + O_2V_2C_2S_{10}
-O_2O_2S_2S_{10} 
- V_2V_2C_2C_{10}\nonumber\\
&&-V_2O_2C_2S_{10} - O_2V_2S_2C_{10}
+O_2O_2C_2C_{10} + V_2V_2S_2S_{10}\big\}C_{16}.
\label{Ghf}
\ea
We have an important comment to make before discussing
the relevant parts of the spectrum.
In the untwisted sector generated by
the $h$ element there are gauge contributions 
($G'_{0h}$, $G''_{0h}$, $G'''_{0h}$) which are multiplied
by massive lattices, not providing any low energy states.  
In the $h$ twisted part these terms can still give a contribution
 to the massless
spectrum (since we rearrange the lattices 
with the transformations (\ref{prop}).
The presence of the terms $G'_{h0}$, $G''_{h0}$, $G'''_{h0}$
and their T transformations will not provide massless states. Thus,
we can neglect these contributions 
when we discuss the relevant part 
of the spectrum. 

\subsection{Torus amplitudes for 
the right and for the left sector}
In the result (\ref{z2xz2}) we have used the 
torus amplitudes defined in terms of the quantities
 $T_{ij}$, $i=0, g, h, f$, providing a simple and compact form
for the partition function.
\ba
T_{i0} &=&  \tau_{i0} +  \tau_{ig} + \tau_{ih} + \tau_{if} \quad , \qquad
T_{ig} =  \tau_{i0} +  \tau_{ig} - \tau_{ih} - \tau_{if} \quad , \nonumber \\
T_{ih} &=&  \tau_{i0} -  \tau_{ig} + \tau_{ih} - \tau_{if} \quad , \qquad
T_{if} =  \tau_{i0} -  \tau_{ig} - \tau_{ih} + \tau_{if} \quad ,
\ea
where the ${\mathbb Z}_2 \times {\mathbb Z}_2$
 characters $\tau_{ij}$ are products
of the four level-one characters, defined explicitly in (\ref{gio}).
The ordering of the four factors refers to the eight transverse 
dimensions of spacetime. The first factor is associated 
to the two transverse space time directions.
For the left sector we have
\ba
G_{i0} &=&  g_{i0} +  g_{ig} + g_{ih} + g_{if} \quad , \qquad
G_{ig} =  g_{i0} +  g_{ig} - g_{ih} - g_{if} \quad , \nonumber \\
G_{ih} &=&  g_{i0} -  g_{ig} + g_{ih} - g_{if} \quad , \qquad
G_{if} =  g_{i0} -  g_{ig} - g_{ih} + g_{if} \quad .
\ea
The content of the above definitions is given in (\ref{gia}).
There we also provide the explicit expressions 
for the gauge contributions  
$G'_{0h}$, $G''_{0h}$, $G'''_{0h}$ for the 
untwisted sector and $G'_{h0}$, $G''_{h0}$ and $G'''_{h0}$ 
for the twisted sector.

\subsection{Massless spectrum}

The formula (\ref{z2xz2}) presents the full modular 
invariant partition function 
for the shift orbifold (\ref{Z-}) 
with ${\mathbb Z}_2 \times {\mathbb Z}_2$ action. 
We notice that 
the only contributions from the \underline{\bf{untwisted spectrum}},
where we have neglected the accented expressions,
come from the combinations
\beq
\tau_{00}g_{00}+\tau_{0g}g_{0g}+\tau_{0h}g_{0h}+\tau_{0f}g_{0f}.
\eeq
Our main interest is as usual the low energy physics of the model 
hence we will present here the massless terms, 
which can be expanded in powers of $q$ by applying the relations
of section \ref{aa33} in Appendix A.
\ba
&&[\overline V_2\overline O_2\overline O_2\overline O_2
-\overline S_2\overline S_2\overline S_2\overline S_2
-\overline C_2\overline C_2\overline C_2\overline C_2]
\times [O_2O_2O_2O_{10}O_{16}],\nonumber\\
&&[\overline O_2 \overline V_2\overline O_2\overline O_2
-\overline C_2\overline C_2\overline S_2\overline S_2
-\overline S_2\overline S_2\overline C_2\overline C_2]
\times [(O_{2}V_2V_2O_{10}+V_{2}O_2O_2V_{10})O_{16}],\nonumber\\
&&[\overline O_2\overline O_2\overline O_2\overline V_2
-\overline C_2\overline S_2\overline S_2\overline C_2
-\overline S_2\overline C_2\overline C_2\overline S_2]
\times [(V_2V_2O_2O_{10}+O_2O_2V_2V_{10})O_{16}],\nonumber\\
&&[\overline O_2\overline O_2\overline V_2\overline O_2
-\overline C_2\overline S_2\overline C_2\overline S_2
-\overline S_2\overline C_2\overline S_2\overline C_2]
\times [(V_2O_2V_2O_{10}+O_2V_2O_2V_{10})O_{16}].\nonumber\\
\label{crz}
\ea
The gauge group of this model is given by 
$G = SO(2)\times SO(2)\times SO(2)\times SO(10)\times SO(16)$
and the representations of the untwisted matter is
provided in the following.
\newpage

$\bullet$ Vectorial supermultiplet:
$$[\overline V_2\overline O_2\overline O_2\overline O_2
-\overline S_2\overline S_2\overline S_2\overline S_2
-\overline C_2\overline C_2\overline C_2\overline C_2]
\times [O_2O_2O_2O_{10}O_{16}]\rightarrow$$
$$[(2,1,1,1)+(1^+,1^+,1^+,1^+)+(1^-,1^-,1^-,1^-)] 
\times(1,1,1,1,1)$$

$\bullet$ Two chiral supermultiplets:
$$[\overline O_2 \overline V_2\overline O_2\overline O_2
-\overline C_2\overline C_2\overline S_2\overline S_2
-\overline S_2\overline S_2\overline C_2\overline C_2]
\times [(O_{2}V_2V_2O_{10}+V_{2}O_2O_2V_{10})O_{16}]\rightarrow$$
$$[(1,2,1,1)+(1^-,1^-,1^+,1^+)+(1^+,1^+,1^-,1^-)]
\times[(1,2,2,1,1)+(2,1,1,10,1)]$$

$\bullet$ Two chiral supermultiplets:
$$[\overline O_2\overline O_2\overline O_2\overline V_2
-\overline C_2\overline S_2\overline S_2\overline C_2
-\overline S_2\overline C_2\overline C_2\overline S_2]
\times [(V_2V_2O_2O_{10}+O_2O_2V_2V_{10})O_{16}]\rightarrow$$
$$[(1,1,1,2)+(1^-1^+1^+1^-)+(1^+1^-1^-1^+)]\times[(2,2,1,1,1)+(1,1,2,10,1)]$$ 

$\bullet$ Two chiral supermultiplets:
$$[\overline O_2\overline O_2\overline V_2\overline O_2
-\overline C_2\overline S_2\overline C_2\overline S_2
-\overline S_2\overline C_2\overline S_2\overline C_2]
\times [(V_2O_2V_2O_{10}+O_2V_2O_2V_{10})O_{16}]\rightarrow$$
$$[(1,1,2,1)+(1^-1^+1^-1^+)(1^+1^-1^+1^-)]
\times[(2,1,2,1,1)+(1,2,1,10,1)].$$

In our notations we indicate with $1^{\pm}$ the two
different chiralities of a spinor in the $SO(2)$ representation.
This model has $N=1$ in four dimensions.

The \underline{\bf{ twisted sector}} gives rise to the only non-vanishing terms
\ba
&&\tau_{g0}g_{g0}+\tau_{gg}g_{gg}+\tau_{gh}g_{gh}+\tau_{gf}g_{gf}+
\tau_{h0}g_{h0}+\tau_{hg}g_{hg}\nonumber\\
&&+\tau_{hh}g_{hh}+\tau_{hf}g_{hf}+
\tau_{f0}g_{f0}+\tau_{fg}g_{fg}+\tau_{fh}g_{fh}+\tau_{ff}g_{ff} ,
\ea
whose
massless contributions have been indicated in the following
\ba
&&[\overline O_2\overline O_2\overline C_2\overline C_2
-\overline C_2\overline S_2\overline O_2\overline O_2]
\times[(C_2O_2O_2C_{10}+V_2C_2C_2O_{10}+O_2S_2S_2V_{10})O_{16}],\nonumber\\
&&[\overline O_2\overline O_2\overline S_2\overline S_2-
\overline S_2\overline C_2\overline O_2\overline O_2]
\times[(S_2O_2O_2S_{10}+V_2S_2S_2O_{10}+O_2C_2C_2V_{10})O_{16}],\nonumber\\
&&[\overline O_2\overline C_2\overline C_2\overline O_2
-\overline C_2\overline O_2\overline O_2\overline S_2]
\times[(C_2C_2V_2O_{10}+S_2S_2O_2V_{10})O_{16}],\nonumber\\
&&[\overline O_2\overline S_2\overline S_2\overline O_2
-\overline S_2\overline O_2\overline O_2\overline C_2]
\times[(S_2S_2V_2O_{10}+C_2C_2O_2V_{10})O_{16}],\nonumber\\
&&[\overline O_2\overline S_2\overline O_2\overline S_2
-\overline S_2\overline O_2\overline C_2\overline O_2]
\times[(O_2S_2O_2S_{10}+S_2V_2S_2O_{10}+C_2O_2C_2V_{10})O_{16}],\nonumber\\
&&[\overline O_2\overline C_2\overline O_2\overline C_2
-\overline C_2\overline O_2\overline S_2\overline O_2]
\times[(O_2C_2O_2C_{10}+C_2V_2C_2O_{10}+S_2O_2S_2V_{10})O_{16}].\nonumber\\
\label{cor2}
\ea

These chiral supermultiplets fall into the representations
presented below.
\newpage
$\bullet$ Three chiral supermultiplets:
$$[\overline O_2\overline O_2\overline C_2\overline C_2
-\overline C_2\overline S_2\overline O_2\overline O_2]
\times[(C_2O_2O_2C_{10}+V_2C_2C_2O_{10}+O_2S_2S_2V_{10})O_{16}]+
$$
$$\,\,[\overline O_2\overline O_2\overline S_2\overline S_2-
\overline S_2\overline C_2\overline O_2\overline O_2]
\times[(S_2O_2O_2S_{10}+V_2S_2S_2O_{10}+O_2C_2C_2V_{10})O_{16}]
\rightarrow$$
$$[(1,1,1^{\pm},1^{\pm})+(1^{\pm},1^{\pm},1,1)]\times [(1^{\pm},1,1,16,1)+
(2,1^{\pm},1^{\pm},1,1)+(1,1^{\pm},1^{\pm},10,1)]
$$

$\bullet$ Two chiral supermultiplets:
$$[\overline O_2\overline C_2\overline C_2\overline O_2
-\overline C_2\overline O_2\overline O_2\overline S_2]
\times[(C_2C_2V_2O_{10}+S_2S_2O_2V_{10})O_{16}]+
$$
$$\,\,\,[\overline O_2\overline S_2\overline S_2\overline O_2
-\overline S_2\overline O_2\overline O_2\overline C_2]
\times[(S_2S_2V_2O_{10}+C_2C_2O_2V_{10})O_{16}]\rightarrow
$$
$$[(1,1^{\pm},1^{\pm},1)+(1^{\pm},1,1,1^{\pm})]\times [(1^{\pm},1^{\pm},2,1,1)+
(1^{\pm},1^{\pm},1,10,1)]
$$

$\bullet$ Three chiral supermultiplets:
$$[\overline O_2\overline S_2\overline O_2\overline S_2
-\overline S_2\overline O_2\overline C_2\overline O_2]
\times[(O_2S_2O_2S_{10}+S_2V_2S_2O_{10}+C_2O_2C_2V_{10})O_{16}]+$$
$$\,\,[\overline O_2\overline C_2\overline O_2\overline C_2
-\overline C_2\overline O_2\overline S_2\overline O_2]
\times[(O_2C_2O_2C_{10}+C_2V_2C_2O_{10}+S_2O_2S_2V_{10})O_{16}]
\rightarrow$$
$$[(1,1^{\pm},1,1^{\pm})+(1^{\pm},1,1^{\pm},1)]\times [(1,1^{\pm},1,16,1)+
(1^{\pm},2,1^{\pm},1,1)+(1^{\pm},1,1^{\pm},10,1)].
$$

We notice that the twisted massless spectrum contains two chiral 
supermultiplets in the spinorial representation of $SO(10)$, plus
few supermultiplets in the fundamental of $SO(10)$ in four dimensions.
This result concludes our chapter.

\chapter{Conclusions}

In this thesis we focus our study on heterotic 
superstring theories and their applications to
particle physics. In particular, we are interested in 
the search of semi-realistic four-dimensional superstring
vacua which can reproduce, at low energy, the Standard Model physics.
Motivated by the $SO(10)$ embedding of matter in heterotic models,
we investigate different schemes of compactification of the $E_8\times E_8$
heterotic string from ten to four dimensions.
A very successful approach is given by free fermionic models.
They give rise to the most realistic three generation string models
to date. Their phenomenology is studied in the effective low energy
field theory by the analysis of supersymmetric flat directions.
In the first example illustrated in chapter 3, the model content
consists of MSSM states in the observable Standard Model sector.
In that model, for the first time, we apply a new general mechanism
that allows the reduction of Higgs content at the string scale
by an opportune choice of asymmetric boundary conditions
for the internal fermions of the theory. An additional result for 
minimal Higgs spectrum models is the fact that 
the supersymmetric moduli space is reduced as well, and this increases
the predictive power of the theory.

A common feature of free fermionic models is the presence of an 
anomalous $U(1)$ which gives rise to a Fayet-Iliopoulos D-term that breaks
supersymmetry at one-loop level in string 
perturbation theory. Supersymmetry is restored
by imposing D and F flatness on the vacuum. Generally, 
it has been assumed that in a given string model 
there should exist a supersymmetric solution
to D and F flatness constraints.
Nevertheless,  
in the second model presented in chapter 3,
such as in the previous example,
no flat solutions
are found after employing the standard analysis for 
flat directions. The Bose-Fermi degeneracy of the spectrum
implies that the cosmological constant vanishes while supersymmetry
remains broken at the perturbative level. This unexpected result
may open new possibilities for the supersymmetry breaking mechanism
in string theory.
By looking at a very different background, the one given by the
orbifold construction, it is possible to obtain complementary
advantages in the understanding of semi-realistic models,
such as a more geometric picture of those.
Moreover, in the case of ${\mathbb Z}_2 \times {\mathbb Z}_2$ for special points
in the compactification space, the correspondence 
with free fermionic models
has been demonstrated. This connection offers interesting
indications in the choice of "good" orbifolds, since the 
number of consistent models is huge and a guiding principle
is needed.
A specific ${\mathbb Z}_2 \times {\mathbb Z}_2$ orbifold with a 
non-factorisable skewed compactification lattice has been
analysed, where the reduction of the number of families
 is realised
and suggests new way of investigating orbifold
compactifications. No semi-realistic models are presented
in this set up yet, nevertheless the possible
combinations of a proper choice for the compactification lattice
plus the presence of suitable Wilson lines provides new chances
in the construction of semi-realistic models.
A challenging outlook in this set up
is the introduction of asymmetric shifts and twists. Indeed, these
elements seem to be related with free fermionic models
where asymmetric boundary conditions are imposed on the compact
dimensions and are responsible for the most successful phenomenological
features of these models.

In the last chapter we present the formalism for the construction
of modular invariant partition functions in heterotic
orbifold models and, among a few examples, the case of
a ${\mathbb Z}_2 \times {\mathbb Z}_2$ shift orbifold model. The study of 
orbifolds with different projections should lighten
the properties of the low energy spectrum and possibly 
provide some selection mechanism for semi-realistic vacua.
For instance, a challenging project would be the realisation
of the Higgs-matter splitting. This mechanism is viable with
an orbifold projection that will allow to obtain string states
uniquely from the untwisted sector and the matter states from the twisted
sectors. This mechanism is already well-known in the free fermionic case.

\appendix
\chapter{}
\section{$\eta$ and $\theta$-functions and modular transformations}
\label{AAA}
The Dedekind $\eta$ function is defined as
\beq
\eta(\tau)=q^{\frac{1}{24}}\prod_{p=1}^{\infty}(1-q^p).
\eeq
We provide the modular transformations of $\eta$ and of the Teichmuller parameter $\tau$ in terms of complex 
function and real components.
\beq
T:\eta(1+\tau)=e^{i\pi/12}\eta(\tau)\quad,\quad
S: \eta(-\frac{1}{\tau})=(-i\tau)^{\frac{1}{2}}\eta(\tau) .
\eeq
\[T:\tau_1\rightarrow \tau_1+1  \quad ;\quad \tau_2 \rightarrow \tau_2
\quad ;\quad d\tau_1 d\tau_2\rightarrow d\tau_1 d\tau_2 .
\]
\[
S:\tau_1+i\tau_2 \rightarrow -\frac{1}{\tau_1 +i\tau_2}=- \frac{\bar\tau}{\tau\bar \tau}
\quad;\quad
 d\tau d\bar\tau\rightarrow 
\frac{d\tau d\bar\tau}{|\tau\bar\tau|^2} .\]
The definition of the $\theta$ function is given in both notations, 
as sum and as product formulae  
\ba
\te\left[\begin{array}{c}  
\a \\
\b
\end{array}\right](0|\theta )&=& \sum_{n=\infty}^{\infty} q^{\frac{1}{2}(n+\a)^2}e^{2\pi i(n+\a)\b}\nonumber\\
&=& e^{2\pi i\a\b}q^{\frac{\a^2}{2}} \prod_{n}^{\infty}(1-q^n)
(1+q^{n+\a-\frac{1}{2}}e^{2\pi i\b})(1+q^{n-\a-\frac{1}{2}}e^{-2\pi i\b})\nonumber\\
\ea
and their modular transformations
\ba
&&T:\te\left[\begin{array}{c}  
\a \\
\b
\end{array}\right](0|\t+1)= e^{-\pi i\a(\a-1)}\,\,\,\te\left[\begin{array}{c}  
\a \\
\b + \a -\frac{1}{2}
\end{array}\right](0|\t),\nonumber\\
&&S:\te\left[\begin{array}{c}  
\a \\
\b
\end{array}\right](0|-\frac{1}{\t})=\,\,\,(-i\t)^{\frac{1}{2}}\,\,\,e^{2\pi i\a\b}\,\,\,\te\left[\begin{array}{c}  
\b\\
-\a 
\end{array}\right](0|\t).
\ea
\newpage
\subsection{$SO(2n)$ characters in terms of $\theta$-functions} 

\beq
O_{2n}=\frac{\te^n_3 +\te^n_4}{2\eta^n}\,\,\,\,;\,\,\,\,
V_{2n}=\frac{\te^n_3 -\te^n_4}{2\eta^n}\,\,\,\,;\,\,\,\,
S_{2n}=\frac{\te^n_2 +i^{-n}\te^n_1}{2\eta^n}\,\,\,\,;\,\,\,\,
C_{2n}=\frac{\te^n_2-i^{-n} \te^n_1}{2\eta^n}\,\,\,\,.\,\,\,\,
\label{Char}
\eeq
It is useful to present the explicit expansions of the previous functions and the $\eta$ function
in terms of powers of $q$, where $q=e^{2i\pi\tau}$
\ba
O_{2n}=&&
\frac{\Pi_{p=1}^{\infty}(1-q^p)^n(1+q^{p-\frac{1}{2}})^{2n}+\Pi_{p=1}^{\infty}(1-q^p)^n(1-q^
{p-\frac{1}{2}})^{2n}}{2q^{\frac{n}{24}}\Pi_{p=1}^{\infty}(1-q^p)^n}\nonumber\\
=&& q^{-\frac{n}{24}}(1+n(2n-1)q+...),\nonumber\\
\nonumber\\
V_{2n}=&&
\frac{\Pi_{p=1}^{\infty}(1-q^p)^n(1+q^{p-\frac{1}{2}})^{2n}-\Pi_{p=1}^{\infty}(1-q^p)^n(1-q^
{p-\frac{1}{2}})^{2n}}{2q^{\frac{n}{24}}\Pi_{p=1}^{\infty}(1-q^p)^n}\nonumber\\
=&& q^{-\frac{n}{24}}(2nq^{\frac{1}{2}}+...),\nonumber\\
\nonumber\\
S_{2n}/C_{2n}=&&q^{\frac{n}{8}}
\frac{\Pi_{p=1}^{\infty}(1-q^p)^n(1+q^{p})^{n}(1+q^{p-1})^n}
{2q^{\frac{n}{24}}\Pi_{p=1}^{\infty}(1-q^p)^n}\nonumber\\
=&& 2^{n-1}q^{\frac{n}{12}}(1+2nq+...),\nonumber\\
\nonumber\\
\frac{1}{\eta^n}=&&q^{-\frac{n}{24}}(1+nq+...) ,
\ea
where the definition of $\theta$-functions and
the binomial expansion below have been applied,
\[
(a+b)^n= \sum_{i=0}^{n}C\left(\begin{array}{c}  
n \\
i\\
\end{array}\right) a^{n-i}b^i=a^n+C\left(\begin{array}{c}  
n \\
i\\
\end{array}\right) a^{n-1}b+...\,.
\]
The decomposition of an $SO(x+y)$ character into the product of
an $SO(x)$ with an $SO(y)$ character 
is given by the expressions below:
\ba
O_{2n}&=& O_xO_y+V_xV_y ,\hspace{1cm} V_{2n}= V_xO_y+O_xV_y ,\nonumber\\
C_{2n}&=& S_xC_y+C_xS_y ,\hspace{1cm} S_{2n}=S_xS_y+C_xC_y ,
\label{a8}
\ea
where $2n=x+y$ and $x, y$ are even. 

In the study of the S transformations of the previous expansions
it can be useful to rearrange (\ref{a8}) with the relations
\ba
aa+bb &=&\frac{1}{2}[(a+b)(a+b)+(a-b)(a-b)]\nonumber\\
aa-bb &=&\frac{1}{2}[(a-b)(a+b)+(a+b)(a-b)]\nonumber\\
ab+ba &=&\frac{1}{2}[(a+b)(a+b)-(a-b)(a-b)]\nonumber\\
ab-ba &=&\frac{1}{2}[(a-b)(a+b)-(a+b)(a-b)]
\label{decomp}
\ea
where $a$ and $b$ can be any of $O_n,V_n,S_n,C_n$.

\subsection{Modular transformations for $SO(2n)$ characters}
\label{SSS}
The modular S and T transformations act on the characters as
\ba
&&\left(\begin{array}{c}
O_{2n}\\
V_{2n}\\
S_{2n}\\
C_{2n}\\
\end{array}\right)\,\,\,\,\,\,\,\,
\underrightarrow{S}
 \,\,\,\,\,\,\,\frac{1}{2}\left(\begin{array}{cccc}  
1&1&1&1 \\
1&1&-1&-1\\
1&-1&i^{-n}&-i^{-n}\\
1&-1&-i^{-n}&i^{-n}\\
\end{array}\right)
\left(\begin{array}{c}
O_{2n}\\
V_{2n}\\
S_{2n}\\
C_{2n}\\
\end{array}\right) ;\nonumber\\
\vspace{0.8cm}
&&\left(\begin{array}{c}
O_{2n}\\
V_{2n}\\
S_{2n}\\
C_{2n}\\
\end{array}\right)\,\,\,\,\,\,\,\,\,
\underrightarrow{T}\,\, e^{-in\pi/12}
\,\,\left(\begin{array}{cccc}  
1 &0&0&0 \\
0&- 1&0&0\\
0&0&e^{in\pi/4}&0\\
0&0&0&e^{in\pi/4}\\
\end{array}\right)
\left(\begin{array}{c}
O_{2n}\\
V_{2n}\\
S_{2n}\\
C_{2n}\\
\end{array}\right).\nonumber\\
\label{S,T}
\ea

\section{Definition of lattice}

The partition function of a compact scalar on a circle of radius $R$ is 

\beq
\La_{m,n}=\frac{1}{\eta\overline \eta} 
\sum_{m,n}q^{\a' p_L^2/4}\overline q^{\a'p_R^2/4} ,
\label{lambda0}
\eeq
where the chiral momenta are defined as
\[ p_{L,R}=\frac{m}{R}\pm \frac{nR}{\a'}.\]
Therefore, if one of the non-compact coordinates of a critical string is replaced with a compact one,
 the continuous integration over internal momenta is replaced by the lattice sum $\frac{1}{\sqrt{\tau_2}\eta(\tau)\overline \eta(\tau)}\rightarrow \La_{m,n}$.

For the case of a d-dimensional torus the eq.(\ref{lambda0}) is generalised to 
\beq
\vec\La_{m,n}=\frac{1}{\eta^d\overline \eta^d} \sum_{m,n}q^{\a' p_L^T g^{-1} p_L/4}
\overline q^{\a'p_R^T g^{-1} p_R/4}
\label{lambda1}
\eeq
where $ p_{L,a}={m_a}+ \frac{1}{\a'}(g_{ab}-B_{ab})n^b$, $
p_{R,a}={m_a}+ \frac{1}{\a'}(g_{ab}+B_{ab})n^b$,
 $g_{ab}$ is the metric on the torus 
and $B_{ab}$ is an antisymmetric NS-NS field. 

\subsection{Definition of shifted lattices}

In this section we present combinations obtained with 
the standard lattice 
$\La_{mn}$ when the shift $\delta:\La_{m,n}\rightarrow(-1)^m\La_{m,n}$ acts on it. Moreover we show 
their main properties whose 
demonstration is given in section \ref{lat}.
\ba
\La_{2m,n}&=& \frac{1+(-1)^m}{2} \La_{m,n},\nonumber\\
\La_{2m+1,n}&=& \frac{1-(-1)^m}{2} \La_{m,n},\nonumber\\
\La_{2m,n+\frac{1}{2}}&=& \frac{1+(-1)^m}{2} \La_{m,n+\frac{1}{2}},\nonumber\\
\La_{2m+1,n+\frac{1}{2}}&=& \frac{1-(-1)^m}{2} \La_{m,n+\frac{1}{2}}.\nonumber\\
\label{lambda}
\ea
\begin{center}
\underline{Transformation properties}
\end{center}
\ba
 &\overbrace{\La_{m,n}}^{\mathit{S\,,\,T\,\, invariant}}\,\,\,;\,\,\,\overbrace{\La_{2m,n}+\La_{2m,n+\frac{1}{2}}}^{\mathit{S\,,\,T\,\, invariant}}&
\nonumber\\
& \overbrace{(-1)^m\La_{m,n}}^{\mathit{T\,\, invariant}}\,\,\,\,\underrightarrow{\mathit{{S}}} \,\,\,\,
\La_{m,n+\frac{1}{2}}\,\,\,\,
\underrightarrow{\mathit{T}} \,\,\,\,\overbrace{(-1)^m\La_{m,n+\frac{1}{2}}}^{\mathit{S\,\, invariant}}&
\nonumber\\
& \overbrace{\La_{2m,n}-\La_{2m,n+\frac{1}{2}}}^{\mathit{T\,\, invariant}}\,\,\,\,\underrightarrow{\mathit{{S}}} \,\,\,\,
\La_{2m+1,n}+\La_{2m+1,n+\frac{1}{2}}\,\,\,\,
\underrightarrow{\mathit{T}} \,\,\,\,\overbrace{\La_{2m+1,n}-\La_{2m+1,n+\frac{1}{2}}}^{\mathit{S\,\, invariant}} & .
\nonumber\\
\label{prop}
\ea

\subsection{Proof for the transformation properties (\ref{prop})}
\label{lat}

In this section we show how to derive some of the properties 
presented in the previous section.
\begin{itemize}
\item[1)] $S$ invariance for $\Lambda_{m,n}$.
\item[2)] $T$ invariance for $\Lambda_{m,n}$ and $(-1)^m\Lambda_{m,n}$.
\item[3)] $(-1)^m\Lambda_{m,n}\quad\underrightarrow{S}\quad\Lambda_{m,n+1/2}$.
\end{itemize}

The other relations shown in (\ref{prop}) can be derived
with the same techniques below.

It is useful to keep in mind the definitions of the 
general lattice (\ref{lambda0}) and the chiral momenta $p_{L,R}$. Moreover
we can rewrite $q$ and $\bar q$ in the convenient way
\[q=e^{2\pi i\tau}=e^{2\pi i(\tau_1+i\tau_2)}=e^{2\pi (i\tau_1-\tau_2)} ,
\quad
\bar q=e^{-2\pi i\bar\tau}=e^{-2\pi i(\tau_1-i\tau_2)}=e^{-2\pi (i\tau_1+\tau_2)}.
\]
The Poisson resummation formula will be applied constantly
in the demonstration of the previous statements, thus we
provide its general expression below
\beq
\sum_{m_i\in Z}e^{-\pi m_i\cdot m_j A_{ij}+\pi B_im_i}=\frac{1}{\sqrt{\text{det} A}}
\sum_{m_k \in Z}e^{-\pi(m_k+\frac{iB_k}{2})(A^{-1})_{kl}(m_l+\frac{iB_l}{2})} 
\eeq
We start by demonstrating \underline{\it{point 1})}.

The best way to proceed is to rewrite the lattice sum in the more 
convenient form
\beq
\Lambda_{m,n}=\sum_{m,n}e^{2\pi(i\tau_1-\tau_2)
\frac{\alpha'}{4}(\frac{m}{R}+\frac{nR}{\alpha'})^2}
e^{2\pi(-i\tau_1-\tau_2)
\frac{\alpha'}{4}(\frac{m}{R}-\frac{nR}{\alpha'})^2}.
\eeq
We notice that the $\frac{1}{\eta\bar\eta}$ factor 
has been dropped for convenience. 

Let us simplify the two exponentials and rewrite
\beq
\Lambda_{m,n}=\sum_{m,n}e^{2\pi i\tau_1mn}
e^{-\pi\tau_2\alpha'(\frac{m^2}{R^2}+\frac{n^2R^2}{\alpha'^2})} .
\label{aa}
\eeq
If we perform a Poisson resummation w.r.t. $m$ we have
\[a=\frac{\alpha'\tau_2}{R^2}\rightarrow \frac{1}{\sqrt{det A}}
=\frac{R}{\sqrt{\alpha'\tau_2}}, \quad b=2i\pi n
\]
then eq.(\ref{aa}) becomes
\beq
\frac{R}{\sqrt{\alpha'\tau_2}}
\sum_{m,n}e^{-\pi( m'-\tau_1n)^2\frac{R^2}{\alpha'\tau_2}}
e^{-\pi\tau_2\frac{n^2R^2}{\alpha'}}.
\label{aaa}
\eeq
We expand the square and we obtain an exponential
with four terms. Two of them can be rewritten as
\[-\frac{\pi R^2}{\alpha'}( \frac{\tau_1^2}{\tau_2}+\tau_2 )n^2
=-\frac{\pi R^2}{\alpha'}\frac{|\tau|^2}{\tau_2}n^2.
\]
We apply now the resummation w.r.t. $n$
\[a=\frac{R^2|\tau^2|}{\alpha'\tau_2} \rightarrow 
\frac{1}{\sqrt{det A}}
=\frac{\sqrt{\alpha'\tau_2}}{R|\tau|}, \quad 
b=\frac{2R^2\tau_1m'}{\tau_2\alpha'}
\]
which transforms (\ref{aaa}) into
\beq
\frac{R}{\sqrt{\alpha' \tau_2}}
\frac{\sqrt{\alpha' \tau_2}}{R\tau_2}
\sum_{m,n} e^{-\pi( m'^2\frac{\pi R^2}{\alpha'\tau_2})}
e^{-\frac{\pi\tau_2\alpha'}{R^2|\tau|^2}(n'+\frac{i m' R^2\tau_1}{\tau_2\alpha'})^2}.
\eeq
We expand the exponents and use the equivalence
\[\frac{\pi R^2m'^2}{\alpha'\tau_2}(-1+\frac{\tau^2_1}{|\tau|^2})
= -\frac{\tau_2\pi R^2m'^2}{|\tau|^2\alpha'}
\]
to get finally
\beq
\Rightarrow \frac{1}{|\tau|}\sum_{m',n'}e^{-2\pi i 
\frac{\tau_1}{|\tau|^2}m'n'} e^{-\frac{\pi \tau_2}{|\tau|^2}(\frac{m'^2R^2}{\alpha'}+
\frac{n'^2\alpha'}{R^2})}.
\label{l}
\eeq

The expression above is equivalent to $\Lambda_{m,n}$ if we redefine
\beq
-\frac{\tau_1}{|\tau|^2}=\tau_1'\quad,\quad \frac{\tau_2}{|\tau|^2}=
\tau_2' ,
\label{sl}
\eeq
 which is in fact the S transformation of $\tau\rightarrow -1/\tau$.
The prefactor $1/|\tau|$ in (\ref{l}) belongs to the transformation of $\eta\overline\eta$ (which we dropped at the beginning),
showing that (\ref{l}) is the $S$ transformation of (\ref{lambda0}).

The explanation for \underline{\it{point 2)}} is very simple since 
the invariance under T is trivial
\[\tau\quad \underrightarrow{\small{T}}\quad \tau+1=(\tau_1+1)+i\tau_2\quad
\rightarrow\quad
\Lambda_{m,n}=\sum_{m,n}e^{2\pi(i\tau_1mn)}\underbrace{e^{2\pi i mn}}_{1} e^{-\pi\tau_2\alpha'(\frac{m^2}{R^2}+\frac{n^2R^2}{\alpha'^2})}.
\]
The quantity $(-1)^m\Lambda_{m,n}$ is obviously
 invariant under $T$ transformation as well.

More algebra is involved for the proof of \underline{\it{point 3)}}.

The main idea here is to show that $(-1)^m\Lambda_{m,n}(\tau)$ 
can be rewritten as $\Lambda_{m,n+1/2}(\tau')$, where $\tau'$
 is given by (\ref{sl}). Let us start with the definition
\beq
(-1)^m\Lambda_{m,n}=
\sum_{m,n}e^{2\pi i m(\tau_1n+1/2)}e^{-\pi\tau_2 \frac{\alpha' m^2}{R^2}}
 e^{-\pi\tau_2\frac{n^2R^2}{\alpha'}}.
\label{b}
\eeq
By applying the Poisson resummation w.r.t. $m$
\[a=\frac{\tau_2\alpha'}{R^2}\quad,\quad b=2 i (\tau_1n+1/2),
\]
eq.(\ref{b})becomes
\beq
\Rightarrow \sum_{m,n}e^{-\pi (m'-(\tau_1 n+1/2))^2\frac{R^2}{\tau_2 \alpha'}}
e^{-\pi\tau_2\frac{n^2R^2}{\alpha'}}.
\eeq
Rearranging the exponential and using the relation
\[-\frac{\pi R^2n^2}{\alpha'}(\frac{\tau_1^2}{\tau_2})=
-\frac{|\tau|^2\pi R^2n^2}{\tau_2\alpha'}
\]
we get
\beq
\frac{R}{\sqrt{\tau_2\alpha'}}\sum_{m',n}
e^{-\pi R^2n^2\frac{|\tau|^2}{\tau_2\alpha'}}
e^{2\pi(m'-1/2)n\frac{\tau_1R^2}{\tau_2\alpha'}}
e^{-\pi(m'-1/2)^2\frac{R^2}{\tau_2\alpha'}}.
\label{bb}
\eeq
A Poisson resummation of (\ref{bb}) w.r.t. $n$, where 
\[a=\frac{R^2|\tau|^2}{\tau_2\alpha'}\quad ,\quad 
b=2(m'-1/2)\frac{\tau_1R^2}{\tau_2\alpha'},
\] 
will provide
\ba
\Rightarrow && \frac{1}{|\tau|}\sum_{m',n'}
e^{-\pi \frac{\tau_2\alpha'}{R^2|\tau|^2}(n'+i(m'-1/2)\frac{\tau_1^2R^2}{\tau_2\alpha'})^2}
e^{-\pi(m'-1/2)^2\frac{R^2}{\tau_2\alpha'}}\nonumber\\
&=& \frac{1}{|\tau|}\sum_{m',n'}
e^{-\pi \frac{\tau_2\alpha'}{R^2|\tau|^2}n'^2}
e^{-\pi(m'-1/2)^2\frac{R^2 \tau_2}{|\tau|^2\alpha'}}
e^{-2i\pi n'(m'-1/2)\frac{\tau_1}{|\tau|^2}}=\Lambda_{n',m'+1/2}.\nonumber\\
\ea
As we said, once redefining $n'\rightarrow n$, $m'\rightarrow m$ and
identifying the transformed $\tau'$ parameter, we have obtained exactly 
the $S$ transformation of the initial (\ref{b}).

\section{Expansion of $SO(2n)$
characters  in powers of q}\label{aa33}
This section presents the explicit expansions of the characters 
used in sections \ref{Z}-\ref{ZZ}
 for the searching of the massless spectrum.
$$
\begin{array}{ccccccc}
V_2&=&q^{-\frac{1}{24}}(2q^{\frac{1}{2}}+...),\quad
O_2&=&q^{-\frac{1}{24}}(1+q+...),\quad
S_2/C_{2}&=&q^{\frac{1}{12}}(1+2q+...),\\
V_4&=&q^{-\frac{1}{12}}(4q^{\frac{1}{2}}+...),\quad
O_4&=&q^{-\frac{1}{12}}(1+6q+...),\quad
S_4/C_{4}&=&2q^{\frac{1}{6}}(1+4q+...),\\
V_{10}&=&q^{-\frac{5}{24}}(10q^{\frac{1}{2}}+...),\quad
O_{10}&=&q^{-\frac{5}{24}}(1+45q+...),\quad
S_{10}/C_{10}&=&2^4q^{\frac{1}{2}}(1+10q+...),\\
V_{12}&=&q^{-\frac{1}{4}}(12q^{\frac{1}{2}}+...),\quad 
O_{12}&=&q^{-\frac{1}{4}}(1+66q+...), \quad
S_{12}/C_{12}&=&2^5q^{\frac{1}{2}}(1+12q+...),\\
V_{16}&=&q^{-\frac{1}{3}}(16q^{\frac{1}{2}}+...),\quad
O_{16}&=&q^{-\frac{1}{3}}(1+120q+...),\quad
S_{16}/C_{16}&=&2^7q^{\frac{2}{3}}(1+16q+...), \\
\label{espan}
\end{array}
$$
The lattice sum contributes to the spectrum as
\ba
\La_{2m,n} &\rightarrow& \overline q^0q^0+...,\nonumber\\
\La_{2m+1,n} &\rightarrow& \text{no massless solutions},\nonumber\\
\La_{2m,n+\frac{1}{2}} &\rightarrow& \text{no massless solutions},\nonumber\\
\La_{2m+1,n+\frac{1}{2}}&\rightarrow& \text{no massless solutions}.
\ea
\chapter{}
\section{Tables for two models with reduced Higgs spectrum}
\addtolength{\oddsidemargin}{-.8in} 
\begin{tabular}{|c|c|c|rrrrrrrr|c|rr|}
\hline
  $F$ & SEC & $SU(3)\times$&$Q_{C}$ & $Q_L$ & $Q_1$ &
   $Q_2$ & $Q_3$ & $Q_{4}$ & $Q_{5}$ & $Q_6$ &
   $SU(2)_{1,..,6}^6$ & $Q_7$ & $Q_{8}$\\
   $$ & $$ & $SU(2)$ & $$ & $$ & $$ & $$ & $$ & $$ & $$ & $$ & $$  & $$ & $$ \\
\hline
   $L_1$ & $b_1$      & $(1,2)$ & $-{3\over2}$ & $0$ &
   $-{1\over2}$ & $0$ & $0$ & $-{1\over2}$ & $0$ & $0$ &
   $(1,1,1,1,1,1)$ & $0$ & $0$ \\
   $Q_1$ &            & $(3,2)$ & $ {1\over2}$ & $0$ &
   $-{1\over2}$ & $0$ & $0$ & $1\over2$ & $0$ & $0$ &
   $(1,1,1,1,1,1)$ & $0$ & $0$ \\
   ${d}_{L_{1}}^c$ &            & $({\bar 3},1)$ & $-{1\over2}$ & $1$ &
   $-{1\over2}$ & $0$ & $0$ & $-{1\over2}$ & $0$ & $0$ &
   $(1,1,1,1,1,1)$ & $0$ & $0$ \\
   ${N}_{L_1}^c $&            & $(1,1)$ & ${3\over2}$ & $-1$ &
   $-{1\over 2}$ & $0$ & $0$ & $-{1\over2}$ & $0$ & $0$ &
   $(1,1,1,1,1,1)$ & $0$ & $0$ \\
   ${u}_{L_1}^c$ &            & $({\bar 3},1)$ & $-{1\over2}$ & $-1$ &
   $-{1\over2}$ & $0$ & $0$ & $1\over2$ & $0$ & $0$ &
   $(1,1,1,1,1,1)$ & $0$ & $0$ \\
   ${e}_{L_1}^c$ &            & $(1,1)$ & ${3\over2}$ & $1$ &
   $-{1\over 2}$ & $0$ & $0$ & $1\over2$ & $0$ & $0$ &
   $(1,1,1,1,1,1)$ & $0$ & $0$ \\
\hline
\end{tabular}\label{matter1}
Table 3.a.
\newpage
\begin{tabular}{|c|c|c|rrrrrrrr|c|rr|}
\hline 
$F$ & SEC & $SU(3)\times $&$Q_{C}$ & $Q_L$ & $Q_1$ &
   $Q_2$ & $Q_3$ & $Q_{4}$ & $Q_{5}$ & $Q_6$ &
   $SU(2)_{1,..,6}^6$ & $Q_{7}$ & $Q_{8}$\\
   $$ & $$ & $SU(2)$ & $$ & $$ & $$ & $$ & $$ & $$ & $$ & $$ & $$  & $$ & $$ \\
\hline
   $T_+$ & $1+b_4$ & $(\bar{3},1)$ & $-{1\over2}$ & $0$ & 
$0$ & $-{1\over2}$ &
   $0$ & $0$ & $-{1\over2}$ & $0$ & $(1,1,1,1,1,1)$ & $0$ &
   $1$ \\
$C_-$ & $+\beta$ & $(1,1)$ & ${3\over2}$ & $0$ & $0$ & $-{1\over2}$ &
   $0$ & $0$ & $-{1\over2}$ & $0$ & $(1,1,1,1,1,1)$ & $0$ &
   $-1$ \\
$C_+$ & $$ & $(1,1)$ & $-{3\over2}$ & $0$ & $0$ & ${1\over 2}$ &
   $0$ & $0$ & $-{1\over2}$ & $0$ & $(1,1,1,1,1,1)$ & $0$ &
   $1$ \\
$T_-$ & $$ & $(3,1)$ & ${1\over2}$ & $0$ & $0$ & ${1\over 2}$ &
   $0$ & $0$ & $-{1\over2}$ & $0$ & $(1,1,1,1,1,1)$ & $0$ &
   $-1$ \\
\hline
   $D_1$ & $b_1+2\gamma$ & $(1,1)$ & $0$ & $0$ & $0$ &
    $-{1\over2}$ & ${1\over 2}$ & $-{1\over2}$ & $0$ & $0$
& $(1,1,2,1,1,2)$ & $0$ &
   $0$ \\
   $S_1$ & $$      & $(1,1)$ & $0$ & $0$ & $0$ &
    $-{1\over2}$ & ${1\over 2}$ & $-{1\over2}$ & $0$ & $0$ & $(1,1,1,1,1,1)$ &
    $-1$ &
    $-1$ \\
\hline
\end{tabular}

\vspace{0.5cm}

Table 3.a continued.
\newpage
\begin{tabular}{|c|c|c|rrrrrrrr|c|rr|}
\hline
  $F$ & SEC & $SU(3)\times $&$Q_{C}$ & $Q_L$ & $Q_1$ &
   $Q_2$ & $Q_3$ & $Q_{4}$ & $Q_{5}$ & $Q_6$ &
   $ SU(2)_{1,..,6}^6$ & $Q_{7}$ & $Q_{8}$\\
   $$ & $$ & $SU(2)$ & $$ & $$ & $$ & $$ & $$ & $$ & $$ & $$ & $$  & $$ & $$ \\
\hline
$D^{(3)}_{--}$ & $b_1+b_3$ & (1,1) & $3\over4$  & $1\over2$ &
   $-{1\over4}$ & $1\over4$ & $-{1\over4}$ &
   $0$ & $-{1\over2}$ & $-{1\over2}$ &
   (1,1,2,1,1,1) & $-{1\over2}$ & $1\over2$  \\
   $D^{(3)}_{+-}$ & $\pm\gamma$ & (1,1) & $3\over4$  & $1\over2$ &
   $-{1\over4}$ & $1\over4$ & $-{1\over4}$ &
   $0$ & $-{1\over2}$ & $1\over2$ &
   (1,1,2,1,1,1) & $-{1\over2}$ & $1\over2$  \\
   $D^{(3)}_{-+}$ & $$ & (1,1) & $-{3\over4}$  & $-{1\over2}$ &
   ${1\over4}$ & $-{1\over4}$ & ${1\over4}$ &
   $0$ & $-{1\over2}$ & $-{1\over2}$ &
   (1,1,2,1,1,1) & ${1\over2}$ & $-{1\over2}$  \\
   $D^{(3)}_{++}$ & $$ & (1,1) & $-{3\over4}$  & $-{1\over2}$ &
   ${1\over4}$ & $-{1\over4}$ & ${1\over4}$ &
   $0$ & $-{1\over2}$ & $1\over2$ &
   (1,1,2,1,1,1) & ${1\over2}$ & $-{1\over2}$  \\
  \hline
$F$ & $1+b_3+\alpha$ & (1,1) & $3\over4$  & $-{1\over2}$ &
   $1\over4$ & $-{1\over4}$ & $-{1\over4}$ &
   $1\over2$ & $0$ & ${1\over2}$ &
   (1,1,2,1,1,1) & $-{1\over2}$ & $-{1\over2}$  \\
$F'$ & $\beta\pm\gamma$ & (1,1) & $3\over4$  & $-{1\over2}$ &
   $1\over4$ & $-{1\over4}$ & $-{1\over4}$ &
   $1\over2$ & $0$ & ${1\over2}$ &
   (1,1,1,1,1,2) & $1\over2$ & ${1\over2}$  \\
\hline
\end{tabular}
\label{matter33}

\vspace{0.5cm}

Table 3.a continued.
\newpage

\begin{tabular}{|c|c|c|rrrrrrrr|c|c|}
\hline
  $F$ & SEC & $SU(3)\times$&$Q_{C}$ & $Q_L$ & $Q_1$ &
   $Q_2$ & $Q_3$ & $Q_{4}$ & $Q_{5}$ & $Q_6$ &
   $SU(2)_{1,..,4} $ & $Q_{H_1}$  \\
   $$ & $$ & $SU(2)$ & $$ & $$ & $$ & $$ & $$ & $$ & $$ & $$ & $\times 
							SU(4)_{H_1}$ & $$  \\
\hline
   $L_1$ & $b_1$      & $(1,2)$ & $-{3\over2}$ & $0$ &
   $-{1\over2}$ & $0$ & $0$ & $1\over2$ & $0$ & $0$ &
   $(1,1,1,1,1)$ & $0$  \\
   $Q_1$ &            & $(3,2)$ & $ {1\over2}$ & $0$ &
   $-{1\over2}$ & $0$ & $0$ & $-{1\over2}$ & $0$ & $0$ &
   $(1,1,1,1,1)$ & $0$  \\
   ${d}_{1}^c$ &            & $({\bar 3},1)$ & $-{1\over2}$ & $1$ &
   $-{1\over2}$ & $0$ & $0$ & $-{1\over2}$ & $0$ & $0$ &
   $(1,1,1,1,1)$ & $0$  \\
   ${N}_{1}^c $&            & $(1,1)$ & ${3\over2}$ & $-1$ &
   $-{1\over 2}$ & $0$ & $0$ & $-{1\over2}$ & $0$ & $0$ &
   $(1,1,1,1,1)$ & $0$   \\
   ${u}_{1}^c$ &            & $({\bar 3},1)$ & $-{1\over2}$ & $-1$ &
   $-{1\over2}$ & $0$ & $0$ & $1\over2$ & $0$ & $0$ &
   $(1,1,1,1,1)$ & $0$   \\
   ${e}_{1}^c$ &            & $(1,1)$ & ${3\over2}$ & $1$ &
   $-{1\over 2}$ & $0$ & $0$ & $1\over2$ & $0$ & $0$ &
   $(1,1,1,1,1)$ & $0$   \\
\end{tabular}
  $\Phi_{45}$ & NS                      & $(1,1)$ & $0$ & $0$ &
   $0$ & $0$ & $0$ & $1$ & $1$ & $0$ &
   $(1,1,1,1,1)$   & $~~0$\\
 $\bar\Phi_{45}$ &                       & $(1,1)$ & $0$ & $0$ &
   $0$ & $0$ & $0$ & $-1$ & $-1$ & $0$ &
   $(1,1,1,1,1)$   &$~~0$\\
   $\Phi_{45}'$ &                       & $(1,1)$ & $0$ & $0$ &
   $0$ & $0$ & $0$ & $1$ & $-1$ & $0$ &
   $(1,1,1,1,1)$   &$~~0$\\
   $\bar\Phi_{45}'$ &                       & $(1,1)$ & $0$ & $0$ &
   $0$ & $0$ & $0$ & $-1$ & $1$ & $0$ &
   $(1,1,1,1,1)$   &$~~0$\\

\hline
   $\Phi^{\alpha\beta}_{1} $ & $b_1+ b_2$      & $(1,1)$ & $0$ & $0$ &
   $0$ & $0$ & $0$ & $-{1\over2}$ & ${1\over2}$ & $0$ &
   $(1,2,1,2,1)$ & $~~0$  \\
   $\bar\Phi^{\alpha\beta}_{1} $ & $\alpha + \beta$  &  $(1,1)$ & $0$ & $0$ &
  $0$ & $0$ & $0$ & ${1\over2}$ & $-{1\over2}$ & $0$ &
   $(1,2,1,2,1)$ & $~~0$  \\
   $\Phi^{\alpha\beta}_{2} $ & $ $  & $(1,1)$ & $0$ & $0$ &
   $0$ & $0$ & $0$ & $-{1\over2}$ & ${1\over2}$ & $0$ &
   $(2,1,2,1,1)$ & $~~0$  \\
   $\bar \Phi^{\alpha\beta}_{2}$ & $ $           & $(1,1)$ & $0$ & $0$ &
   $0$ & $0$ & $0$ & ${1\over2}$ & $-{1\over2}$ & $0$ &
   $(2,1,2,1,1)$ & $~~0$  \\
\hline
 \end{tabular}

\vspace{0.5cm}

Table 3.b.
\newpage
\begin{tabular}{|c|c|c|rrrrrrrr|c|c|}
\hline
  $F$ & SEC & $SU(3)\times$&$Q_{C}$ & $Q_L$ & $Q_1$ &
   $Q_2$ & $Q_3$ & $Q_{4}$ & $Q_{5}$ & $Q_6$ &
   $SU(2)_{1,..,4} $ & $Q_{H_1}$  \\
   $$ & $$ & $SU(2)$ & $$ & $$ & $$ & $$ & $$ & $$ & $$ & $$ & $\times
							SU(4)_{H_1}$  & $$  \\
\hline
   $V_7 $ & $b_3+ 2 \gamma$      & $(1,1)$ & $0$ & $0$ &
   $-{1\over2}$ & $-{1\over2}$ & $0$ & $0$ & $0$ & $1\over2$ &
   $(1,1,1,1,6)$ & $~~0$  \\
   $V_8 $ &            & $(1,1)$ & $0$ & $0$ &
   $-{1\over2}$ & $-{1\over2}$ & $0$ & $0$ & $0$ & $-{1\over2}$ &
   $(1,1,1,1,1)$ & $-2$  \\
   $V_9 $ &            & $(1,1)$ & $0$ & $0$ &
   $-{1\over2}$ & $-{1\over2}$ & $0$ & $0$ & $0$ & $-{1\over2}$ &
   $(1,1,1,1,1)$ & $~~2$  \\
\hline
  
   $V_{10} $ & $1+b_2+ $      & $(1,1)$ & $0$ & $0$ &
   $0$ & $-{1\over2}$ & $1\over2$ & $-{1\over2}$ & $0$ & $0$ &
   $(1,2,2,1,1)$ & $~~0$   \\
 $V_{11} $ & $b_3+2\gamma $      & $(1,1)$ & $0$ & $0$ &
   $0$ & $-{1\over2}$ & $1\over2$ & $-{1\over2}$ & $0$ & $0$ &
   $(2,1,1,2,1)$ & $~~0$   \\
\hline
   $V_{12} $ & $1+b_1+ $      & $(1,1)$ & $0$ & $0$ &
   $-{1\over2}$ & $0$ & $1\over2$ & $0$ & $1\over2$ & $0$ &
   $(2,1,1,2,1)$ & $~~0$   \\
\end{tabular}
   $H_3 $ & $b_3\pm \gamma$      & $(\bar{3},1)$ & $1\over4$ & $-{1\over2}$ &
   $1\over4$ & $1\over4$ & $-{1\over4}$ & $0$ & $0$ & $-{1\over2}$ &
   $(1,1,1,1,1)$ & $~~1$  \\
   $H_{4} $ &            & $(1,2)$ & $-{3\over4}$ & $1\over2$ &
   $1\over4$ & $1\over4$ & $-{1\over4}$ & $0$ & $0$ & $-{1\over2}$ &
   $(1,1,1,1,1)$ & $~~1$  \\
   $H_{5} $ &            & $(1,1)$ & $-{3\over4}$ & $-{1\over2}$ &
   $-{3\over4}$ & $1\over4$ & $-{1\over4}$ & $0$ & $0$ & $-{1\over2}$ &
   $(1,1,1,1,1)$ & $~~1$  \\
   $H_{6}$  &            & $(1,1)$ & $-{3\over4}$ & $-{1\over2}$ &
   $1\over4$ & $-{3\over4}$ & $-{1\over4}$ & $0$ & $0$ & $-{1\over2}$ &
   $(1,1,1,1,1)$ & $~~1$  \\ 
$H_{7} $ &            & $(1,1)$ & $-{3\over4}$ & $-{1\over2}$ &
   $1\over4$ & $1\over4$ & $3\over4$ & $0$ & $0$ & $-{1\over2}$ &
   $(1,1,1,1,1)$ & $~~1$  \\
   $\bar{H}_{3} $ & $ $      & $( 3,1)$ & $-{1\over4}$ & $1\over2$ &
   $-{1\over4}$ & $-{1\over4}$ & $1\over4$ & $0$ & $0$ & ${1\over2}$ &
   $(1,1,1,1,1)$ & $-1$  \\
  $\bar{H}_{4} $ &            & $(1,2)$ & $3\over4$ & $-{1\over2}$ &
   $-{1\over4}$ & $-{1\over4}$ & $1\over4$ & $0$ & $0$ & ${1\over2}$ &
   $(1,1,1,1,1)$ & $-1$  \\
   $\bar{H}_{5} $ &            & $(1,1)$ & $3\over4$ & $1\over2$ &
   ${3\over4}$ & $-{1\over4}$ & $1\over4$ & $0$ & $0$ & ${1\over2}$ &
   $(1,1,1,1,1)$ & $-1$  \\
\hline
\end{tabular}

\vspace{0.5cm}

Table 3.b continued.
{\addtolength{\oddsidemargin}{.5in} 
\newpage
\begin{tabular}{|r||r||rrrrrrrrrrrrrr|}
\hline 
\hline
FD &$\frac{Q^{(A)}}{15}$& $\Phi_{46}$ & $\Phi^{'}_{45}$ & 
	$\bar{\Phi}^{'}_{56}$&$V_3$&$V_2$&
	$V_6$&$V_5$&$V_9$&$V_8$&$N^{c}_{1}$&$N^{c}_{2}$&
			$N^{c}_{3}$&$\Phi^{'}_{46}$    &$\Phi_{45}$\\
       &  &$\bar{\Phi}_{56}$&$e^c_1$& $e^c_2$& $e^c_3$ &$H_7$& $H_6$ & 
						$H_5$ &&&&&&&\\
\hline
\hline
${\cal{D}}^{'}_1$&   1&  1& -2&  1&  0&  0&  0&  0&  0&  3&  0&  0&  0&  
				0&  0\\&&   0& 0&  0&  3&  2&  2&  2&&&&&&&\\
${\cal{D}}^{'}_2$&   2&  2& -1& -1&  0&  6&  0&  0&  0&  0&  0&  0&  0&
				0&  0\\&&   0& 0&  0&  6&  1&  4&  7&&&&&&&\\
${\cal{D}}^{'}_3$&   2& -1& -1&  2&  0&  0&  0&  6&  0&  0&  0&  0&  0&
			  	0&  0\\&&   0& 0&  0&  6&  1&  7&  4&&&&&&&\\
${\cal{D}}^{'}_4$&   0&  1& -1&  0&  0&  0&  0&  0&  0&  0&  0&  0&  0&
				0&  0\\&&   0& 0&  2& -2&  1& -1&  0&&&&&&&\\
${\cal{D}}^{'}_5$&   0& -1&  1&  0&  0&  0&  0&  0&  0&  0&  0&  2& -2&  
				0&  0\\&&   0& 0&  0&  0&  1& -1&  0&&&&&&&\\
${\cal{D}}^{'}_6$&   0&  0& -1&  1&  0&  0&  0&  0&  0&  0&  0&  0&  0&  
				0&  0\\&&   0& 2&  0& -2&  1&  0& -1&&&&&&&\\
${\cal{D}}^{'}_7$&   0&  1&  0& -1&  0&  0&  0&  0&  0&  0&  0&  0&  0&  
				0&  1\\&&   0& 0&  0&  0&  0&  0&  0&&&&&&&\\
${\cal{D}}^{'}_8$&   0&  1& -1&  0&  0&  0&  0&  0&  0&  0&  0&  0&  0&  
				0&  0\\&&   1& 0&  0&  0&  0&  0&  0&&&&&&&\\
${\cal{D}}^{'}_9$&   0&  0& -1&  1&  0&  0&  0&  0&  0&  0&  0&  0&  0&  
				1&  0\\&&   0& 0&  0&  0&  0&  0&  0&&&&&&&\\
${\cal{D}}^{'}_{10}$&0&  0&  0&  0&  0&  0&  0&  0&  1&  0&  0&  0&  0&  
				0&  0\\&&   0& 0&  0& -1&  0& -1& -1&&&&&&&\\
${\cal{D}}^{'}_{11}$&0&  0&  1& -1&  0&  0&  0&  0&  0&  0&  2&  0& -2&  
				0&  0\\&&   0& 0&  0&  0&  1&  0& -1&&&&&&&\\
${\cal{D}}^{'}_{12}$&0& -1&  1&  0&  0&  0&  2&  0&  0&  0&  0&  0&  0&  
				0&  0\\&&   0& 0&  0& -2& -1& -1&  -2&&&&&&&\\
${\cal{D}}^{'}_{13}$&0&  0&  1& -1&  2&  0&  0&  0&  0&  0&  0&  0&  0&  
				0&  0\\&&   0& 0&  0& -2& -1& -2& -1&&&&&&&\\
\hline
\hline
\end{tabular}

\vspace{0.5cm}

Table 3.c. $D$-Flat direction basis of non-abelian singlet fields.
Column 2 specifies the anomalous charge and columns 3 through
16 specify the norm-square VEV components of each basis direction.
The six fields $e^c_i$ and $H_{5,6,7}$ carry hypercharge, 
the remaining do not.
A negative component indicates the vector partner of a field (if it exists) 
must take on VEV rather than the field.
 \newpage
\begin{tabular}{|r||r|}
\hline 
\hline
FD & VEV\\
\hline
\hline
${\cal{D}}^{'}_1$& $V_8$\\
${\cal{D}}^{'}_2$& $V_2$\\
${\cal{D}}^{'}_3$& $V_5$\\
${\cal{D}}^{'}_4$& $e^c_2$\\
${\cal{D}}^{'}_5$& $N^{c}_{2}$\\
${\cal{D}}^{'}_6$& $e^c_1$\\
${\cal{D}}^{'}_7$& $\Phi_{45}$\\
${\cal{D}}^{'}_8$& $\bar{\Phi}_{56}$\\
${\cal{D}}^{'}_9$& $\Phi^{'}_{46}$\\
${\cal{D}}^{'}_{10}$& $V_9$\\
${\cal{D}}^{'}_{11}$& $N^{c}_{1}$\\
${\cal{D}}^{'}_{12}$& $V_6$\\
${\cal{D}}^{'}_{13}$& $V_3$\\
\hline
\hline
\end{tabular}

\vspace{0.5cm}

Table 3.d. Unique VEV associated with each non-abelian singlet field 
$D$-Flat basis direction.
\newpage
\begin{tabular}{|r||r||rrrrrrrrrrrrrr|}
\hline 
\hline
FD    &$\frac{Q^{(A)}}{15}$& $\Phi_{46}$ & $\Phi^{'}_{45}$ & $\bar{\Phi}^{'}_{56}$	&
	$V_3$&$V_2$&
	$V_6$&$V_5$&$V_9$&$V_8$&$N^{c}_{1}$&$N^{c}_{2}$&$N^{c}_{3}$&
  	$\Phi^{'}_{46}$      &$\Phi_{45}$\\
& &$\bar{\Phi}_{56}$&$e^c_1$& $e^c_2$& $e^c_3$ &$H_7$& $H_6$ & $H_5$ &&&&&&&\\
& &$\bar{\Phi}^{\alpha\beta}_{1,2}$&$H_{11}$&
		$H_{10}$ & $\bar{H}_{9}$& $\bar{H}_{8}$ & $H_{16}$
& $H_{14}$ & $H_{12}$ & $V_{1}$ & $V_{4}$ & $V_{7}$ & $H_{17}$ 
		                       & $H_{15}$ & $H_{13}$ \\
& & $V_{15}$& $V_{14}$ & $V_{13}$ & $V_{12}$ & $V_{10}$ & $V_{11}$&&&&&&&&\\
&& $Q_1$ & $Q_2$ & $Q_3$ & $d^c_1$ & $d^c_2$ & $d^c_3$ & $u^c_1$ & $u^c_2$ & 
$u^c_3$ & $H_3$ & $h$   & $L_1$ & $L_2$   & $L_3 $  \\
&& $H_4$&$\bar{H}_1$ & $\bar{H}_2$ &&&&&&&&&&&\\                 
\hline
\hline
${\cal{D}}_{1}$&   -2& 4& 1& 1& 0& 0& 0& 0& 0& 0& 6& 0& 0& 0& 0\\&& 
                       0& 0& 0&-6&-1&-4&-7&  &  &  &  &  &  &  \\
\hline
\hline
\end{tabular}

\vspace{0.5cm}

Table 3.e $D$-Flat direction basis of all fields.
\newpage
\begin{tabular}{|r||r||rrrrrrrrrrrrrr|}
\hline 
\hline
FD  &$\frac{Q^{(A)}}{15}$& $\Phi_{46}$ & $\Phi^{'}_{45}$ & $\bar{\Phi}^{'}_{56}$&$V_3$&$V_2$&
$V_6$&$V_5$&$V_9$&$V_8$&$N^{c}_{1}$&$N^{c}_{2}$&$N^{c}_{3}$&$\Phi^{'}_{46}$  
				&$\Phi_{45}$\\
&         &$\bar{\Phi}_{56}$&$e^c_1$& $e^c_2$& $e^c_3$ &$H_7$& $H_6$ & 
						$H_5$ &&&&&&&\\
& & $\bar{\Phi}^{\alpha\beta}_{1,2}$ & $H_{11}$ & $H_{10}$ & $\bar{H}_{9}$ & 
	$\bar{H}_{8}$ & $H_{16}$ & $H_{14}$ & $H_{12}$ & $V_{1}$ & $V_{4}$ & 
	$V_{7}$ & $H_{17}$ & $H_{15}$ & $H_{13}$ \\
& & $V_{15}$& $V_{14}$ & $V_{13}$ & $V_{12}$ & $V_{10}$ & $V_{11}$&&&&&&&&\\
&& $Q_1$ & $Q_2$ & $Q_3$ & $d^c_1$ & $d^c_2$ & $d^c_3$ & $u^c_1$ & $u^c_2$ & 
$u^c_3$  & $H_3$ & $h$   & $L_1$ & $L_2$   & $L_3 $  \\
&& $H_4$&$\bar{H}_1$ & $\bar{H}_2$ &&&&&&&&&&&\\                  
\hline
\hline
${\cal{D}}_{7}$&   -1& 1& 0& 1& 0& 0& 0& 0& 0& 0& 0& 0& 0& 0& 0\\&&
                       0& 0& 0&-2&-1&-2&-2&  &  &  &  &  &  &  \\
\hline
\hline
\end{tabular}

\vspace{0.5cm}

Table 3.e continued.
\newpage
\begin{tabular}{|r||r||rrrrrrrrrrrrrr|}
\hline 
\hline
FD    &$\frac{Q^{(A)}}{15}$& $\Phi_{46}$ & $\Phi^{'}_{45}$ & $\bar{\Phi}^{'}_{56}$&$V_3$&
$V_2$& $V_6$&$V_5$&$V_9$&$V_8$&$N^{c}_{1}$&$N^{c}_{2}$&$N^{c}_{3}$&$
\Phi^{'}_{46}$      &$\Phi_{45}$\\
& &$\bar{\Phi}_{56}$&$e^c_1$& $e^c_2$& $e^c_3$ &$H_7$& $H_6$ & $H_5$ &&&&&&&\\
& & $\bar{\Phi}^{\alpha\beta}_{1,2}$ & $H_{11}$ & $H_{10}$ & $\bar{H}_{9}$ &
$\bar{H}_{8}$ & $H_{16}$ & $H_{14}$ & $H_{12}$ & $V_{1}$ & $V_{4}$ & $V_{7}$ 
				& $H_{17}$ & $H_{15}$ & $H_{13}$ \\
& & $V_{15}$& $V_{14}$ & $V_{13}$ & $V_{12}$ & $V_{10}$ & $V_{11}$&&&&&&&&\\
&& $Q_1$ & $Q_2$ & $Q_3$ & $d^c_1$ & $d^c_2$ & $d^c_3$ & $u^c_1$ & $u^c_2$ &
$u^c_3$ & $H_3$ & $h$   & $L_1$ & $L_2$   & $L_3 $  \\
&& $H_4$&$\bar{H}_1$ & $\bar{H}_2$ &&&&&&&&&&&\\                   
\hline
\hline
${\cal{D}}_{13}$   &0& 0& 0&-1& 0& 0& 0& 0& 0& 0& 0& 0& 0& 0& 0\\&&
                       0& 0& 0& 0& 0& 0& 0&  &  &  &  &  &  &  \\
\hline
\hline
\end{tabular}

\vspace{0.5cm}

Table 3.e continued.
\newpage

\begin{tabular}{|r||r||rrrrrrrrrrrrrr|}
\hline 
\hline
FD    &$\frac{Q^{(A)}}{15}$& $\Phi_{46}$ & $\Phi^{'}_{45}$ & $\bar{\Phi}^{'}_{56}$&
	$V_3$&$V_2$& $V_6$&$V_5$&$V_9$&$V_8$&$N^{c}_{1}$&$N^{c}_{2}$&$N^{c}_{3}$&
	$\Phi^{'}_{46}$      &$\Phi_{45}$\\
& &$\bar{\Phi}_{56}$&$e^c_1$& $e^c_2$& $e^c_3$ &$H_7$& $H_6$ & $H_5$ &&&&&&&\\
& & $\bar{\Phi}^{\alpha\beta}_{1,2}$ & $H_{11}$ & $H_{10}$ & $\bar{H}_{9}$ & 
$\bar{H}_{8}$ & $H_{16}$ & $H_{14}$ & $H_{12}$ & $V_{1}$ & $V_{4}$ & $V_{7}$ & 
$H_{17}$ & $H_{15}$ & $H_{13}$ \\
& & $V_{15}$& $V_{14}$ & $V_{13}$ & $V_{12}$ & $V_{10}$ & $V_{11}$&&&&&&&&\\
&& $Q_1$ & $Q_2$ & $Q_3$ & $d^c_1$ & $d^c_2$ & $d^c_3$ & $u^c_1$ & $u^c_2$ & 
$u^c_3$ & $H_3$ & $h$   & $L_1$ & $L_2$   & $L_3 $  \\
&& $H_4$&$\bar{H}_1$ & $\bar{H}_2$ &&&&&&&&&&&\\                          
\hline
\hline
${\cal{D}}_{18}$&   0&-1& 1& 0& 0& 0& 2& 0& 0& 0& 0& 0& 0& 0& 0\\&&
                       0& 0& 0&-2&-1&-1&-2&  &  &  &  &  &  &  \\
\hline
\hline
\end{tabular}

\vspace{0.5cm}

Table 3.e continued.
\newpage
\begin{tabular}{|r||r||rrrrrrrrrrrrrr|}
\hline 
\hline
FD    &$\frac{Q^{(A)}}{15}$& $\Phi_{46}$ & $\Phi^{'}_{45}$ & $\bar{\Phi}^{'}_{56}$&
	$V_3$&$V_2$& $V_6$&$V_5$&$V_9$&$V_8$&$N^{c}_{1}$&$N^{c}_{2}$&$N^{c}_{3}$&
	$\Phi^{'}_{46}$ &$\Phi_{45}$\\
& &$\bar{\Phi}_{56}$&$e^c_1$& $e^c_2$& $e^c_3$ &$H_7$& $H_6$ & $H_5$ &&&&&&&\\
& & $\bar{\Phi}^{\alpha\beta}_{1,2}$ & $H_{11}$ & $H_{10}$ & $\bar{H}_{9}$ & 
$\bar{H}_{8}$ & $H_{16}$ & $H_{14}$ & $H_{12}$ & $V_{1}$ & $V_{4}$ & $V_{7}$ & 
$H_{17}$ & $H_{15}$ & $H_{13}$ \\
& & $V_{15}$& $V_{14}$ & $V_{13}$ & $V_{12}$ & $V_{10}$ & $V_{11}$&&&&&&&&\\
&& $Q_1$ & $Q_2$ & $Q_3$ & $d^c_1$ & $d^c_2$ & $d^c_3$ & $u^c_1$ & $u^c_2$ & 
$u^c_3$ & $H_3$ & $h$   & $L_1$ & $L_2$   & $L_3 $  \\
&& $H_4$&$\bar{H}_1$ & $\bar{H}_2$ &&&&&&&&&&&\\                           
\hline
\hline
${\cal{D}}_{24}$&   0& 0&-1& 1& 0& 0& 0& 0& 0& 0& 0& 0& 0& 0& 0\\&&
                       0& 2& 0&-2& 1& 0&-1&  &  &  &  &  &  &  \\
\hline
\hline
\end{tabular}

\vspace{0.5cm}

Table 3.e continued.
\newpage
\begin{tabular}{|r||r||rrrrrrrrrrrrrr|}
\hline 
\hline
FD    &$\frac{Q^{(A)}}{15}$& $\Phi_{46}$ & $\Phi^{'}_{45}$ & $\bar{\Phi}^{'}_{56}$&$V_3$&
	$V_2$&$V_6$&$V_5$&$V_9$&$V_8$&$N^{c}_{1}$&$N^{c}_{2}$&$N^{c}_{3}$&
	$\Phi^{'}_{46}$      &$\Phi_{45}$\\
& &$\bar{\Phi}_{56}$&$e^c_1$& $e^c_2$& $e^c_3$ &$H_7$& $H_6$ & $H_5$ &&&&&&&\\
& & $\bar{\Phi}^{\alpha\beta}_{1,2}$ & $H_{11}$ & $H_{10}$ & $\bar{H}_{9}$ & 
	$\bar{H}_{8}$ & $H_{16}$ & $H_{14}$ & $H_{12}$ & $V_{1}$ & $V_{4}$ & 
	$V_{7}$ & $H_{17}$ & $H_{15}$ & $H_{13}$ \\
& & $V_{15}$& $V_{14}$ & $V_{13}$ & $V_{12}$ & $V_{10}$ & $V_{11}$&&&&&&&&\\
&& $Q_1$ & $Q_2$ & $Q_3$ & $d^c_1$ & $d^c_2$ & $d^c_3$ & $u^c_1$ & $u^c_2$ & 
$u^c_3$ & $H_3$ & $h$   & $L_1$ & $L_2$   & $L_3 $  \\
&& $H_4$&$\bar{H}_1$ & $\bar{H}_2$ &&&&&&&&&&&\\             
\hline
\hline
${\cal{D}}_{30}$&   1& 4& 1&-2& 0& 0& 0& 0& 0& 0& 0& 0& 0& 0& 0\\&&
                       0& 0& 0& 6&-4& 5&-1&  &  &  &  &  &  &  \\
\hline
\hline
\end{tabular}

\vspace{0.5cm}

Table 3.e continued.

\begin{tabular}{|r||r||rrrrrrrrrrrrrr|}
\hline 
\hline
FD  &$\frac{Q^{(A)}}{15}$& $\Phi_{46}$ & $\Phi^{'}_{45}$ & $\bar{\Phi}^{'}_{56}$&$V_3$&$V_2$&
$V_6$&$V_5$&$V_9$&$V_8$&$N^{c}_{1}$&$N^{c}_{2}$&$N^{c}_{3}$&$\Phi^{'}_{46}$
      &$\Phi_{45}$\\
      & &$\bar{\Phi}_{56}$&$e^c_1$& $e^c_2$& $e^c_3$ &$H_7$& $H_6$ & 
	$H_5$ &&&&&&&\\
& & $\bar{\Phi}^{\alpha\beta}_{1,2}$ & $H_{11}$ & $H_{10}$ & $\bar{H}_{9}$ & 
	$\bar{H}_{8}$ & $H_{16}$ & $H_{14}$ & $H_{12}$ & $V_{1}$ & $V_{4}$ & 
	$V_{7}$ & $H_{17}$ & $H_{15}$ & $H_{13}$ \\
& & $V_{15}$& $V_{14}$ & $V_{13}$ & $V_{12}$ & $V_{10}$ & $V_{11}$&&&&&&&&\\
&& $Q_1$ & $Q_2$ & $Q_3$ & $d^c_1$ & $d^c_2$ & $d^c_3$ & $u^c_1$ & $u^c_2$ & 
$u^c_3$ & $H_3$ & $h$   & $L_1$ & $L_2$   & $L_3 $  \\
&& $H_4$&$\bar{H}_1$ & $\bar{H}_2$ &&&&&&&&&&&\\             
\hline
\hline
${\cal{D}}_{36}$&   1&-2& 1& 1& 0& 0& 0& 0& 0& 0& 0& 0& 0& 0& 0\\&&
                       0& 0& 0& 0&-1& 2&-1&  &  &  &  &  &  &  \\
\hline
\hline
\end{tabular}

\vspace{0.5cm}

Table 3.e continued.
\newpage
\begin{tabular}{|r||r||rrrrrrrrrrrrrr|}
\hline 
\hline
FD &$\frac{Q^{(A)}}{15}$& $\Phi_{46}$ & $\Phi^{'}_{45}$ & $\bar{\Phi}^{'}_{56}$&$V_3$&$V_2$&
$V_6$&$V_5$&$V_9$&$V_8$&$N^{c}_{1}$&$N^{c}_{2}$&$N^{c}_{3}$&$\Phi^{'}_{46}$
      &$\Phi_{45}$\\
& &$\bar{\Phi}_{56}$&$e^c_1$& $e^c_2$& $e^c_3$ &$H_7$& $H_6$ & $H_5$ &&&&&&&\\
& & $\bar{\Phi}^{\alpha\beta}_{1,2}$ & $H_{11}$ & $H_{10}$ & $\bar{H}_{9}$ & 
	$\bar{H}_{8}$ & $H_{16}$ & $H_{14}$ & $H_{12}$ & $V_{1}$ & $V_{4}$ & 
	$V_{7}$ & $H_{17}$ & $H_{15}$ & $H_{13}$ \\
& & $V_{15}$& $V_{14}$ & $V_{13}$ & $V_{12}$ & $V_{10}$ & $V_{11}$&&&&&&&&\\
&& $Q_1$ & $Q_2$ & $Q_3$ & $d^c_1$ & $d^c_2$ & $d^c_3$ & $u^c_1$ & $u^c_2$ &
	$u^c_3$ & $H_3$ & $h$   & $L_1$ & $L_2$   & $L_3 $  \\
&& $H_4$&$\bar{H}_1$ & $\bar{H}_2$ &&&&&&&&&&&\\             
\hline
\hline
${\cal{D}}_{42}$&   1&-2& 1& 4& 0& 0& 0& 0& 0& 0& 0& 0& 0& 0& 0\\&&
                       0& 0& 0& 6&-4&-1& 5&  &  &  &  &  &  &  \\
\hline
\hline
\end{tabular}

\vspace{0.5cm}

Table 3.e continued.
\newpage

\begin{tabular}{|r||r||rrrrrrrrrrrrrr|}
\hline 
\hline
FD &$\frac{Q^{(A)}}{15}$& $\Phi_{46}$ & $\Phi^{'}_{45}$ & $\bar{\Phi}^{'}_{56}$&$V_3$&$V_2$&
$V_6$&$V_5$&$V_9$&$V_8$&$N^{c}_{1}$&$N^{c}_{2}$&$N^{c}_{3}$&$\Phi^{'}_{46}$ &
		$\Phi_{45}$\\
& &$\bar{\Phi}_{56}$&$e^c_1$& $e^c_2$& $e^c_3$ &$H_7$& $H_6$ & $H_5$ &&&&&&&\\
& & $\bar{\Phi}^{\alpha\beta}_{1,2}$ & $H_{11}$ & $H_{10}$ & $\bar{H}_{9}$ & 
	$\bar{H}_{8}$ & $H_{16}$ & $H_{14}$ & $H_{12}$ & $V_{1}$ & $V_{4}$ & 
	$V_{7}$ & $H_{17}$ & $H_{15}$ & $H_{13}$ \\
& & $V_{15}$& $V_{14}$ & $V_{13}$ & $V_{12}$ & $V_{10}$ & $V_{11}$&&&&&&&&\\
&& $Q_1$ & $Q_2$ & $Q_3$ & $d^c_1$ & $d^c_2$ & $d^c_3$ & $u^c_1$ & $u^c_2$ & 
$u^c_3$ & $H_3$ & $h$   & $L_1$ & $L_2$   & $L_3 $  \\
&& $H_4$&$\bar{H}_1$ & $\bar{H}_2$ &&&&&&&&&&&\\                         
\hline
\hline
${\cal{D}}_{48}$&   2& 0& 1&-3& 0& 0& 0& 0& 0& 0& 0& 0& 0& 0& 0\\&& 
                       0& 0& 0& 2& 3& 4& 1&  &  &  &  &  &  &  \\
\hline
\hline
\end{tabular}

\vspace{0.5cm}

Table 3.e continued.
\newpage

\begin{tabular}{|r|r|||r|r|||r|r|||r|r|||r|r|}
\hline 
\hline
FD & VEV & FD & VEV & FD & VEV & FD & VEV & FD & VEV\\
\hline
\hline
\hline
\bda{1} & $N^{c}_{1}$  & \bda{11}& $\bar{H}_{2}$  & \bda{21}& ${u}^{c}_{3}$  &
	\bda{31}& $V_{ 4}$  & \bda{41}&$V_{1}$\\
\bda{2} & $N^{c}_{2}$  & \bda{12}& $\bar{H}_{1}$  & \bda{22}& $h$            &
	 \bda{32}& $H_{13}$  & \bda{42}&$H_{17}$\\
\bda{3} & $N^{c}_{3}$  & \bda{13}& $H_{10}$       & \bda{23}& ${e}^{c}_{2}$  &
	 \bda{33}& $V_{10}$  & \bda{43}&$V_{7}$\\
\bda{4} & $H_{16}$     & \bda{14}& $\Phi^{'}_{46}$  
                                                  & \bda{24}& ${e}^{c}_{1}$  &
	 \bda{34}& $V_{15}$  & \bda{44}&$V_{11}$\\
\bda{5} & $H_{14}$     & \bda{15}& $\bar{\Phi}_{56}$
                                                  & \bda{25}& $V_{ 9}$       &
	 \bda{35}& $V_{14}$  & \bda{45}&$L_{1}$\\
\bda{6} & $H_{12}$     & \bda{16}& $\bar{\Phi}^{\alpha\beta}_{1,2}$ 
                                                  & \bda{26}& $V_{ 3}$       &
	 \bda{36}& $V_{13}$  & \bda{46}&$V_{2}$\\
\bda{7} & $H_{ 3}$     & \bda{17}& $\bar{H}_{8}$  & \bda{27}& ${u}^{c}_{1}$  &
	 \bda{37}& $V_{12}$  & \bda{47}&$V_{5}$\\
\bda{8} & $\bar{H}_{9}$& \bda{18}& $V_{ 6}$       & \bda{28}& $Q_{ 2}$       &
	 \bda{38}& ${d}^{c}_{3}$ &\bda{48}&${d}^{c}_{1}$\\
\bda{9} & $H_{11}$     & \bda{19}& $Q_{ 1}$       & \bda{29}& ${u}^{c}_{2}$  &
	 \bda{39}& ${L}_{3}$  &\bda{49}&${d}^{c}_{2}$\\
\bda{10}& $\Phi_{45}$  & \bda{20}& $Q_{3}$  & \bda{30}& $H_{15}$       &
		 \bda{40}& $V_{ 8}$  & \bda{50}&$L_{2}$\\

\hline
\end{tabular}

\vspace{0.5cm}

Table 3.f. Unique VEV associated with each $D$-Flat basis direction.
\newpage

\noindent Quintic superpotential: 
\beqn
W_5&=&  
     Q_{1}  H_{3}   L_{1}  \bar{H}_{5}   \xi_{2}
+    Q_{2}  H_{3}   L_{2}  \bar{H}_{6}   \xi_{1}
+    Q_{3}   u^{c}_{3}  \bar{H}_{1}  \bar{H}_{7}  H_{10}
+    Q_{3}   u^{c}_{3}  H_{2}  \bar{H}_{7}  \bar{H}_{8}
\nonumber\\&+&
     d^{c}_{1}   u^{c}_{1}  H_{3}  \bar{H}_{5}   \xi_{2}
+    d^{c}_{1}  H_{3}  H_{3}  \Phi_{46}  V_{2}
+    d^{c}_{2}   u^{c}_{2}  H_{3}  \bar{H}_{6}   \xi_{1}
+    d^{c}_{2}  H_{3}  H_{3}  \bar{\Phi}^{'}_{56}  V_{5}
\nonumber\\&+&
     H_{3}  \bar{H}_{4}  \bar{H}_{1}   \bar{H}_{3}  H_{10}
+    H_{3}  \bar{H}_{4}  H_{2}   \bar{H}_{3}  \bar{H}_{8}
+    H_{3}  \bar{H}_{1}  \bar{H}_{2}   \bar{H}_{3}  \bar{\Phi}^{\alpha\beta}_{2}
+    H_{3}  \bar{H}_{3}  \Phi^{\alpha\beta}_{1}  H_{11}  H_{9}
\nonumber\\&+&
     H_{3}   \bar{H}_{3}  \bar{\Phi}^{\alpha\beta}_{2}  \bar{H}_{8}  \bar{H}_{10}
+    L_{3}  \bar{H}_{1}   N^{c}_3  \bar{H}_{7}  H_{10}
+    L_{3}  H_{2}   N^{c}_3 \bar{H}_{7}  \bar{H}_{8}
+    H_{4}  H_{4} \bar{\Phi}^{'}_{46}  H_{8}  H_{8}
\nonumber\\&+&
     H_{4} H_{4} \Phi_{46}   N^{c}_{1} V_{2}
+    H_{4} H_{4} \bar{\Phi}^{'}_{56}   N^{c}_{2}  V_{5}
+    H_{4} H_{4} \bar{\Phi}^{'}_{56}  \bar{H}_{10}  \bar{H}_{10}
+    H_{4} \bar{H}_{4}  \bar{H}_{4}  \bar{H}_{1}  H_{10}
\nonumber\\&+&
     H_{4} \bar{H}_{4}  \bar{H}_{4}  H_{2}  \bar{H}_{8}
+    H_{4} \bar{H}_{4}  \bar{H}_{1}  \bar{H}_{2}  \bar{\Phi}^{\alpha\beta}_{2}
+    H_{4} \bar{H}_{4}  \Phi^{\alpha\beta}_{1}  H_{11}  H_{9}
+    H_{4} \bar{H}_{4}  \bar{\Phi}^{\alpha\beta}_{2}  \bar{H}_{8}  \bar{H}_{10}
\nonumber\\&+&
     H_{4} H_{1}   \xi_{2}  H03  H_{8}
+    H_{4} H_{2}  \bar{\Phi}^{'}_{56}  \Phi^{\alpha\beta}_{2}  \bar{H}_{10}
+   \bar{H}_{4}  \bar{H}_{4}  \Phi^{'}_{46}  \bar{H}_{8}  \bar{H}_{8}
+   \bar{H}_{4}  \bar{H}_{4}  \Phi^{'}_{56} H_{10}  H_{10}
\nonumber\\&+&
    \bar{H}_{4}  \bar{H}_{1}  \bar{H}_{1}  H_{1}  H_{10}
+   \bar{H}_{4}  \bar{H}_{1}  H_{1}  H_{2}  \bar{H}_{8}
+   \bar{H}_{4}  \bar{H}_{1}  \bar{H}_{2}  H_{2}  H_{10}
+   \bar{H}_{4}  \bar{H}_{1}  H_{7}  \bar{H}_{7}  H_{10}
\nonumber\\&+&
    \bar{H}_{4}  \bar{H}_{1}  H_{6}  \bar{H}_{6}  H_{10}
+   \bar{H}_{4}  \bar{H}_{1}  H_{5}  \bar{H}_{5}  H_{10}
+   \bar{H}_{4}  \bar{H}_{1} \xi_{2} \bar{\Phi}^{\alpha\beta}_{2}  \bar{H}_{8}
+   \bar{H}_{4}  \bar{H}_{1} \Phi^{\alpha\beta}_{1} \bar{\Phi}^{\alpha\beta}_{1} H_{10}
\nonumber\\&+&
   \bar{H}_{4}  \bar{H}_{1} \Phi^{\alpha\beta}_{2} \bar{\Phi}^{\alpha\beta}_{2} H_{10}
+  \bar{H}_{4}  \bar{H}_{1}  H_{11}  H_{10}  \bar{H}_{11}
+   \bar{H}_{4} \bar{H}_{1}  H_{10}  H_{10}  \bar{H}_{10}
+   \bar{H}_{4} \bar{H}_{1}  H_{10}  \bar{H}_{9}  H_{9}
\nonumber\\&+&
    \bar{H}_{4}  \bar{H}_{1}  H_{10}  \bar{H}_{8}  H_{8}
+   \bar{H}_{4}  H_{1}  H_{10}  H_{16}  H_{17}
+   \bar{H}_{4}  \bar{H}_{2}  H_{2}  H_{2}  \bar{H}_{8}
+   \bar{H}_{4}  \bar{H}_{2}  \Phi^{'}_{56} \bar{\Phi}^{\alpha\beta}_{2}  H_{10}
\nonumber\\&+&
    \bar{H}_{4} H_{2} H_{7}  \bar{H}_{7}  \bar{H}_{8}
+   \bar{H}_{4} H_{2} H_{6}  \bar{H}_{6}  \bar{H}_{8}
+   \bar{H}_{4} H_{2} H_{5}  \bar{H}_{5}  \bar{H}_{8}
+   \bar{H}_{4} H_{2} \Phi^{\alpha\beta}_{1} \bar{\Phi}^{\alpha\beta}_{1} 
							\bar{H}_{8}
\nonumber\\&+&
    \bar{H}_{4} H_{2}\Phi^{\alpha\beta}_{2} \bar{\Phi}^{\alpha\beta}_{2}
							\bar{H}_{8}
+   \bar{H}_{4} H_{2} H_{11}  \bar{H}_{8}  \bar{H}_{11}
+   \bar{H}_{4} H_{2} H_{10}  \bar{H}_{8}  \bar{H}_{10}
+   \bar{H}_{4} H_{2}\bar{H}_{9}  \bar{H}_{8}  H_{9}
\nonumber\\&+&
    \bar{H}_{4}  H_{2}  \bar{H}_{8}  \bar{H}_{8}  H_{8}
+   \bar{H}_{1}  \bar{H}_{1}  H_{1}  \bar{H}_{2}  \bar{\Phi}^{\alpha\beta}_{2}
+   \bar{H}_{1}  \bar{H}_{1}  \bar{H}_{2}  \bar{H}_{2}  \Phi^{'}_{45}
+    \bar{H}_{1}  \bar{H}_{1}  \bar{\Phi}^{'}_{46}  \bar{\Phi}^{\alpha\beta}_{1}
                                                   \bar{\Phi}^{\alpha\beta}_{1}
\nonumber\\&+&
   \bar{H}_{1}  \bar{H}_{1}  \bar{\Phi}^{'}_{46}  \bar{\Phi}^{\alpha\beta}_{2}
                                                   \bar{\Phi}^{\alpha\beta}_{2}
+   \bar{H}_{1}  H_{1}  \Phi^{\alpha\beta}_{1}  H_{11}  H_{9}
+   \bar{H}_{1}  H_{1}  \bar{\Phi}^{\alpha\beta}_{2}  \bar{H}_{8}  \bar{H}_{10}
+   \bar{H}_{1}  \bar{H}_{2}  \bar{H}_{2}  H_{2}  \bar{\Phi}^{\alpha\beta}_{2}
\nonumber\\&+&
    \bar{H}_{1}  \bar{H}_{2}  \Phi^{'}_{45}  \bar{H}_{8}  \bar{H}_{10}
+   \bar{H}_{1}  \bar{H}_{2}  H_{7}  \bar{H}_{7}  \bar{\Phi}^{\alpha\beta}_{2}
+   \bar{H}_{1}  \bar{H}_{2}  H_{6}  \bar{H}_{6}  \bar{\Phi}^{\alpha\beta}_{2}
+   \bar{H}_{1}  \bar{H}_{2}  H_{5}  \bar{H}_{5}  \bar{\Phi}^{\alpha\beta}_{2}
\nonumber\\&+&
   \bar{H}_{1}  \bar{H}_{2}  \Phi^{\alpha\beta}_{1}  \bar{\Phi}^{\alpha\beta}_{1}
                                                     \bar{\Phi}^{\alpha\beta}_{2}
+   \bar{H}_{1}  \bar{H}_{2}  \Phi^{\alpha\beta}_{2}  \bar{\Phi}^{\alpha\beta}_{1}
                                                     \bar{\Phi}^{\alpha\beta}_{1}
+   \bar{H}_{1}  \bar{H}_{2}  \Phi^{\alpha\beta}_{2}  \bar{\Phi}^{\alpha\beta}_{2}
                                                     \bar{\Phi}^{\alpha\beta}_{2}
+   \bar{H}_{1}  \bar{H}_{2}  \bar{\Phi}^{\alpha\beta}_{2}  H_{11}  \bar{H}_{11}
\nonumber\\&+&
   \bar{H}_{1}  \bar{H}_{2}  \bar{\Phi}^{\alpha\beta}_{2}  H_{10}  \bar{H}_{10}
+   \bar{H}_{1}  \bar{H}_{2}  \bar{\Phi}^{\alpha\beta}_{2}  \bar{H}_{9}  H_{9}
+   \bar{H}_{1}  \bar{H}_{2}  \bar{\Phi}^{\alpha\beta}_{2}  \bar{H}_{8}  H_{8}
+     H_{1}  H_{1}  H_{2}  H_{2}  \bar{\Phi}^{'}_{45}
\nonumber\\&+&
    H_{1}  H_{1}  \Phi^{'}_{46}  \Phi^{\alpha\beta}_{1}  \Phi^{\alpha\beta}_{1}
+    H_{1}  H_{1}  \Phi^{'}_{46}  \Phi^{\alpha\beta}_{2}  \Phi^{\alpha\beta}_{2}
+    H_{1}  H_{1}  \Phi^{'}_{46}  H_{12}  H_{13}
+     H_{1}  H_{1}  \Phi_{45}  H_{14}  H_{15}
\nonumber\\&+&
    H_{1}  \bar{H}_{2}  \bar{\Phi}^{\alpha\beta}_{2}  H_{16}  H_{17}
+    H_{1}  H_{2}  \bar{\Phi}^{'}_{45}  H_{10}  H_{8}
+    \bar{H}_{2}  \bar{H}_{2}  \Phi^{'}_{45}  H_{16}  H_{17}
+     \bar{H}_{2}  \bar{H}_{2}  \Phi^{'}_{56} \bar{\Phi}^{\alpha\beta}_{1}
                                               \bar{\Phi}^{\alpha\beta}_{1}
\nonumber\\&+&
    \bar{H}_{2}  \bar{H}_{2}  \Phi^{'}_{56} \bar{\Phi}^{\alpha\beta}_{2}
                                               \bar{\Phi}^{\alpha\beta}_{2}
+    \bar{H}_{2}  H_{2}  \Phi^{\alpha\beta}_{1}  H_{11}  H_{9}
+    \bar{H}_{2}  H_{2}  \bar{\Phi}^{\alpha\beta}_{2}  \bar{H}_{8}  \bar{H}_{10}
+   H_{2}  H_{2}\bar{\Phi}^{'}_{56}\Phi^{\alpha\beta}_{1}\Phi^{\alpha\beta}_{1}
\nonumber\\&+&
   H_{2}  H_{2}\bar{\Phi}^{'}_{56}\Phi^{\alpha\beta}_{2}\Phi^{\alpha\beta}_{2}
+  H_{2}  H_{2}  \bar{\Phi}^{'}_{56}  H_{12}  H_{13}
+ \bar{\Phi}^{'}_{46}\bar{\Phi}^{\alpha\beta}_{1} 
		\bar{\Phi}^{\alpha\beta}_{1}H_{16} H_{17}
+ \bar{\Phi}^{'}_{46}\bar{\Phi}^{\alpha\beta}_{2} 
		\bar{\Phi}^{\alpha\beta}_{2}H_{16}  H_{17}
\nonumber\\&+&
    \Phi_{46}   N^{c}_3 V_{9}  \bar{H}_{9}  \bar{H}_{9}
+    \Phi_{46}   N^{c}_3 V_{8}  \bar{H}_{8}  \bar{H}_{8}
+    \Phi^{'}_{45}  N^{c}_{2}  V_{6}  \bar{H}_{9}  \bar{H}_{9}
+    \Phi^{'}_{45}   N^{c}_{2}  V_{5}  \bar{H}_{8}  \bar{H}_{8}
\nonumber\\&+&
    \Phi^{'}_{45}  \bar{H}_{9}  \bar{H}_{9}  \bar{H}_{11}  \bar{H}_{11}
+    \Phi^{'}_{45}  \bar{H}_{8}  \bar{H}_{8}  \bar{H}_{10}  \bar{H}_{10}
+    \bar{\Phi}^{'}_{45}  H_{11}  H_{11}  H_{9}  H_{9}
+    \bar{\Phi}^{'}_{45}  H_{10}  H_{10}  H_{8}  H_{8}
\nonumber\\&+&
     \Phi_{45}   N^{c}_{1} V_{3}  H_{11}  H_{11}
+    \Phi_{45}   N^{c}_{1} V_{2}  H_{10}  H_{10}
+    \Phi_{56}   N^{c}_3 V_{9}  H_{11}  H_{11}
+    \Phi_{56}   N^{c}_3 V_{8}  H_{10}  H_{10}
\nonumber\\&+&
     \Phi_{56}  \bar{\Phi}^{\alpha\beta}_{1}  \bar{\Phi}^{\alpha\beta}_{1}  
		H_{14}  H_{15}
+    \Phi_{56}  \bar{\Phi}^{\alpha\beta}_{2}  \bar{\Phi}^{\alpha\beta}_{2}  
		H_{14}  H_{15}
+    N^{c}_{2}  V_{5}  \bar{\Phi}^{\alpha\beta}_{2}  H_{10}  \bar{H}_{8}
+    N^{c}_{2}  \bar{\Phi}^{\alpha\beta}_{2}  H_{11}  \bar{H}_{8}  V_{12}
\nonumber\\&+&
     H_{7}  \bar{H}_{7}  \Phi^{\alpha\beta}_{1}  H_{11}  H_{9}
+    H_{7}  \bar{H}_{7}  \bar{\Phi}^{\alpha\beta}_{2}  \bar{H}_{8}  \bar{H}_{10}
+    H_{6}  \bar{H}_{6}  \Phi^{\alpha\beta}_{1}  H_{11}  H_{9}
+    H_{6}  \bar{H}_{6}  \bar{\Phi}^{\alpha\beta}_{2}  \bar{H}_{8}  \bar{H}_{10}
\nonumber\\&+&
     H_{5}  \bar{H}_{5}  \Phi^{\alpha\beta}_{1}  H_{11}  H_{9}
+    H_{5}  \bar{H}_{5}  \bar{\Phi}^{\alpha\beta}_{2}  \bar{H}_{8}  \bar{H}_{10}
+   \Phi^{\alpha\beta}_{1}  \Phi^{\alpha\beta}_{1}  \bar{\Phi}^{\alpha\beta}_{1}  
		H_{11}  H_{9}
+   \Phi^{\alpha\beta}_{1}  \Phi^{\alpha\beta}_{2}  \bar{\Phi}^{\alpha\beta}_{2}  
		H_{11}  H_{9}
\nonumber\\&+&
    \Phi^{\alpha\beta}_{1}  \bar{\Phi}^{\alpha\beta}_{1}  
		\bar{\Phi}^{\alpha\beta}_{2} \bar{H}_{8}  \bar{H}_{10}
+   \Phi^{\alpha\beta}_{1}  H_{11}  H_{11}  \bar{H}_{11}  H_{9}
+    \Phi^{\alpha\beta}_{1}  H_{11}  H_{10}  \bar{H}_{10}  H_{9}
+     \Phi^{\alpha\beta}_{1}  H_{11}  \bar{H}_{9}  H_{9}  H_{9}
\nonumber\\&+&
    \Phi^{\alpha\beta}_{1}  H_{11}  \bar{H}_{8}  H_{9}  H_{8}
+   \Phi^{\alpha\beta}_{2}  \Phi^{\alpha\beta}_{2}  
		\bar{\Phi}^{\alpha\beta}_{1}  H_{11}  H_{9}
+   \Phi^{\alpha\beta}_{2}  \bar{\Phi}^{\alpha\beta}_{1}  
		\bar{\Phi}^{\alpha\beta}_{1} \bar{H}_{8}  \bar{H}_{10}
+   \Phi^{\alpha\beta}_{2}  \bar{\Phi}^{\alpha\beta}_{2}  
		\bar{\Phi}^{\alpha\beta}_{2} \bar{H}_{8}  \bar{H}_{10}
\nonumber\\&+&
     \bar{\Phi}^{\alpha\beta}_{1}  H_{11}  H_{12}  H_{9}  H_{13}
+    \bar{\Phi}^{\alpha\beta}_{2}  H_{11}  \bar{H}_{8}  \bar{H}_{11}  \bar{H}_{10}
+    \bar{\Phi}^{\alpha\beta}_{2}  H_{10}  \bar{H}_{8}  \bar{H}_{10}  \bar{H}_{10}
+    \bar{\Phi}^{\alpha\beta}_{2}  \bar{H}_{9}  \bar{H}_{8}  \bar{H}_{10}  H_{9}
\nonumber\\&+&
     \bar{\Phi}^{\alpha\beta}_{2}  \bar{H}_{8}  \bar{H}_{8}  \bar{H}_{10}  H_{8}.
\label{w5all}\\
\nonumber
\eeqn

\chapter{}
\section{Weight roots of $E_6$ representations in the twisted sector $\theta_2$ of the $SO(4)^3$ model}

\begin{center}
\begin{tabular}{|c|c|}
\hline
 $p_{sh}= p - V_2 $ & $p_{sh, DL_(E_6)}$ \\
\hline
\hline
 $(1, -1/2, -1/2, 0^5)$   & $ (0, 0, 0, 0, 0, 0)$\\
 $(-1, -1/2, -1/2, 0^5)$   &  $(1, 0, -1, 0, 0, 1)$\\
 $(0, 1/2, 1/2,1, 0^4)$   &  $(0, -1, 1, 0, 0, -1)$\\
 $(0, 1/2, 1/2,-1, 0^4)$   &  $(-1, 1, 0, 0, 0, 0)$\\
 $(0, 1/2, 1/2,0,1, 0^3)$   & $ (0, 1, 0, 0, 0, -1)$\\
 $(0, 1/2, 1/2,0,-1, 0^3)$   &  $(0, -1, 1, 0, 0, 0)$\\
 $(0, 1/2, 1/2,0^2,1, 0^2)$   &  $(-1, 0, 0, 1, -1, 0)$\\
 $(0, 1/2, 1/2, 0^2,-1,0^2)$   &  $(0, 0, 1, -1, 1, 0)$\\
 $(0, 1/2, 1/2, 0^3,1,0)$   & $ (-1, 0, 1, -1, 0, 0)$\\
 $(0, 1/2, 1/2, 0^3,-1,0)$   &  $(0, 0, 0, 1, 0, -1)$\\
 $(0, 1/2, 1/2, 0^4,1)$   &  $(0, 0, 1, 0, -1, -1)$\\
 $(0, 1/2, 1/2, 0^4,-1)$   &  $(-1, 0, 0, 0, -1, 0)$\\
$(-1/2,0,0, 1/2, 1/2, 1/2,1/2,1/2)$     &  $(0, 0, 0, 0, -1, 0)$\\
$(-1/2,0,0, -1/2, -1/2, 1/2,1/2,1/2)$   &  $(0, 0, 0, 0, -1, 0)$\\
$(-1/2,0,0, -1/2, 1/2, -1/2,1/2,1/2)$   &  $(0, 1, 0, -1, 0, 0)$\\
$(-1/2,0,0, -1/2, 1/2, 1/2,-1/2,1/2)$   &  $(0, 1, -1, 1, -1, 0)$\\
$(-1/2,0,0, -1/2, 1/2, 1/2,1/2,-1/2)$   &  $(-1, 1, -1, 0, 0, 1)$\\
$(-1/2,0,0, 1/2,- 1/2, -1/2,1/2,1/2)$   &  $(1, -1, 1, -1, 0, 0)$\\
$(-1/2,0,0, 1/2, -1/2, 1/2,-1/2,1/2)$   &  $(1, -1, 0, 1, 0, 0)$\\
$(-1/2,0,0, 1/2, -1/2, 1/2,1/2,-1/2)$   &  $(0, -1, 0, 0, 0, 1)$\\
$(-1/2,0,0, 1/2, 1/2,- 1/2,-1/2,1/2)$   &  $(1, 0, 0, 1, -1, -1)$\\
$(-1/2,0,0, 1/2, 1/2, -1/2,1/2,-1/2)$   &  $(0, 0, 0, -1, 0, 0)$\\
$(-1/2,0,0, 1/2, 1/2, 1/2,-1/2,-1/2)$   &  $(0, 0, -1, 1, 0, 0)$\\
$(-1/2,0,0, -1/2, -1/2,- 1/2,-1/2,1/2)$   &  ${\bf{(1, 0, 0, 0, 0, 0)}}$\\
$(-1/2,0,0, -1/2, -1/2,- 1/2,1/2,-1/2)$   &  $(0, 0, 0, -1, 1, 1)$\\
$(-1/2,0,0, -1/2,- 1/2, 1/2,-1/2,-1/2)$   &  $(0, 0, -1, 1, 0, 1)$\\
$(-1/2,0,0, -1/2, 1/2, -1/2,-1/2,-1/2)$   &  $(0, 1, -1, 0, 0, 0)$\\
$(-1/2,0,0, 1/2, -1/2, -1/2,-1/2,-1/2)$   &  $(1, -1, 0, 0, 0, 0)$\\
\hline
\end{tabular}
\end{center}

Table c.1 contains the 28 roots which fulfil
the massless equation for the twisted sector $\theta_2$ for the fixed
torus $T_2$. The solutions $p_{sh}$, shifted by $V_2$, are
showed in the first column. In the second column the roots are 
rewritten in Dynkin labels with respect to $E_6$. 
The first root is a singlet of $E_6$. The other 27 belong 
to the same multiplet and form in fact the ${\bf{27}}$ of $E_6$. 
The highest weight
of the ${\bf{27}}$ representation is ${\bf{(1, 0, 0, 0, 0, 0)}}$.
These states are singlets under the hidden $E_8'$ gauge group.

\vspace{0.5cm}

The simple roots of $E_6$ are given below :

\ba
 \a_1&=&(-1/2,-1/2,-1/2, 1/2, -1/2, -1/2,-1/2, 1/2)\nonumber\\
 \a_2&=&(0,0,0,-1,1,0,0,0)\nonumber\\
 \a_3&=&(1/2,1/2,1/2, 1/2, -1/2, -1/2,1/2, 1/2)\nonumber\\
 \a_4&=&(0,0,0,0,0,1,-1,0)\nonumber\\
 \a_5&=&(0,0,0,0,0,-1,0,-1)\nonumber\\
 \a_6&=&(-1/2,-1/2,-1/2,- 1/2, -1/2, 1/2,1/2,-1/2) .
\label{appc}
\ea

\chapter{}
{\addtolength{\oddsidemargin}{0.5in} 
\section{Total amplitude contributions of the shift orbifold in eq.(\ref{Z-})}
\label{D1}
\ba
\mathcal{Z}_{o,ab}&=&(\overline{V_8} -\overline{S_8})\La_1\La_2\La_{m',n'}
\La_{m,n}[(O_{16}-S_{16})(O_{16}-S_{16})]\underrightarrow{S}\nonumber\\
\mathcal{Z}_{ab,o}&=& (\overline{V_8} -\overline{S_8})\La_1\La_2\La_{m',n'}
\La_{m,n}[(V_{16}+C_{16})(V_{16}+C_{16})]\underrightarrow{T}\nonumber\\
\mathcal{Z}_{ab,ab}&=& (\overline{V_8} -\overline{S_8})\La_1\La_2\La_{m',n'}
\La_{m,n}[(V_{16}-C_{16})(V_{16}-C_{16})]\rightarrow \text{S invariant}\nonumber\\
\mathcal{Z}_{o,a}&=&(\overline{V_8} -\overline{S_8})\La_1\La_2\La_{m',n'}
(-1)^{m}\La_{m,n}[(O_{16}-S_{16})(O_{16}+S_{16})]\underrightarrow{S}\nonumber\\
\mathcal{Z}_{a,o}&=& (\overline{V_8} -\overline{S_8})\La_1\La_2\La_{m',n'}
\La_{m,n+1/2}[(V_{16}+C_{16})(O_{16}+S_{16})]\underrightarrow{T}\nonumber\\
\mathcal{Z}_{a,a}&=& (\overline{V_8} -\overline{S_8})\La_1\La_2\La_{m',n'}
(-1)^{m}\La_{m,n+1/2}[(-V_{16}+C_{16})(O_{16}+S_{16})]\rightarrow \text{S invariant}\nonumber\\
\mathcal{Z}_{o,b}&=&(\overline{V_8} -\overline{S_8})\La_1\La_2\La_{m',n'}
(-1)^{m}\La_{m,n}[(O_{16}+S_{16})(O_{16}-S_{16})]\underrightarrow{S}\nonumber\\
\mathcal{Z}_{b,o}&=& (\overline{V_8} -\overline{S_8})\La_1\La_2\La_{m',n'}
\La_{m,n+1/2}[(O_{16}+S_{16})(V_{16}+C_{16})]\underrightarrow{T}\nonumber\\
\mathcal{Z}_{b,b}&=& (\overline{V_8} -\overline{S_8})\La_1\La_2\La_{m',n'}
(-1)^{m}\La_{m,n+1/2}[(O_{16}+S_{16})(-V_{16}+C_{16})]\rightarrow \text{S invariant}\nonumber\\
\mathcal{Z}_{a,b}&=&(\overline{V_8} -\overline{S_8})\La_1\La_2\La_{m',n'}
(-1)^{m}\La_{m,n+1/2}[(V_{16}+C_{16})(O_{16}-S_{16})]
\underrightarrow{T}\nonumber\\
\mathcal{Z}_{a,ab}&=&(\overline{V_8} -\overline{S_8})\La_1\La_2\La_{m',n'}
\La_{m,n}[(C_{16}-V_{16})(O_{16}-S_{16})]\underrightarrow{S}\nonumber\\
\mathcal{Z}_{ab,a}&=& (\overline{V_8} -\overline{S_8})\La_1\La_2\La_{m',n'}
(-1)^{m}
\La_{m,n+1/2}[(-V_{16}+C_{16})(V_{16}+C_{16})]\underrightarrow{T}\nonumber\\
\mathcal{Z}_{ab,b}&=& (\overline{V_8} -\overline{S_8})\La_1\La_2\La_{m',n'}
(-1)^{m}\La_{m,n+1/2}[(V_{16}+C_{16})(-V_{16}+C_{16})]
\rightarrow \text{S}\nonumber\\
\mathcal{Z}_{b,ab}&=&(\overline{V_8} -\overline{S_8})\La_1\La_2\La_{m',n'}
\La_{m,n+1/2}[(O_{16}-S_{16})(C_{16}-V_{16})]\underrightarrow{T}\nonumber\\
\mathcal{Z}_{b,a}&=&
(\overline{V_8} -\overline{S_8})\La_1\La_2\La_{m',n'}
(-1)^{m}\La_{m,n}[(O_{16}-S_{16})(V_{16}+C_{16})].\nonumber\\
\ea

\section{Left amplitudes of ${\mathbb Z}_2 \times {\mathbb Z}_2$ orbifold model in eq.(\ref{z2xz2})}\label{gia}
In this sector we assume that the first three elements of each product
correspond to the compact space, hence they feel 
the action of the ${\mathbb Z}_2 \times {\mathbb Z}_2$ orbifold.
 
\underline{Untwisted}
\ba
g_{00}&=&(O_{2}O_2O_2O_{10}+V_{2}V_2V_2V_{10})O_{16} +
(S_2S_2S_2C_{10} + C_2C_2C_2S_{10})C_{16};\nonumber\\
g_{0g}&=&(O_2V_2V_2O_{10}+V_2O_2O_2V_{10})O_{16}+
(S_2C_2C_2C_{10} + C_2S_2S_2S_{10})C_{16} ;\nonumber\\
g_{0h}&=&(V_2V_2O_2O_{10}+O_2O_2V_2V_{10})O_{16}+
(C_2C_2S_2C_{10} + S_2S_2C_2S_{10})C_{16} ;\nonumber\\
g_{0f}&=&(V_2O_2V_2O_{10}+O_2V_2O_2V_{10})O_{16}+
(C_2S_2C_2C_{10} + S_2C_2S_2S_{10})C_{16} .\nonumber\\
\ea
\ba
G'_{oh}&=& (S_2S_2S_2S_{10}+C_2C_2S_2S_{10}
-C_2S_2C_2S_{10}-S_2C_2C_2S_{10}-C_2S_2S_2C_{10}\nonumber\\
&&-S_2C_2S_2C_{10}+S_2S_2C_2C_{10}+C_2C_2C_2C_{10})S_{16}\nonumber\\
&&+(-V_2O_2O_2O_{10}-O_2V_2O_2O_{10}
+O_2O_2V_2O_{10}+V_2V_2V_2O_{10}+O_2O_2O_2V_{10}\nonumber\\
&&+V_2V_2O_2V_{10}-V_2O_2V_2V_{10}-O_2V_2V_2V_{10})V_{16};\nonumber\\
G''_{oh}&=&(S_2S_2S_2C_{10}+C_2C_2S_2C_{10}
-C_2S_2C_2C_{10}-S_2C_2C_2C_{10}-C_2S_2S_2S_{10}\nonumber\\
&&+S_2S_2C_2S_{10}+C_2C_2C_2S_{10}-S_2C_2S_2S_{10})O_{16}\nonumber\\
&&+(O_2O_2O_2O_{10}+V_2V_2O_2O_{10}
-V_2O_2V_2O_{10}-O_2V_2V_2O_{10}-V_2O_2O_2V_{10}\nonumber\\
&&-O_2V_2O_2V_{10}+O_2O_2V_2V_{10}+V_2V_2V_2V_{10})C_{16};\nonumber\\
G'''_{oh}&=&(S_2S_2S_2S_{10}+C_2C_2S_2S_{10}
-C_2S_2C_2S_{10}-S_2C_2C_2S_{10}-C_2S_2S_2C_{10}\nonumber\\
&&-S_2C_2S_2C_{10}+S_2S_2C_2C_{10}+C_2C_2C_2C_{10})V_{16}\nonumber\\
&&+(-V_2O_2O_2O_{10}-O_2V_2O_2O_{10}
+O_2O_2V_2O_{10}+V_2V_2V_2O_{10}+O_2O_2O_2V_{10}\nonumber\\
&&+V_2V_2O_2V_{10}-V_2O_2V_2V_{10}-O_2V_2V_2V_{10})S_{16};\nonumber\\
\label{mm}
\ea
where each of these (\ref{mm}) contributions do not play a role in the 
massless untwisted spectrum, beside they can contribute in the 
twisted massless sector.

\underline{Twisted sector h}

\ba
g_{h0}=(S_2C_2O_2O_{10}+C_2S_2V_2V_{10})O_{16}+
(V_2O_2S_2C_{10} + O_2V_2C_2S_{10})C_{16};\nonumber\\
g_{hg}=(S_2S_2V_2O_{10}+C_2C_2O_2V_{10})O_{16}+
(O_2O_2S_2S_{10} + V_2V_2C_2C_{10})C_{16} ;\nonumber\\
g_{hh}=(C_2S_2O_2O_{10}+S_2C_2V_2V_{10})O_{16}+
 (V_2O_2C_2S_{10} + O_2V_2S_2C_{10})C_{16} ;\nonumber\\
g_{hf}=(C_2C_2V_2O_{10}+S_2S_2O_2V_{10})O_{16}+
(O_2O_2C_2C_{10} + V_2V_2S_2S_{10})C_{16} .\nonumber\\
\ea

\underline{Twisted sector g}
\ba
g_{g0}&=&(O_2C_2S_2O_{10}+C_2V_2O_2S_{10}+
S_2O_2V_2C_{10}+V_2S_2C_2V_{10})(O_{16}+C_{16})\nonumber\\
&+&(O_2C_2C_2O_{10} + C_2V_2O_2C_{10}+
S_2O_2V_2S_{10} + V_2S_2S_2V_{10})S_{16}\nonumber\\
&+&(S_2O_2O_2C_{10} + V_2S_2C_2O_{10}+
C_2V_2V_2S_{10} + O_2C_2S_2V_{10})V_{16};\nonumber\\
g_{gg}&=&(O_2S_2C_2O_{10}+S_2V_2O_2C_{10}+
C_2O_2V_2S_{10}+V_2C_2S_2V_{10})(O_{16}+C_{16})\nonumber\\
&+&(O_2S_2S_2O_{10} + S_2V_2O_2S_{10}+
C_2O_2V_2C_{10} + V_2C_2C_2V_{10})S_{16}\nonumber\\
&+&(C_2O_2O_2S_{10} + V_2C_2S_2O_{10}+
S_2V_2V_2C_{10} + O_2S_2C_2V_{10})V_{16};\nonumber\\
g_{gh}&=&(S_2O_2O_2S_{10}+V_2S_2S_2O_{10}+
C_2V_2V_2C_{10}+O_2C_2C_2V_{10})(O_{16}+C_{16})\nonumber\\
&+&(S_2O_2O_2C_{10} + V_2S_2C_2O_{10}+
C_2V_2V_2S_{10} + O_2C_2S_2V_{10})S_{16}\nonumber\\
&+&(O_2C_2C_2O_{10} + C_2V_2O_2C_{10}+
S_2O_2V_2S_{10} + V_2S_2S_2V_{10})V_{16};\nonumber\\
g_{gf}&=&(C_2O_2O_2C_{10}+V_2C_2C_2O_{10}+
S_2V_2V_2S_{10}+O_2S_2S_2V_{10})(O_{16}+C_{16})\nonumber\\
&+&(C_2O_2O_2S_{10} + V_2C_2S_2O_{10}+
S_2V_2V_2C_{10} + O_2S_2C_2V_{10})S_{16}\nonumber\\
&+&(O_2S_2S_2O_{10} + S_2V_2O_2S_{10}+
C_2O_2V_2C_{10} + V_2C_2C_2V_{10})V_{16}.\nonumber\\
\ea

\underline{Twisted sector f}
\ba
g_{f0}&=&(C_2O_2S_2O_{10}+V_2C_2O_2S_{10}+
O_2S_2V_2C_{10}+S_2V_2C_2V_{10})(O_{16}+C_{16})\nonumber\\
&+&(C_2O_2C_2O_{10} + V_2C_2O_2C_{10}+
O_2S_2V_2S_{10} + S_2V_2S_2V_{10})S_{16}\nonumber\\
&+&(O_2S_2O_2C_{10} + S_2V_2C_2O_{10}+
V_2C_2V_2S_{10} + C_2O_2S_2V_{10})V_{16};\nonumber\\
g_{fg}&=&(O_2C_2O_2C_{10}+C_2V_2C_2O_{10}+
V_2S_2V_2S_{10}+S_2O_2S_2V_{10})(O_{16}+C_{16})\nonumber\\
&+&(O_2C_2O_2S_{10} + C_2V_2S_2O_{10}+
V_2S_2V_2C_{10} + S_2O_2C_2V_{10})S_{16}\nonumber\\
&+&(S_2O_2S_2O_{10} + V_2S_2O_2S_{10}+
O_2C_2V_2C_{10} + C_2V_2C_2V_{10})V_{16};\nonumber\\
g_{fh}&=&(O_2S_2O_2S_{10}+S_2V_2S_2O_{10}+
V_2C_2V_2C_{10}+C_2O_2C_2V_{10})(O_{16}+C_{16})\nonumber\\
&+&(O_2S_2O_2C_{10} + S_2V_2C_2O_{10}+
V_2C_2V_2S_{10} + C_2O_2S_2V_{10})S_{16}\nonumber\\
&+&(C_2O_2C_2O_{10} + V_2C_2O_2C_{10}+
O_2S_2V_2S_{10} + S_2V_2S_2V_{10})V_{16};\nonumber\\
g_{ff}&=&(S_2O_2C_2O_{10}+V_2S_2O_2C_{10}+
O_2C_2V_2S_{10}+C_2V_2S_2V_{10})(O_{16}+C_{16})\nonumber\\
&+&(S_2O_2S_2O_{10} + V_2S_2O_2S_{10}+
O_2C_2V_2C_{10} + C_2V_2C_2V_{10})S_{16}\nonumber\\
&+&(O_2C_2O_2S_{10} + C_2V_2S_2O_{10}+
V_2S_2V_2C_{10} + S_2O_2C_2V_{10})V_{16}.\nonumber\\
\ea
\newpage
For completeness we present also the twisted amplitudes which 
do not contribute to the low energy spectrum
\ba
G'_{h0}&=& (C_2C_2O_2O_{10}+S_2S_2O_2O_{10}
+S_2C_2V_2O_{10}+C_2S_2V_2O_{10}+S_2C_2O_2V_{10}\nonumber\\
&&+C_2S_2O_2V_{10}+C_2C_2V_2V_{10}+S_2S_2V_2V_{10})V_{16}\nonumber\\
&&+(O_2O_2S_2C_{10}+O_2O_2C_2S_{10}
+V_2O_2C_2C_{10}+V_2O_2S_2S_{10}+O_2V_2C_2C_{10}\nonumber\\
&&+O_2V_2S_2S_{10}+V_2V_2S_2C_{10}+V_2V_2C_2S_{10})S_{16};\nonumber\\
G''_{h0}&=&(O_2O_2C_2C_{10}+O_2O_2S_2S_{10}
+V_2O_2S_2C_{10}+V_2O_2C_2S_{10}+O_2V_2S_2C_{10}\nonumber\\
&&+O_2V_2C_2S_{10}+V_2V_2C_2C_{10}+V_2V_2S_2S_{10})O_{16}\nonumber\\
&&+(S_2C_2O_2O_{10}+C_2S_2O_2O_{10}
+C_2C_2V_2O_{10}+S_2S_2V_2O_{10}+C_2C_2O_2V_{10}\nonumber\\
&&+S_2S_2O_2V_{10}+S_2C_2V_2V_{10}+C_2S_2V_2V_{10})C_{16};\nonumber\\
G'''_{h0}&=&(O_2O_2S_2C_{10}+O_2O_2C_2S_{10}
+V_2O_2C_2C_{10}+V_2O_2S_2S_{10}+O_2V_2C_2C_{10}\nonumber\\
&&+O_2V_2S_2S_{10}+V_2V_2S_2C_{10}+V_2V_2C_2S_{10})V_{16}\nonumber\\
&&+(C_2C_2O_2O_{10}+S_2S_2O_2O_{10}
+S_2C_2V_2O_{10}+C_2S_2V_2O_{10}+S_2C_2O_2V_{10}\nonumber\\
&&+C_2S_2O_2V_{10}+C_2C_2V_2V_{10}+S_2S_2V_2V_{10})S_{16}.\nonumber\\
\label{mmmm}
\ea
One obtains analogous expressions by applying T transformations
on each of the previous amplitudes, giving rise to
 $G'_{hh}$, $G''_{hh}$ and $G'''_{hh}$ respectively.
\newpage
\section{Right amplitudes of ${\mathbb Z}_2 \times {\mathbb Z}_2$ orbifold model in eq.(\ref{z2xz2})}\label{gio}
We assume that the first element of the following products
corresponds to spacetime degrees of freedom, hence the action 
of the ${\mathbb Z}_2 \times {\mathbb Z}_2$ orbifold applies on the last three elements.
\ba
\tau_{00}&=&V_2O_2O_2O_2+O_2V_2V_2V_2-S_2S_2S_2S_2-C_2C_2C_2C_2 \ , \nonumber \\
\tau_{0g}&=&O_2V_2O_2O_2+V_2O_2V_2V_2-C_2C_2S_2S_2-S_2S_2C_2C_2 \ , \nonumber \\
\tau_{0h}&=&O_2O_2O_2V_2+V_2V_2V_2O_2-C_2S_2S_2C_2-S_2C_2C_2S_2 \ , \nonumber \\
\tau_{0f}&=&O_2O_2V_2O_2+V_2V_2O_2V_2-C_2S_2C_2S_2-S_2C_2S_2C_2 \ , \nonumber \\
\tau_{g0}&=&V_2O_2S_2C_2+O_2V_2C_2S_2-S_2S_2V_2O_2-C_2C_2O_2V_2 \ , \nonumber \\
\tau_{gg}&=&O_2V_2S_2C_2+V_2O_2C_2S_2-S_2S_2O_2V_2-C_2C_2V_2O_2 \ , \nonumber \\
\tau_{gh}&=&O_2O_2S_2S_2+V_2V_2C_2C_2-C_2S_2V_2V_2-S_2C_2O_2O_2 \ , \nonumber \\
\tau_{gf}&=&O_2O_2C_2C_2+V_2V_2S_2S_2-S_2C_2V_2V_2-C_2S_2O_2O_2 \ , \nonumber \\
\tau_{h0}&=&V_2S_2C_2O_2+O_2C_2S_2V_2-C_2O_2V_2C_2-S_2V_2O_2S_2 \ , \nonumber \\
\tau_{hg}&=&O_2C_2C_2O_2+V_2S_2S_2V_2-C_2O_2O_2S_2-S_2V_2V_2C_2 \ , \nonumber \\
\tau_{hh}&=&O_2S_2C_2V_2+V_2C_2S_2O_2-S_2O_2V_2S_2-C_2V_2O_2C_2 \ , \nonumber \\
\tau_{hf}&=&O_2S_2S_2O_2+V_2C_2C_2V_2-C_2V_2V_2S_2-S_2O_2O_2C_2 \ , \nonumber \\
\tau_{f0}&=&V_2S_2O_2C_2+O_2C_2V_2S_2-S_2V_2S_2O_2-C_2O_2C_2V_2 \ , \nonumber \\
\tau_{fg}&=&O_2C_2O_2C_2+V_2S_2V_2S_2-C_2O_2S_2O_2-S_2V_2C_2V_2 \ , \nonumber \\
\tau_{fh}&=&O_2S_2O_2S_2+V_2C_2V_2C_2-C_2V_2S_2V_2-S_2O_2C_2O_2 \ , \nonumber \\
\tau_{ff}&=&O_2S_2V_2C_2+V_2C_2O_2S_2-C_2V_2C_2O_2-S_2O_2S_2V_2 \ ,
\ea
where for brevity we dropped the bar which labels the supersymmetric sector.

\bibliographystyle{h-physrev3}
\bibliography{thesis}
\addcontentsline{toc}{chapter}{Bibliography}



\end{document}